\documentclass[smallextended]{svjour3}       

\usepackage{graphicx}
\usepackage{subfigure}
\usepackage{amssymb}

\usepackage[colorlinks=true]{hyperref}

\usepackage[authoryear]{natbib}
\setcitestyle{authorcat}

\newcommand{\eprint}[2][]{\href{https://arxiv.org/abs/#2}{arXiv:~\nolinkurl{#2}}}
\newcommand{\doi}[2][]{\href{http://dx.doi.org/#2}{DOI:~\nolinkurl{#2}}}

\usepackage{lmodern}

\newcommand{\vs}{{\it vs.}}
\newcommand{\eg}{{\it e.g.}}
\newcommand{\ie}{{\it i.e.}}

\newcommand{\apriori}{{\it a priori}}
\newcommand{\scimm}[2]{#1\times10^{#2}}
\newcommand{\sci}[2]{$\scimm{#1}{#2}$}
\newcommand{\msolar}{M_\odot}
\newcommand{\D}{\mathcal{D}}
\newcommand{\TwoF}{2\mathcal{F}}
\newcommand{\fstatistic}{$\mathcal{F}$-statistic}
\newcommand{\fabstatistic}{$\mathcal{F}_{\rm AB}$-statistic}
\newcommand{\gstatistic}{$\mathcal{G}$-statistic}
\newcommand{\jstatistic}{$\mathcal{J}$-statistic}

\newcommand{\mcoh}{m_{\rm coh}}
\newcommand{\msemicoh}{m_{\rm semi-coh}}
\newcommand{\weave}{Weave}
\newcommand{\tdf}{TD-$\mathcal{F}$-statistic}
\newcommand{\gctf}{GCT-$\mathcal{F}$-statistic}
\newcommand{\Om}{\mathcal{O}}
\newcommand{\C}{\mathcal{C}}
\newcommand{\M}{\mathcal{M}}
\newcommand{\Mij}{\mathcal{M}_{ij}}
\newcommand{\fdot}{\dot f}
\newcommand{\fdotdot}{\ddot f}
\newcommand{\Tcoh}{T_{\rm coh}}
\newcommand{\Ncoh}{N_{\rm coh}}
\newcommand{\Tobs}{T_{\rm obs}}
\newcommand{\Tseg}{T_{\rm seg}}
\newcommand{\fgw}{f_{\rm GW}}
\newcommand{\tssb}{\tau}

\newcommand{\fgwdot}{\dot{f}_{\rm GW}}
\newcommand{\fgwddot}{\ddot{f}_{\rm GW}}
\newcommand{\frot}{f_{\rm rot}}
\newcommand{\fkepler}{f_{\rm Kepler}}
\newcommand{\fbase}{f_{\rm base}}
\newcommand{\rmode}{$r$-mode}
\newcommand{\rmodes}{$r$-modes}
\newcommand{\alfven}{Alfv\'en}
\newcommand{\Izz}{I_{\rm zz}}
\newcommand{\fsig}{f_{\rm sig}}
\newcommand{\Sh}{S_h}
\newcommand{\Shi}{S_{h_{i}}}
\newcommand{\Shki}{S_{h_{i}}^k}
\newcommand{\etaki}{\eta_{i}^k}
\newcommand{\Di}{\tilde D_i}
\newcommand{\Pki}{\tilde P^{(k)}_i}
\newcommand{\Pssi}{P^{(k)}_{i(\rm SS)}}
\newcommand{\Pss}{\mathcal{P}_{i(\rm SS)}}
\newcommand{\NDFT}{N_{\rm DFT}}
\newcommand{\Dki}{\tilde D^{(k)}_i}
\newcommand{\Dkidemod}{\tilde D^{(k)}_{i(\rm demod)}}
\newcommand{\Donei}{\tilde D_{1,i}}
\newcommand{\Dtwoi}{\tilde D_{2,i}}
\newcommand{\DIi}{\tilde D_{I,i}}
\newcommand{\DIj}{\tilde D_{I,j}}
\newcommand{\DJi}{\tilde D_{J,i}}
\newcommand{\DJj}{\tilde D_{J,j}}
\newcommand{\phionei}{\phi_{1,i}}
\newcommand{\phitwoi}{\phi_{2,i}}
\newcommand{\expphionei}{e^{-i\phi_{1,i}}}
\newcommand{\expphitwoi}{e^{-i\phi_{2,i}}}

\newcommand{\expphiIiJj}{e^{-i(\phi_{I,i}-\phi_{J,j})}}
\newcommand{\Nseg}{N_{\rm seg}}
\newcommand{\Nlag}{N_{\rm lag}}
\newcommand{\Nsample}{N_{\rm sample}}
\newcommand{\fsample}{f_{\rm sample}}
\newcommand{\imag}{{\rm i}}
\newcommand{\rhoi}{\rho_i}
\newcommand{\rhocc}{\rho_{\rm CC}}
\newcommand{\rhostari}{\rho_i^*}
\newcommand{\Rpf}{R_{\rm PF}}
\newcommand{\Ri}{R_i}
\newcommand{\Wi}{W_i}
\newcommand{\wi}{w_i}
\newcommand{\numi}{n_i}
\newcommand{\nbar}{\bar n}
\newcommand{\etastar}{\eta^*}
\newcommand{\nth}{n_{\rm th}}
\newcommand{\Fi}{F_i(\iota,\psi)}
\newcommand{\Fplusi}{F_i^+}
\newcommand{\Fcrossi}{F_i^\times}
\newcommand{\Rstari}{R_i^*}
\newcommand{\erfc}{{\rm erfc}}
\newcommand{\hbayes}{h_0^{95\%\>{\rm UL}}}
\newcommand{\hnf}{h_0^{95\%}}
\newcommand{\mm}{\mu}
\newcommand{\lambdabar}{{\mkern0.75mu\mathchar '26\mkern -9.75mu\lambda}}
\newcommand{\sinc}{{\rm sinc}}
\newcommand{\dlam}{\Delta\lambda}
\newcommand{\vlam}{\vec\lambda}
\newcommand{\vlamp}{\vec\lambda'}
\newcommand{\dvlam}{\Delta\vec\lambda}
\newcommand{\gkell}{g_{k\ell}}
\newcommand{\tauconst}{\tau_{\rm SD}}
\newcommand{\Omegar}{\Omega_r}
\newcommand{\hsd}{h_{\rm spin-down}}
\newcommand{\casa}{Cas A}
\newcommand{\vela}{Vela Jr.}
\def\figpermission#1#2{Image reproduced with permission from #1, copyright by #2}

\journalname{Living Reviews in Relativity}

\begin{document}

\title{Searches for Continuous-Wave Gravitational Radiation}


\author{Keith Riles
}


\institute{K. Riles \at
              Physics Department \\
              University of Michigan\\
              Ann Arbor, MI 48109, USA\\
              Tel.: +1-734-764-4652\\
              Fax: +1-734-936-6529\\
              \email{kriles@umich.edu}          
}

\date{Received: 13 June 2022 / Accepted: 20 February 2023}

\maketitle

\begin{abstract}
  Now that detection of gravitational wave signals from the coalescence of extra-galactic compact
  binary star mergers has become nearly routine, it is intriguing to consider
  other potential gravitational wave signatures. Here we examine the prospects for
  discovery of continuous gravitational waves from fast-spinning neutron stars in
  our own galaxy and from more exotic sources. Potential continuous-wave sources are reviewed, search methodologies
  and results presented and prospects for imminent discovery discussed. 
\end{abstract}

\tableofcontents

\section{Introduction}
\label{intro}

  The LIGO~\citep{bib:aligodetector1} and Virgo~\citep{bib:avirgodetector} gravitational wave detectors have made historic discoveries
  over the last seven years. The first direct detection in September 2015 of gravitational waves marked a
  milestone in fundamental science~\citep{bib:GW150914}, confirming a longstanding prediction of
  Einstein's General Theory of Relativity~\citep{bib:EinsteinGW1,bib:EinsteinGW2}. That the detection came from the
  first observation of a binary black hole merger provided a bonus not only
  in verifying detailed predictions of General Relativity, but in establishing
  unambiguously that stellar-mass black holes exist in the Universe.
  More than 80 binary black hole (BBH) systems have been observed since
  GW150914~\citep{bib:GW151226,bib:GW170104,bib:GW170608,bib:GW170814,bib:GWTC1,bib:GWTC2,bib:GWTC3}.
  Merging binary neutron star (BNS) systems~\citep{bib:GW170817,bib:GW190425} have also been observed,
  including GW170817~\citep{bib:GW170817}, which was accompanied by a multitude
  of electromagnetic observations~\citep{bib:GW170817MMA}. Those observations confirmed the association of at least
  some short gamma ray bursts with binary neutron star mergers~\citep{bib:GW170817GRB} and the onset of kilonovae in
  BNS mergers that contribute substantially to the heavy element production in the Universe~\citep{bib:GW170817MMA}.
  More recently came detections of merging neutron star -- black hole (NSBH) systems~\citep{bib:NSBH}.
  These discoveries of transient gravitational wave signals have ignited the field of gravitational wave astronomy.

  This review concerns a quite different and as-yet-undiscovered gravitational wave signal type, one defined by stability and near-monochromaticity
  over long time scales, namely {\it continuous waves}. CW signals with strengths detectable by current and imminent
  ground-based gravitational wave interferometers could originate from relatively nearby galactic sources, such as
  fast-spinning neutron stars exhibiting non-axisymmetry~\citep{bib:Thorne300}, or more exotically, from strong extra-galactic sources,
  such as super-radiant Bose-Einstein clouds surrounding black holes~\citep{bib:axiverseArvanitaki1}.

  We already know from prior LIGO and Virgo searches that the strengths of CW signals must be exceedingly weak [$\sim$$10^{-24}$ or less], which
  is consistent with theoretical expectation, from which we expect plausible CW strain amplitudes to be orders of magnitudes
  lower than the amplitudes of the transient signals detected to date [$\sim$$10^{-21}$]. This disparity in signal strength holds despite
  the much nearer distance of galactic neutron stars ($\sim$kpc) compared to the compact binary mergers ($\sim$40 Mpc to multi-Gpc) seen to date.
  In fact, it is only their long-lived nature that
  gives us any hope of detecting CW signals through integration over long data spans, so as to achieve a statistically
  viable signal-to-noise (SNR) ratio. As discussed below, however, that SNR increases, at best, as the square root of observation time,
  but for most CW searches, increases with an even lower power of observation time, while computational cost increases with much higher powers.
  These different scalings of signal sensitivity and cost have led to a variety of approaches in targeting signals, depending on
  the size of signal parameter space searched.

  The search for continuous gravitational radiation has been under way since the 1970's, using
  data from interferometers~\citep{bib:LevineStebbins} and bars~\citep{bib:EarlyBarLimits,bib:Suzuki}, including from early
  prototypes~\citep{bib:LivasArticle}
  for the large gravitational wave detectors to come later. This review focuses primarily on the most
  recent searches from the Advanced LIGO and Virgo detectors, although summaries of search algorithm developments in the
  initial LIGO and Virgo era (and before) provide some historical context. For reference, the Advanced LIGO and Virgo
  runs to date comprise (with selected highlighted detections):
  \begin{itemize}
  \item The O1 observing run (LIGO only): September 12, 2015 -- January 12, 2016 -- First detection of gravitational waves from a BBH merger: GW150914~\citep{bib:GW150914}.
  \item The O2 observing run (LIGO joined by Virgo in last month): November 30, 2016 -- August 25, 2017 -- First detection of gravitational waves from a BNS merger: GW170817~\citep{bib:GW170817}.
  \item The O3 observing run (LIGO and Virgo): April 1, 2019 -- March 27, 2020 -- First detection of gravitational waves from the formation of an intermediate-mass black hole: GW190521~\citep{bib:GW190521} and the first detection of NSBH mergers. The run was divided into a 6-month ``O3a'' epoch (April 1, 2019 -- October 1, 2019) and ``O3b'' (November 1 -- March 27, 2020) by a 1-month commissioning break. Many initial publications focused on results from the O3a data.
  \end{itemize}

  In the following, section~\ref{sec:sources} reviews both conventional and exotic potential sources
  of CW gravitational radiation. Section~\ref{sec:searches} describes a wide variety of search methodologies
  being used to address the challenges of detection.
  Section~\ref{sec:results} presents
  results (so far only upper limits) from searches based on these algorithms, with an emphasis on the most recent results from
  the Advanced LIGO and Virgo detectors.
  Finally, section~\ref{sec:outlook} discusses the outlook for discovery in the coming years, including the
  prospects for electromagnetic observations of the continuous gravitational wave sources.
  This review focuses on CW radiation potentially detectable with current-generation and next-generation
  ground-based gravitational wave interferometers, which are sensitive to gravitational frequencies in the human-audible band for sound.
  Past and future searches for lower-frequency CW radiation from supermassive black hole binaries
  at $\sim$nHz frequencies using pulsar timing arrays~\citep{bib:ptareview} or from stellar-mass galactic binaries
  at $\sim$mHz frequencies using the space-based LISA~\citep{bib:LISA} are not discussed here. 

  Textbooks addressing gravitational waves, their detection and their analysis include~\cite{bib:MTWtext,bib:Schutztext,bib:Maggioretext1,bib:Maggioretext2,bib:Saulsontext,bib:CreightonAndersontext,bib:JaranowskiKrolaktext,bib:AnderssonText}. Review articles and volumes on gravitational wave science
  include~\cite{bib:Thorne300,bib:Blairbook1,bib:lrre-SathyaprakashSchutz,bib:lrre-Pitkinetal,bib:lrre-FreiseStrain,bib:Blairbook2,bib:RilesPPNP,bib:lrre-RomanoCornish}.
  This review is a substantial expansion upon a briefer previous article~\citep{bib:RilesMPLA}. Other reviews of CW search methodology
  include~\cite{bib:Prixreview,bib:palombareview,bib:Laskyreview,bib:SieniawskaBejgerreview,bib:TenorioKeitelSintesreview,bib:PiccinniReview}.

\section{Potential sources of CW radiation}
\label{sec:sources}

In the frequency band of present 
ground-based detectors, the canonical sources of continuous gravitational waves are
galactic, non-axisymmetric neutron stars spinning fast enough to produce
gravitational waves in the LIGO and Virgo detectable band (at 1$\times$, $\sim$4/3$\times$ or 2$\times$ rotation frequency,
depending on the generation mechanism).
These nearby neutron stars
offer a ``conventional'' source of CW radiation -- as astrophysically extreme as
such objects are.

A truly exotic postulated source
is a ``cloud'' of bosons, such as QCD axions, surrounding a fast-spinning
black hole, bosons that can condense in gargantuan numbers to a small
number of discrete energy levels, enabling coherent
gravitation wave emission from boson annihilation or from level transitions.
Attention here focuses mainly on the conventional neutron stars, but
the exotic boson cloud scenario is also discussed. 

\subsection{Fast-spinning neutron stars}
\label{sec:neutronstarsources}

The following subsections give an overview of neutron star formation, structure, observables and populations, present
the phenomenology of neutron-star spin-down, discuss potential sources of non-axisymmetry in neutron stars, and
consider a number of particular GW search targets of interest. Although neutron stars were first postulated by
Baade \&\ Zwicky~\citep{bib:BaadeZwicky} and their basic properties worked out by Oppenheimer and Volkoff~\citep{bib:OppenheimerVolkoff},
the first definitive establishment of their existence came with the discovery of the first
radio pulsar~\citep{bib:Hewishetal} PSR B1919+21 in 1967 with prior theoretical support for neutron star radiation
contributing to supernova remnant shell energetics~\citep{bib:Pacini_1967} and rapid theoretical follow-up to explain the
pulsation mechanism~\citep{bib:Gold,bib:GoldreichJulian,bib:RudermanSutherland}. 

\subsubsection{Neutron star formation, structure, observables and populations}
\label{sec:nsbasics}

As background, this section surveys at a basic level the fundamentals of neutron star formation, structure, observables and populations.
Much more detailed information can be found in the following review articles or volumes on neutron stars~\citep{bib:LattimerPrakash,bib:lrre-ChamelHaensel,bib:NSandPulsars-Becker,bib:OzelFreire},
pulsars~\citep{bib:LorimerKramer,bib:LyneGrahamSmith,bib:lrre-Lorimer},
and rotating relativistic stars~\citep{bib:lrre-PaschalidisStergioulas}.
  
Neutron stars are the final states of stars too massive to form white dwarfs upon collapse after fuel consumption and too light to form black holes, having progenitor
masses in the approximate range 6--15 $\msolar$~\citep{bib:LyneGrahamSmith,bib:CerdaduranEliasrosa,bib:StockingerEtal}. These remarkably dense objects, supported by neutron degeneracy pressure,
boast near-nuclear densities in their crusts and well-beyond-nuclear densities in their cores. The range of densities and associated total stellar
masses and radii depend on an equation of state that is not experimentally accessible in terrestrial laboratories because of the combination of high density
and (relatively) low temperature. A variety of equations of state have been proposed~\citep{bib:LattimerPrakash}, with a small subset disfavored by
the measurement of neutron stars greater than two solar masses~\citep{bib:EOSexclusiontwoSM}, by radii of
approximately ten kilometers~\citep{bib:NICERzero,bib:NICERIII,bib:NICERanother,bib:NICERIV}
and by the absence of severe tidal deformation effects
in the gravitational waveforms measured for the BNS merger GW170817~\citep{bib:GW170817,bib:GW170817EOS,bib:LimHolt,bib:EssickEtal}.
The detection of a $\sim$2.6-$\msolar$ object in the GW190814 merger~\citep{bib:GW190814} poses a challenge to the nuclear equation of state
if the object is indeed a neutron star instead of a light black hole. 

In broad summary, a neutron star is thought to have a crust with outer radius between 10 and 15 km and a thickness of
$\sim$1 km~\citep{bib:ShapiroTeukolsky},
composed near the top of a tight lattice of neutron-rich heavy nuclei, permeated by 
neutron superfluid. Deeper in the star, as pressure and density increase, the nuclei may become distorted and elongated, forming a
``nuclear pasta'' of ordered nuclei and gaps~\citep{bib:nuclearpasta,bib:CaplanHorowitzPasta}.
Still deeper, the pasta gives way to a hyperdense neutron fluid and perhaps
undergoes phase transitions involving hyperons,
perhaps to a quark-gluon plasma, or even perhaps to
a solid strange-quark core~\citep{bib:ShapiroTeukolsky,bib:LattimerPrakash}.  

Uncertainties in equation of state lead directly to uncertainties in the expected maximum mass and radius of a neutron star~\citep{bib:LattimerPrakash},
but theoretical prejudice is consistent with the absence of observation in binary systems of neutron star masses much higher than two solar masses~\citep{bib:OzelFreire,bib:DemorestEtalJ1614,bib:ArzoumanianEtalJ1614,bib:AntoniadisJ0348,bib:CromartieEtalJ0740,bib:FonsecaEtalJ0740}.
Neutron star radii are especially challenging
to measure directly, with older measurements coming from X-ray measurements, where inferences are drawn from brightness of
the radiation, its temperature and distance to the source, assuming black-body radiation, with corrections for the strong space-time curvature
affecting the visible surface area~\citep{bib:OzelFreire,bib:DegenaarSuleimanov}. New measurements from the NICER X-ray satellite are improving upon the precision
with which mass and radius can be determined simultaneously from individual stars,
constraining more tightly the allowed equations of state~\citep{bib:NICERzero,bib:NICERIII,bib:NICERI,bib:NICERII,bib:NICEREOSIII,bib:NICEREOS,bib:NICERV,bib:NICERanother,bib:NICERIV}.

Measurements of the gravitational waveform from the binary neutron star merger GW170817 have also provided new constraints and disfavor
very stiff equations of state that lead to large neutron star radii~\citep{bib:GW170817EOS}. Detection of additional binary neutron star mergers
in the coming years should improve these constraints. Broadly, one expects average neutron star densities of $\sim$\sci{7}{14} g cm$^{-3}$, well
above the density of nuclear matter ($\sim$\sci{3}{14})~\citep{bib:LorimerKramer}, with densities at the core likely above $10^{15}$ g cm$^{-3}$~\citep{bib:ShapiroTeukolsky}.
See~\citep{bib:YunesMiller} for a recent review of what has been learned
about the neutron star equation of state from gravitational wave and X-ray observations.
A recent Bayesian combined analysis~\citep{bib:CombinedEOSAnalysis} of predictions from chiral effective field theory of QCD, measured BNS gravitational waveforms,
NICER X-ray observations and measurements from heavy ion (gold) collisions indicate a somewhat stiffer equation of state than previously favored and hence larger allowed radii of neutron stars.

Given the immense pressure on the nuclear matter, one expects a neutron star to assume a highly spherical shape in the limit of no rotation
and, with rotation, to become an axisymmetric oblate spheroid. True axisymmetry would preclude emission of quadrupolar gravitational waves from rotation alone. Hence
CW searchers count upon a small but detectable mass (or mass current) non-axisymmetry, discussed in detail in section~\ref{sec:nonaxisymmetry}.

During the collapse of their slow-spinning stellar progenitors, neutron stars can acquire an impressive rotational speed as angular momentum
conservation spins up the infalling matter. Even the two slowest-rotating known pulsars spin on their axes every 76 seconds~\citep{bib:LongPeriodPulsar}
and 24 seconds~\citep{bib:SlowestRadioPulsar,bib:ATNFdb}, implying rotational
kinetic energies greater than $\sim$$10^{35}$ J, and other young pulsars with spin frequencies of 10's of Hz have rotational energies of $\sim$$10^{43}$ J.
Recycled millisecond pulsars acquire even higher spins via accretion from a binary companion star, leading to measured spin frequencies above
700 Hz~\citep{bib:HesselsEtal,bib:BassaEtal} and a rotational energy of $\sim$$10^{45}$ J, or several percent of the magnitude of the gravitational bound energy of the star.
This immense reservoir of rotational energy might appear to bode well for supporting detectable gravitational wave emission, but
vast energy is required to create appreciable distortions in highly rigid space-time.
From the perspective of gravitational-wave energy density~\citep{bib:MTWtext},
one can define an effective, frequency-dependent Young's
modulus $Y_{\rm eff} \sim {c^2\fgw^2\over G}$ ($\sim$$10^{31}$ Pa for $\fgw\approx100$ Hz,
or 20 orders of magnitude higher than steel). As a result, one must tap a significant fraction of the reservoir's energy loss rate in order
to produce detectable radiation, as quantified below.

Most of the $\sim$3300 known neutron stars in the galaxy are pulsars, detected via pulsed electromagnetic emission, primarily in the radio band,
but also in X-rays and $\gamma$-rays (with a small number detected optically)~\citep{bib:LyneGrahamSmith,bib:ATNFdb}.
Pulses are typically observed at
the rotation frequency of the star, as a beam of radiation created by curvature radiation~\citep{bib:BuschauerBenfordCurvatureRadiation}
from particles that are flung out in a plasma from the magnetic poles
(misaligned with the spin axis) and accelerated transversely by the magnetic field,
sweeps across the Earth once per rotation (see~\citep{bib:MelroseEtal}, however, for a critique of this model).
A subset of neutron stars presumed to have
magnetic poles tilted nearly 90 degrees from the spin axis display two distinct
pulses.

\begin{figure}[t!]
\begin{center}
\includegraphics[width=13.cm]{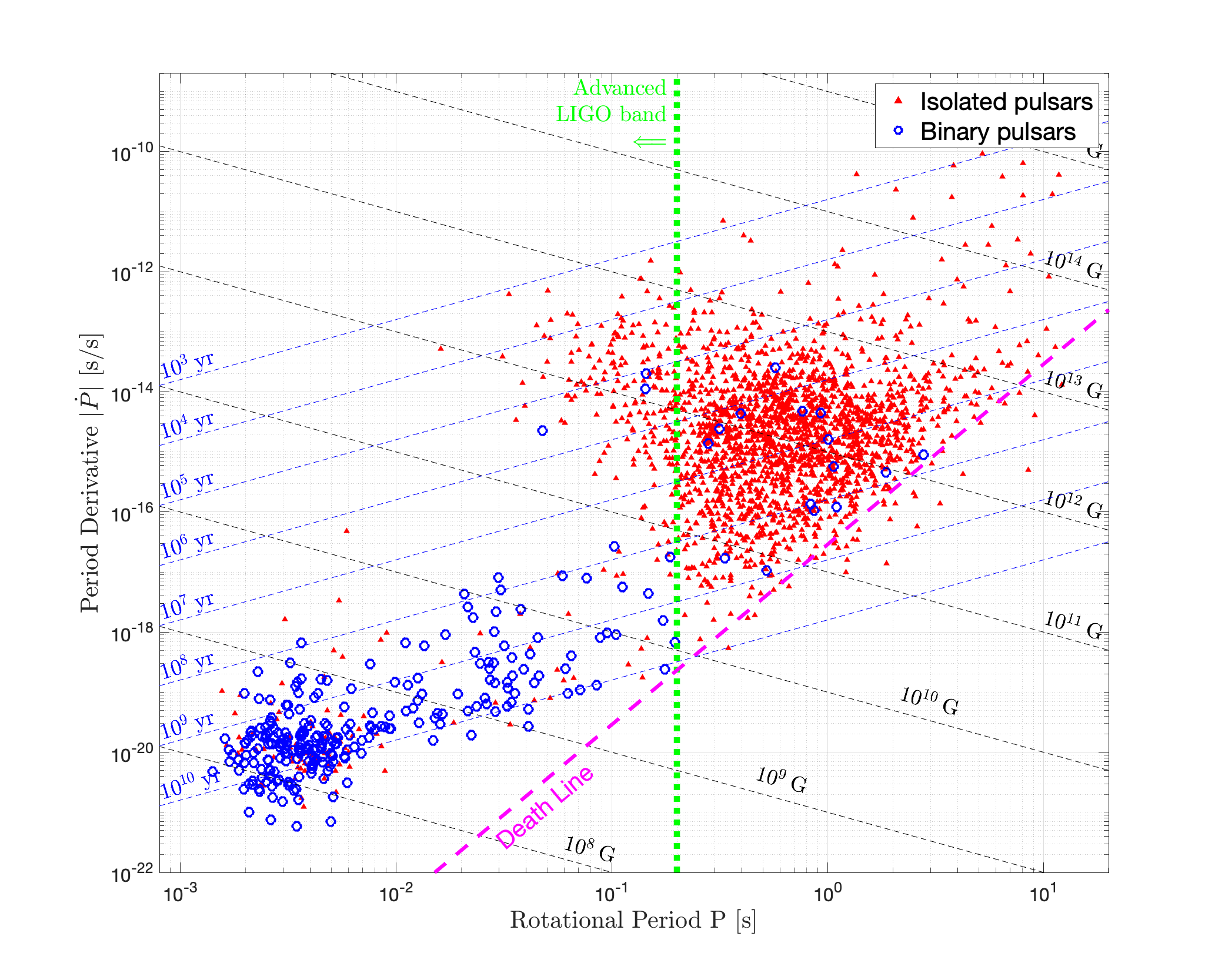}
\caption{Measured rotational periods and period derivatives for known
pulsars. Closed red triangles indicate isolated stars. Open blue circles
indicate binary stars. The vertical dotted line denotes the
approximate sensitivity band for Advanced LIGO at design sensitivity
($\fgw>10$ Hz, assuming $\fgw=2\frot$). A similar band applies to
design sensitivities of the Advanced Virgo and KAGRA detectors~\citep{bib:obsscenario}.}
\label{fig:pvspdot}
\end{center}
\end{figure}

Other neutron stars are known from detection of X-rays from thermal emission (heat from formation
and perhaps from magnetic field decay),
particularly at sites consistent with the birth locations
and times of supernova remnants~\citep{bib:LyneGrahamSmith}.
Still other neutron stars are inferred from accretion X-rays observed in binary systems, particularly
low-mass X-ray binaries with accretion disks~\citep{bib:LyneGrahamSmith},
although some accreting binaries with compact stars contain black
holes, such as the high-mass X-ray binary Cygnus X-1. Figure~\ref{fig:pvspdot} shows nearly the entire population of currently known
pulsars~\citep{bib:ATNFdb} with spin period $P$ shorter than 20
seconds\footnote{The longest known pulsation period is 76f seconds from the recently discovered PSR J0901-4046~\citep{bib:LongPeriodPulsar}, which
also displays unusual pulse length and variability and which may represent a new pulsar class.}
in the $P$--$\dot P$ plane, where
$\dot P$ is the first time derivative of the period. Red triangles show isolated pulsars, and blue circles show binary pulsars. 
  
Neutron stars have strong magnetic field intensities as a natural result of their collapse. If the magnetic flux is approximately conserved, the reduction
of the outer surface of the star to a radius of $\sim$10 km ensures a static surface field
far higher than achievable in a terrestrial laboratory~\citep{bib:Pacini_1967},
with inferred values (see below) ranging from $10^8$ G to more than $10^{15}$ G~\citep{bib:LyneGrahamSmith}.
The strongest fields are seen in
so-called ``magnetars,'' young neutron stars with extremly rapid spin-down, for which dynamo generation is also
likely relevant~\citep{bib:MagnetarDynamo1,bib:MagnetarDynamo2}. Both in young pulsars and in binary millisecond pulsars, there is
reason to believe that stronger magnetic fields are ``buried'' in the star from accreting plasma~\citep{bib:PayneMelatos},
although the burial mechanism is not confidently understood~\citep{bib:ChevalierAccretion,bib:GeppertEtalAccretion,bib:LyneGrahamSmith,bib:BernalEtalAccretion}.
It has been suggested there is evidence in at least some pulsars for slowly re-emerging (strengthening) magnetic
field~\citep{bib:HoAccretion,bib:BfieldEmergence}. Energy density deformation from a potentially non-axisymmetric buried field is another
potential source of GW emission~\citep{bib:BonazzolaGourgoulhon}. See \citep{bib:CrucesReiseneggerTauris} for a discussion of magnetic field decay preceding the accretion stage.

In principle, there should be $\sim$$10^{8-9}$ neutron stars in our galaxy,~\citep{bib:nspopulation,bib:TrevesEtal}.
That only a small fraction have been detected is expected, for several reasons.
Radio pulsations require high magnetic field and rotation frequency.
Early studies~\citep{bib:GoldreichJulian,bib:Sturrock,bib:RudermanSutherland,bib:LorimerKramer}
implied the relation
\begin{equation}
  B\cdot \frot^2>\scimm{1.7}{11}\>{\rm G}\cdot({\rm Hz})^2,
  \label{eqn:ChenRuderman}
\end{equation}
based on a model of radiation dominated by electron-positron pair creation in the stellar magnetosphere, a model
broadly consistent with empirical observation, although the resulting ``death line'' (see Figure~\ref{fig:pvspdot})
in the plane of period and period derivative is perhaps better understood
to be a valley~\citep{bib:ChenRuderman,bib:ZhangHardingMuslimov,bib:BeskinLitvinovDeathLine,bib:BeskinIstominDeathValley}.

The death line can be understood qualitatively from the
following argument. The rotating 
magnetic field of a neutron star creates a strong electric field that pulls charged particles from the star,
forming a plasma with charge density $\rho_0$ that satisfies (SI units):~\citep{bib:ChenRuderman}
\begin{eqnarray}
  \rho_0 & = & -\epsilon_0\nabla \cdot \left[(\vec\Omega\times\vec r)\times\vec B(\vec r)\right] \\
  & \approx & -2\epsilon_0\,\vec\Omega\cdot\vec B(\vec r),
\end{eqnarray}
\noindent where  $\vec\Omega$ is the angular velocity of the star, and
$\vec B$ is the local magnetic field at location $\vec r$ with respect to the star's center.
In steady-state equilibrium, one expects $\vec E\cdot\vec B \approx 0$ since free charges can
move along $B$-field lines. In so-called ``gaps,'' however, where the plasma density is low,
a potential difference large enough to produce spontaneous electron-positron pair production
can lead to radio-frequency synchrotron radiation as the accelerated particles encounter curved magnetic fields.
This emission is thought to account for most radio pulsations~\citep{bib:LyneGrahamSmith}, where
an ``inner gap'' refers to a region just outside the magnetic poles above the star's surface, and an ``outer gap'' refers to
a region where a nominally dipolar magnetic field is approximately perpendicular to the rotation direction,
separating regions of proton and electron flow from the star to the region beyond the ``light cylinder,''
defined by the cylindrical radius at which a co-rotating particle in the magnetosphere must travel at the
speed of light. For the inner gap to have a voltage drop high enough to induce an amplifying cascade of
pair production leading to coherent radio wave emission imposes a minimum value on the
gap potential difference $\Delta V$ which, in general, can be approximated by
(SI units):~\citep{bib:GoldreichJulian,bib:Sturrock,bib:RudermanSutherland,bib:ChenRuderman}
\begin{equation}
  \Delta V \sim {B\Omega^2R^3\over2c},
\end{equation}
\noindent where $R$ is the neutron star radius, 
leading (in a more detailed calculation) to Eqn.~\ref{eqn:ChenRuderman} and
via magnetic dipole emission assumptions (see section~\ref{sec:spindown}) to the death line
shown in Figure~\ref{fig:pvspdot} (but see \citep{bib:SmithRevisedDeathline} for evidence of
selection effects and \citep{bib:Petri} for a discussion of potentially important effects from
higher order multipoles). 
Presumably, the vast majority of neutron stars created in the galaxy's existence to date are now to the right of
the line. Additional negative-sloped dashed lines in the figure indicate different nominal magnetic
dipole field strengths and positive-sloped dashed lines indicate different nominal ages, based on observed
present-day periods and period derivatives $P/(2\dot P)$ (see section~\ref{sec:spindown}).

Two distinct major pulsar populations are apparent in Figure~\ref{fig:pvspdot}, defined by location
in the diagram. The bulk of the population lies above and to the right of the line corresponding
to $B\sim10^{11}$ G. The bulk also lies above and to the left of the line corresponding to
ages younger than $\sim$$10^8$ years. Assuming a star's magnetic field strength is stable, stars are expected
to migrate down to the right along the $B$-field contours.
Isolated pulsars seem to have typical pulsation lifetimes of $\sim$$10^7$ years~\citep{bib:LyneGrahamSmith},
after which they become increasingly difficult to observe in radio.
On this timescale, they also cool to where thermal X-ray emission is difficult
to detect~\citep{bib:NSCooling}. There remains the possibility of X-ray emission
from steady accretion of interstellar medium (ISM)~\citep{bib:OstrikerReesSilk,bib:BlaesMadauAccretion},
but it appears that the
kick velocities from birth highly suppress such accretion~\citep{bib:bondi1,bib:bondi2}
which depends on the inverse cube of the star's velocity through
the ISM, and steady-state X-ray emission from accretion onto even slow-moving neutron stars can
be highly suppressed, consistent with non-observation to date of such accretion~\citep{bib:PopovISMaccretion}.

The remaining population, in the lower left of the figure, is characterized by shorter periods and smaller
period derivatives. These are so-called ``millisecond pulsars'' (MSPs), thought to arise from ``recycling''
of rotation speed due to accretion of matter from a binary companion. MSPs are stellar zombies, brought
back from the dead with immense rotational energies imparted by infalling matter~\citep{bib:AlparEtalRecycling,bib:RadhakrishnanSrinivasanRecycling}.
The rotation frequencies achievable through this spin-up are
impressive -- the fastest known rotator is PSR J1748$-$2446ad at 716 Hz~\citep{bib:HesselsEtal}.
One progenitor class for MSPs is the set of low mass X-ray binaries (LMXBs)
in which the neutron star ($\sim$1.4 $\msolar$) has a much
lighter companion ($\sim$0.3 $\msolar$)~\citep{bib:LyneGrahamSmith} that
overfills its Roche lobe, spilling material onto an accretion
disk surrounding the neutron star or possibly spilling material
directly onto the star, near its magnetic polar caps. 
When the donor companion star eventually shrinks and decouples
from the neutron star, the neutron star can retain a large
fraction of its maximum angular momentum and rotational energy.
Because the neutron star's magnetic field decreases during
accretion (through processes that are not well understood),
the spin-down rate after decoupling can be very small.
The minority of MSPs that are isolated are thought to have lost their one-time companions via consumption and
ablation. A bridging class called ``black widows'' and ``redbacks'' refer to binary systems with
actively ablating companions, such as B1957+20~\citep{bib:blackwidowfirst,bib:redbackcollection,bib:blackwidowredback}, where
black widows denote the extreme subclass with companion masses below 0.1 $\msolar$~\citep{bib:blackwidowredback}.

A nice confirmation of the link between LMXBs and recycled MSPs comes from ``transitional millisecond pulsars'' (tMSPs)
in which accreting LMXB behavior alternates with detectable radio pulsations. The first tMSP found was
PSR J1023$+$0038~\citep{bib:tMSPfirst1,bib:tMSPfirst2,bib:tMSPfirst3}, with two more systems since detected~\citep{bib:tMSPsystems}.
The nominal ages of MSPs extend beyond 10$^{10}$ years, that is, some have
apparent ages greater than that of the galaxy (or even that of the Universe).
One possible explanation of this anomaly is
reverse-torque spin-down during the Roche decoupling phase~\citep{bib:taurisrldp},
although a recent numerical study suggests a more complex frequency evolution
before and during the decoupling~\citep{bib:BhattacharyyaTwoModes}.

An obvious pattern in Figure~\ref{fig:pvspdot}, consistent with the recycling model,
is the higher fraction of binary systems at lower periods. For example, binary systems
account for 3/4 of the lowest 200 pulsar periods (below $\sim$4 ms).

Aside from the disappearance of stars from this diagram as they evolve toward the lower right
and cease pulsations, there are also strong selection effects that suppress the visible population.
We observe pulsars only if their radiation
beams cross the Earth, only if that radiation is bright enough to be seen in the observing band, and
only if the radiation is not sufficiently absorbed, scattered or frequency-dispersed to prevent
detection with current radio telescopes. When the Square Kilometer Array project comes to fruition in the late 2020's,
it is estimated that the current known population of pulsars will grow tenfold~\citep{bib:SKAIncrease}.

\subsubsection{Neutron star spin-down phenomenology and mechanisms}
\label{sec:spindown}

Nearly every known pulsar is observed to be spinning down, that is, to have a negative
rotational frequency time derivative, implying loss of rotational kinetic energy.
As discussed below in detail, there are many physical mechanisms, electromagnetic and
gravitational, that can lead to this energy loss. For CW signal detection we want a
gravitational wave component, but there is good reason to believe that electromagnetic
processes dominate for nearly every known pulsar.

A convenient and commonly used phenomenological model for spin-down is a power law:
\begin{equation}
  \fdot = K f^n\>,
  \label{eqn:spindownpowerlaw}
\end{equation}
\noindent where $f$ is the star's instantaneous frequency (rotational $\frot$ or gravitational: $\fgw\propto\frot$),
$\fdot$ is the first time derivative, and
$K$ is a negative constant for all but a handful of stars (thought to
be experiencing large acceleration toward us because of nearness to a deep gravitational
well, such as in the core of a globular cluster). The exponent $n$ depends on the spin-down
mechanism and is known as the {\it braking index}. The four most common theoretical braking indices discussed in
the literature are the following:
\begin{itemize}
\item $n=1$ -- ``Pulsar wind'' (extreme model)
\item $n=3$ -- Magnetic dipole radiation
\item $n=5$ -- Gravitational mass quadrupole radiation (``mountain'')
\item $n=7$ -- Gravitational mass current quadrupole radiation (\rmodes).
\end{itemize}
\noindent In principle, other oscillation modes that can generate gravitational waves are also possible, but the
$n\!=\!5$ and $n\!=\!7$ modes discussed below are thought to be the most promising.

Assuming the same power law has applied since the birth of the star, the age
$\tau$ of the star can be related to its birth rotation frequency $f_0$ and
current frequency $f$ by ($n\ne1$):
\begin{equation}
\label{eqn:ageindex}
\tau \quad = \quad -\left[{f\over (n-1)\,\dot f}\right]\,\left[1-\left({f\over f_0}\right)^{(n-1)}\right],
\end{equation}
and in the case that $f\ll f_0$, 
\begin{equation}
\label{eqn:approxageindex}
\tau \quad \approx \quad -\left[{f\over (n-1)\,\dot f}\right]\,.
\end{equation}

A common baseline assumption in radio pulsar astronomy is that the braking index is $n=3$ from which the nominal magnetic dipole age of a star
can be defined
\begin{equation}
  \tau_{\rm mag} \equiv -{f\over2\dot f},
  \label{eqn:approxagemagnetic}
\end{equation}
again, under the assumption $f\ll f_0$.

From the more generic power-law spin-down model (Eqn.~\ref{eqn:spindownpowerlaw}), the 2nd frequency derivative can be written:
\begin{equation}
  \ddot f = nKf^{n-1}\dot f = nK^2f^{2n-1}\>,
    \label{eqn:fdoubledot}
\end{equation}
from which the current braking index can be determined if the spin frequency's 2nd time derivative
can be measured reliably:
\begin{equation}
  \label{eqn:brakingindex}
  n \quad = \quad {f\ddot f \over \dot f^2}\,.
\end{equation}
Before examining the empirical measurements of the braking indices, which are mostly inconsistent with $n=3$,
let's briefly review spin-down mechanisms with well defined braking indices, when dominant. For
GW radiation spin-down dominance, related ``spin-down'' limits on strain amplitude will also be presented.

\paragraph{``Pulsar wind'' ($n=1$).} \leavevmode\\

Early on in pulsar astronomy~\citep{bib:Michel,bib:MichelTucker} it was recognized that the streaming
of relativistic particles (electrons and positrons mainly, with some ions)
away from the magnetosphere of a fast-spinning neutron star would lead
to a spin-down torque that could, in principle, rival that from magnetic dipole radiation, in addition to distorting
the shape of the magnetic field lines and affecting the dipole radiation~\citep{bib:GaenslerSlane}.
In this perhaps too-simple model, the spin-down is
dominated by a braking torque from a return current (predominantly counter-flowing electrons and
positrons) crossing magnetic field lines in the polar cap regions of the star~\citep{bib:ContopoulosKazanasFendt},
leading to a braking index of one.
A more recent study of magnetar
spin-down~\citep{bib:HardingContopoulosKazanas} considered a model with sporadic high winds following bursts, with
magnetic dipole emission dominating spin-down between bursts.
In the steady state, however, considering the interaction of the magnetic field and the plasma of the
magnetosphere, both magnetic dipole emission and pulsar wind contributions tend to
yield a braking index of about three~\citep{bib:MichelLi,bib:Spitkovsky}, discussed next.
A phenomenological model~\citep{bib:Melatos} that is a variant of the vacuum dipole mode, featuring an inner magnetosphere strongly coupled to the star,
accounts successfully for the braking indices of the Crab and other young pulsars with $n<1$.
  
\paragraph{Magnetic dipole ($n=3$).} \leavevmode\\

The radiation energy loss due to a rotating magnetic dipole moment is~\citep{bib:Pacini_1968}
\begin{equation}
\left({dE\over dt}\right)_{\rm mag} \quad = \quad -{\mu_0M_\perp^2\omega_{\rm rot}^4\over6\pi c^3},
\end{equation}
where $\omega_{\rm rot}$ is the rotational angular speed and
$M_\perp$ is the component of the star's magnetic dipole
moment perpendicular to the rotation axis (taken to be the $z$ axis): $M_\perp=M\sin(\alpha)$,
with $\alpha$ the angle between the axis and north magnetic pole.

In a pure dipole moment model, the magnetic pole field strength
at the surface is $B_0 = \mu_0M\,/\,2\pi R^3$.
Equating the radiation energy loss to that of the (Newtonian) rotational
energy ${1\over2}\Izz\omega_{\rm rot}^2$ leads to the prediction:
\begin{equation}
{d\omega_{\rm rot}\over dt} \quad = \quad -{2\pi\over3} {R^6 \over \mu_0c^3\Izz}B_\perp^2\omega_{\rm rot}^3.
\end{equation}
Hence the magnetic dipole spin-down  rate is proportional to the square of $B_\perp=B_0\sin(\alpha)$
and to the cube of the rotation frequency, giving $n=3$.

\paragraph{Gravitational mass quadrupole (``mountain'', $n=5$).} \leavevmode\\

Let's now consider the gravitational radiation one might
expect from these stars.
It is conventional to characterize a star's mass quadrupole
asymmetry by its equatorial ellipticity:
\begin{equation}
\label{eqn:ellipticity}
\epsilon \quad \equiv \quad {|I_{xx}-I_{yy}|\over \Izz}.
\end{equation}
An oblate spheroid naturally has a polar ellipticity,
but in the absence of precession\footnote{Free precession of an oblate
neutron star can lead to gravitational radiation at the rotation frequency~\citep{bib:zimmermannszedenits}, but there is little empirical evidence
for such precession in pulsars and good reason to expect that such precession would be rapidly damped by internal
dissipation~\citep{bib:JonesAnderssonPrecession2}.},
such a deformation does not lead to GW emission.
Henceforth ``ellipticity'' will
refer to equatorial ellipticity, often attributed to a ``mountain''.
For a star at a distance $d$ away and spinning about the approximate symmetry axis of rotation ($z$),
(assumed optimal -- pointing toward the Earth), then the expected intrinsic strain amplitude $h_0$ is 
\begin{eqnarray}
  \label{eqn:hexpected}
h_0 & = & {4\,\pi^2G\epsilon\Izz\fgw^2\over c^4d}  \\
    & = & (\scimm{1.1}{-24})\left({\epsilon\over10^{-6}}\right)\left({\Izz\over I_0}\right)\left({\fgw\over1\>{\rm kHz}}\right)^2
\left({1\>{\rm kpc}\over d}\right),
\end{eqnarray}
where $I_0=10^{38}$ kg$\cdot$m$^2$ (10$^{45}$ g$\cdot$cm$^2$) is a nominal moment of inertia of
a neutron star used throughout this article, and the gravitational radiation is emitted at frequency $\fgw=2\,\frot$.
The total power emission in gravitational waves from
the star (integrated over all angles) is 
\begin{eqnarray}
\label{eqn:powerloss}
{dE\over dt} & = & - {32\over5} {G\over c^5}\,\Izz^2\, \epsilon^2\, \omega_{\rm rot}^6  \\
             & = & - (\scimm{1.7}{33}\>{\rm J/s})\left({\Izz\over I_0}\right)^2 \left({\epsilon\over10^{-6}}\right)^2 \left({\fgw\over1\>{\rm kHz}}\right)^6.
\end{eqnarray}

Equating this loss to the reduction of rotational kinetic energy ${1\over2}\Izz\omega_{\rm rot}^2$ leads to the spin-down relation:
\begin{eqnarray}
  \label{eqn:GWbraking}
  \fgwdot & = & -{32\,\pi^4\over5}{G\over c^5}\Izz\epsilon^2\fgw^5\\
          & = & -(\scimm{1.7}{-9}\> {\rm Hz/s}) \left({\epsilon\over10^{-6}}\right)^2 \left({\fgw\over1{\rm\>kHz}}\right)^5,
\end{eqnarray}
\noindent in which the braking index of 5 is apparent.

For an observed neutron star of measured $f$ and $\dot f$,
one can define the ``spin-down limit'' on maximum allowed
strain amplitude by equating the power loss in Eqn.~(\ref{eqn:powerloss})
to the time derivative of the (Newtonian) rotational kinetic
energy: ${1\over2}\Izz\omega_{\rm rot}^2$, as above for magnetic dipole radiation. 
One finds:
\begin{eqnarray}
\label{eqn:spindownlimit}
h_{\rm spin-down} & = & {1\over d}\sqrt{-{5\over2}{G\over c^3}\Izz{\dot\fgw\over\fgw}} \nonumber \\
& = &  (\scimm{2.6}{-25})
\left[ {1\>{\rm kpc}\over d} \right]\!
  \left[
    \left({1\>{\rm kHz}\over\fgw}\right)\!
    \left({-\dot \fgw \over10^{-10}\>{\rm Hz/s}}\right)\!
    \left({\Izz\over I_0}\right)\!
    \right]^{1\over2}\!\!.\quad\>\>\>
\end{eqnarray}
Hence for each observed pulsar with a measured frequency, spin-down and
distance $d$,
one can determine whether or not energy conservation even permits detection
of gravitational waves in an optimistic scenario. Unfortunately,
nearly all known pulsars have strain spin-down limits below what
can be detected by the LIGO and Virgo detectors at current
sensitivities, as detailed below. 

\paragraph{Gravitational mass current quadrupole (\rmodes, $n=7$).} \leavevmode\\

Different frequency scalings apply to mass quadrupole and mass current quadrupole emission.
The most promising source of the mass current non-axisymmetry in neutron stars is thought
to be ``\rmodes,'' due to fluid motion of neutrons (or protons) in the crust or core
of the star. Like jet streams in the Earth's atmosphere that manifest Rossby waves, these currents are
deflected by Coriolis forces, giving rise to spatial oscillations~~\citep{bib:rmodes1,bib:rmodes2,bib:rmodes3,bib:rmodes4}.
These \rmodes\ can be inherently unstable, arising from azimuthal
interior currents that are retrograde in the star's rotating frame, but
which are prograde in an external reference frame. As a result, the
quadrupolar gravitational wave emission due to these currents
leads to an {\it increase} in the strength of the current. This
positive-feedback loop leads to a potential intrinsic (Chandrasekhar-Friedman-Schutz~\citep{bib:cfs1,bib:cfs2}) instability.
The frequency of such emission is expected to be a bit more than approximately 4/3 the rotation 
frequency~\citep{bib:rmodes1,bib:rmodes2,bib:rmodes3,bib:rmodes4,bib:rmodes5,bib:CarideIntaOwenRajbhandari}. 

Following the notation of Owen~\citep{bib:Owenalpha,bib:CarideIntaOwenRajbhandari}, the mass current can be treated
as due to a velocity field perturbation $\delta v_j$, integration over which leads to the following expression
for the intrinsic strain amplitude seen at a distance $d$:
\begin{eqnarray}
  \label{eqn:rmode1}
  h_0 & = & \sqrt{512\,\pi^7\over5} {G\over c^5}{1\over d}\fgw^3\alpha MR^3\tilde J \\
  & = & \scimm{3.6}{-26} \left({1\>{\rm kpc}\over d}\right) \left({\fgw\over100\>{\rm Hz}}\right)^3 \left({\alpha\over10^{-3}}\right)
  \label{eqn:rmode2}
    \left({R\over11.7\>{\rm km}}\right)^3,
\end{eqnarray}
\noindent where $\alpha$ is the dimensionless \rmode\ amplitude, $M$ is the stellar mass,
$R$ its radius, and $\tilde J$ is a dimensionless functional of the stellar equation of state,
which for a Newtonian polytrope with index 1 gives $\tilde J\approx .0164$~\citep{bib:Owenalpha},
assumed in the fiducial Eqn.~\ref{eqn:rmode2}.

The energy loss in this model is~\citep{bib:ThorneMultipoles,bib:Owenalpha}
\begin{eqnarray}
  {dE\over dt} & = & -{1024\,\pi^9\over 25} {G\over c^7} \fgw^8 \alpha^2 M^2 R^6 \tilde J^2.
\end{eqnarray}
Equating this loss to the reduction of rotational kinetic energy ${1\over2}\Izz\omega_{\rm rot}^2$, as above, leads to the spin-down relation:
\begin{eqnarray}
  \fgwdot & = & -{4096\,\pi^7\over225} {G\over c^7} {M^2R^6\tilde J^2\over\Izz} \alpha^2\fgw^7 \\
          & = & -\scimm{9.0}{-14}\>{\rm Hz/s}\left({R\over11.7\>{\rm km}}\right)^6 \left({\alpha\over10^{-3}}\right)^2 \left({\fgw\over100\>{\rm Hz}}\right)^7 ,
\end{eqnarray}
\noindent in which the braking index of 7 is apparent.

As before, one can define a spin-down limit, but one based on pure \rmode\ radiation:
\begin{equation}
h_{\rm spin-down}  =  {1\over r}\sqrt{-{45\over8}{G\over c^3}\Izz{\fgwdot\over\fgw}}\>,
\end{equation}
\noindent where the ratio of this spin-down limit to the one given in Eqn.~\ref{eqn:spindownlimit} is 3/2,
which arises simply from the different ratios of GW signal frequency to spin frequency for mass
quadrupole \vs\ mass current quadrupole
radiation\footnote{The 4/3 ratio assumed here for $\fgw/\frot$ is a slow-rotation approximation in Newtonian gravity; the ratio changes by tens of percent for fast rotation
in General Relativity~\citep{bib:IdrisyOwenJones,bib:CarideIntaOwenRajbhandari} (see section~\ref{sec:directedisolated}).}. 

\paragraph{Measured braking indices.} \leavevmode\\

\begin{figure}[t!]
\begin{center}
\includegraphics[width=13.cm]{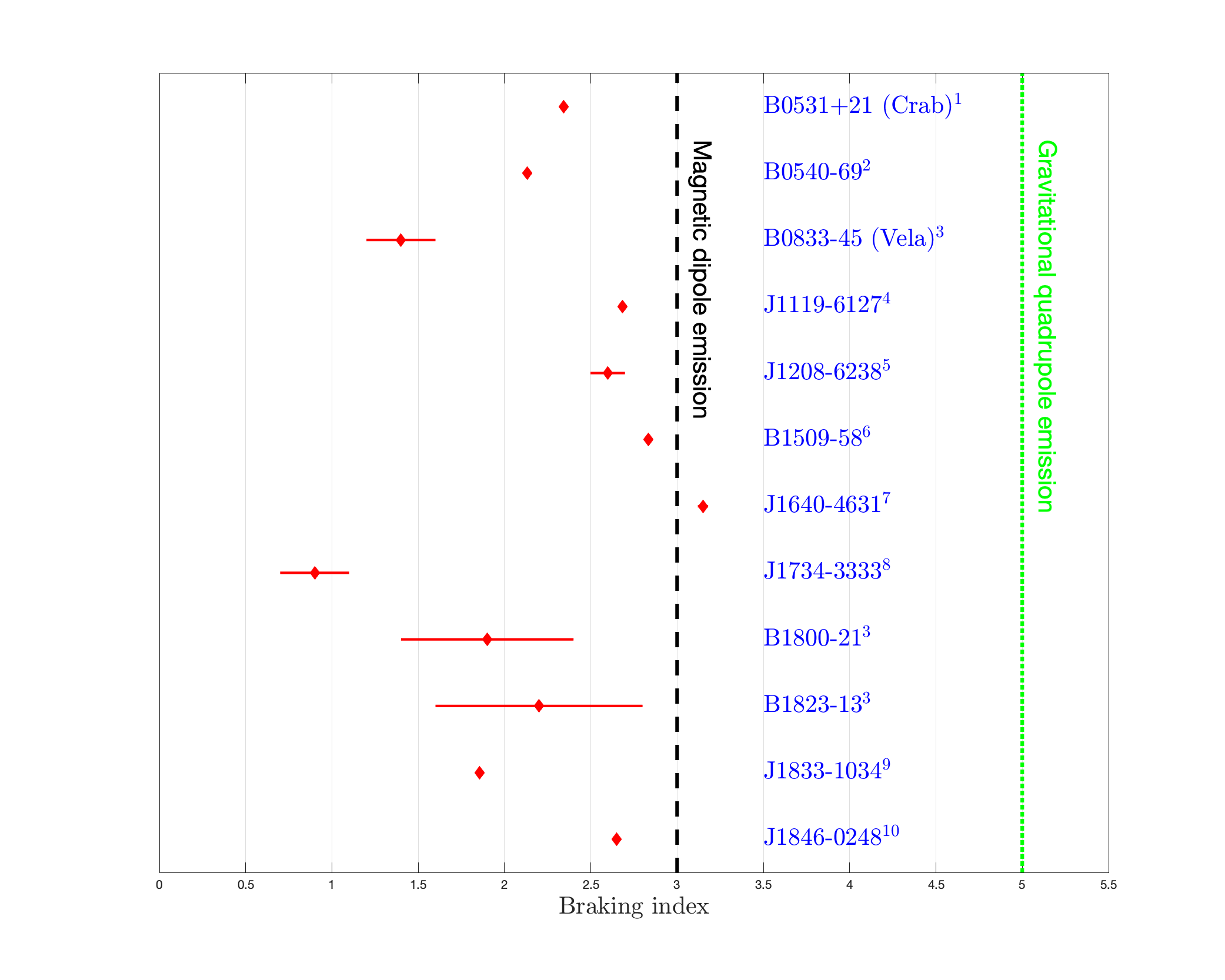}
\caption{Measured braking indices inferred from frequency derivatives of young pulsars with
  rotation frequencies greater than 10 Hz.
  For frequently glitching pulsars, such as Vela, the braking index is computed as a long-term
  average~\citep{bib:EspinozaEtal_2016}. Horizontal bars indicate uncertainties and are smaller than the
  plot markers for several pulsars. Vertical lines at braking indices of 3 and 5 denote the nominal
  expectations for magnetic dipole and gravitational quadrupole emission, respectively.
  References: 1~\citep{bib:LyneEtal_2015}, 2~\citep{bib:FerdmanEtal_2015}, 3~\citep{bib:EspinozaEtal_2016},
  4~\citep{bib:WeltevredeEtal_2011}, 5~\citep{bib:ClarkEtal_2016}, 6~\citep{bib:Livingstone_Kaspi_2011},
  7~\citep{bib:ArchibaldEtal_2016}, 8~\citep{bib:BfieldEmergence}, 9~\citep{bib:RoyEtal_2012},
  10~\citep{bib:LivingstoneEtal_2007}.}

\label{fig:brakingindices}
\end{center}
\end{figure}

Figure~\ref{fig:brakingindices} shows the distribution of 12 reliably measured braking indices from a recent snapshot of the $\sim$3300
pulsars listed in the ATNF catalog (release V1.66 -- January 10, 2022~\citep{bib:ATNFdb}). Nearly all have values below the nominal value of 3 for a magnetic
dipole radiator, although several have large uncertainties.

\begin{figure}[t!]
\begin{center}
\includegraphics[width=7.95cm]{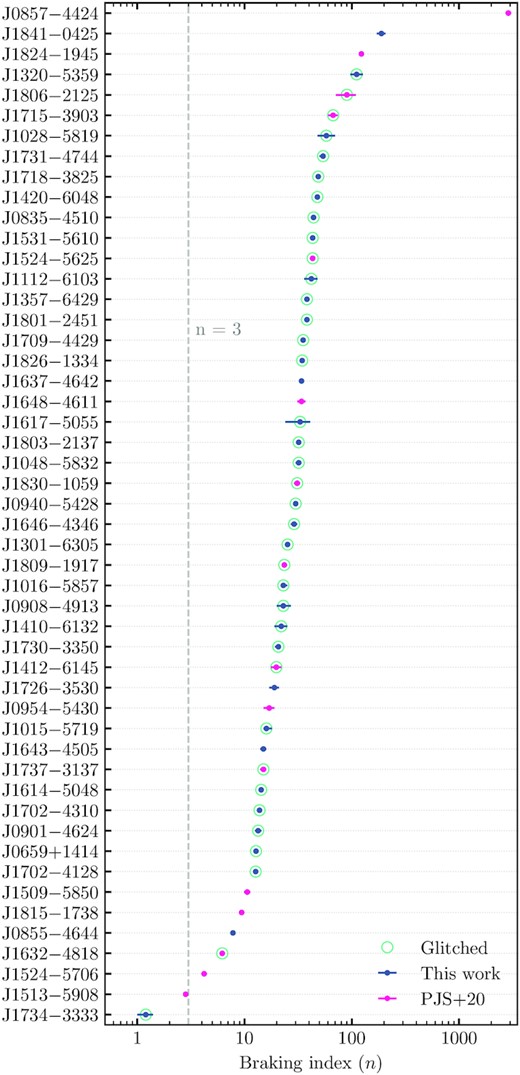}
\caption{Measured braking indices inferred from frequency derivatives of pulsars compiled
  in (\cite{bib:LowerEtal} -- ``this work'', including from \citep{bib:ParthasarathyEtalBrakingIndicesII} -- PJS).
  An ensemble of glitching (open green circles) and non-glitching pulsars are included.
  For most of the glitching stars, the braking indices are representative of their
  average inter-glitch braking, not their long-term evolution.}
\label{fig:brakingindiceslarge}
\end{center}
\end{figure}

This distribution suggests
that the model of a neutron star spinning down with
constant magnetic field is, most often, inaccurate~\citep{bib:LyneGrahamSmith}.
All measured values for this collection lie below 3.0, except X-ray pulsar PSR J1640$-$4631
with a measured index of 3.15$\pm$0.03~\citep{bib:ArchibaldEtal_2016}.
It is possible that for many stars the departure of the measured braking index from the nominal value is due to an admixture
of magnetic dipole radiation and other steady-state processes~\citep{bib:Melatos},
although secular mechanisms may also play a role.
See \citep{bib:palomba1,bib:palomba2} for discussions of spin-down evolution
in the presence of both gravitational wave and electromagnetic
torques. Other suggested mechanisms for less-than-3 braking indices are
decaying magnetic fields~\citep{bib:RomaniUnifiedBfieldModel},
re-emerging buried magnetic fields~\citep{bib:HoAccretion},
a changing inclination angle between the magnetic dipole axis 
the spin axis~\citep{bib:MiddleditchEtalStarquakes,bib:tauriskonar,bib:emergingBfield,bib:LyneEtal_2015,bib:decayinginclination}, and
a changing superfluid moment of inertia~\citep{bib:HoAnderssonSupercooling}.

An interesting observation of the aftermath of two
short GRBs noted indirectly inferred braking indices near or equal to
three~\citep{bib:magnetarbrakingindex},
suggesting the rapid spin-down of millisecond magnetars, possibly born
from neutron star mergers. (No direct gravitational wave evidence of a such a post-merger remnant
has been observed from GW170817~\citep{bib:Postmerger1,bib:Postmerger2}.)
Similarly, a recent analysis of X-ray afterglows of gamma-ray bursts~\citep{bib:SarinLaskyAshton}
argues that at least some have millisecond magnetar remnants powering their emission,
with GRB 061121 yielding a braking index $n=4.85^{+0.11}_{-0.15}$, consistent with gravitational radiation dominance
(albeit with large required ellipticity~\citep{bib:HoGRB,bib:KashiyamaEtalGRB}).
See~\cite{bib:StrangEtal}, however, for an alternative study in which radiation driven from a millisecond magnetar
can account for short GRB X-ray afterglows. See \cite{bib:DallOssoStellMillisecondMagnetar} for a recent brief review of millisecond magnetars,
including evidence of their serving as central engines to create GRBs, and see~\cite{bib:Jordana-MitjansEtalGRBRemnant}
for evidence of a protomagnetar remnant in the aftermath of GRB 180618A.

It has been argued that the inter-glitch evolution of spin for the
X-ray pulsar PSR J0537$-$6910 displays behavior consistent with a braking index of 7,~\citep{bib:rmodeJ0537-6910,bib:HoEtalJ0537}
consistent with \rmode\ emission, while the long-term trends points to an underlying braking index of
$-$1.25$\pm$0.01~\citep{bib:HoEtalJ0537}.
When intepreting the generally low values of well measured braking indices, one
must bear in mind the potential for selection bias.
Baysesian analysis of the spin evolution of 19 young pulsars~\citep{bib:ParthasarathyEtalBrakingIndicesI,bib:ParthasarathyEtalBrakingIndicesII},
taking into account timing noise and extracting
the long-term behavior from short-term, glitch-driven fluctuations, leads to braking indices
{\it much larger} than 3. A similar follow-up analysis of an ensemble of glitching and
non-glitching pulsars~\citep{bib:LowerEtal} confirmed that braking indices exceeding
100 are observed (see Figure~\ref{fig:brakingindiceslarge}), albeit for stars in
which a simple power-law spin-down is clearly inappropriate. 

\paragraph{The gravitar model and associated figures of merit.}\leavevmode\\

\begin{figure}[t!]
\begin{center}
\includegraphics[width=13.cm]{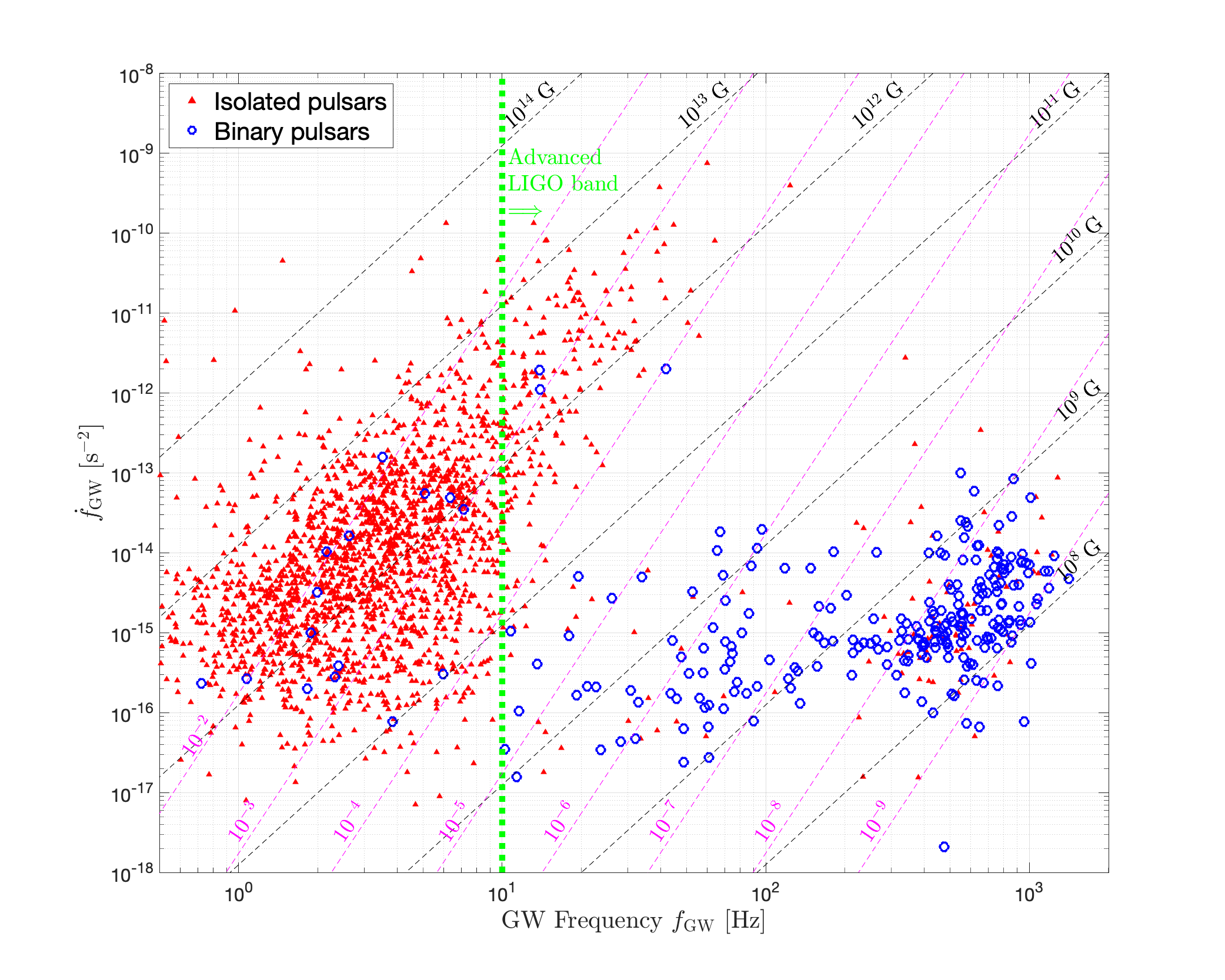}
\caption{Nominal expected GW frequencies and frequency derivatives for known 
pulsars. Closed triangles indicate isolated stars. Open circles
indicate binary stars. Contours are shown for constant magnetic fields
(ellipticities) for spin-down dominated by magnetic dipole (gravitational mass quadrupole) emissions.
In this figure and in Figs.~\ref{fig:fvshspindown}--\ref{fig:fvsdist2}, the frequency derivatives have been
corrected for the Shklovskii effect~\citep{bib:Shklovskii} (apparent negative frequency derivative
due to proper motion orthogonal to the line of sight).
The vertical dotted line denotes the
approximate sensitivity band for Advanced LIGO at design sensitivity.
A similar band applies to
design sensitivities of the Advanced Virgo and KAGRA detectors~\citep{bib:obsscenario}.}
\label{fig:fvsfdot}
\end{center}
\end{figure}

Gravitars refer to neutron stars with spin-down dominated by gravitational wave energy loss~\citep{bib:palomba2}.
Although there is good reason to believe that most known pulsars are {\it not} gravitars,
nonetheless the model is useful in bounding expectation on what is {\it possibly} detectable.
Figure~\ref{fig:fvsfdot} shows a subset of the pulsars from Figure~\ref{fig:pvspdot}, now
graphed in the $\fgw$--$\fgwdot$ plane, under the assumption that $\fgw = 2\,\frot$. Again,
isolated and binary stars are denoted by closed circles and open triangles, respectively.
A vertical dashed line bounds the approximate detection
bandwidth for Advanced LIGO at design sensitivity ($\sim$10 Hz and above).
The same approximate frequency boundary applies to the design sensitivities of
the Advanced Virgo and KAGRA detectors~\citep{bib:obsscenario}.
As in Figure~\ref{fig:pvspdot},
contours are shown for constant magnetic field, assuming spin-down dominated by magnetic
dipole emission ($n=3$). In addition, contours of higher slope are shown for 
constant ellipticity.
An intriguing deficit of millisecond pulsars with extremely low period derivatives appears consistent~\citep{bib:WoanEtalMSP}
with a population of sources with a minimum ellipticity of about $\sim$$10^{-9}$ with additional spin-down losses from
magnetic dipole radiation (see near absence of sources in figure~\ref{fig:fvsfdot} to the right of the
$\epsilon=10^{-9}$ line). At the other extreme are lower-frequency, younger pulsars with high spin-downs,
the highest of which is \sci{7.6}{-10} Hz/s (Crab pulsar). 

\begin{figure}[t!]
\begin{center}
\includegraphics[width=13.cm]{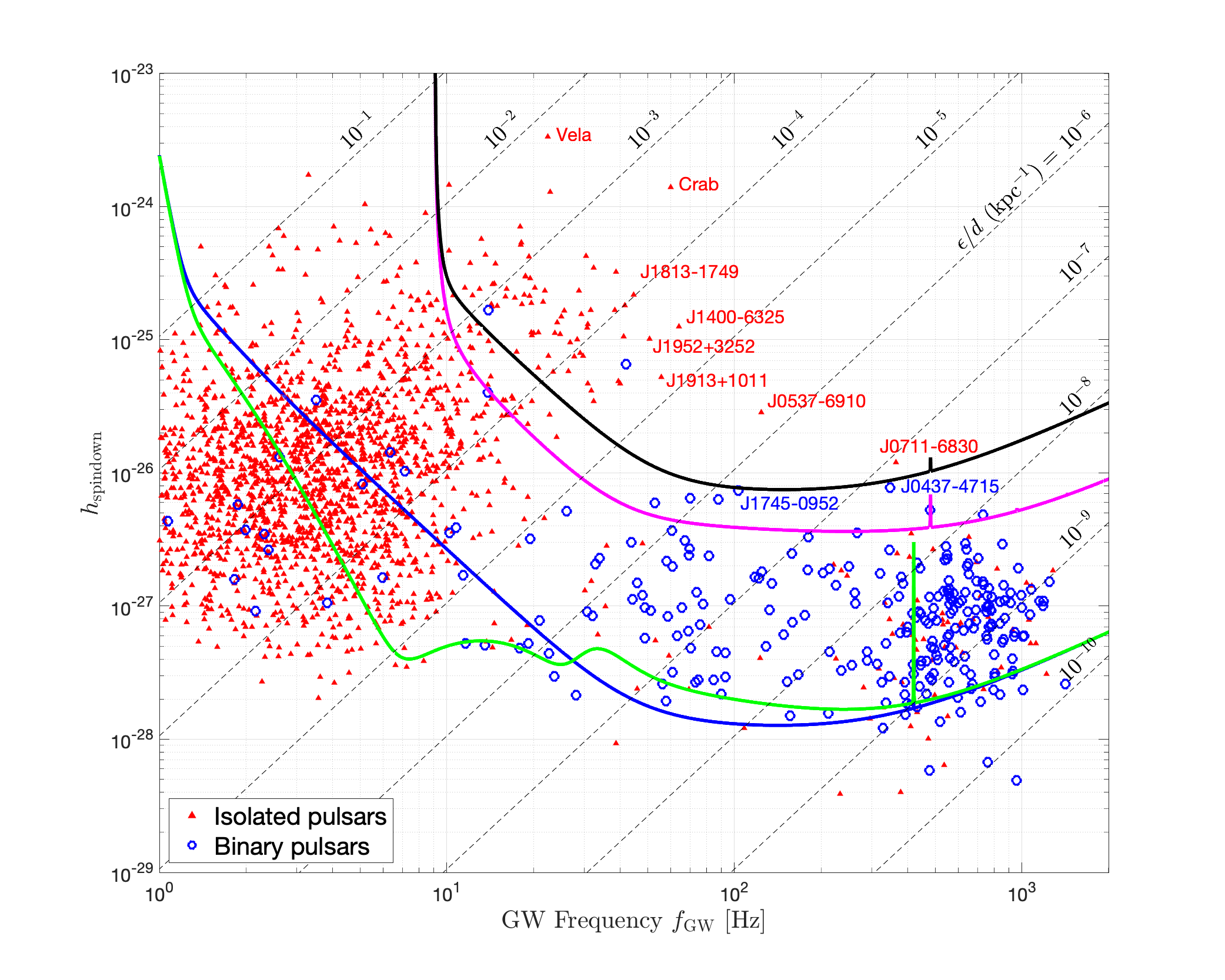}
\caption{Nominal expected GW frequencies and nominal strain spin-down limits for known
pulsars. Closed triangles indicate isolated stars. Open circles
indicate binary stars. The solid curves indicate the nominal (idealized) strain noise sensitivity for the
O3 observing run (black), and expected sensitivities for
2-year advanced detector data run at design sensitivity (magenta) and a 5-year Einstein Telescope data run
for two different detector designs: ETB (blue) and ETC (green). Dashed diagonal lines correspond to particular
quotients of ellipticity over distance.
A subset of pulsars of particular interest are labeled on the figure.}
\label{fig:fvshspindown}
\end{center}
\end{figure}

Using Eqn.~\ref{eqn:spindownlimit}, these known pulsars can be mapped onto a plane
of $\fgw$--$h_0$ {\it under the gravitar assumption}, indicated in Figure~\ref{fig:fvshspindown}. That is, the spin-down strain limit (for $n=5$)
is shown on the vertical axis.  Also shown are corresponding contours of constant implied values of $\epsilon/d$, under the gravitar assumption,
where $d$ is the distance to the star. In addition, detector network sensitivities are
shown for advanced detectors at design sensitivity~\citep{bib:obsscenario} and
for two proposed configurations of the ``3rd-generation'' Einstein Telescope (ET)~\citep{bib:EinsteinTelescope}
(ETB and ETC, for three detectors for five observing years).
Another 3rd-generation proposal is for
the ``Cosmic Explorer''~\citep{bib:CosmicExplorer} 
which would have performance comparable to that of ET,
being more sensitive at frequencies above $\sim$10 Hz
and less sensitive at lower frequencies.
To avoid clutter in these figures, only the ET sensitivities are shown.

In Fig.~\ref{fig:fvshspindown} and in succeeding figures,
the ``advanced detector'' sensitivities are represented by those computed for
two Advanced LIGO detectors running continuously for two observing years,
henceforth designated as the ``O4/O5 run''. Although the O4 run scheduled to start near the start of 2023
will likely run for only $\sim$1 year~\citep{bib:obsscenario} and may not quite reach the original Advanced LIGO
design sensitivity, the succeeding O5 run in the ``A$^+$'' configuration
is expected to exceed Advanced LIGO sensitivity significantly
and to last for more than a year, making the detector sensitivities assumed here conservative, in principle.
Including Advanced Virgo and KAGRA into the network sensitivity would improve these sensitivities still further.
On the other hand, the O4/O5 observing time assumed here does not account for realistic deadtime losses, which
can be substantial ($\sim$25\%\ per detector~\citep{bib:DavisEtalDetchar}). 
The detection sensitivities shown in Fig.~\ref{fig:fvshspindown}
assume a {\it targeted search} (discussed below) using known pulsar ephemerides.
If a star is marked above a sensitivity curve, then it is at least possible to detect it
if its spin-down makes it a gravitar. Note, however, that Eqn.~\ref{eqn:spindownlimit} has been
applied with a nominal moment of inertia $\Izz$, but the uncertainty in $\Izz$ is of order a factor
of two, depending on equation of state and stellar mass~\citep{bib:UncertainIzz}.

\begin{figure}[t!]
\begin{center}
\includegraphics[width=13.cm]{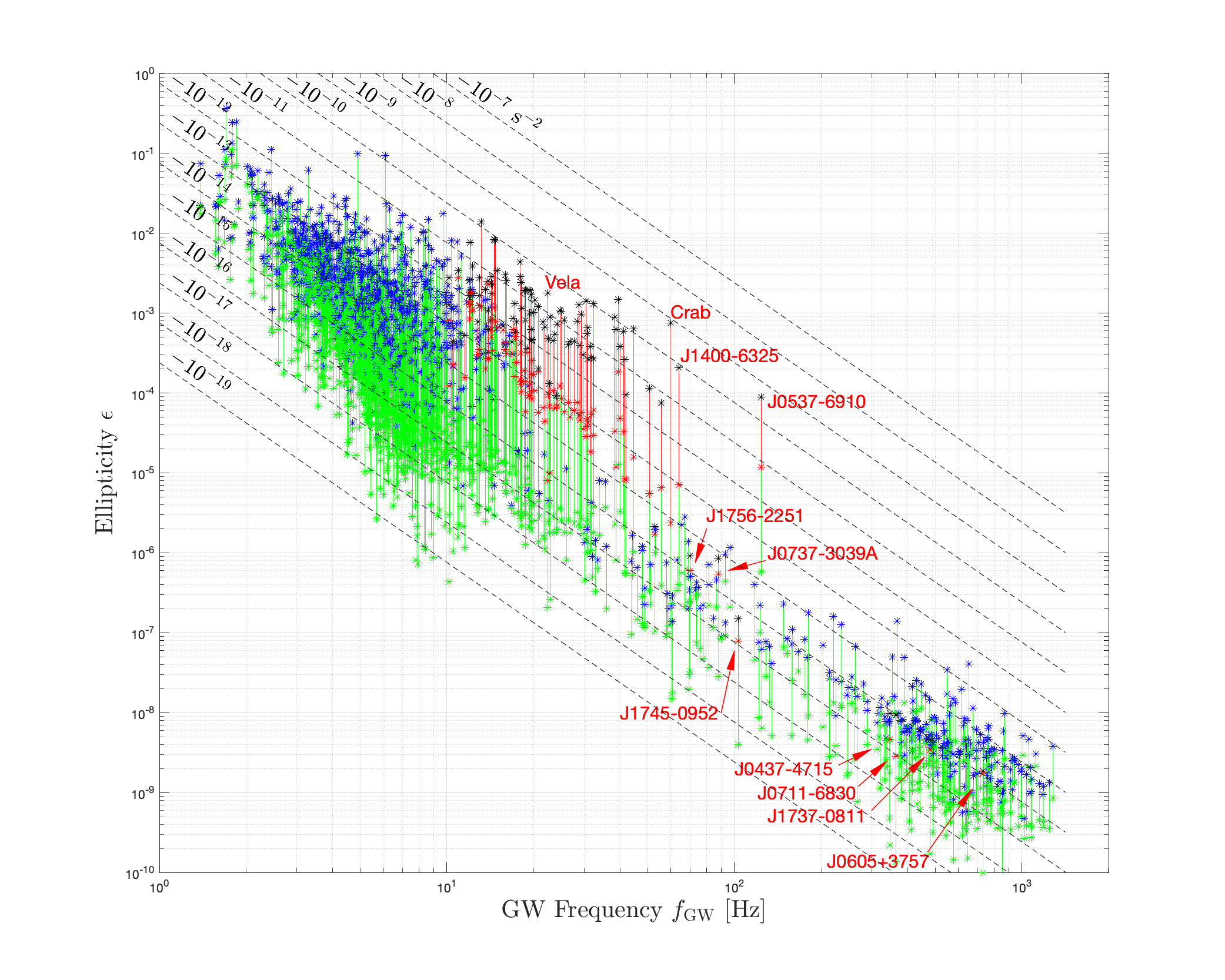}
\caption{Nominal expected GW frequencies and maximum allowed ellipticities for known
  pulsars.  Black or blue asterisks indicate ellipticities
  accessible with advanced detectors or ETC sensitivities (3 detectors, 5 years), respectively, using targeted searches,
  where red vertical lines terminated by red asterisks indicate ellipticity sensitivity range
  for advanced detectors, and green vertical lines and green asterisks indicate additional ellipticity sensitivity range
  for ETC. A selection of pulsars accessible with advanced detector sensitivity are labeled in red.
  Diagonal dashed lines correspond to corresponding $\fgwdot$ values under the gravitar model.}
\label{fig:fvsepsilon}
\end{center}
\end{figure}

Another take on the pulsars with accessible spin-down limits is shown in Figure~\ref{fig:fvsepsilon},
where accessible ellipticity $\epsilon$ values are shown for advanced detector and
Einstein Telescope (ETC) sensitivities. Each vertical bar represents a range of ellipticities detectable
for that star (red = accessible to advanced detectors, green = accessible to Einstein Telescope), where the
asterisk at the top of the each bar is the ellipticity corresponding to that star's spin-down limit,
given its $\fgw$, $\fgwdot$ and distance $d$ values, while the depth to which the bar falls indicates
the lowest detectable ellipticity. Straight dashed lines of negative slope depict corresponding $\fgwdot$ values
{\it under the mass quadrupole gravitar model}. The actual $\fgwdot$ may be significantly higher because of
the spin-down mechanisms discussed earlier. A striking feature of this figure is that sensitivities to very low
ellipticities come almost entirely from the highest-frequency stars (as a reminder from Eqn.~\ref{eqn:hexpected}, $h_0\propto \epsilon\fgw^2$).
For example, no known pulsar with an ellipticity below $10^{-6}$ and that is accessible to advanced detectors  has a $\fgw$ value
lower than 70 Hz, and no ellipticity below $10^{-8}$ is accessible to advanced detectors below 300 Hz.

\begin{figure}[t!]
\begin{center}
\includegraphics[width=13.cm]{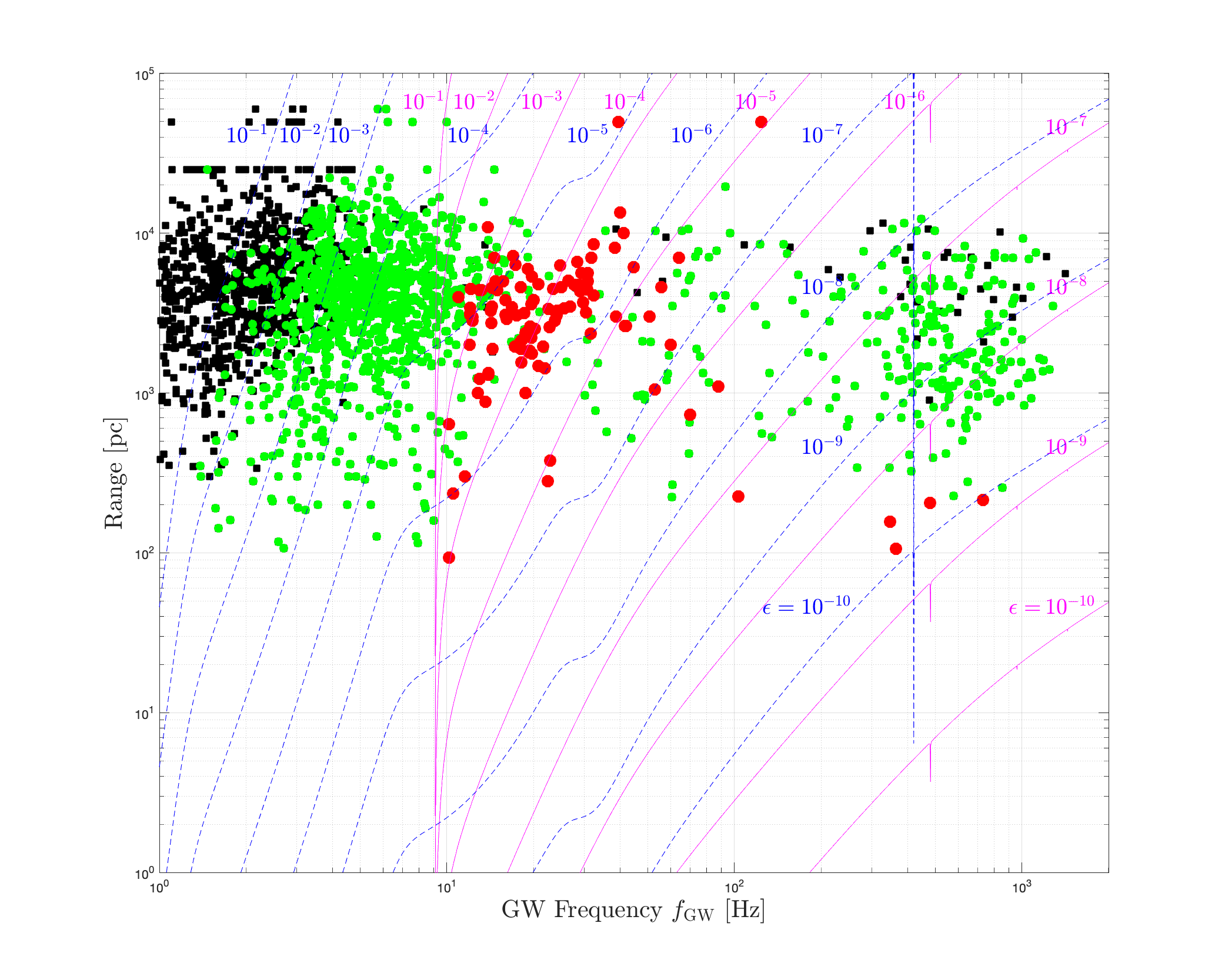}
\caption{Maximum allowed targeted-search ranges for gravitars vs GW frequencies for different assumed
  ellipticities for advanced detector sensitivity (solid magenta curves) and corresponding ranges for ETC sensitivity (dashed blue curves). Known pulsar
  distances are shown vs the expected GW frequencies, where red dots indicate pulsars with accessible spin-down limits for
  advanced detector sensitivity, and smaller green dots indicated pulsars with accessible spin-down limits for ETC sensitivity.
  Known pulsars in distinct horizontal bands (common distance) arise from stars in clusters or from distance capping
  in the galactic electron density model~\citep{bib:YMWModel} used in the ATNF catalog~\citep{bib:ATNFdb}.}
\label{fig:fvsdist1}
\end{center}
\end{figure}

Another figure of merit is the distance to which searches can detect sources of a particular ellipticity.
Figure~\ref{fig:fvsdist1} shows the estimated distances to known pulsars over the detection frequency band.
Also shown are solid contours of advanced detector sensitivity range for different ellipticity values and dashed contours
for Einstein Telescope. Pulsars with spin-down limits accessible to advanced detectors are shown in red,
and those accessible to Einstein Telescope are shown in green. Only a handful of pulsars within 500 pc are
accessible to advanced detectors with ellipticities below 10$^{-8}$.
On the other hand, to reach the galactic center ($\sim$8.5 kpc) at a
signal frequency of 1 kHz requires an ellipticity larger than $\sim$\sci{3}{-8}, and at 100 Hz requires
an ellipticity greater than $\sim$\sci{3}{-6}.

\begin{figure}[t!]
\begin{center}
\includegraphics[width=13.cm]{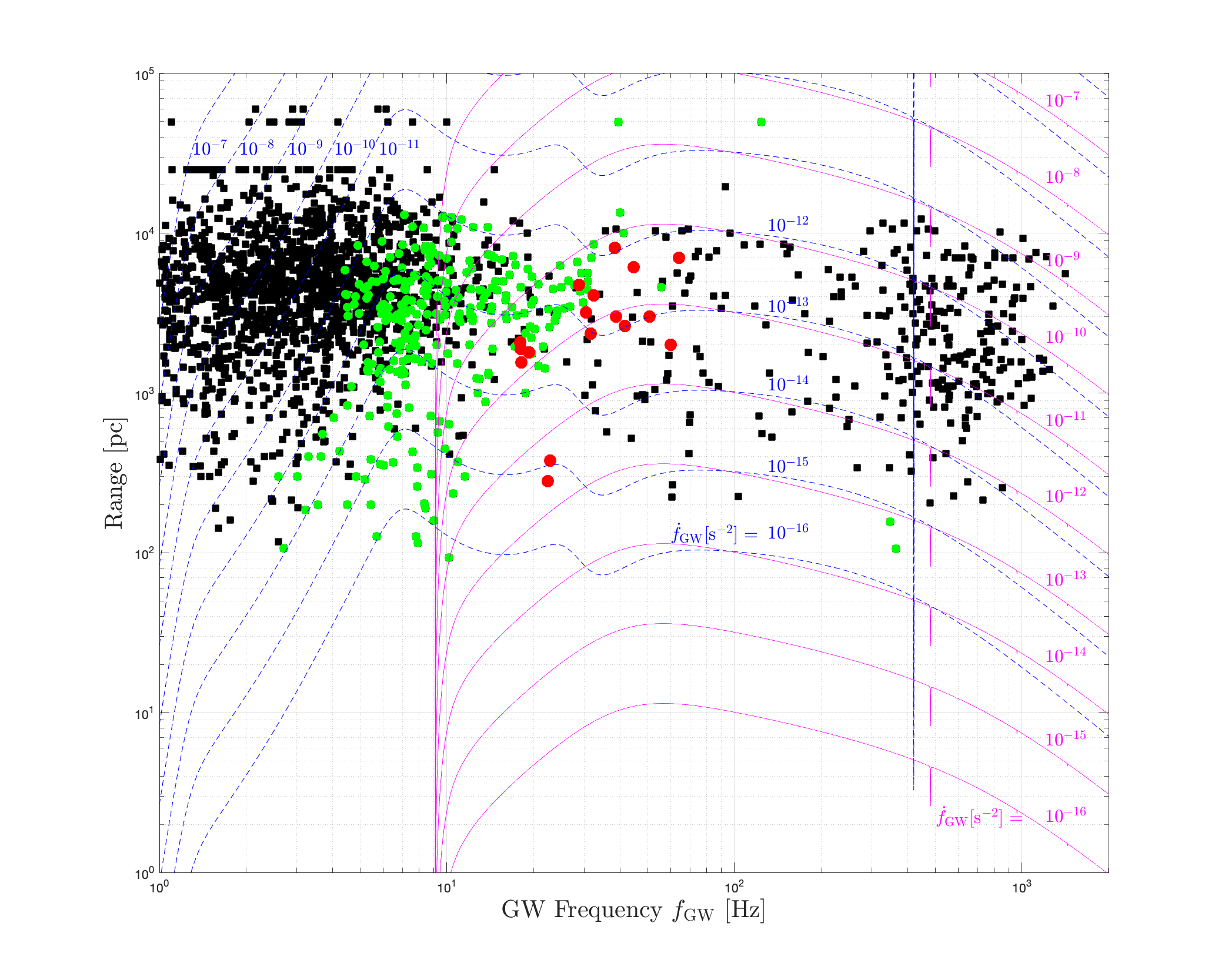}
\caption{Maximum allowed (optimistic) all-sky-search ranges for gravitars vs GW frequencies for different assumed
  spin-down derivatives  for advanced detector sensitivity (solid magenta curves) and corresponding ranges for ETC sensitivity (dashed green curves). Known pulsar
  distances are shown vs the expected GW frequencies, where red dots indicate pulsars with accessible spin-down limits for
  advanced detector sensitivity, and smaller green dots indicated pulsars with accessible spin-down limits for ETC sensitivity.
  These all-sky search ranges assume an optimistic sensitivity depth of 50 Hz$^{-1/2}$ (see section~\ref{sec:depth}).}
\label{fig:fvsdist2}
\end{center}
\end{figure}

As discussed in detail below, all-sky searches for unknown neutron stars necessarily have reduced sensitivity, such that
the ranges shown for targeted searches using known pulsar timing do not apply. Figure~\ref{fig:fvsdist2} shows another
range vs frequency plot, but for which (optimistic) advanced detector and Einstein Telescope all-sky sensitivities are assumed.
For reference, the all-sky strain sensitivity is taken to be about 20 times worse than its targeted-search sensitivity for
advanced detector and the corresponding ratio about 40 times worse for Einstein Telescope\footnote{Because targeted-search strain sensitivity improves
as the square root of time, while all-sky search sensitivity improves, at best, as only the fourth root of time (see
section~\ref{sec:challenges}), the disparity between targeted and all-sky sensitivity increases for longer observing times.}.
Consequently, the all-sky range contours
corresponding to those in Figure~\ref{fig:fvsdist1} would be reduced by the same ratios. Alternatively, to obtain the same ranges
in the all-sky search would require ellipticities higher by the same ratios.
The all-sky ranges in Figure~\ref{fig:fvsdist2}, in contrast, are shown as contours for different assumed $\fgwdot$ values under the gravitar assumption. These contours
are useful in assessing all-sky searches, since those searches are defined, in part, by their maximum spin-down range,
which affects computational cost. Once again, known pulsars for which this search technique can reach the spin-down limit
are shown in red for advanced detector and in green for
Einstein Telescope. We see that for the advanced detectors to reach the galactic center 
at a signal frequency of 1 kHz requires a minimum spin-down magnitude greater than $10^{-9}$ Hz/s (minimum because another mechanism,
such as magnetic dipole emission, may contribute to a higher spin-down magnitude), and at 100 Hz requires a minimum spin-down
magnitude just less than $10^{-10}$ Hz/s. The corresponding required ellipticities at those frequencies are $\sim$$\scimm{8}{-5}$ and
$\sim$$\scimm{8}{-7}$, respectively.

A simple steady-state argument by Blandford~\citep{bib:Thorne300} led to 
an early estimate of the maximum detectable strain amplitude expected from a population of
isolated gravitars of a few times 10$^{-24}$, independent of typical ellipticity values,
in the optimistic scenario that most neutron stars become gravitars. 
A later detailed numerical simulation~\citep{bib:knispelallen}
revealed, however, that the steady-state assumption does not generally hold
for mass quadrupole radiation, leading
to ellipticity-dependent expected maximum amplitudes that can be 2-3 orders
of magnitude lower in the LIGO/Virgo/KAGRA band for ellipticities as low as 10$^{-9}$ and a few
times lower for ellipticity of about $10^{-6}$. Mass current quadrupole (\rmode) emission, however,
would spin stars down faster, leading back to more optimistic maximum amplitudes~\citep{bib:Owenalpha}.
A more detailed simulation including both electromagnetic and gravitational wave spin-down
demonstrated the potential for setting joint constraints on natal neutron star magnetic fields
and ellipticities~\citep{bib:Wadeetal}.
A recent population simulation study~\citep{bib:ReedEtal} estimated fractions of neutron stars probed by
previous CW searches for different assumed ellipticities and concluded that the greatest potential gain
from improving detector sensitity 
in accessing more neutron stars of plausible ellipticity comes at higher frequencies.

The spin-down limit on strain defined in Eqn.~\ref{eqn:spindownlimit} for known pulsars requires knowing the frequency $\fgw$, its
first derivative $\fgwdot$ and the distance $d$ to the star. There are other neutron stars for which no pulsations are observed, hence for
which neither $\fgw$ nor $\fgwdot$ is known, but for which the distance and the age of the star are known with some precision. For
such stars one can define an ``age-based'' limit -- under the assumption of gravitar behavior since the neutron star's birth in
a supernova event. 
Using Eqn.~\ref{eqn:approxageindex} and a braking index of 5 for mass quadrupole radiation gives the
gravitar age:
\begin{equation}
  \tau_{\rm gravitar} = -{\frot\over 4\,\dot\frot}\>.
\end{equation}

Therefore, if one knows the distance and the age of the star, \eg, from
the expansion rate of its visible nebula, then
under the assumption that the star has been
losing rotational energy since birth primarily
due to gravitational wave emission, then one
has the following {\it frequency-independent}
age-based limit on strain~\citep{bib:cwcasamethod}:
\begin{equation}
h_{\rm age} =  (\scimm{2.3}{-24})\left({1\>{\rm kpc}\over r}\right)\sqrt{\left({1000\>{\rm yr}\over\tau}\right)
\left({\Izz\over I_0}\right)}\>,
  \label{eqn:agebasedlimit}
\end{equation}
\noindent along with a corresponding frequency-dependent but distance-independent ellipticity upper limit~\citep{bib:cwcasamethod}:
\begin{equation}
  \epsilon_{\rm age} = (\scimm{2.2}{-4}) \left({100\>{\rm Hz}\over\fgw}\right)^2 \sqrt{ \left({1000\>{\rm yr}\over\tau}\right) \left({I_0\over\Izz}\right)}.
\end{equation}

The corresponding calculation for \rmode\ emission leads to the age-based strain limit relation~\citep{bib:Owenalpha}:
\begin{equation}
h_{\rm age}^{r{\rm-mode}} =  (\scimm{1.9}{-24})\left({1\>{\rm kpc}\over r}\right)\sqrt{\left({1000\>{\rm yr}\over\tau}\right)
  \left({\Izz\over I_0}\right)}\>,
\label{eqn:agebasedlimitrmode}
\end{equation}
\noindent along with a corresponding frequency-dependent but distance-independent \rmode\ amplitude upper limit~\citep{bib:cwcasamethod}:
\begin{equation}
  \alpha_{\rm age} = 0.076 \left({1000\>{\rm yr}\over\tau}\right)^{1/2} \left({100\>{\rm Hz}\over\fgw}\right)^2 .
\end{equation}

Yet another empirically determined strain upper limit can be defined
for accreting neutron stars in binary systems,
such as Scorpius X-1. The X-ray luminosity from
the accretion is a measure of mass accumulation rate at
the surface. As the material rains down on the surface
it can add angular momentum to the star, which in 
equilibrium may be radiated away in gravitational waves.
Hence one can derive a torque-balance limit~\citep{bib:wagoner,bib:papapringle,bib:rmodes2} in
the form~\citep{bib:WattsEtal}:
\begin{eqnarray}
\label{eqn:torquebalance1}
h_{\rm torque} & \sim & (\scimm{3}{-27})
\left({R\over\mathrm{10\>km}}\right)^{3\over4}
\left({\msolar\over M}\right)^{1\over4} \nonumber \\
& & \times \left({1000\>{\rm Hz}\over\frot}\right)^{1\over2}
\left({\mathcal{F}_{\rm x}\over10^{-8}\>{\rm erg/cm}^2/{\rm s}}\right)^{1\over2},
\end{eqnarray}
where $\mathcal{F}_{\rm x}$ is the observed energy flux at the Earth of
X-rays from accretion, $M$ is the neutron star mass and $R$ its radius.
Taking nominal values of $R$ = 10 km, $M = 1.4 \msolar$ and reformulating
in terms of the gravitational wave frequency $\fgw$ (benchmarked to 600 Hz), one obtains:
\begin{equation}
\label{eqn:torquebalance2}
h_{\rm torque}\quad \sim \quad(\scimm{5}{-27})
\left({600\>{\rm Hz}\over\fgw}\right)^{1\over2}
\left({\mathcal{F}_{\rm x}\over10^{-8}\>{\rm erg/cm}^2/{\rm s}}\right)^{1\over2}.
\end{equation}

Equations~\ref{eqn:torquebalance1}-\ref{eqn:torquebalance2} assume the radius at which the
accretion torque is applied is the stellar surface. If one assumes the torque lever arm is
the \alfven\ radius because of the coupling between the stellar rotation and the magnetosphere,
then the implied equilibrium strain is $\sim$2.4 times higher~\citep{bib:cwdirectedO2ScoX1Viterbi}.
This limit is independent of the distance to the star. In general, variations in
accretion inferred from X-ray flux fluctuations suggest similar (slower) fluctuations
in the equilibrium frequency, which could degrade GW detection sensitivity for coherent
searches that assume exact equilibrium. A first attempt to address these potential
frequency fluctuations for Scorpius X-1 may be found in~\citep{bib:spinwandering}.
See~\cite{bib:SerimSerimBaykalAccretingPulsarTiming} for a recent compilation of
timing fluctuations of seven accretion-powered pulsars, providing evidence that accretion fluctuations
indeed dominate timing noise. 

\subsubsection{Assessing potential sources of neutron star non-axisymmetry}
\label{sec:nonaxisymmetry}

From the above, it is clearly {\it possible} for neutron stars in our galaxy to produce
continuous gravitational waves detectable by current ground-based detectors, but
is it {\it likely} that putative emission mechanisms are strong enough to give us
a detection in the next few years. Let's look more critically at those mechanisms~\footnote{This section
  draws, in part, from a recent review~\citep{bib:GandG} of gravitational wave emission from neutron stars.}.

Isolated neutron stars may exhibit intrinsic non-axisymmetry from residual crustal
deformation (\eg, from ``starquakes'' due to cooling \&\ cracking of the crust~\citep{bib:crustdeformation,bib:KerinMelatosCrustalFailure}
or due to changing centrifugal stress induced by stellar spin-down~\citep{bib:Ruderman,bib:BaymEtal,bib:FattoyevHorowitzLu,bib:GilibertiCambiotti}), from
non-axisymmetric distribution of magnetic field energy trapped
beneath the crust~\citep{bib:ZimmermannBuriedBfield,bib:CutlerToroidalBFields} or from a pinned neutron superfluid component
in the star's interior~\citep{bib:JonesSuperfluid,bib:MelatosEtalSuperfluid,bib:HaskellEtalPinnedSuperfluid}.
See~\citep{bib:HaskellEtal,bib:SinghHaskellMukherjeeBulik} for a discussion of emission from
magnetic and thermal ``mountains'' and~\citep{bib:Laskyreview,bib:GandG} for recent, comprehensive reviews of
GW emission mechanisms from neutron stars.

Maximum allowed asymmetries depend on the
neutron star equation of state~\citep{bib:JohnsonMcdanielOwen,bib:KrastevLi} and on the breaking strain of
the crust. Detailed molecular dynamics simulations borrowed from condensed matter theory have suggested in
recent years that the breaking strain may be an order of magnitude higher than previously thought
feasible~\citep{bib:HorowitzKadau,bib:CaplanHorowitzPasta}. Analytic treatments~\citep{bib:BaikoChugunov}
indicate, however, that anisotropy may be important and caution that simulations based on relatively
small numbers of nuclei may not capture effects due to a polycrystalline structure in the crust.
A recent cellular automaton-based simulation~\citep{bib:KerinMelatosCrustalFailure} of a spinning-down
neutron star used
nearest-neighbour tectonic interactions involving strain redistribution and thermal dissipation.
That study found the resulting annealing led to emitted gravitational strain amplitudes too low to be detected by present-generation
detectors. 

A recent revisiting of the mountain-building scenario~\citep{bib:GittinsAnderssonJones} finds
systematically lower ellipticities to be realistic.
It is argued in~\citep{bib:WoanEtalMSP} that a possible minimum ellipticity in millisecond pulsars
may arise from asymmetries of buried internal magnetic field $B_i$~\citep{bib:CutlerToroidalBFields,bib:LanderEtal2011,bib:Lander2014}
of order of~\citep{bib:WoanEtalMSP}
\begin{equation}
  \epsilon \sim 10^{-8}\>\left({<\!B_i\!>\over10^{12}\>\mathrm{G}}\right) \left({<\!H_c\!>\over10^{16}\>\mathrm{G}}\right),
\end{equation}
\noindent where $H_c$ is the lower critical field for superconductivity (protons in the stellar core are assumed
to form a Type II superconductor). Hence, a buried toroidal (equatorial) field of
$\sim$10$^{11}$ G could yield an ellipticity at the 10$^{-9}$ level. It has been argued, on the other hand, that
an explicit model of braking dynamics with non-axisymmetry due to magnetic field non-axisymmetry
leads to still smaller ellipticities,
based on observed braking indices of younger pulsars~\citep{bib:DeAraujoCoelhoCosta1,bib:DeAraujoCoelhoCosta2},
where the magnetic contribution to the ellipticity depends quadratically on the field
strength~\citep{bib:BonazzolaGourgoulhon,bib:KonnoObataKojima,bib:RegimbauDeFreitasPacheco}.
An analysis~\citep{bib:OsborneJones} of internal magnetic field contributions to non-axisymmetric temperature distributions
in the neutron star crust finds that high field strengths ($>10^{13}$ G) are needed in an accreting system for
GW emission to halt spin-up from the accretion, four orders of magnitude higher than is
expected for surface fields in LMXBs. A follow-up study~\citep{bib:HutchinsJones} finds more
optimistically large thermal asymmetries to be possible deeper in a star,
and another study~\citep{bib:MoralesHorowitz} finds a maximum allowed ellipticity of $\sim$$\scimm{7.4}{-6}$.

\rmodes\ (mass current quadrupole, see section~\ref{sec:spindown}) offer an intriguing alternative GW emission source~\citep{bib:MytidisEtalRmode}.
Serious concerns have been raised,~\citep{bib:rmodesdoubts,bib:GandG} however, about the 
detectability of the emitted radiation for young 
isolated neutron stars, for which mode saturation appears to occur at low \rmode\ amplitudes because
of various dissipative effects~\citep{bib:Owenalpha}.
Another study,~\citep{bib:alfordschwenzeryoungpulsar} though,
is more optimistic about newborn neutron stars. The same authors, on the other hand, find that \rmode\ emission
from millisecond pulsars is likely to be undetectable by advanced detectors~\citep{bib:alfordschwenzerMSP}.

The notion of a runaway rotational instability was
first appreciated for high-frequency $f$-modes,~\citep{bib:cfs1,bib:cfs2} (Chandrasekhar-Friedman-Schutz
instability), but realistic viscosity effects seem likely to 
suppress the effect in conventional neutron star production~\citep{bib:cfskiller1,bib:cfskiller2}.
Moreover, \citep{bib:HoEtalJ0952} set limits on the \rmodes\ amplitude $\alpha$ for J0952$-$0607 below $10^{-9}$ based
on the absence of heating observed in its X-ray spectrum, despite its high rotation frequency (707 Hz) which
places it in the nominal \rmodes\ instability window.
Similarly,~\citep{bib:BoztepeGogusGuverSchwenzer} set limits on $\alpha$ as low
as $\scimm{3}{-9}$, based on observations of two other millisecond pulsars
(PSR J1810$+$1744 and PSR J2241$-$5236) which also sit in the instability window.
Another potential source of \rmodes\ dissipation is from the interaction of
``ordinary'' and superfluid modes, leading to a stabilization window for
LMXB stars~\citep{bib:GusakovChugunovKantor,bib:KantorGusakovDommes}.
The $f$-mode stability could play an important role, however, for
a supramassive neutron star formed as the remnant
of a binary neutron star merger~\citep{bib:donevaetal} (spinning too fast to collapse immediately
despite exceeding the nominal maximum allowed neutron star mass).

In addition, as discussed below, a binary neutron star may experience direct non-axisymmetry
from non-isotropic accretion~\citep{bib:OwenElastic,bib:UshomirskyEtal,bib:MelatosPayne} (also possible for an isolated 
young neutron star that has experienced fallback accretion shortly
after birth), or may exhibit \rmodes\ induced by accretion spin-up.

Given the various potential mechanisms for generating continuous gravitational waves
from a spinning neutron star, detection of the waves should yield valuable information
on neutron star structure and on the equation of state of nuclear matter 
at extreme pressures, especially when combined with electromagnetic observations
of the same star.

The notion of gravitational wave torque equilibrium is potentially important,
given that the maximum observed rotation frequency of neutron
stars in LMXBs is substantially lower than one might expect from
calculations of neutron star breakup rotation speeds ($\sim$1400 Hz)~\citep{bib:breakupspeed}.
It has been suggested~\citep{bib:speedlimit} that there is a ``speed limit''
due to gravitational wave emission that governs the maximum
rotation rate of an accreting star. In principle, the distribution
of frequencies could have a quite sharp upper frequency cutoff,
since the angular momentum emission is proportional to the 
5th power of the frequency for mass quadrupole radiation. For example, for 
an equilibrium frequency corresponding to a particular accretion rate,
doubling the accretion rate would increase the equilibrium frequency
by only about 15\%. For \rmode\ GW emission, with a braking index of 7, the cutoff
would be still sharper.

Note, however, that a non-GW speed limit may well arise
from interaction between the neutron star's magnetosphere
and an accretion disk~\citep{bib:ghoshlamb,bib:HaskellPatruno,bib:PatrunoHaskellDangelo}.
It has also been argued~\citep{bib:ErtanAlpar} that correlation between the accretion rate
and the frozen surface dipole magnetic field resulting from Ohmic diffusion through the neutron star crust
in the initial stages of accretion in low mass X-ray binaries can explain a minimum rotation period
well above the naive expectation.

A number of mechanisms have been proposed by which the accretion
leads to gravitational wave emission in binary systems. The simplest is localized accumulation
of matter, \eg, at the magnetic poles (assumed offset from the rotation axis), 
leading to a non-axisymmetry.
One must remember, however, that matter can and will diffuse into
the crust under the star's enormous gravitational field. This diffusion of
charged matter can be slowed by the also-enormous magnetic fields in
the crust, but detailed calculations~\citep{bib:vigeliusmelatos} indicate the
slowing is not dramatic. Relaxation via thermal conduction is considered in \citep{bib:SuvorovMelatos}.

Another proposed mechanism is excitation of
\rmodes\ in the fluid interior of the star,~\citep{bib:rmodes1,bib:rmodes2,bib:rmodes3,bib:rmodes4}
with both steady-state emission and cyclic spin-up/spin-down 
possible~\citep{bib:rmodeslmxb,bib:HeylLMXB,bib:rmodesdoubts}. Intriguing,
sharp lines consistent with expected \rmode\ frequencies were reported 
in the accreting millisecond X-ray pulsar XTE J1751$-$305~\citep{bib:strohmayermahmoodifar1}
and in a thermonuclear burst of neutron star 4U 1636$-$536~\citep{bib:strohmayermahmoodifar2}.
The inconsistency of the observed stellar spin-downs for these sources with ordinary \rmode\
emission, however, suggests that a different type of oscillation is being observed~\citep{bib:AnderssonJonesHo}
or that the putative r-modes are restricted to the neutron star crust and hence gravitationally
much weaker than core r-modes~\citep{bib:lee2014}. Another recent study~\citep{bib:PatrunoHaskellAndersson}
suggests that spin frequencies observed in accreting LMXB's are consistent with two sub-populations, where
the narrow higher-frequency component ($\sim$575 Hz with standard deviation of $\sim$30 Hz) may signal an equilibrium
driven by gravitational wave emission.
It has been suggested~\citep{bib:HaskellPatrunoJ1023} that the transitional
millisecond pulsar PSR J1023$+$0038 (for which spin-down has been measured in both accreting and non-accreting states)
shows evidence for mountain building (or \rmodes) during the accretion state, based on different spin-downs observed
in accreting \vs\ non-accreting states.
It has also been argued~\citep{bib:Bhattacharyya} that
J1023$+$0038 shows evidence for a permanent ellipticity in the range $\scimm{0.48-0.93}{-9}$.
An analysis~\citep{bib:Chen} of
three transitional millisecond pulsars and ten redbacks concluded their ellipticities
ranged over $\scimm{0.9-23.4}{-9}$.

A recent analysis~\citep{bib:DeLilloEtal} based on the absence of evidence of a stochastic gravitational wave background
emitted by a population of neutron stars with a rotational frequency distribution similar to that of known pulsars inferred that
the average ellipticity of the galactic population is less than $\sim$$\scimm{2}{-8}$.

\subsubsection{Particular GW targets}
\label{sec:targets}

In the following, particular neutron star targets for gravitational wave searches are discussed in the
following categories: known young pulsars with high spin-down rates;
known high-frequency millisecond pulsars; neutron stars in supernova remnants,
neutron stars in low-mass X-ray binary systems;
and particular directions on the sky.

\paragraph{Known young pulsars with high spin-down rates}\leavevmode\\

A young pulsar with a high spin-down rate presents an attractive target. Its age
offers the hope of a star not yet annealed into smooth axisymmetry, a hope
strengthened by the prevalence of observed timing glitches among young stars.
A high spin-down rate not only makes it more likely that the spin-down limit
is accessible, but also suggests a star with a reservoir of magnetic energy,
some of which could give rise to non-axisymmetry. From the Advanced LIGO / Virgo
O1, O2 and O3 data sets more than 20 pulsars were spin-down accessible~\citep{bib:cwtargetedO2,bib:cwtargetedO3}
(see section~\ref{sec:targeted}), but most correspond to ellipticities
of $\sim$$10^{-4}$--$10^{-3}$. A small number are highlighted here, for which
ellipticities below $10^{-5}$ are accessible already or with a 2-year data run
at advanced detector design sensitivity (``O4/O5 run'').

\begin{itemize}
\item {\bf Crab (PSR J0534$+$2200)} -- This pulsar, created in a 1054 A.D. supernova
  observed by Chinese astronomers and discovered in 1968~\citep{bib:staelinreifenstein}, has received more attention from LIGO / Virgo analysts than any other.
  Its spin-down limit was first beaten in the initial LIGO data set S5~\citep{bib:cwtargetedcrabS5}, and now has been beaten (O3 data) by a factor
  of $\sim$100~\citep{bib:cwtargetedO3} (see section~\ref{sec:targeted}), leading to a 95\%\ upper limit on ellipticity of \sci{1.0}{-5}.
  For a 2-year O4/O5 run, this sensitivity reaches $\sim$$\scimm{2}{-6}$. Spinning at just below 30 Hz, its nominal $\fgw$ is just below 60 Hz, making
  the detector spectrum susceptible to power mains contamination (including non-linear upconversion, see section~\ref{sec:lines}) in the LIGO and KAGRA interferometers, but not in the Virgo interferometer, which uses 50 Hz power mains.
  Its inferred rotational kinetic energy loss rate based on its spin-down is $dE/dt \sim \scimm{-5}{38}$ erg s$^{-1}$, assuming the
  nominal $\Izz = 10^{38}$~kg~m$^2$ (10$^{45}$~g~cm$^{2}$).
\item {\bf Vela (PSR J0835$-$4510)} -- Although older and lower in frequency than the Crab with a higher ellipticity spin-down limit (\sci{1.9}{-3}),
  the Vela pulsar, discovered in 1968~\citep{bib:largeetal}, is nonetheless interesting, given its frequent glitches~\citep{bib:velaglitches,bib:AshtonLaskyGraberPalfreyman}.
  Its O4/O5 ellipticity sensitivity reaches $~\sim$$\scimm{8}{-6}$. Spinning at just above 11 Hz, its nominal $\fgw$ is about 22 Hz, where detector
  noise is several times higher than at the Crab frequency.   Its inferred $dE/dt \sim \scimm{-7}{36}$ erg s$^{-1}$.
\item {\bf PSR J0537$-$6910} -- This pulsar, observed to pulse only in X-rays, is distant ($\sim$50 kpc in the Large Magellanic Cloud).
  With a rotation frequency of $\sim$62 Hz, its nominal GW frequency of 124 Hz
  is quite high for a young pulsar (magnetic dipole spin-down age $\sim$ 5000 yr), and its spin-down energy loss is comparable to the Crab's.
  It is also extremely glitchy ($\sim$1 per 100 days)~\citep{bib:Antonopoulouetal,bib:FerdmanEtal} and as noted above, may show evidence
  of \rmode\ emission between glitches~\citep{bib:rmodeJ0537-6910,bib:HoEtalJ0537} (which would imply a GW frequency at $\sim$90 Hz).  Its inferred $dE/dt \sim \scimm{-5}{38}$ erg s$^{-1}$.
\item {\bf PSR J1400$-$6325} -- This relatively recently discovered X-ray pulsar~\citep{bib:RenaudEtal} lies in a supernova remnant 7-10 kpc away and 
  displays a spin-down energy about 1/10 of the Crab pulsar's, but may be younger than 1000 yr. With a spin frequency of
  $\sim$32 Hz, its nominal $\fgw$ is 64 Hz, comparable to the Crab's, but farther from the troublesome 60 Hz power mains.
  Its inferred $dE/dt \sim \scimm{-5}{37}$ erg s$^{-1}$.
\item {\bf PSR J1813$-$1749} -- First detected as a TeV $\gamma$-ray source~\citep{bib:HESSJ1813}, this star was found to
  exhibit non-thermal X-ray emission and to have a tentative association with a radio supernova remnant G12.8-0.0~\citep{bib:BroganEtalJ1813} suggesting
  a distance greater than 4 kpc and an age perhaps younger than 1000 years. X-ray pulsations detected still later
  with a period of 44 ms confirmed a pulsar source and posited an association with a young star cluster at 4.7 kpc~\citep{bib:GotthelfHalpern},
  while yielding a nominal pulsar spin-down age of 3.3-7.5 kyr.
  A more recent detection of highly dispersed radio pulsations, however, suggest a distance of 6 or 12 kpc~\citep{bib:CamiloEtalJ1813},
  depending on electron dispersion model, casting doubt on the association with the star cluster.
  The spin frequency of 22 Hz yields a nominal GW
  frequency of $\sim$45 Hz, and the frequency derivative imply $dE/dt \sim \scimm{-6}{37}$ erg s$^{-1}$.
\end{itemize}

\paragraph{Known high-frequency millisecond pulsars}\leavevmode\\

Because nearly all millisecond pulsars are old, with some characteristic ages greater than 10 billion years,
they can be assumed to retain little asymmetry from their initial formation or from the accretion that
spun them up. Thus one sees low spin-down for this population in Fig.~\ref{fig:fvshspindown}
and hence low inferred maximum ellipticities in Fig.~\ref{fig:fvsepsilon}. On the other hand, the vast energy reservoirs in
their rotation and the quadratic dependence of $h_0$ on frequency still makes these stars potentially intriguing.
As noted above, there may be empirical evidence for a minimum ellipticity of order $\sim$$10^{-9}$~\citep{bib:WoanEtalMSP}.
Highlighted below are particular millisecond pulsars of interest in the coming years.

\begin{itemize}
\item {\bf PSR J0711$-$6830} This isolated star at a distance of 0.11 kpc, with a nominal  $\fgw \sim$ 364 Hz, a spin-down upper limit of \sci{1.2}{-26} and
  corresponding maximum ellipticity of \sci{9.4}{-9}, is the first MSP to have its spin-down limit beaten (in early O3 data, see section~\ref{sec:targeted}).
\item {\bf PSR J0437$-$4715} -- This binary star at a distance of 0.16 kpc, with a nominal $\fgw \sim$ 347 Hz, a spin-down upper limit of \sci{7.8}{-27} and
  corresponding maximum ellipticity of \sci{9.7}{-9} also had its spin-down limit beaten (in the full O3 data).
\item {\bf PSR J1737$-$0811} This binary star at a distance of 0.21 kpc, with a nominal  $\fgw \sim$ 479 Hz, a spin-down upper limit of \sci{5.3}{-27} and
  corresponding maximum ellipticity of \sci{4.6}{-9}, will likely have its spin-down limit beaten by the O4/O5 data set.
\item {\bf PSR J1231$-$1411} This binary star at a distance of 0.42 kpc, with a nominal  $\fgw \sim$ 543 Hz, a spin-down upper limit of \sci{2.8}{-27} and
  corresponding maximum ellipticity of \sci{3.8}{-9}, will likely have its spin-down limit beaten by the O4/O5 data set.
\item {\bf PSR J2124$-$3358} This binary star at a distance of $\sim$0.4 kpc, with a nominal  $\fgw \sim$ 406 Hz, a spin-down upper limit of \sci{2.3}{-27} and
  corresponding maximum ellipticity of \sci{5.6}{-9}, will likely have its spin-down limit beaten by the O4/O5 data set.
\item {\bf PSR J1643$-$1224} This binary star at a distance of 0.79 kpc\footnote{A recent parallax measurement~\citep{bib:ReardonEtalParkes},
though, finds a distance for J1643$-$1224 of 1.2$^{+0.4}_{-0.3}$ kpc.}, with a nominal  $\fgw \sim$ 433 Hz, a spin-down upper limit of \sci{2.1}{-27} and
  corresponding maximum ellipticity of \sci{8.0}{-9}, may {\it not} have its spin-down limit beaten by the O4/O5 data set, but as noted by~\citep{bib:WoanEtalMSP},
  would have the highest GW SNR of any known star if its ellipticity were $10^{-9}$.
\end{itemize}

\paragraph{Central compact objects and Fomalhaut b}\leavevmode\\

Not every neutron star of interest has been detected to pulsate. Central compact objects (CCOs) at the heart of
supernova remnants present especially intriguing targets, especially those in remnants inferred from their
size and expansion rate to be young~\citep{bib:DelucaCCOs}. There may be direct evidence of a neutron star, such as from thermal X-rays emitted
from a hot surface or from X-rays due to interstellar accretion, or there may be indirect evidence from a pulsar wind
nebula driven by a fast-spinning star at the core. Most  GW searches to date for a CCO lacking detected pulsations
have focused on the particularly promising source, Cassiopeia A, but in recent years, such searches have also been carried out for as
many as 15 supernova remnants~\citep{bib:cwdirectedSNRO1,bib:cwdirectedO3aSNRs}. Highlighted below are particular supernova remnants
(``G'' naming terminology based on the Green Catalog~\citep{bib:GreenSNRCatalog}, see also~\citep{bib:FerrandSafiHarbSNRCatalog}) with known
or suspected central compact objects, in addition to an object, Fomalhaut b, originally thought to
be an exoplanet, but which may be a nearby neutron star. Results from searches for these targets are presented further below
in section~\ref{sec:directedisolated}. 

  \begin{itemize}
  \item {\bf Cassiopeia A} -- Cas~A (G111.7$-$2.1) is perhaps the most promising example of gravitational wave CCO source in a supernova remnant.
    Its birth aftermath may have been observed by Flamsteed~\citep{bib:casabirth} $\sim$340 years ago in
    1680, and the expansion of the visible shell is consistent
    with that date~\citep{bib:FesenEtalCasA}. Hence Cas~A, which is visible in X-rays~\citep{bib:TananbaumCasA,bib:HoEtalCasA}
    but shows no pulsations~\citep{bib:HalpernGotthelfCasA}, is almost certainly a very young 
    neutron star at a distance of about 3.3 kpc~\citep{bib:ReedEtalCasA,bib:AlarieEtalCasA}. From Eqn.~\ref{eqn:agebasedlimit},
    one finds an age-based strain limit of $\sim$\sci{1.2}{-24}, which is readily accessible to
    LIGO and Virgo detectors in their most sensitive band.
  \item {\bf Vela~Jr.} -- This star (G266.2$-$1.2) is observed in
    X-rays~\citep{bib:PavlovEtalVelaJr,bib:KargaltsevEtalVelaJr,bib:BeckerEtalVelaJr} and is potentially
    quite close ($\sim$0.2 kpc) and young (690 yr)~\citep{bib:IyudinEtalVelaJr}, but searches have also conservatively
    assumed more a more pessimistic distance (0.9 kpc) and age (5100 yr), based on other measurements~\citep{bib:AllenEtalVelaJr}.
    The optimistic age and distance assumptions lead to an age-based strain limit of $\sim$\sci{1.4}{-23}, even more accessible than
    the Cas~A limit. Even the pessimistic age-base limit of \sci{1.1}{-24} is only slightly lower than that of Cas~A.
    It has been argued~\citep{bib:MingEtalOptimization} that a search over multiple CCOs, optimized for most likely detection
    success given fixed computing resources, favors focusing those resources on Vela~Jr. over other CCOs, including Cas~A.
  \item {\bf G347.3$-$0.5} -- An X-ray source~\citep{bib:SlaneEtal,bib:HoEtalCasA} is consistent with the core of this supernova remnant,
    the nearness ($\sim$0.9 kpc) and youth (1600 yr) of which
    make a search aimed at the remnant's center intriguing, as they yield an age-based strain limit of $\sim$\sci{2.0}{-24} -- 
    higher than that of Cas~A.
  \item {\bf G1.9$+$0.3} This supernova remnant, the youngest in the galaxy at 100 yr~\citep{bib:ReynoldsEtal}, has no detected
    CCO at its core, which is consistent with a Type IA supernova's having left no neutron star behind. Nonetheless,
    its youth make it interesting despite this doubt and its distance (8.5 kpc), yielding an age-based strain limit of $\sim$\sci{8.4}{-25}.
  \item {\bf Fomalhaut b} This object was assumed to be an extrasolar planet~\citep{bib:KalasEtal} until \citep{bib:NeuhauserEtalFomalhautb} noted
    that the absence of detected infrared radiation could indicate the object is a remarkably nearby neutron star ($\sim$0.01 kpc).
    The absence of attempted X-ray detection with {\it Chandra} observations~\citep{bib:PoppenhaegerEtal}, however, disfavors its being a young, hot neutron star.
    More recent evidence~\citep{bib:GasparRiekeFomalhautb} argues, in fact, that the optical observations point to a planetesimal collision.
  \end{itemize}
  
\paragraph{Neutron stars in low-mass X-ray binary systems}\leavevmode\\

Because of its high X-ray flux ($\mathcal{F}_{\rm x}\sim\scimm{3.9}{-7}$ erg cm$^{-2}$ s$^{-1}$~\citep{bib:WattsEtal}) and the torque-balance
relation for low-mass X-ray binaries [Eqn.~(\ref{eqn:torquebalance1})],
Scorpius X-1 is thought to be an especially promising search
target for advanced detectors and has been the subject
of multiple searches in initial and Advanced gravitational wave detector data. From Eqn.~(\ref{eqn:torquebalance2}),
one expects a strain amplitude limited by~\citep{bib:cwfstatS2,bib:ScoX1MDC1}
\begin{equation}
  h \quad \sim \quad (\scimm{3}{-26})\,\left({600\>{\rm Hz}\over\fgw}\right)^{1\over2}.
  \label{eqn:torquebalance3}
\end{equation}
While Sco X-1's rotation frequency
remains unknown~\citep{bib:GalaudageEtal}, its orbital period is well measured,~\citep{bib:scox1period,bib:WangEtalScoX1}
which allows substantial reduction in search space. 
A similar but less bright LMXB system is Cygnus X-2~\citep{bib:GallowayCygX2Source} at
a distance of 7 kpc and an average flux $\mathcal{F}_{\rm x} = \scimm{11}{-9}$ erg/cm$^2$ s$^{-1}$~\citep{bib:GallowayCygX2Flux},
yielding a torque-balance strain limit about 20 times lower than that of Sco X-1.
Unlike Sco X-1 which is assumed but not known to contain a
neutron star (as opposed to a black hole with an accretion disk),
Cyg X-2 has displayed thermonuclear bursts, confirming the presence of a neutron surface.

Another interesting class contains ``accreting X-ray millisecond pulsars'' (AXMPs) which are fast-spinning neutron stars in LMXBs that
show sporadic outbursts during accretion episodes (when ``active'') from which rotation frequencies can be determined. When active,
the frequencies can increase or decrease, while frequencies between outbursts (when ``quiescent'') generally decrease.
One could hope to detect CW radiation from either active or quiescent phases. Although the limited durations of bursts
and their stochastic nature constrain potential search sensitivity, it is during such outbursts when one might expect
the largest generation of non-axisymmetries or excitation of \rmodes. The fastest-spinning stars, such as IGR J00291$+$5934
at $\frot\sim$ 599 Hz and a distance of $\sim$4 kpc~\citep{bib:IGRJ00291Outburst,bib:IGRJ00291Later}, offer deeper probing of equatorial
ellipticity and \rmode\ amplitude. Current search sensitivities to strain amplitude~\citep{bib:cwAXMPO3}
remain an order of magnitude or more away from inferred spin-down limits ($\sim$10$^{-28}$-10$^{-27}$), but improvements in detector sensitivity,
search methodology and potential future electromagnetic observations make this type of source potentially intriguing in the coming years.

\paragraph{Particular sky directions}\leavevmode\\

In addition to known (or suspected) neutron stars, there are other localized sky regions or points where
a directed search might yield a continuous gravitational wave detection. Listed below are possibilities that
have attracted attention in recent years.

\begin{itemize}

\item {\bf Galactic center} -- The vicinity of the galactic center (Sgr A*) is particularly interesting~\citep{bib:cwdirectedgalacticcenterS5}, as
  an active, star-forming region with known pulsars~\citep{bib:DenevalEtal}. Moreover, it is highly likely that only a small fraction
  of pulsars near the galactic center have been detected to date, since there is extreme dispersion and scattering of radio
  signals along the propagation line to the Earth~\citep{bib:LazioCordes}. The inference of there being many hidden pulsars is supported by $\sim$20 pulsar wind nebula candidates
  detected within 20 pc of Sgr A*~\citep{bib:MunoEtal}. In addition, searches for dark-matter annihilation signals have detected an excess of
  high-energy gamma ray emission from the galactic center region above what is
  expected from conventional models of diffuse gamma-ray emission and catalogs of known gamma-ray sources~\citep{bib:AckermannEtal},
  a tension which may be resolved by the existence of a hidden population of millisecond pulsars~\citep{bib:Abazajian}.
  A systematic radio survey of the central 1 parsec of Sgr A* at a frequency of 15 GHz~\citep{bib:MacquartEtal}, high enough to reduce dispersion and
  scattering substantially, yielded no detections, but the rapidly falling spectrum of most pulsars makes detection at 15 GHz at that distance difficult.
  This survey obtained a 90\%\ CL upper limit of 90 on the number of pulsars within 1 parsec of Sgr A*, assuming the population there is similar
  to known pulsars. Unfortunately,
  the $\sim$8.5 kpc distance to the galactic center makes CW searches challenging with present detector sensitivities. Only stars with
  extreme ellipticities are accessible to advanced detectors at design sensitivity
  (see Figure~\ref{fig:fvsdist2}). At the same time, however, young neutron stars are those most likely
  to exhibit such ellipticities.
\item {\bf Globular cluster cores} -- One normally associates globular clusters with ancient stellar populations and might
  expect, at best, to see only pulsars that are themselves ancient -- recycled and well annealed millisecond pulsars.
  Indeed many MSPs are seen in globular clusters~\citep{bib:FreireReview}. For example, Tucanae 47 is known to host at least 25 MSPs~\citep{bib:FreireEtal}.
  Nonetheless, not all observed pulsars in globular clusters seem to be old~\citep{bib:FreireReview}. A plausible explanation is that
  the dense core of a globular cluster leads to multibody exchange interactions in which a previously recycled but decoupled neutron star
  acquires a close new companion that proceeds  to overflow its Roche lobe, leading to new accretion. Another, related mechanism is possible
  debris accretion triggered by multibody interactions, given that some pulsars are known to host debris disks and even planets~\citep{bib:S6NGC6544}.
  The well localized core of a globular cluster makes a deep, directed search tractable. 
\item {\bf High-latitude Fermi sources} The Fermi satellite's LAT experiment has detected $\sim$100 previously unknown gamma ray pulsars
  since observing began in 2008. Gamma ray pulsars tend to be sources with low variability and relatively low spectral cutoffs,
  and most lie near the galactic plane, as expected. Fermi-LAT point sources well outside the galactic plane tend to be extagalactic, \eg, active galactic nuclei,
  but an intriguing possiblity is that a source with high galactic latitude could be a galactic neutron star, in which case the high latitude favors a nearby source~\citep{bib:SandersThesis},
  consistent with a scale height of $\sim$600 pc with respect to the galactic plane observed for known pulsars~\citep{bib:LyneGrahamSmith}.
  Arguing against this possibility, however, are extensive searches for gamma-ray pulsations from pulsar-like Fermi-LAT sources
  (see, \eg., \citep{bib:fermipulsarsearchexample}), based in part on algorithms developed for CW gravitational wave searches~\citep{bib:fermifirsteathdetection}.
  On the other hand, such searches are challenged to probe binary sources with large accelerations, suggesting that CW searches directed at such sources include
  algorithms sensitive to binary sources~\citep{bib:NeunzertThesis}.
\end{itemize}

In between all-sky searches and directed searches for single sky points reside ``spotlight'' searches, in which
a patch of sky is searched more deeply than in all-sky searches (with increased computational cost), but less deeply
than is computationally feasible for a single sky location. Such spotlights have been applied in searches for
a broad star-forming region along two directions of the Orion spur of the local galactic spiral arm~\citep{bib:CWOrionSpur} and toward
the galactic center region, including the globular cluster Terzan 5~\citep{bib:AEIGCTerzan5}.

\subsection{Axion clouds bound to black holes}
\label{sec:axions}

An intriguing potential connection between gravitational waves and
the still-unknown missing dark matter of the Universe comes from the
possibility that the dark matter is composed of ultralight, electromagnetically invisible
bosons, such as axions. One novel idea is that these bosons
could be disproportionately found in the vicinity of rapidly spinning black
holes~\citep{bib:axiverseArvanitaki1,bib:axiverseArvanitaki2}.
The ultralight particles could, in principle, be spontaneously created via energy extraction from the
black hole's rotation~\citep{bib:penrose1,bib:penrose2} and form a Bose-Einstein ``cloud'' with
nearly all of the quanta occupying a relatively small number of energy levels. For a cloud bound to a black hole,
the approximate inverse-square law attraction outside the Schwarzchild radius ($r_{\rm Schwarz.}\equiv{2\,GM_{\rm BH}\over c^2}$)
leads to an
energy level spacing directly analogous to that of the hydrogen atom~\citep{bib:axiverseArvanitaki1,bib:BaumannChiaStoutTerHaar}. 
The number of quanta occupying the low-lying levels can be amplified enormously by the phenomenon of superradiance
in the vicinity of a rapidly spinning black hole (with angular momentum that is a signficant fraction of the
maximum value allowed in General Relativity). The bosons in a non-$s$ ($\ell>0$) negative-energy
state can be thought of as propagating
in a well formed between an $\ell$-dependent centrifugal barrier at $r>r_{\rm Schwarz.}$ 
and a potential rising toward zero as $r\rightarrow\infty$;
wave function penetration into the black hole ergosphere permits transfer of energy
from the black hole spin~\citep{bib:superradiance1,bib:superradiance2,bib:superradiance3} into the creation of new quanta.

Two particular gravitational wave emission modes of interest here can arise in the axion scenario, both
potentially leading to intense coherent radiation~\citep{bib:axionArvanitaki}.
In one mode, axions can annihilate with each other to produce gravitons with frequency double that
corresponding to the axion mass: $f_{\rm graviton} = 2m_{\rm axion}c^2/h$. In another mode,
emission occurs from level transitions of quanta in the cloud.
This Bose condensation is most pronounced when
the reduced Compton wavelength of the axion is comparable to but larger than the scale of the black hole's Schwarzchild radius:
\begin{eqnarray}
 \lambdabar \equiv {\lambda\over2\,\pi} & = & {\hbar\over m_{\rm axion}c} \gtrapprox {2\,GM_{\rm BH}\over c^2} \\
  \Rightarrow \qquad m_{\rm axion} & \lessapprox & (\scimm{7}{-11} {\>{\rm eV/c^2}}) {\msolar\over M_{\rm BH}} ,
\end{eqnarray}
\noindent where $\hbar$ is the reduced Planck constant and $G$ is Newton's gravitational constant.
A key parameter governing detectability
is a parameter analogous to the electromagnetic fine structure constant:
\begin{equation}
  \label{eqn:alphadef}
  \alpha \equiv {Gm_{\rm axion}M_{\rm BH}\over\hbar c},
\end{equation}
\noindent where both the growth rate of a cloud upon black hole formation and the amplitude of gravitational wave emission due
to axion annihilation depend on high powers of $\alpha$. Hence small $\alpha$ impedes detection; at the same time,
superradiance itself requires~\citep{bib:IsiEtalBosons}:
\begin{equation}
  \label{eqn:alphalimit}
  \alpha < {1\over2}m\chi\left(1+\sqrt{1-\chi^2}\right)^{-1} < {m\over2},
\end{equation}
\noindent where $\chi$ is the dimensionless black hole spin proportional to its total angular momentum magnitude $J$:
$\chi = {cJ\over GM_{\rm BH}^2}$, and $m$ is the quantum number corresponding to the axion's orbital angular momentum
projection along the spin axis of the black hole (the first level to be populated in a newborn black hole is $m=1$~\citep{bib:IsiEtalBosons}). Hence the range of $\alpha$ (and therefore axion mass) for which
a particular black hole produces superradiance may be narrow. In general, more massive black holes produce stronger
signals over wider ranges in axion mass. Clouds composed of ultralight vector or tensor bosons would lead to stronger, but
shorter-lived signals~\citep{bib:SiemonsenEast,bib:BritoEtalTensorBosons}.
Nominal limits on axion masses can be placed based on the existence of high-spin binary black holes
in our galaxy~\citep{bib:axionBBHmerger,bib:CardosoEtal}, but those limits are subject to uncertainties in inferred
black hole spins~\citep{bib:Reynolds,bib:McClintockEtal} and may be invalidated by tidal disruption effects from the
companion star~\citep{bib:CardosoDuqueIkeda}. Constraints have also been inferred from spin measurements in the population of binary black
hole merger detections~\citep{bib:NgEtalGWTC2BosonConstraints}.

Given the many orders of magnitude of
uncertainty in, for example, axion masses that could account for dark matter~\citep{bib:darkmatterreview},
the relatively narrow mass window accessible to currently feasible CW searches (1-2 orders of magnitude) makes
searching for such an emission a classic example of ``lamppost'' physics, where one can only
hope that nature places the axion in this lighted area of a vast parameter space. 

In principle, searching for these potential CW sources requires
no fundamental change in the search methods described below, but search optimization can be
refined for the potentially very slow (and positive) frequency evolution expected during annihilation emission
(as the relative magnitude of the axion field's binding energy decreases). In addition, for a known black hole
location, a directed search can achieve better sensitivity than an all-sky search.
For string axiverse models, however,
the axion cloud~\citep{bib:axiverseArvanitaki1,bib:axiverseArvanitaki2,bib:YoshinoKodama2014,bib:YoshinoKodama2015}
can experience significant self-interactions which can
lead to appreciable frequency evolution of the signal and to uncertainty in that evolution,
a complication less important for the postulated QCD axion~\citep{bib:axionArvanitaki}.
In an optimistic scenario with many galactic black holes producing individually detectable
signals,~\citep{bib:ZhuEtalBosonCWSignal} points out that the signals would all lie in a very narrow
band, complicating CW searches, which typically implicitly assume no more than one detectable signal
in narrow bands. A later study~\citep{bib:PieriniEtalSignalConfusion}, however, finds that 
semi-coherent searches can be robust with respect to potential signal confusion.

Until recently, most published searches have not been tailored for a black hole axion cloud source,
but instead existing (non-optimized) limits on neutron star CW emission could be reinterpreted as limits on
such emission~\citep{bib:axionArvanitaki,bib:FalconPaper,bib:PalombaEtalAxion}.
More recently, though, searches have been carried out that exploit the narrow spin-up parameter space
expected for such sources~\citep{bib:ViterbiCygX1,bib:cwallskyO3BosonCloud}.

One interesting suggestion includes the possibility that a black hole formed from the detected merger of binary black
holes or neutron stars could provide a natural target for follow-up CW searches~\citep{bib:axionBBHmerger,bib:GhoshEtalSuperradiance,bib:IsiEtalBosons}.
Recent studies~\citep{bib:brito1,bib:brito2,bib:TsukadaEtal2019,bib:TsukadaEtal2021} argue that 
the lack of detection of a stochastic gravitational radiation background from the superposition of
extragalactic black holes already places significant limits on axion masses relevant to CW searches.
Another recent study~\citep{bib:IsiEtalBosons} examined in detail the prospects for detecting superradiance from
both post-merger black hole remnants and known black holes in galactic X-ray binaries, such 9as Cygnus X-1.

\section{Continuous Wave Search Methods}
\label{sec:searches}

Being realistic, we must acknowledge that the first discovered CW signal will be exceedingly weak compared to
the transient signals detected to date, an assumption borne out by many unsuccessful CW searches to
date. One must integrate the signal over a long duration to observe it
with statistical significance. Those long integrations in noise that is instantaneously much higher in
amplitude require application of assumed signal templates to the data. In general, the more restrictive
is the model, the better is the achievable signal-to-noise ratio, as one can search over a smaller
volume of source parameter space. The following sections discuss the challenges faced in searching over
larger parameter space volumes, a common classification of general search methods, and 
specific algorithms devised to meet the challenges.

\subsection{Challenges in CW signal detection and types of searches}
\label{sec:challenges}

At first glance, it may seem puzzling that a signal due to a rapidly spinning neutron star is challenging to find.
One might expect a simple discrete Fourier transform of the data stream to reveal a sharp spike at the nominal frequency.
There are several severe complications, however, for most CW searches. For concreteness, imagine that a signal is
weak enough to require a coherent, phase-preserving 1-year integration time $\Tcoh$. The nominal frequency resolution from a
discrete Fourier transform (DFT\footnote{CW search literature frequently refers to SFTs, ``Short'' discrete Fourier transforms,
where short is relative to the span of an observing run, but which may correspond to coherence times as long as hours.})
is then 1/yr $\sim$ 30 nHz. In order for the signal's central frequency to remain in
the same DFT bin (integer index into the transform result, see Eqn.~\ref{eqn:DFTdefinition} below) during that year, its first derivative $\fdot$ would need to satisfy $\fdot\Tcoh \lesssim 1/\Tcoh$,
or $\fdot \lesssim 10^{-15}$ Hz/s and its second derivative $\fdotdot \lesssim \scimm{6}{-23}$ Hz/s$^2$.
In practice, not only are Doppler modulations of detected frequency due to the Earth's motion much larger than these values, as discussed below,
but the frequency derivative of a detectable source is typically also much larger, in order for its rotational
kinetic energy loss to be compatible with detection (detectable spin-down limit). 
If the precise frequency evolution of the source is known already from radio or gamma-ray pulsar timing (assuming a fixed EM/GW phase relation),
then one can make corrections for that evolution via {\it barycentering}, discussed below, without SNR degradation as long
as the uncertainties in frequency derivatives are well below the above constraints.

For sources with large frequency uncertainties, however, especially those with unknown frequencies,
correcting for intrinsic source frequency evolution and for modulations due to the Earth's motion incurs
a substantial computing cost for searching over parameter space. Because of these costs, it is useful
to categorize CW searches broadly into three categories~\citep{bib:Prixreview} (while recognizing there are special cases that
fall near the boundaries). 
\begin{enumerate}
  \item {\it Targeted} searches in which
    the star's position and rotation frequency are known, \ie, known 
    radio, X-ray or $\gamma$-ray pulsars; 
  \item {\it Directed} searches in which the star's position is known, but rotation
    frequency is unknown, \eg, a non-pulsating X-ray source at the
    center of a supernova remnant; and
  \item {\it All-sky} searches for unknown
    neutron stars.
\end{enumerate}

The volume of parameter space over which to search increases
in large steps as one progresses through these categories. In each
category a star can be isolated or binary. For 2) and 3) any unknown binary
orbital parameters further increase the search volume, making a subclassification
helpful, as discussed below. In general, the greater the \apriori\
knowledge of sources parameters, the more computationally feasible it is to integrate data
coherently for longer time periods in order to improve strain amplitude sensitivity.

To illustrate, consider a directed search for a source of known location but with unknown frequency and
unknown frequency derivatives, where the signal phase is expanded in truncated Taylor form in the
source frame time $\tau$ with respect to a reference time $\tau_0$:
\begin{equation}
\label{eqn:phaseevolution}
\Phi(\tau) \quad \approx \quad \Phi_0 + 2\,\pi\left[f_s(\tau-\tau_0) + {1\over2}\dot f_s(\tau-\tau_0)^2 + 
{1\over6}\ddot f_s(\tau-\tau_0)^3\right].
\end{equation}

Using the phase evolution model of Eqn.~\ref{eqn:phaseevolution},
if we wish to preserve phase fidelity to a tolerance $\Delta\Phi$ over a coherence time $\Tcoh\approx \tau-\tau_0$, then
we need (in a naive estimate) to know the frequency and derivatives to a tolerance better than
\begin{eqnarray}
  \label{eqn:ftolerance}
  \Delta\fgw & \approx & {\Delta\Phi\over2\,\pi}{1\over\Tcoh}, \\
  \label{eqn:fdottolerance}
  \Delta\fgwdot & \approx & {\Delta\Phi\over2\,\pi}{2\over\Tcoh^2}, \\
  \label{eqn:fddottolerance}
  \Delta\fgwddot & \approx & {\Delta\Phi\over2\,\pi}{6\over\Tcoh^3}.
\end{eqnarray}
Hence the numbers of steps to take in $\fgw$, $\fgwdot$, and $\fgwddot$ to cover
a given range in the parameters are proportional
to $\Tcoh$, $\Tcoh^2$ and $\Tcoh^3$, respectively -- if it's necessary to step at all in
those derivatives. Naively, for a search over a long enough coherence time to require
multiple steps in $\fgwddot$, one has a template count proportional to $\Tcoh^6$ and,
presumably, pays a price proportional to another factor of $\Tcoh$ in computational cost
in processing the associated data volume. In principle, then, the computational cost
of a coherent search scales as the 7th power of the coherence time used, although, in
practice the scaling tends not to be as extreme because the numbers of steps needed for
$\fgwddot$ can be small integers that take on new discrete values only slowly with increased
$\Tcoh$. In practice, these considerations for a 2nd frequency derivative come into play
for only directed searches or for the deep follow-up of outliers from all-sky searches,
when segment coherence times exceed several days.
Section~\ref{sec:templates} will discuss more quantitatively the placement of
search templates in parameter space to maintain acceptable phase tolerance.

In carrying out {\it all-sky} searches for unknown neutron stars, 
the computational considerations grow worse. The corrections for
Doppler modulations and antenna pattern modulation due to the Earth's 
motion must be included, as for the targeted and directed searches,
but the corrections are sky-dependent, and the spacing of the
demodulation templates is dependent upon the inverse of
the coherence time of the search. Specifically, for a coherence time $T_{\rm coh}$
the required angular resolution is~\citep{bib:cwallskyS4}
\begin{equation}
\label{eqn:angres}
\delta\theta \quad \approx \quad {0.5\, {\rm c}\, \delta f\over f\,[v\sin(\theta)]_{\rm max}},
\end{equation}
where $\theta$ is the angle between the detector's velocity relative
to a nominal source direction, where the maximum relative frequency shift 
$[v\sin(\theta)]_{\rm max}/c\approx10^{-4}$, and where $\delta f$
is the size of the frequency bins in the search. For $\delta f=1/T_{\rm coh}$,
one obtains:
\begin{equation}
\delta\theta \quad \approx \quad \scimm{9}{-3}\>{\rm rad}\>\left({30\>{\rm minutes}\over T_{\rm coh}}\right)
\left({300\>{\rm Hz}\over f_s}\right),
\end{equation}
where $T_{\rm coh}$ = 30 minutes has been used in several all-sky searches to date.
Because the number of required distinct points on the sky scales like $1/(\delta\theta)^2$,
the number of search templates scales like $(T_{\rm coh})^2(f_s)^2$ for a fixed signal frequency $f_s$.
Now consider attempting a search with a coherence time of 1 year for a
signal frequency $f_s=1$ kHz. One obtains $\delta\theta\sim0.3$ $\mu$rad and
a total number of sky points to search of $\sim$$10^{14}$ -- again, for a fixed
frequency. Adding in the degrees of freedom to search over ranges in 
$f_s$, $\dot f_s$ and $\ddot f_s$ (and higher-order derivatives, as needed)
makes a brute-force, fully coherent 1-year all-sky
search hopelessly impractical, given the Earth's present total computing capacity.

As a result, tradeoffs in sensitivity must be made to achieve tractability
in all-sky searches. The simplest tradeoff is to reduce the observation
time to a computationally acceptable coherence time.
It can be more attractive, however, to reduce the coherence time still further
to the point where the total observation time is divided into $N=T_{\rm obs}/T_{\rm coh}$,
segments, each of which is analyzed coherently and the results added incoherently
to form a detection statistic. One sacrifices intrinsic sensitivity per 
segment in the hope of compensating (partially) with the 
increased statistics from being able to use more total data.
In practice, for realistic data observation spans (weeks or longer), the semi-coherent
approach gives better sensitivity for fixed computational cost and hence has been used
extensively in both all-sky and directed searches~\citep{bib:PrixShaltev}. One finds a
strain sensitivity (threshold for detection) that scales approximately as the inverse fourth root 
of $N$~\citep{bib:cwallskyS2}. Hence, for a fixed observation time, the strain sensitivity degrades
roughly as $N^{1\over4}$ as $T_{\rm coh}$ decreases (see~\citep{bib:WetteEstimation} for a discussion
of variations from this scaling). This degradation is a price one pays
for not preserving phase coherence over the full observation 
time, in order to make the search computationally tractable. An important virtue of
semi-coherent searches methods, however, is robustness with respect to deviations of a signal from an assumed coherent model.

In general, fully coherent search methods are potentially the most sensitive, but their applicability depends
on several considerations, perhaps the most important being sheer computational tractability.
Even when tractable for a particular search, moreover, a fully coherent initial search stage
may incur a statistical trials factor large enough to make a putative detection questionable,
because of the necessarily ultra-fine search needed to probe coherently a multi-dimensional signal parameter space.
That is, the statistical significance of a nominally ``loud'' detection statistic must account for
the number of independent trials carried out in the search. 
Applying {\it a priori} constraints instead, when available from electromagnetic observations or theoretical expectation,
can reduce the parameter space volume and hence trials factor, making a detection more convincing.
For example, a ``5 $\sigma$'' detection of a signal from a known pulsar in a targeted search
might not qualify as even a weak outlier in an all-sky search, much less as a discovery.
\citep{bib:loosecoherence} discusses the tradeoff between fully templated and ``loose coherence''
methods (see section~\ref{sec:longlagloose}) in a broad parameter space, arguing against brute-force
template application.

In the next sections, a variety of general approaches and specific algorithms will be
presented, methods that attempt to achieve tradeoffs best suited to particular CW search types.

\subsection{Signal model}

CW searches must account for large phase modulations (or, equivalently, frequency modulations) of the source signal due
to  detector motion and potentially due to source motion (expecially for binary sources).
The precision of the applied modulation corrections must be high in the case of {\it targeted} searches, which
use measured ephemerides from radio, optical, X-ray or $\gamma$-ray
observations valid over the gravitational wave observation time. The precision must also be high
in following up outliers from {\it directed} or {\it all-sky} searches, while much less
precision is needed in the first stage of hierarchical searches. This section describes the intrinsic signal
model assumed, along with the expected modulations due to detector motion.

For the Earth's motion, one has
a daily relative frequency modulation of $v_{\rm rot}/c\approx10^{-6}$ and a much
larger annual relative frequency modulation of $v_{\rm orb}/c\approx10^{-4}$. 
The pulsar astronomy community has developed a powerful and mature software infrastructure for
measuring ephemerides and applying them in measurements, using the TEMPO 2 program~\citep{bib:tempo}.
The same physical corrections for the Sun's, Earth's and Moon's motions (and for the motion of other planets),
along with general relativistic effects
including gravitational redshift in the Sun's potential and Shapiro delay for
waves passing near the Sun, have been incorporated into the LIGO and Virgo software
libraries~\citep{bib:lal,bib:cwexplorer2}. 

Consider an isolated, rotating rigid triaxial ellipsoid (conventional model for a GW-emitting neutron star), for which
the strain waveform detected by an interferometer can be written as
\begin{equation}
\label{eqn:cwhdefinition}
h(t) \quad=\quad F_+(t,\psi)\,h_0{1+\cos^2(\iota)\over2}\,\cos(\Phi(t)) 
\>+\> F_\times(t,\psi)\,h_0\,\cos(\iota)\,\sin(\Phi(t)),
\end{equation}
where $\iota$ is the angle between the star's spin direction and the propagation
direction $\hat k$ of the waves (pointing toward the Earth).
$F_+$ and $F_\times$ are the (real) detector antenna pattern response factors
($-1 \le F_+,F_\times \le 1)$ to the $+$ and $\times$ polarizations. $F_+$ and $F_\times$ 
depend on the orientation of the detector and the source, and on 
the polarization angle $\psi$~\citep{bib:cwtargetedS1}. Here, $\Phi(t)$ is
the phase of the signal, which can often usefully be Taylor-expanded as in Eqn.~\ref{eqn:phaseevolution},
in the solar system barycenter (SSB) time $\tssb$ with apparent frequency derivatives with respect to detector-frame time arising
from source motion. A more general signal model with GW emission at both once and twice the rotation frequency is
considered in \citep{bib:JKS}, with effects of free precession addressed in
\citep{bib:JonesAnderssonPrecession1,bib:JonesAnderssonPrecession2,bib:VanDenBroeck,bib:GaoEtal},
and a convenient reparametrization is presented in \citep{bib:JonesParametrization}.

Explicitly, the time of arrival of a signal at the solar system barycenter, $\tssb(t)$, can be written
in terms of the signal time of arrival $t$ at the detector:
\begin{equation}
\label{eqn:phasedefinition}
\tssb(t) \quad \equiv \quad t + \delta t \quad = \quad t - {\vec r_d\cdot\hat k\over c}+ \Delta_{E\odot}+\Delta_{S\odot},
\end{equation}
\noindent where
$\vec r_d$ is the position
of the detector with respect to the SSB, and $\Delta_{E\odot}$ and $\Delta_{S\odot}$ 
are solar system Einstein and Shapiro time delays, respectively~\citep{bib:taylorssb,bib:tempo}.

Equation~\ref{eqn:phasedefinition} implicitly assumes planar gravitational wavefronts and neglects proper
motion of the source (transverse to the line of sight), corrections for which are common in radio pulsar astronomy~\citep{bib:LorimerKramer,bib:LyneGrahamSmith}.
In principle, long-duration (multi-year) fully coherent observations of a near-enough ($\sim$100 pc), high-frequency ($\sim$1 kHz) CW source would allow
inference of its distance from determination of the wavefront curvature~\citep{bib:SetoWavefrontCurvature}.
Similarly, multi-year coherent observations of a high-frequency source would need to account for significant proper motions ($\sim$50 mas/year, typical of
known pulsars)~\citep{bib:CovasProperMotion}. Both wavefront curvature and proper motion have been neglected
in CW searches for unknown sources to date because the coherence times used in the searches don't require those corrections, but
in the happy event of a future detection and subsequent extended observations, these corrections may become relevant.

Existing gravitational wave detectors are far from isotropic in their response functions. In the long-wavelength limit,
Michelson interferometers have an antenna pattern sensitivity with polarization-dependent maxima
normal to their planes and nodes along the bisectors of the arms. As the Earth rotates at angular velocity $\Omegar$ with respect to
a fixed source, the antenna pattern modulation is quite large and polarization dependent via
the functions $F_+(t)$ and $F_\times(t)$ which
depend on the orientation of the detector and the source.

A commonly used parametrization of these amplitude response modulations is defined in~\citep{bib:JKS}.
\begin{eqnarray}
  \label{eqn:jksone}
  F_+(t)     & = & \sin(\zeta)\left[ a(t)\cos(2\psi)+b(t)\sin(2\psi)\right], \\
  F_\times(t) & = & \sin(\zeta)\left[ b(t)\cos(2\psi)-a(t)\sin(2\psi)\right] ,
\end{eqnarray}
\noindent where $\zeta$ is the angle between the arms of the interferometer (nearly or precisely 90 degrees for all major
ground-based interferometers), and where $\psi$ defines the polarization angle of the source wave frame
(\eg, angle between neutron star spin axis projected onto the plane of the sky and
local Cartesian coordinates aligned with its right ascension and declination directions).
The antenna pattern functions $a(t)$ and $b(t)$ depend on the position and orientation of the interferometer on
the Earth's surface, the source location and sidereal time: 
\begin{eqnarray}
  \label{eqn:aoftdef}
  a(t) & = & {1\over16}\sin(2\gamma)(3-\cos(2\lambda))(3-\cos(2\delta))\cos[2(\alpha-\phi_r-\Omegar t)] \nonumber \\
  & & -{1\over4}\cos(2\gamma)\sin(\lambda)(3-\cos(2\delta))\sin[2(\alpha-\phi_r-\Omegar t)] \nonumber \\
  & & +{1\over4}\sin(2\gamma)\sin(2\lambda)\sin(2\delta)\cos[(\alpha-\phi_r-\Omegar t] \nonumber \\
  & & -{1\over2}\cos(2\gamma)\cos(\lambda)\sin(2\delta)\sin[(\alpha-\phi_r-\Omegar t] \nonumber \\
  & & +{3\over4}\sin(2\gamma)\cos^2(\lambda)\cos^2(\delta), \\
  b(t) & = & \cos(2\gamma)\sin(\lambda)\sin(\delta) \cos[2(\alpha-\phi_r-\Omegar t)] \nonumber \\
  & & + {1\over4}\sin(2\gamma)(3-\cos(2\lambda))\sin(\delta)\sin[2(\alpha-\phi_r-\Omegar t)] \nonumber \\
  & & + \cos(2\gamma)\cos(\lambda)\cos(\delta)\cos[\alpha-\phi_r-\Omegar t] \nonumber \\
  & & + {1\over2}\sin(2\gamma)\sin(2\lambda)\cos(\delta)\sin[\alpha-\phi_r-\Omegar t].
  \label{eqn:boftdef}
  \label{eqn:jkstwo}
\end{eqnarray}
\noindent Specifically, in these equations, $\lambda$ is the interferometer's latitude, and $\gamma$ is the counterclockwise angle between the
bisector of its arms and the eastward direction. The source direction is specified by right ascension $\alpha$ and declination $\delta$, while
$\phi_r$ is a deterministic phase defined implicitly by the interferometer's longitude.
These functions reveal amplitude modulations with periods of 1/2 and 1 sidereal day, and in the case of $a(t)$, a constant term independent
of time. As a result, the interferometer's response to
a monochromatic source in the Earth center's reference frame will, in general, display five distinct frequency
components, corresponding to the ``carrier'' frequency and two pairs of positive and negative sidebands, with a
splitting between adjacent frequencies of ${\Omegar\over2\,\pi} \approx \scimm{1.16}{-5}$ Hz.

Searches for CW signals must take into account the phase/frequency modulations embodied in Eqn.~\ref{eqn:phasedefinition} due
to detector translational motion and
the antenna pattern modulations embodied in Eqns.~\ref{eqn:jksone}-\ref{eqn:jkstwo} due to detector orientation changes.

\begin{figure}[t!]
\begin{center}
\includegraphics[width=13.cm]{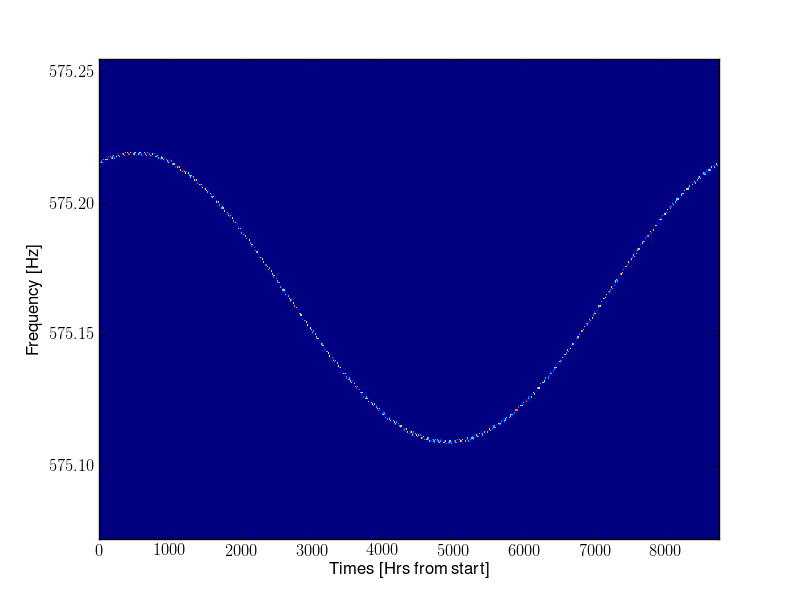}
\caption{Sample signal spectrogram for a LIGO ``hardware injection'' (negligible noise), where the pixel dimensions are 0.5 hours by 0.556 mHz.}
  \label{fig:samplesignalspectrogram}
\end{center}
\end{figure}

\begin{figure}[t!]
\begin{center}
\includegraphics[width=13.cm]{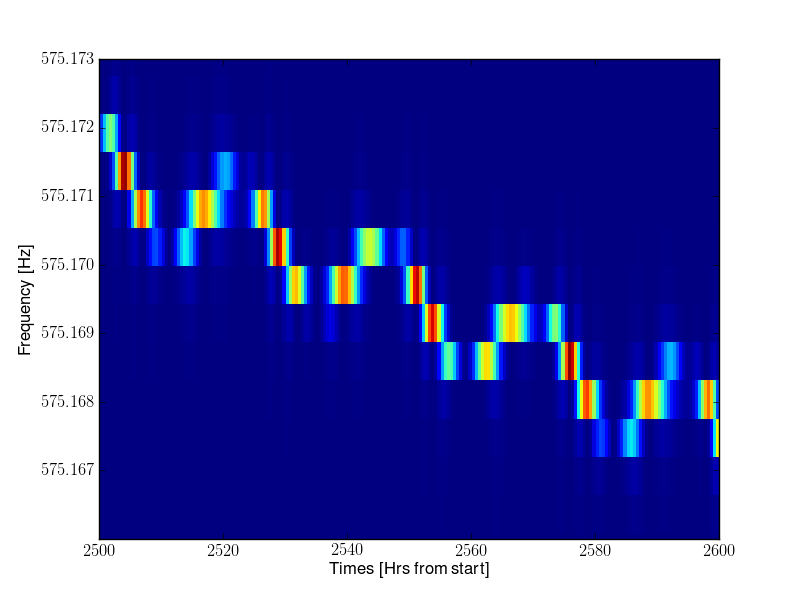}
\caption{Zoomed-in sample signal spectrogram for a LIGO ``hardware injection'' (100 hours from spectrogram in Figure~\ref{fig:samplesignalspectrogram}).
The sidereal Doppler modulations ($\sim$24 hrs) of frequency and amplitude modulations ($\sim$12 and 24 hrs) are more apparent. }
  \label{fig:samplesignalspectrogramzoom}
\end{center}
\end{figure}

Figure~\ref{fig:samplesignalspectrogram} shows a sample spectrogram of a pure signal simulation using
one of the so-called ``hardware injections'' used in LIGO data runs. These signal simulations are used to verify
end-to-end the detector's response to a CW signal, including sustained phase coherence over long durations.
The simulations are injected via ``photon calibrators,'' which are auxiliary lasers shining on mirrors with
a modulated intensity. The imposed relative motion of the mirror mimics (in the long-wavelength regime) the response
of the interferometer to a gravitational wave. Various such signals, ranging in nominal frequency from 12 Hz to 2991 Hz were
injected into the LIGO detectors over the O1, O2 and O3 observation runs~\citep{bib:hwinjectionpaper}.
In the example shown, a signal (``Pulsar 2'')
with source frequency 575.163573 Hz (reference time = November 1, 2003 00:00 UTC) and spin-down \sci{-1.37}{-13} Hz/s is
simulated at a sky location of right ascension $\alpha$ = 3.75692884 radians (14h 21m 1.48s)
and declination $\delta$ = 0.060108958 radians ($3^\circ$ 26' 38.36''). Its orientation is defined by
inclination angle $\iota$ = 2.76 radians (158. deg), and polarization angle $\psi$ = $-$0.222 radians ($-$12.7 deg).
The simulation shown (with negligible noise for clarity) in Figure~\ref{fig:samplesignalspectrogram}
applies over a duration of the calendar year 2019 UTC. One can see the annual modulation from the
Earth's orbit imposed on an imperceptible decrease in the intrinsic frequency. A zoom of 100-hour duration is also
shown in Figure~\ref{fig:samplesignalspectrogramzoom}, to indicate the
much smaller frequency modulation (by $\sim$2 orders of magnitude), along with the intensity modulation.

\begin{figure}[t!]
\begin{center}
\includegraphics[width=13.cm]{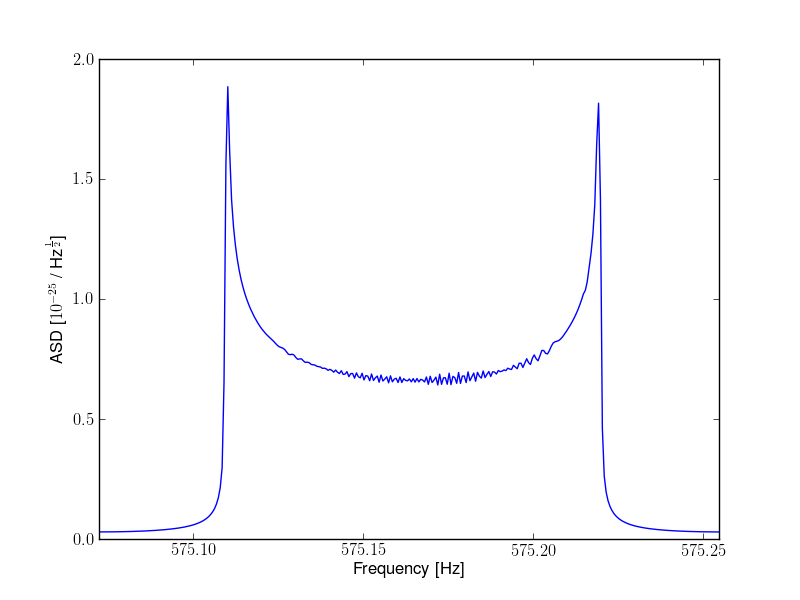}
\caption{Sample signal amplitude spectral density for a LIGO ``hardware injection'' (same injection as for spectrograms in
  Figure~\ref{fig:samplesignalspectrogram}).}
  \label{fig:samplesignalspectrum}
\end{center}
\end{figure}

As seen from the spectrogram, the frequency modulations lead to stationary bands at the turning points of the modulation. As a result,
the spectrum averaged over the signal duration peaks at the turning points, as shown in Figure~\ref{fig:samplesignalspectrum}. These
``horns'' are a characteristic spectral signature of expected signals, where the relative heights of the horns depend on the duration
of the observation and on the Earth's orbital phase at the start. For a signal with negligible spin-down and a duration equal to a multiple
of a year, the horns are approximately symmetric, but in the general case that includes observations of a few months or less,
one or both horns may not be apparent. A more detailed analysis of the Fourier transform of a CW signal can be found
in~\citep{bib:CWFourierTransform}.

In the following, we discuss in more detail the implementations of a selection of these
methods developed in searches of the initial LIGO and Virgo data sets (2001-2011) and
that have been further refined for searches of advanced detector data.
Section~\ref{sec:results} presents the results of each type, from
searches in the data of Advanced LIGO's and Virgo's observing runs O1, O2 and O3. 

The volume of parameter space over which to search increases
in large steps as one progresses through these categories. In each
category a star can be isolated or binary. Any unknown binary
orbital parameters further increase the search volume.
In all cases we expect (and have now verified from unsuccessful searches to date)
that source strengths are very small. Hence one must integrate data over long
observation times to have any chance of signal detection. How much one knows about
the source governs the nature of that integration. In general, the greater that
knowledge, the more computationally feasible it is to integrate data
coherently (preserving phase information) over long observation times, for reasons
explained below.

\subsection{Broad approaches in CW searches}
\label{sec:algorithms}

Computational cost depends 
critically upon the search method used, which in turn, depends on the \apriori\ knowledge one
has about the source. In the following, a broad overview is given of a few key search methods
used in published searches to date. More details of implementation are presented in
sections~\ref{sec:results}, where a selection of these methods is applied to
particular classes of potential CW sources.

In this overview, a simplified ``toy model'' will be used to
illustrate scaling relations. Methods specific to correcting for modulations will be addressed further below
in the presentation of particular search implementations. For now, amplitude modulation of the signal strength due to rotation
of the GW detectors with respect to the source is ignored in the following, along with Doppler modulations.

\subsubsection{Fully coherent methods}
\label{sec:fullycoherent}

When applicable, as discussed in section~\ref{sec:challenges},
fully coherent methods provide the best sensitivity. 
Explicit search methods, taking into account source frequency and amplitude evolution along
with detector noise non-stationarity,
are discussed in section~\ref{sec:targeted}. Here, though, let's consider the highly simplified
problem of detecting a sinuosidal signal of amplitude $h_0$ and known frequency $\fsig$, but with
unknown phase constant,
in random Gaussian noise.

Imagine the data
observation is continuous of duration $\Tobs$ and has a one-sided power spectral noise density function
$\Sh(\fgw)$, and for convenience, assume $\fsig$ is an integer multiple of $1/\Tobs$ (see~\citep{bib:AllenPapaSchutz} for
a didactic treatment of the more general case).
The DFT bin $\Di$ is defined by the following sum over a real time series of length $\Nsample = \fsample\Tobs$ with values $d_j$ ($j$ = 0...$\Nsample$) where
$\fsample$ is the sampling frequency of the data stream:
\begin{equation}
  \Di = \sum_{j=0}^{\Nsample-1}d_je^{-\imag2\pi ji/\Nsample}.
  \label{eqn:DFTdefinition}
\end{equation}
\noindent From the DFT, one can define the one-sided power spectral noise density estimate $\Shi$:
\begin{equation}
  \Shi = {2 \left< \!\left[ ( \Re\{\Di\})^2 + (\Im\{\Di\})^2 \right]\!\right>\Tobs\over \Nsample^2} = {2\left<\!|\Di|^2\!\right>\Tobs\over\Nsample^2},
\end{equation}
\noindent for $0<i<\Nsample/2$  and where ``$\langle$ $\rangle$'' indicates an expectation value in the absence of signal
(\eg, determined from an average over many nearby bins {\it and excluding the bin $i$ itself}).
Then one can construct a dimensionless detection statistic $\rhoi^2$ using the measured strain power
in the DFT bin $\Di$ corresponding to a signal frequency $\fsig$:
\begin{equation}
  \rhoi^2 = 4{|\Di|^2\Tobs\over\Nsample^2\Shi},
  \label{eqn:rhodefinition}
\end{equation}
\noindent which follows a
non-central $\chi^2$ distribution with two degrees of freedom and a non-centrality parameter $\lambda(h_0) = {h_0^2\Tobs\over\Shi}$,
which implies an expectation value $2+{h_0^2\Tobs\over\Shi}$ and variance $4+4{h_0^2\Tobs\over\Shi}$. In Gaussian noise one
expects a $\chi^2$ distribution with two degrees of freedom from summing the squares of the normally distributed real and imaginary 
DFT coefficients.

In the absence of a signal, one can define a threshold value $\rhostari$ corresponding to a
false alarm probability $\alpha$ such that 
the cumulative density probability function satisfies:
\begin{equation}
  {\rm CDF}_{\rm noise}[{\rhostari}^2] \equiv \int_0^{{\rhostari}^2} {p_{\rm noise}(\rhoi^2;2)}\,d(\rhoi^2) \equiv 1-\alpha\>,
      \label{eqn:cdfrelationnoisecoherent}
\end{equation}
\noindent where the probability density function is ($\chi^2$ with two degrees of freedom)
\begin{equation}
  p_{\rm noise}(x;2) = {1\over2}e^{-x/2}.
  \label{eqn:probnoisecoherent}
\end{equation}

From this threshold and a desired false dismissal rate $\beta$, one can then determine
the corresponding signal amplitude $h_0^{1-\beta}$  from
\begin{equation}
  {\rm CDF}_{\rm signal+noise}[{\rhostari}^2] \equiv \int_0^{{\rhostari}^2} {p_{\rm signal+noise}(\rhoi^2;2,\lambda(h_0^{1-\beta}))}\,d(\rhoi^2) = \beta\>,
      \label{eqn:cdfrelationsignalnoisecoherent}
\end{equation}
\noindent where the probability density function is (non-central $\chi^2$ with two degrees of freedom)
\begin{equation}
  p_{\rm signal+noise}(x;2,\lambda) = {1\over2}e^{-(x+\lambda)/2}I_0(\sqrt{\lambda x})\>,
  \label{eqn:probsignalnoisecoherent}
\end{equation}
\noindent and where $I_0(y)$ is a modified Bessel function of the first kind:
\begin{equation}
  I_0(y) = \sum_{j=0}^\infty {(y^2/4)^j\over(j!)^2}\>.
\end{equation}

Choosing a 1\%\ false alarm probability ($\alpha$ = 0.01) leads to ${\rhostari}^2\approx9.21$, from which numerical
evaluation of Equation~\ref{eqn:cdfrelationsignalnoisecoherent} for a false dismissal probability $\beta$ = 5\%\
leads to an expected sensitivity $\hnf$ of
\begin{equation}
  \hnf \approx 4.54\, \sqrt{\Shi\over\Tcoh},
  \label{eqn:coherentsinesens}
\end{equation}
which can be taken as a proxy for the expected 95\%\ confidence level upper limit on signal amplitude
based simply on an observation that an observed $\rhoi$ does not exceed $\rhostari$. In practice, many CW search upper
limits are based on the loudest statistic found in a search, (\eg, largest $\rho^2$ value for multiple computations 
at different frequencies), regardless of whether or not a pre-defined
threshold has been exceeded. \citep{bib:TenorioEtalLoudestCandidate} discusses the statistics of
loudest candidates using extreme value theory.

The sensitivity expression in Eqn.~\ref{eqn:coherentsinesens} is shown as the leftmost point of the lower blue curve in Figure~\ref{fig:sinusoidsensitivity}, where it can be compared
to sensitivities from other methods, discussed below.

If simultaneous data sets from two independent detectors of identical sensitivity are added coherently (and a phase correction applied,
to account for detector separation and source direction), 
one can construct a combined averaged detection statistic $\rho_{i,{\rm comb}}^2$ using the measured power
from the square of the sum of the simultaneous DFT bin coefficients $\Donei$ and $\Dtwoi$ containing $\fsig$:
\begin{equation}
  \rho_{i,{\rm comb}}^2 = {1\over2}\left[ 4{|\Donei+\Dtwoi|^2\,\Tobs\over\Nsample^2\Shi}\right]\,
  \label{eqn:twodetectorDFTsum}
\end{equation}
\noindent which has an expectation value of $2+{2\,h_0^2\Tobs\over\Shi}$ and a variance of $4+{8\,h_0^2\Tobs\over\Shi}$, that is, it follows a
non-central $\chi^2$ distribution with two degrees of freedom and a non-centrality parameter $\lambda(h_0) = {2\,h_0^2\Tobs\over\Shi}$.
Applying the same methodology as above for a single detector, one obtains an expected sensitivity of
\begin{equation}
  \hnf \approx {1\over\sqrt{2}} \times 4.54\, \sqrt{\Shi\over\Tcoh},
\end{equation}
\noindent indicating an improvement by $\sqrt2$ with respect to a single detector. This sensitivity is shown as the
leftmost point of the green curve in Figure~\ref{fig:sinusoidsensitivity}. More generally, $N$ identical
detectors with simultaneous data sets\footnote{Simultaneity is not strictly required, if phase corrections for time offsets between detectors
can be computed precisely enough.}, for which phase corrections are known, gain a sensitivity improvement of $\sqrt{N}$
with respect to a single detector, as one would naively expect. Put another way, combining the $N$ sets gives
a sensitivity equal to that of a single detector with an amplitude spectral noise density of $\sqrt{\Shi/N}$.
Again, this is a simplified model. In practice, detectors
have different, frequency-dependent $\Shi$ noise levels and unequal live times of observing, in addition to different
orientations affecting antenna pattern sensitivity, ignored here.

\subsubsection{Semi-coherent methods}
\label{sec:semicoherent}

Fully coherent methods are computationally costly when covering a large parameter space because
of the fine steps needed to avoid missing a signal as coherence times increase. A crude solution is simply
to reduce the coherence time and suffer the reduction in strain sensitivity, by approximately $\sqrt{\Tcoh}$. A better
solution is to apply a {\it semi-coherent} method in which the observation time is divided into $\Nseg$ 
segments of equal length $\Tcoh = \Tobs/\Nseg$, where the detection statistic is constructed from an incoherent sum of powers
from the individual segments. This method sacrifices the constraint of phase consistency among the different segments and
hence is less sensitive than a fully coherent method, but is more sensitive than analysis of a single segment alone.
As shown below, the strain sensitivity scales with $1/\Nseg^{1\over4}$ for fixed coherence time and large $\Nseg$.

For illustration, consider a detection statistic constructed from the sum of $\Nseg$ individual DFT powers covering
the observation time $\Tobs$, and once again, each normalized by its power spectral density (taken to be stationary here,
for convenience):
\begin{equation}
  \label{eqn:Ridef}
  \Ri \equiv \sum_{k=1}^{\Nseg} 4{|\Dki|^2\Tcoh\over\Nsample^2\Shi},
\end{equation}
\noindent where $\Nseg=1$ yields $\Ri=\rhoi^2$ in Eqn.~\ref{eqn:rhodefinition}. The underlying statistical distribution of $\Ri$ is that of
a non-central $\chi^2$ with $2\Nseg$ degrees of freedom and a non-centrality parameter of 
$\lambda(h_0) = \Nseg {h_0^2\Tcoh\over\Shi}$.

In the absence of signal a false alarm probability $\alpha$ implies a threshold $\Rstari$, 
found from requiring that the cumulative probability density function satisfy:
\begin{equation}
  {\rm CDF}_{\rm noise}[\Rstari] \equiv \int_0^{\Rstari} {p_{\rm noise}(\Ri;2)}\,d\Ri \equiv 1-\alpha\>,
      \label{eqn:cdfrelationnoisesemicoherent}
\end{equation}
\noindent where the probability density function is
\begin{equation}
  p_{\rm noise}(x;2\Nseg) = {x^{\Nseg-1}e^{-x/2}\over2^{\Nseg}\Gamma(\Nseg)},
\end{equation}
\noindent which reduces to Eqn.~\ref{eqn:probnoisecoherent} for $\Nseg=1$. $\Rstari$ can be
determined numerically for arbitrary $\Nseg$, but in the limit of large $\Nseg$, $p_{\rm noise}$ reduces
to a normal distribution with a mean of $2\Nseg$ and variance $4\Nseg$, in which approximation
$\Rstari(\alpha=0.01) \approx 2\Nseg + 4.65\sqrt{\Nseg}$.

In the presence of a signal, 
the $\hnf$ value can be obtained numerically from the cumulative probability density function:
\begin{equation}
  {\rm CDF}_{\rm signal+noise}[\Rstari] \equiv \int_0^{\Rstari} {p_{\rm signal+noise}(\Ri;2,h_0)}\,d\Ri = \beta,
      \label{eqn:cdfrelationsignalnoisesemicoherent}
\end{equation}
\noindent where the probability density function is
\begin{eqnarray}
  p_{\rm signal+noise}(x;2\Nseg,h_0) & = & {1\over2} e^{-{1\over2}\left(x+{h_0^2\Tobs\over\Sh}\right)}
  \left({x\Sh\over h_0^2\Tobs}\right)^{{\Nseg-1\over2}} \nonumber \\ 
  & &    I_{\Nseg-1}\left(h_0\sqrt{{x\Tobs\over\Sh}}\right)
\end{eqnarray}
\noindent which reduces to Eqn.~\ref{eqn:probsignalnoisecoherent} for $\Nseg=1$, where $\Tobs=\Nseg\Tcoh$ has been used.

In the limit of large $\Nseg$ and weak signal, however, the distribution approaches that of a Gaussian
with variance $4\Nseg$, from which an approximate expression for $\hnf$ can be obtained:
\begin{equation}
 h_0^{1-\beta} \approx \sqrt{2}\,\left[\sqrt{2}\,(\erfc^{-1}(2\alpha)+\erfc^{-1}(2\beta))\right]^{1/2} \Nseg^{1\over4} \sqrt{{\Sh\over\Tobs}},
 \label{eqn:h095approxsemicoherent}
\end{equation}
where $\erfc$ is the inverse complementary error function.
This scaling of sensitivity with $\Nseg^{1\over4}$ for fixed observation time is a universal result in semi-coherent searches with large
$\Nseg$~\citep{bib:houghmethod,bib:stackslideimplementation,bib:PrixShaltev,bib:WetteEstimation}. It can be understood
qualitatively from the SNR of an approximately Gaussian detection statistic (large $\Nseg$) scaling with
$\sqrt{\Nseg\Tcoh} = \sqrt{\Tobs/\Nseg}$ and the direct dependence of that detection statistic on the squared signal amplitude.

\begin{figure}[t!]
\begin{center}
\includegraphics[width=13.cm]{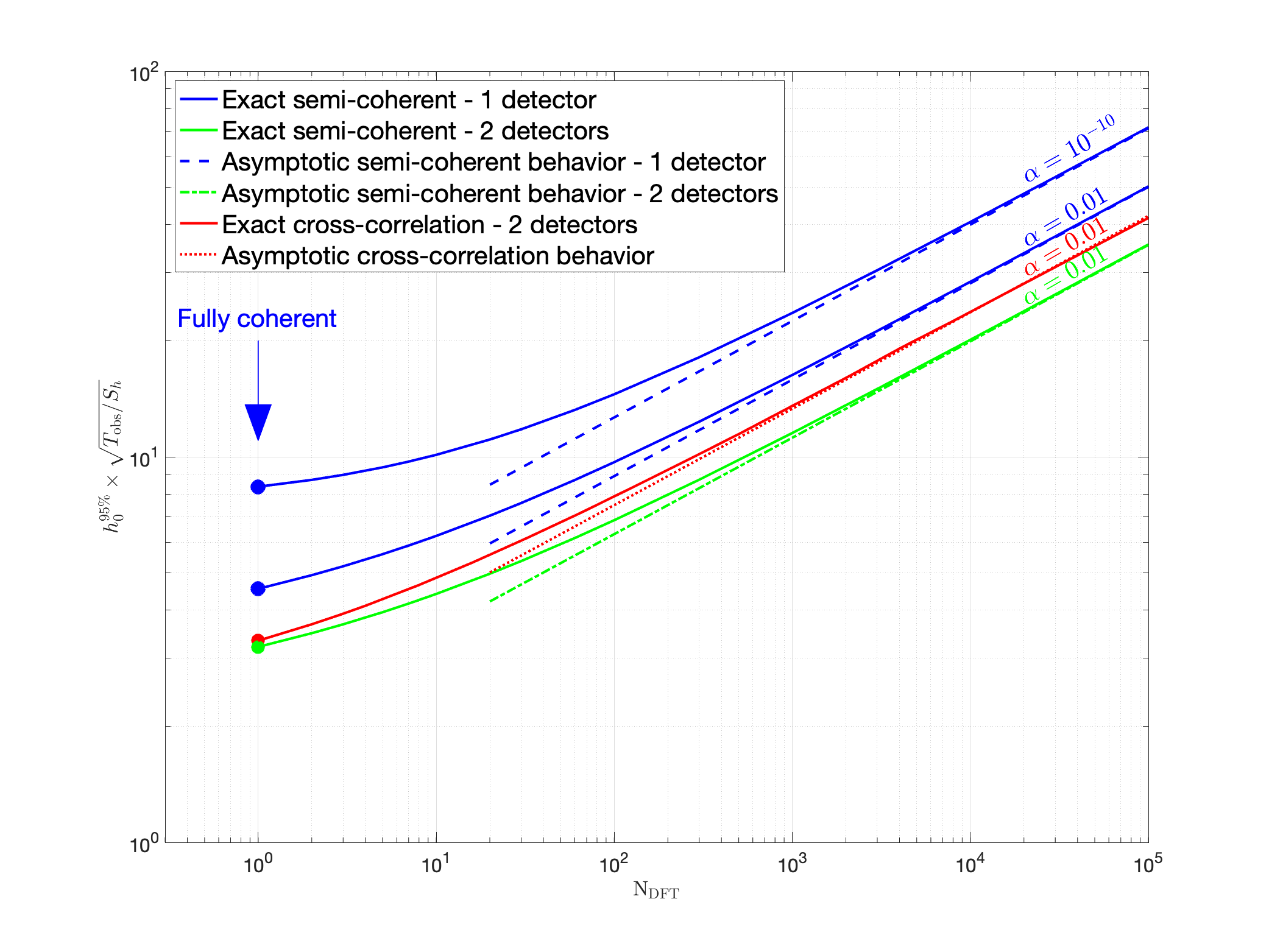}
\caption{Strain sensitivities (defined by $\alpha=0.01$ or $\alpha=10^{-10}$; and $\beta=0.05$)
  of various search methods to a bin-centered sinusoidal signal in noise of power spectral density $\Sh$
  \vs\ the number of DFT segments into
  which a fixed observing time $\Tobs$ is divided. Curves are shown for $\hnf$ for $\alpha=0.01$ and
  $\alpha = 10^{-10}$ for semi-coherent searches in a single detector's data (blue) and
  for $\alpha=0.01$ in two-detector searches (green). Asymptotic expressions based on the large-$\Nseg$
  Gaussian approximation are shown as dashed lines.
  The points at the bottom left of each curve
  represent the fully coherent search sensitivities for 1 and 2 identical detectors.
  Semi-coherent curves for two detectors assumed coherent summing of simultaneous DFTs for the 2 detectors.
  Sensitivities for cross-correlation of simultaneous data from two detectors are also shown (red).
  }
  \label{fig:sinusoidsensitivity}
\end{center}
\end{figure}

For $\alpha=0.01$ and $\beta=0.05$,
Equation~\ref{eqn:h095approxsemicoherent} yields $\hnf\approx2.82\Nseg^{1\over4}\sqrt{{\Sh/\Tobs}}$. This expression
does {\it not} agree with Eqn.~\ref{eqn:coherentsinesens} when $\Nseg=1$ because the Gaussian approximation breaks down for small $\Nseg$.
For the much lower
false alarm probability $\alpha=10^{-10}$, Eqn.~\ref{eqn:h095approxsemicoherent} yields $\hnf\approx4.00\Nseg^{1\over4}\sqrt{{\Sh/\Tobs}}$.
These asymptotic approximations are shown as dashed lines in Figure~\ref{fig:sinusoidsensitivity},
together with numerically evaluated values from Eqn.~\ref{eqn:cdfrelationsignalnoisesemicoherent}.

As is the case for combining data in a fully coherent search from two identical detectors with simultaneous data sets,
there is a gain of $\sqrt{2}$ in sensitivity for combining simultaneous DFTs from two detectors in a semi-coherent search -- as long as the
DFT coefficients are combined coherently for each segment (otherwise, the gain is only $2^{1/4}$ from semi-coherent combination
of DFT powers).
Figure~\ref{fig:sinusoidsensitivity} shows exact (solid green curve) and asymptotic (dashed green line) results for two detectors.

An important variation on semi-coherent methods, used in Hough transform methods described below,
applies {\it thresholding} to DFT powers prior
to summing. By applying a threshold corresponding to a relatively high false alarm rate (and relatively
high false dismissal rate for weak signals), one can reduce the data volume in processing, to reduce
computational cost. In addition, by adding integer counts (or, optimally, pre-computed weights) instead
of measured DFT powers, one's detection statistic is less susceptible to distortions from transient,
non-Gaussian outliers from instrumental contamination. The optimum false alarm probability for thresholding
in this idealized sine-wave detection analysis (weak-signal limit) is $\approx$20\%~\citep{bib:AllenPapaSchutz}.

\subsubsection{Cross-correlation methods}
\label{sec:crosscorrelation}

Another attractive approach uses cross correlation between independent (and ideally, simultaneous) data streams. The canonical example is cross correlation between coincident data sets taken with the nearly aligned Hanford and Livingston interferometers, but cross correlation can also be used with poorly aligned detector pairs and with non-coincident data streams -- if sufficient signal coherence can be established over longer time scales.
In this section, only truly coincident data sets will be considered, for simplicity.

Once again, let's use the artificial but informative toy model of a bin-centered, constant-amplitude sinusoidal signal in Gaussian noise. Let's also assume two identically oriented detectors (approximation to Hanford-Livingston, which have normal vectors to their planes only 27.3 degrees apart~\citep{bib:AlthouseJonesLazzarini}, in addition to a 90-degree relative rotation about the normal, leading to a sign flip in GW response). Also assume that the relative positions of the detectors can be accounted for via a signal phase correction for any given source direction. In the following, that phase correction is assumed to have been applied.

Using the notation from section~\ref{sec:semicoherent}, we can combine the two independent data streams
via DFT coefficients in a narrow band of interest to define a new detection statistic:
\begin{equation}
  \rhocc = {2\,\sqrt{2}\,\Re\left\{\Donei\Dtwoi^*\right\}\Tobs\over\Nsample^2\Shi}.
\end{equation}
\noindent
In the absence of a signal, this statistic has an expectation value of zero and (by construction) a variance of one.
The underlying statistical distribution is far from Gaussian, however. In the absence of
a signal, one has the sum of two normal product distributions
(from $\Re\left\{\Donei\Dtwoi^*\right\} = \Re\left\{\Donei\right\}\Re\left\{\Dtwoi\right\} + \Im\left\{\Donei\right\}\Im\left\{\Dtwoi\right\}$),
which can be obtained analytically, using characteristic functions.

Specifically, a single normal production distribution for
the product of two zero-mean normal distributions of variance $\sigma_1^2$ and $\sigma_2^2$ is
\begin{equation}
  p_{\rm 1\> normal\> product}(x) = {1\over\pi\sigma_1\sigma_2}K_0\left({|x|\over\sigma_1\sigma_2}\right),
\end{equation}
\noindent where $K_0$ is a modified Bessel function of the second kind, and for which the characteristic function is~\citep{bib:McNolty}
\begin{equation}
  {\rm CF}[K_0\left({|x|\over\sigma_1\sigma_2}\right)] = {1\over\sqrt{1+\sigma_1^2\sigma_2^2t^2}}.
\end{equation}
\noindent Inverting the product of characteristic functions gives the
following probability distribution for the sum of two such (signal-free) normal product distributions:
\begin{equation}
  p_{\rm 2\> normal\> products}(x) = {1\over2\sigma_1\sigma_2}e^{-{|x|\over\sigma_1\sigma_2}}.
\end{equation}

For identical detectors ($\sigma_1=\sigma_2\equiv\sigma$) of power spectral density $\Shi$ with a signal
present of amplitude $h_0$, the expectation value of $\rhocc$ is
${h_0^2\Tobs\over\sqrt{2}\,\Shi}$, and the variance is $1 + {h_0^2\Tobs\over\Shi}$.

In the presence of a common signal $h_0$ in both data streams, one can evaluate numerically the value $h_0^{1-\beta}$ for
which the false dismissal rate is $\beta$ for a threshold on the detection statistic corresponding to the signal-free false alarm probability $\alpha$.
In the absence of a signal, the threshold on $\rhocc$ for a false alarm probability $\alpha = 0.01$ is approximately 2.766.
For $\beta=0.05$  one then finds 
$\hnf\approx 3.335\sqrt{{\Shi/\Tobs}}$, which is only slightly higher than that obtained for a fully coherent
search using data from two identical detectors (see Figure~\ref{fig:sinusoidsensitivity}).

As with semi-coherent searches, discussed in section~\ref{sec:semicoherent}, one typically finds it necessary in wide-parameter searches
to segment the data. As before, consider dividing the observation time $\Tobs$ into $\Nseg$ equal-duration segments of
coherence time $\Tcoh$. In the presence of a signal of amplitude $h_0$, the following detection statistic,
\begin{equation}
\rhocc^{\Nseg}  = {1\over\Nseg}\sum_{i=1}^{\Nseg} \rhocc^i,
  \label{eqn:crosscorrelation}
\end{equation}
\noindent has a mean value of  ${h_0^2\Tcoh\over\sqrt{2}\Shi} = {1\over\Nseg}{h_0^2\Tobs\over\sqrt{2}\Shi}$
and variance ${1\over\Nseg}\left[1 + {h_0^2\Tobs\over\Shi}\right]$.

In the regime of large $\Nseg$ and weak signal, the underlying probability distribution approaches that of a Gaussian
for which one expects:
\begin{equation}
 h_0^{1-\beta} \approx \sqrt{2}\,\left[\sqrt{2}\,(\erfc^{-1}(2\alpha)+\erfc^{-1}(2\beta))\right]^{1/2} \left[{\Nseg\over2}\right]^{1\over4} \sqrt{{\Sh\over\Tobs}}. 
 \label{eqn:h095approxcorrelation}
\end{equation}
\noindent This asymptotic expectation is shown as a dotted red line in Figure~\ref{fig:sinusoidsensitivity}, together with results from numerical simulation over a range of $\Nseg$ values (solid red curve). This detection statistic
is $2^{1/4}$ more sensitive than the asymptotic 1-detector semicoherent sensitivity (lower dashed blue curve),
equally sensitive to the asymptotic 2-detector semicoherent sensitivity for which powers from separate detectors are added (not shown), 
and $2^{1/4}$ less sensitive than the asymptotic 2-detector semicoherent behavior with coherent summing of simultaneous DFTs from the 2 detectors
before computing power, as in Eqn.~\ref{eqn:twodetectorDFTsum} (green dash-dotted curve). 

One practical consideration to keep in mind for these comparisons is that while coherent summing or cross-correlation of simultaneous DFTs provides
improved sensitivity where possible, those gains are limited by achievable livetimes of interferometers that operate near their technological limits.

One can compute nominal signal-to-noise ratios for given signal strengths for the coherent, semi-coherent and cross-correlations
methods from the noise-only variances and the expectation value
dependences on signal $h_0$ presented above. Those SNRs allow sensible direct comparisons among semi-coherent and cross-correlation
methods when $\Nseg$ is large
enough for the noise-only detection statistics to exhibit approximate Gaussian behavior over the range of interest,
but for small $\Nseg$, including especially the fully coherent case of $\Nseg$ = 1, the underlying statistics are highly non-Gaussian, requiring
care in making comparisons.

\subsubsection{Long-lag cross-correlation and loose coherence}
\label{sec:longlagloose}

Fully coherent, long integrations seem distinctly different from the multiple-short-segment searches based on semi-coherent and cross-correlation
described (in simplified form) above, and in fact, using fully coherent methods to follow up on outliers produced by the latter methods is challenging
because of the typical mismatch in parameter space fineness. Nonetheless, between these extremes exist bridges that offer the possibility of both
systematic follow-up of outliers and more sensitive initial stages for multi-segment searches. The first method is long-lag correlation~\citep{bib:xcorrmethod1},
and the second method is known as loose coherence~\citep{bib:loosecoherence}.

Each method benefits from coherently summing DFT coefficients from data segments offset in time, which can be motivated by considering
the segmentation of a single (continuous) Fourier transform of a strain signal $h(t)$ into $\Nseg$ segments:
\begin{eqnarray}
  F(\omega;[t_A,t_B]) & \equiv & {1\over T} \int_{t_A}^{t_B} h(t) e^{-i\omega (t-t_A)}dt \\
  & = & {e^{i\omega t_A}\over\Nseg} \sum_{i=1}^{\Nseg}{e^{-i\omega t_{i-1}}\over\Tseg}\int_{t_{i-1}}^{t_i} h(t)e^{-i\omega (t-t_{i-1})}dt \\
  & = & {e^{i\omega t_A}\over\Nseg} \sum_{i=1}^{\Nseg} e^{-i\omega t_{i-1}} F(\omega;[t_{i-1},t_i]) \\
  & = & {e^{i\omega t_A}\over\Nseg} \sum_{i=1}^{\Nseg} e^{-i\phi_{i}} F_i(\omega),
\end{eqnarray}
\noindent where $\Tseg = (t_B-t_A)/\Nseg$, $t_i = t_A + i{\Tseg}$ and $\phi_i = \omega t_{i-1}$. Hence the Fourier transform for the
full data span is proportional to the sum of transforms for the individual segment transforms with phase corrections $e^{-i\phi_i}$.

Now consider once again the artificial but informative special case of a monochromatic signal detected via its strength in
DFT bins from two detectors 1 and 2 with identical observation periods segmented into $\NDFT$ epochs for which
DFT are computed. For simplicity, assume the signal is bin-centered for each epoch's DFT and that the detector noise
is both stationary and identical for the two detectors. Guided by the relation above, one can then define a detection statistic based on the coherent
sum of {\it all} DFTs from both detectors:
\begin{eqnarray}
  P & = & \left|\sum_{i=1}^{\NDFT} \left[ \Donei\expphionei+\Dtwoi\expphitwoi \right] \right|^2 \\
    & = & \sum_{I,J=1}^2\sum_{i,j=1}^{\NDFT} \DIi\DJj^*\expphiIiJj,
\end{eqnarray}
\noindent where $\phionei$ and $\phitwoi$ account for the signal phase evolution for each detector for each DFT $i$ and for any geometric offset between the detectors
relative to the source direction (see Eqn.~\ref{eqn:phasedefinition}). This full double sum is computationally costly to evaluate explicitly, not only
because of the additional operations, but more important, because the implicit full coherence requires a fine stepping in parameter space. The form, however,
makes more clear the relations between full coherence and both semi-coherence and cross-correlations~\citep{bib:xcorrmethod1}, which can be viewed as subsets of the double sum.
A semi-coherent sum of powers from individual detectors can be represented by
\begin{equation}
  P_{\rm semi-coherent} = \sum_{I=1}^2\sum_{i=1}^{\NDFT} \DIi\DIi^*,
\end{equation}
\noindent while cross-correlation of simultaneous amplitudes (see equation~\ref{eqn:crosscorrelation}) from the two detectors is proportional to
\begin{equation}
  P_{\rm cross-correlation} = \sum_{(I\ne J)=1}^2\sum_{i=1}^{\NDFT} \DIi\DJi^*\expphiIiJj.
\end{equation}

This last relation suggests the possibility of following up an interesting search outlier from the first stage of a simultaneous-segment cross-correlation
search by increasing the number of terms kept from the full double sum, allowing non-simultaneous cross terms and allowing self-correlation terms.
For example, allowing an offset of up to
$\Nlag$ segment durations would yield:
\begin{equation}
  P_{\rm cross-correlation(\Nlag)} = \sum_{I,J=1}^2\sum_{\begin{array}{c}i,j=1;\\|i-j|\le\Nlag\end{array}}^{\NDFT} \DIi\DJj^*\expphiIiJj.
\end{equation}
\noindent This approach can also be used at the first stage if rapid frequency evolution of the source, such as in a short-period
binary system or in young object, argues for short DFT coherence times.

Another, related approach is to define a ``loosely coherent'' detection statistic as a subset of the original sum for which phase correlation
between nearby (small-lag) DFT coefficients is favored (as opposed to the completely random relation allowed by semi-coherent sums).
To illustrate\footnote{For simplicity, the initial implementation of loose coherence is described here. Later
refinements~\citep{bib:LooseCoherenceWellModeledSignals,bib:LooseCoherenceMediumScale,bib:FalconPaper}
led to substantial performance improvements.},
consider for simplicity a sum restricted to a single detector. Assume the phase correction applied to a product
of DFT coefficients for segments separated by a single segment lag is taken to be unknown but uniformly distributed in probability between
$-\delta$ and $+\delta$. A useful detection statistic can then be formed by~\citep{bib:loosecoherence}
\begin{eqnarray}
  P_{\rm loose-coherence} & = & {1\over(2\delta)^{\Nseg-1}}\int_{-\delta}^{+\delta}d\Delta\phi_{1,2}\int_{-\delta}^{+\delta}d\Delta\phi_{2,3}...\int_{-\delta}^{+\delta}d\Delta\phi_{\Nseg-2,\Nseg-1} \nonumber\\
 & &  \sum_{i,j}^{\Nseg}\DIi\DIj^*e^{-i(\Delta\phi_{i',i'+1}+\Delta\phi_{i'+1,i'+2}+...+\Delta\phi_{j'-1,j'})},
\end{eqnarray}
\noindent where $i' (j') = {\rm min} ({\rm max}) \{i,j\}$ and $\Delta\phi_{m,n}=\phi_n-\phi_m$. Evaluation of the integrals leads to
\begin{equation}
  P_{\rm loose-coherence} =   \sum_{i,j}^{\Nseg}\DIi\DIj^*\left({\sin(\delta)\over\delta}\right)^{|i-j|},
\end{equation}
where the factor $\left({\sin(\delta)\over\delta}\right)^{|i-j|}$ weights adjacent-lag products more heavily than those with longer lags (and yields unity for $i=j$).
The single-detector semicoherent power sum is recovered by setting $\delta=\pi$.
In practice, to reduce computational cost, the factor is replaced by a discrete kernel function that truncates terms for which the
weight contribution is too small to warrant the additional operations.
A generalized version of this detection statistic, including more than one detector
and non-integer segment lags, has been used in multi-stage searches with decreasing $\delta$ at each stage, \eg,
$\delta=\pi \rightarrow \pi/2 \rightarrow \pi/4 \rightarrow \pi/8$. Each reduction in $\delta$ brings
an increase in computational cost per parameter space volume as more terms are retained in the sum and the
parameter space is searched more finely (``zooming in'').

\subsection{Barycentering and coherent signal demodulation}

Taking into account the phase/frequency modulations of Eqn.~\ref{eqn:phasedefinition} due
to detector translational motion and
the antenna pattern modulations embodied in Eqns.~\ref{eqn:jksone}-\ref{eqn:jkstwo} due to detector orientation changes
requires an accurate model of relevant solar system motions. As noted above, the gravitational wave community has
adopted techniques of pulsar astronomy researchers, with many LIGO data searches using the TEMPO 2 program~\citep{bib:tempo} as a guide and for
cross-checking~\citep{bib:lal,bib:cwtargetedS1}. Correction for the Earth's and a detector's motions with respect
to the solar system barycenter is called barycentering. Independently, Virgo analysts developed another barycentering
software package~\citep{bib:VirgoBarycentering}, also checked against TEMPO 2 and the LIGO software. These packages choose
steps in time fine enough to allow reliable interpolation of detector motion between sampled times.

Several approaches to incorporating the corrections have been developed for continuous gravitational wave searches. These include
time-domain heterodyning, Fourier-domain decomposition and hybrid techniques, as discussed below.
More recently, techniques have been developed for more computationally efficient barycentering
for use in targeted searches~\citep{bib:PitkinBarycentering} and
all-sky searches~\citep{bib:SauterBarycentering}.

\subsubsection{Heterodyne method}

Since CW signals are inherently quite narrowband with respect to deviations from idealized models,
a heterodyning procedure using a base frequency near the nominal signal frequency,
followed by a low-pass filter allows a large reduction in the number of data samples required to capture the signal modulations.
Conceptually, if one has a pure signal $h(t)$ that can be expressed as a slowly varying amplitude function $A(t)$ times a
sinusoid of base frequency $\fbase$, namely, $h(t) = \Re\left\{A(t)e^{\imag(2\,\pi\fbase \tssb(t)+\phi_0)}\right\}$,
one can apply the following heterodyne for a base frequency $\fbase$:
\begin{equation}
  H_{\fbase}(t) \equiv h(t)\>e^{-\imag2\,\pi\fbase \tssb(t)} = A(t)\,e^{\imag\phi_0},
\end{equation}
\noindent where $\tssb(t$) relates the SSB time to detector time.
This approach allows the heterodyned function to have a low effective bandwidth. Applying a low-pass filter and then downsampling
allows a large reduction in data volume while preserving signal fidelity.

In practice, the heterodyne used in targeted CW searches applies not a pure sinusoid factor, but rather a slowly modulated sinusoidal
phase $\phi_{\rm model}(t)$ dependent on the topocentric (observatory-centric) time $t$, a model
that includes the effects of Eqns.~\ref{eqn:phaseevolution} and~\ref{eqn:phasedefinition} on signal frequency evolution and propagation delays:
\begin{equation}
  H_{\rm model}(t) \equiv d(t)\>e^{-\phi_{\rm model}(t)},
\end{equation}
\noindent where it is assumed the signal is well approximated by the model: $h(t) = \Re\left\{f(t)e^{\imag(\phi_{\rm model}(t))}\right\}$,
and the data stream $d(t)$ contains $h(t)$ and a (much larger-amplitude) random noise $n(t)$.
The resulting heterodyne product $H_{\rm model}(t)$ can then be interrogated for consistency with noise in addition to
a signal amplitude function subject to antenna pattern modulations.
Small residual deviations from the model (``timing noise'') measured empirically from electromagnetic pulsation observations
can also be taken into account straightforwardly.

This technique is well suited to searches for known pulsars, for which the nominal frequency is precisely
known from ephemeris measurements. For example, it has been customary in many LIGO and Virgo targeted searches to
heterodyne, low-pass filter and then downsample to 1 data sample per minute, starting from a raw data
stream of 16384 Hz. This technique assumes the residual intrinsic bandwidth of the signal following the heterodyne is no greater than
the Nyquist frequency of 8.3 mHz, which is an excellent approximation for effects due to Earth / detector motion.
This specific implementation is not as well suited to wide-parameter searches, for which the bandwidth must be increased or
many distinct heterodynes be carried out.

The resulting heterodyned data samples have had frequency / phase modulations due to detector motion removed, but
they retain antenna pattern due to detector rotation about the Earth's spin axis. Section~\ref{sec:targeted} below
presents a Bayesian analysis method for such samples~\citep{bib:DupuisWoan}.

\subsubsection{Resampling methods}

An alternative barycentering technique to heterodyning is to ``resample'' the detector data in order to transform
it into SSB time~\citep{bib:SchutzBlairbook,bib:JKS,bib:stackslide1}. A mundane but difficult nuisance is that data samples uniformly in detector frame time is {\it not}
uniformly sampled in SSB time, making it difficult to apply conventional discrete Fourier transforms to the
SSB samples. Two distinct methods have been used to date in CW analysis, to make the SSB samples uniform in time.
The first~\citep{bib:PatelResampling,bib:CrossCorrResampling} uses spline interpolation of the non-uniformly sampled data to create uniformly
sampled data. In practice, this interpolation is carried out on heterodyned subbands, much wider than those used
in targeted searches, but much narrower than the full bandwidth of the original data collected.
Another method~\citep{bib:VirgoResampling,bib:SinghalEtalResampling} is based on selective data sample deletions and duplications, where
narrow bands of data are temporarily upsampled to much higher frequencies, allowing smaller errors when extra samples
are deleted or duplicated as SSB time appears to run faster or slower than detector-frame time (as defined by
successive gravitational wavefronts), depending on the relative velocity of the detector with respect to the source.
As for the heterodyne method, the result in both methods is a data stream for which detector translational motion has been
corrected, but which still contains antenna pattern modulations from daily detector rotation.

\subsubsection{Dirichlet kernel method}

An alternative method can be applied in the Fourier domain by breaking the observation time into segments of
short-enough duration that the signal frequency has negligible evolution during that duration, that is,
the frequency change during the time $\Tseg$ is small relative to intrinsic frequency resolution of a
discrete Fourier transform over that duration: ${1\over\Tseg}$. In a templated search for a particular signal,
the frequency for that segment is known, and a Dirichlet filter~\citep{bib:Dirichlet} can be applied to the DFT coefficients in
a narrow band surrounding the nominal frequency (\eg, $\pm$4 DFT bins), using the expected weights for those
bins for the nominal central frequency. 

\begin{figure}[t!]
\begin{center}
\includegraphics[width=13.cm]{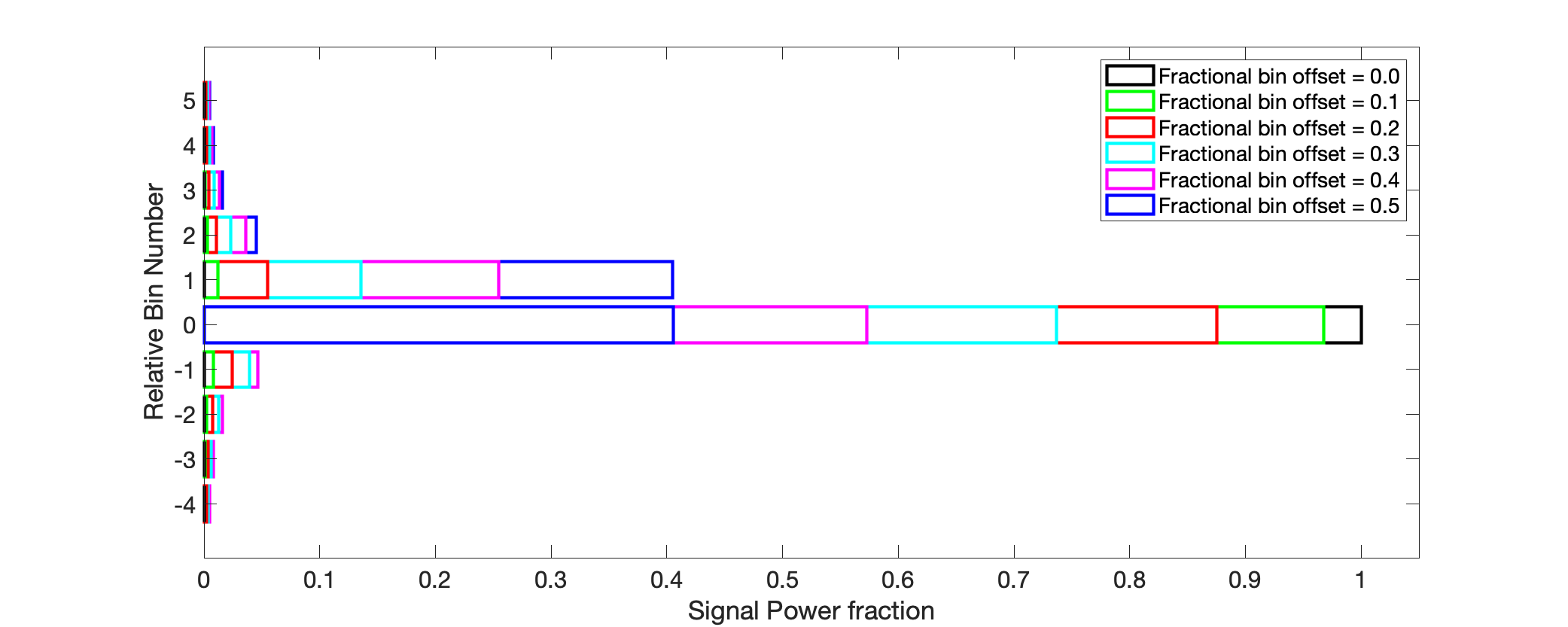}
\caption{Visual representation of the fractional power values listed in Table~\ref{tab:Dirichlet}.}
\label{fig:Dirichlet}
\end{center}
\end{figure}

For a bin-centered signal and rectangular windowing, the filter would
be a Kronecker delta, but in general, spectral leakage favors use of a handful of neighboring bins, to
recover the full signal strength. By coherently combining the resulting extracted complex coefficients from
the observed segments, one can achieve a full, coherent demodulation of the signal\footnote{An algorithm
commonly used is known as LALDemod~\citep{bib:WilliamsSchutzDemod,bib:PrixFstatNote}.}.
Table~\ref{tab:Dirichlet}
shows example power fractions in adjacent DFT bins (using rectangular windowing)
for a monochromatic signal that is bin-centered or offset positively from the bin center by bin fractions
of 0.1-0.5 in increments of 0.1. One sees that the bulk of signal power can be recovered by a modest number
of neighboring bins. Figure~\ref{fig:Dirichlet} shows a visual representation of the values in Table~\ref{tab:Dirichlet}.

\def\colone#1{\color{black}#1}
\def\coltwo#1{\color{green}#1}
\def\colthree#1{\color{red}#1}
\def\colfour#1{\color{cyan}#1}
\def\colfive#1{\color{magenta}#1}
\def\colsix#1{\color{blue}#1}
\begin{table}
  \begin{center}
  \begin{tabular}{|r|c|c|c|c|c|c|}
    \hline
    & \multicolumn{6}{c|}{Fractional bin offset of signal from bin center} \\
    & \multicolumn{1}{c}{\bf \colone{0.0}}   & \multicolumn{1}{c}{\bf \coltwo{0.1}} & \multicolumn{1}{c}{\bf \colthree{0.2}} & \multicolumn{1}{c}{\bf \colfour{0.3}} & \multicolumn{1}{c}{\bf \colfive{0.4}} & {\bf \colsix{0.5}}  \\
    \hline
    Bin +5:        &  \colone{0}          & \coltwo{0.0004}    & \colthree{0.0015}    & \colfour{0.0030}    & \colfive{0.0043}    & \colsix{0.0050} \\
    Bin +4:         & \colone{0}          & \coltwo{0.0006}    & \colthree{0.0024}    & \colfour{0.0048}    & \colfive{0.0071}    & \colsix{0.0083} \\
    Bin +3:         & \colone{0}          & \coltwo{0.0012}    & \colthree{0.0045}    & \colfour{0.0091}    & \colfive{0.0136}    & \colsix{0.0162} \\
    Bin +2:         & \colone{0}          & \coltwo{0.0027}    & \colthree{0.0108}    & \colfour{0.0229}    & \colfive{0.0358}    & \colsix{0.0450} \\
    Bin +1:         & \colone{0}          & \coltwo{0.0119}    & \colthree{0.0547}    & \colfour{0.1353}    & \colfive{0.2546}    & \colsix{0.4053} \\
    {\bf Bin$\>\>$ 0:}   & \colone{1}          & \coltwo{0.9675}    & \colthree{0.8751}    & \colfour{0.7368}    & \colfive{0.5728}    & \colsix{0.4053} \\
    Bin $-$1:         & \colone{0}          & \coltwo{0.0080}   & \colthree{0.0243}    & \colfour{0.0392}    & \colfive{0.0468}    & \colsix{0.0450} \\
    Bin $-$2:         & \colone{0}          & \coltwo{0.0022}   & \colthree{0.0072}    & \colfour{0.0125}    & \colfive{0.0159}    & \colsix{0.0162} \\
    Bin $-$3:         & \colone{0}          & \coltwo{0.0010}   & \colthree{0.0034}    & \colfour{0.0061}    & \colfive{0.0079}    & \colsix{0.0083} \\
    Bin $-$4:         & \colone{0}          & \coltwo{0.0006}   & \colthree{0.0020}    & \colfour{0.0036}    & \colfive{0.0047}    & \colsix{0.0050} \\
    \hline
    {\bf Sum:}     &  \colone{1.000}      & \coltwo{0.9982}    & \colthree{0.9933}    & \colfour{0.9874}    & \colfive{0.9825}    & \colsix{0.9807} \\
    \hline
  \end{tabular}
  \caption{Fractional powers in neighboring DFT bins (rectangularly windowed) for a monochromatic signal
    with a frequency that ranges from bin-centered (bin 0 of the 10 bins shown) to a positive offset of a half-bin.
    \label{tab:Dirichlet}  
    The last row gives the total fractional power in these 10 bins.}
  \end{center}
\end{table}

\subsubsection{Time-domain parameter extraction}
\label{sec:timedomainPE}

After frequency demodulation for detector translational motion, one has a highly reduced
band-limited, data stream (time-domain for heterodyne or resampling,
Fourier-domain for the Dirichlet filter) implicitly containing the amplitude modulations embodied in
Eqns.~\ref{eqn:jksone}-\ref{eqn:jkstwo}, which can be considered deterministic functions of time,
dependent upon the putative signal parameters. In the case of a source of known position and phase evolution but unknown orientation, as for
many known pulsars, the unknown source parameters can be taken to be the strain amplitude $h_0$, the signal phase constant $\phi_0$,
the inclination angle $\iota$ and the polarization angle $\psi$.
The data stream can then be analyzed to extract those parameters.

For direct time-domain analysis, the principal method used to date for LIGO and Virgo data analysis has been a
Bayesian inference~\citep{bib:cwtargetedS1,bib:DupuisWoan}. In brief, the heterodyned data samples $\{B_k\}$ can
be expressed as complex quantities, with signal template expectations: (adopting the notation of~\citep{bib:cwtargetedS1}):
\begin{eqnarray}
  y(t_k,\vec a) & = & {1\over4}F_+(t_k;\psi)h_0(1+\cos^2(\iota))e^{\imag2\phi_0} \nonumber \\
                &   & -{\imag\over2}F_\times(t_k;\psi)h_0\cos(\iota)e^{\imag2\phi_0},
\end{eqnarray}
\noindent where $\vec a$ is a vector with components $(h_0,\iota,\psi,\phi_0)$ and $t_k$ is the time stamp
of the $k$th sample. 

With a set of priors on the $\vec a$ parameters
one can extract a joint posterior probability density function for these parameters:
\begin{eqnarray}
  p(\vec a|\{B_k\}) & \propto &
  p(\vec a) \exp\left[-\sum_k {\Re\left\{B_k-y(t_k;\vec a)\right\}^2\over2\,\sigma_{\Re\{B_k\}}^2}\right] \nonumber \\
                  &  & \times \exp\left[-\sum_k {\Im\left\{B_k-y(t_k;\vec a)\right\}^2\over2\,\sigma_{\Im\{B_k\}}^2}\right],
\end{eqnarray}
\noindent where $p(\vec a)$ is the prior on $\vec a$ (uniform for $\cos(\iota)$, $\psi$ and $\phi_0$ and $h_0$),
and $\sigma_{\Re\{B_k\}}^2$ and $\sigma_{\Im\{B_k\}}^2$ are the variances on the real and imaginary parts of $B_k$. This posterior
distribution can be examined for evidence of a signal present. In the absence of a signal, an upper limit on strain amplitude
$h_0$ can be found via marginalization over the other three signal parameters to obtain a marginalized posterior:
\begin{equation}
  p(h_0|\{B_k\}) \propto \int\!\!\!\!\int\!\!\!\!\int p(\vec a|\{B_k\})\,d\iota\, d\psi\, d\phi_0, 
\end{equation}
normalized so that $\int_0^\infty p(h_0|\{B_k\})dh_0 = 1$. Unlike a frequentist confidence level, the resulting curve vs
$h_0$ represents the distribution of degree of belief in any particular value of $h_0$, given the signal model, the
parameter priors and the data observations $\{B_k\}$. One can derive 
a 95\%\ credible Bayesian upper limit $\hbayes$ for which the probability lies below $\hbayes$ via
\begin{equation}
  0.95 = \int_0^{\hbayes} p(h_0|\{B_k\})dh_0.
\end{equation}

The combined posterior distribution from multiple, independent detectors
can be obtained via the product of the individual likelihoods~\citep{bib:DupuisWoan}.
In the event that estimates of $\iota$ and $\psi$ can be inferred from electromagnetic measurements of the source,
\eg, from images of jets assumed to be emitted along the spin axis of a star, then the precision on $h_0$ can
be improved by assigning much narrower priors to the parameters.

\subsubsection{Five-vector method}
\label{sec:fivevectormethod}

The so-called ``Five-vector'' method exploits the property that the complexity in Eqns.~\ref{eqn:cwhdefinition} and~\ref{eqn:jksone}-\ref{eqn:jkstwo}
can be distilled down to five terms~\citep{bib:fivevector}
\begin{equation}
  h(t) = h_0\vec A\cdot\vec We^{\imag(\omega_0 t+\phi_0)},
\end{equation}
\noindent where $\omega_0$ is the signal frequency in the SSB frame,
where $\vec A$ can be decomposed into plus- and cross-polarized terms that depend on complex amplitudes $H_+$ and $H_\times$:
\begin{equation}
  \vec A = H_+\vec A^+ + H_\times \vec A^\times,
\end{equation}
\noindent and where $\vec A^+$ and $\vec A^\times$ can be expressed in terms of trigonometic functions, using
Eqns.~\ref{eqn:jksone}-\ref{eqn:jkstwo} (see~\citep{bib:fivevector} for detailed expressions). The vector $\vec W$
is a five-component set of basis functions, indexed by $k = [-2, -1, 0, 1, 2]$:
\begin{equation}
  \vec W_k = e^{-\imag k\Theta},
\end{equation}
\noindent where $\Theta$ is the detector's local sidereal time in radians.

The data stream $x(t)$ too can be decomposed using these basis functions:
\begin{equation}
  \vec X = \int_Tx(t)\vec W e^{-\imag\omega_0 t}dt.
\end{equation}
One can then construct a detection statistic using a weighted sum of the squared projections:
\begin{equation}
  S = c_+|\hat h_+|^2 +c_\times|\hat h_\times|^2,
\end{equation}
where the projections are defined by
\begin{equation}
  \hat h_+ = {\vec X\cdot\vec A^+\over|\vec A^+|^2};  \qquad \hat h_\times = {\vec X\cdot\vec A^\times\over|\vec A^\times|^2}.
\end{equation}
Empirically~\citep{bib:fivevector}, it is found that best performance for known $\iota$, $\psi$ can be
obtained with the weightings: $c_{+,\times} = |\vec A^{+,\times}|^4$, while estimation of signal amplitude can be
obtained from
\begin{equation}
  \hat h_0 = \sqrt{|\hat h_+|^2+|\hat h_\times|^2}.
\end{equation}

\subsubsection{The \fstatistic}
\label{sec:fstatistic}

The most pervasive detection statistic used in broadband CW searches can also be used for targeted searches, namely
the \fstatistic~\citep{bib:JKS}. 
As above, the \fstatistic\ is constructed to take into account not only the frequency / phase modulation of the
detector's translational motion (using time-domain or frequency-domain techniques), but also the amplitude modulation
from daily detector rotation.

It is constructed from a general maximum likelihood approach, where the data is taken to be a sum of random noise $n(t)$ and
a signal $h(t)$:
\begin{equation}
  x(t) = n(t) + h(t),
\end{equation}
where $h(t)$ from Eqns.~\ref{eqn:jksone}-\ref{eqn:jkstwo} can be written\footnote{In the original \fstatistic\ article~\citep{bib:JKS},
  a two-component signal model is assumed, corresponding to frequencies at once and twice the source rotation frequency. 
  Only the component at twice the rotational frequency is considered here
  where the wobble angle $\theta$ in~\citep{bib:JKS} is taken to be $\pi/2$ for a triaxial ellipsoid, allowing a simplification of notation.}
\begin{equation}
  \label{eqn:hoftdef}
h(t) = \sum_{i=1}^4 A_i h_i(t),
\end{equation}
\noindent where the coefficients $A_i$ are inferred from Eqns.~\ref{eqn:jksone}-\ref{eqn:jkstwo}: 
\begin{eqnarray}
  A_1 & = & h_0\sin(\zeta)\Biggl[{1\over2}(1+\cos^2(\iota)\cos(2\psi)\cos(2\,\Phi_0) \nonumber\\
      &   & \qquad\qquad -\cos(\iota)\sin(2\psi)\sin(2\Phi_0)\Biggr], \\
  A_2 & = & h_0\sin(\zeta)\Biggl[{1\over2}(1+\cos^2(\iota)\sin(2\psi)\cos(2\,\Phi_0) \nonumber\\
      &   & \qquad\qquad +\cos(\iota)\cos(2\psi)\sin(2\Phi_0)\Biggr], \\
  A_3 & = & h_0\sin(\zeta)\Biggl[-{1\over2}(1+\cos^2(\iota)\cos(2\psi)\sin(2\,\Phi_0) \nonumber\\
      &   & \qquad\qquad -\cos(\iota)\sin(2\psi)\cos(2\Phi_0)\Biggr], \\
  A_4 & = & h_0\sin(\zeta)\Biggl[-{1\over2}(1+\cos^2(\iota)\sin(2\psi)\sin(2\,\Phi_0) \nonumber\\
      &   & \qquad\qquad +\cos(\iota)\cos(2\psi)\cos(2\Phi_0)\Biggr],
\end{eqnarray}
\noindent and the time-dependent functions $h_i$ have the form:
\begin{eqnarray}
  \label{eqn:hidefinitions1}
  h_1 = a(t)\cos(2\Phi(t)), &  \qquad & h_2(t) = b(t)\cos(2\Phi(t)) \\
  \label{eqn:hidefinitions2}
  h_3 = a(t)\sin(2\Phi(t)), &  \qquad & h_4(t) = b(t)\sin(2\Phi(t)),
\end{eqnarray}
\noindent where $\Phi(t)$ is the phase of the signal, including modulations.

\noindent A log-likelihood function $\log(\Lambda)$ is constructed via:
\begin{equation}
\log(\Lambda) = (x|h) - {1\over2}(h|h),
\end{equation}
\noindent where the scalar product $(\>|\>)$ is defined by a filter matched to the detection noise spectrum:
\begin{equation}
  (x|y) := 4\Re\left\{ \int_0^\infty{\tilde x(f)\tilde y^*(f) \over S_h(f) }df \right\},
\end{equation}
\noindent where $\tilde{ }$ denotes a Fourier transform, $*$ is the complex conjugation, and $S_h$ is the one-sided power spectral density.

Following~\citep{bib:JKS}, the narrowband signal allows, in principle, conversion of the scalar product to a time-domain expression:
\begin{equation}
  (x|h) \approx {2\over S_h(f_0)}\int_0^{\Tobs} x(t)h(t)dt,
\end{equation}
\noindent where stationarity of the noise over the observation period $\Tobs$ is implicitly assumed, which unfortunately, is rarely a good assumption
for interferometers at the frontier of technology. Nonetheless, practical implementations of the \fstatistic\ are not limited by this assumption.
Defining a time-domain scalar product:
\begin{equation}
(x||y) := {2\over\Tobs}\int_0^{\Tobs}x(t)y(t)dt,
\end{equation}
\noindent the log-likelihood function can be approximated via
\begin{equation}
  \label{eqn:loglikelihood}
  \log(\Lambda) \approx {\Tobs\over S_h(f_0)}\left[(x||h)-{1\over2}(h||h)\right],
\end{equation}
\noindent which is proportional to a normalized log-likelihood $\log(\Lambda')$:
\begin{equation}
  \log(\Lambda') = (x||h) - {1\over2}(h||h),
\end{equation}
\noindent which does not depend explicitly on the spectral noise density. The signal depends linearly
on the four amplitudes $A_i$ and can, in principle, be extracted from a likelihood maximization:
\begin{equation}
  {\partial \log\Lambda'\over\partial A_i} = 0,
\end{equation}
\noindent from which a set of linear algebraic equations can be derived:
\begin{equation}
  \sum_{i=1}^4\Mij A_j = (x||h_i),
\end{equation}
\noindent where the components of the matrix $\Mij$ are given by
\begin{equation}
  \Mij := (h_i||h_j). 
\end{equation}
Cross-terms of the $\cos(\Phi(t))$ and $\sin(\Phi(t))$ terms in Eqns.~\ref{eqn:hidefinitions1}-\ref{eqn:hidefinitions2}
can be neglected in the time integrations. The surviving terms can be expressed:
\begin{eqnarray}
  (h_1|h_1) \approx & (h_3|h_3) & \approx {1\over2}A, \nonumber\\
  (h_2|h_2) \approx & (h_1|h_4) & \approx {1\over2}B, \nonumber\\
  (h_1|h_2) \approx & (h_3|h_4) & \approx {1\over2}C, 
\end{eqnarray}
\noindent where $A:=(a||a)$, $B:=(b||b)$ and $C:=(a||b)$. With these approximations, the matrix $\M$ becomes
\begin{equation}
  \M = \left(\begin{array}{cc} \C & \Om \\ \Om & \C \\ \end{array} \right),
\end{equation}
\noindent where $\Om$ is a zero 2 $\times$ 2 matrix, and $\C$ is
\begin{equation}
  \C = {1\over2}\left(\begin{array}{cc} A & C \\ C & B \\ \end{array} \right),
\end{equation}
\noindent from which maximum-likelihood estimators $\tilde A_i$ of the true amplitudes $A_i$ can be obtained:
\begin{eqnarray}
  \tilde A_1 & = & 2{B(x||h_1)-C(x|h_2)\over D}, \nonumber \\
  \tilde A_2 & = & 2{A(x||h_2)-C(x|h_1)\over D}, \nonumber \\
  \tilde A_3 & = & 2{B(x||h_3)-C(x|h_4)\over D}, \nonumber \\
  \tilde A_4 & = & 2{A(x||h_4)-C(x|h_3)\over D},
\end{eqnarray}
\noindent where $D = AB-C^2$. Substituting these expressions into Eqns.~\ref{eqn:loglikelihood} leads to
the \fstatistic\ (denoted by $\TwoF$): 
\begin{eqnarray}
  \TwoF & = {\Tobs\over S_h(f_0)} \Biggl[
    & {B(x||h_1)^2+A(x||h_2)^2-2C(x||h_1)(x||h_2)\over D} \nonumber \\
     &   & + {B(x||h_3)^2+A(x||h_4)^2-2C(x||h_3)(x||h_4) \over D} \Biggr].
\end{eqnarray}
\noindent The quantity $\TwoF$ has a probability distribution of a chi-squared with
four degrees of freedom in the absence of a signal and that of a non-central chi-squared
with a non-centrality parameter:
\begin{equation}
  \label{eqn:lambdadefinition}
  \lambda \equiv d^2 = (h|h)
\end{equation}
\noindent where $d$ is proportional to signal amplitude~\citep{bib:JKS}. The probability distributions $p_{\rm noise}(\TwoF)$ and $p_{\rm signal+noise}(\TwoF;d)$ are hence:
\begin{eqnarray}
  p_{\rm noise}(\TwoF) & = & {1\over4}(\TwoF) e^{-(\TwoF)/2}, \\
  p_{\rm signal+noise}(\TwoF;d) & = & {\left(\TwoF\right)^{1\over2}\over d}I_1\left(d\sqrt{(\TwoF)}\right) e^{-{1\over2}(\TwoF)-{1\over2}d^2},
\end{eqnarray}
\noindent where $I_1$ is a modified Bessel function of the first kind (order 1).

As discussed in section~\ref{sec:algorithms}, $\TwoF$ can be used as a detection statistic, where
a threshold $\TwoF_0$ can be chosen to satisfy a desired false alarm probability:
\begin{eqnarray}
  {\rm CDF}_{\rm noise}[\TwoF_0] & = & \int_0^{\TwoF_0} {p_{\rm noise}(\TwoF)}\,d(\TwoF) = 1-\alpha, \\
                             & = & 1 - \left(1+\TwoF_0+{1\over2}\TwoF_0^2+{1\over6}\TwoF_0^3\right)e^{-\TwoF_0},
\end{eqnarray}
\noindent and where the probability of detection for a given $d$ is
\begin{equation}
  P_{\rm detection}(d,\TwoF_0) = \int_{\TwoF_0}^\infty p_{\rm signal+noise}(\TwoF;d)\,d(\TwoF).
\end{equation}

The formalism above describes a time-domain implementation~\citep{bib:JKS,bib:tdfstatistic},
but a narrowband frequency implementation~\citep{bib:PrixFstatNote} has been used extensively in LIGO searches.

In searches for known pulsars for which optical or X-ray observations of pulsar wind nebulae allow inference
of $\iota$ and $\psi$, a modified version of the \fstatistic\ known as the \gstatistic\ can be applied
to gain slightly in sensitivity, depending on the stellar orientation~\citep{bib:gstatisticmethod}.

Although originally derived in a frequentist, log-likelihood framework, the \fstatistic\
can also be obtained in a Bayesian approach~\citep{bib:PrixKrishnan} with an unphysical prior (non-isotropic in stellar orientation),
an alternative framework that has received additional
study~\citep{bib:PrixGiampanisMessenger,bib:lineveto1,bib:WhelanEtalCoordinates,bib:DhurandharKrishnanWillis,bib:BeroWhelan,bib:WetteGeometry}.

\subsection{Semi-coherent signal demodulation}

Let's now consider a coarser demodulation, in which phase fidelity is not required for the full observation time. Instead,
the observation is broken into discrete segments of coherence time $\Tcoh$ which need not be contiguous with each other.
The segmentation reduces the fineness with which the parameter space (\eg, frequency, frequency derivatives, sky location) must be sampled,
leading to often dramatic reduction in computing cost to search a given parameter space volume, albeit with a degradation of achievable
strain sensitivity.

\subsubsection{The stack-slide method}
\label{sec:stackslide}

\begin{figure}[t!]
\begin{center}
\includegraphics[width=13.cm]{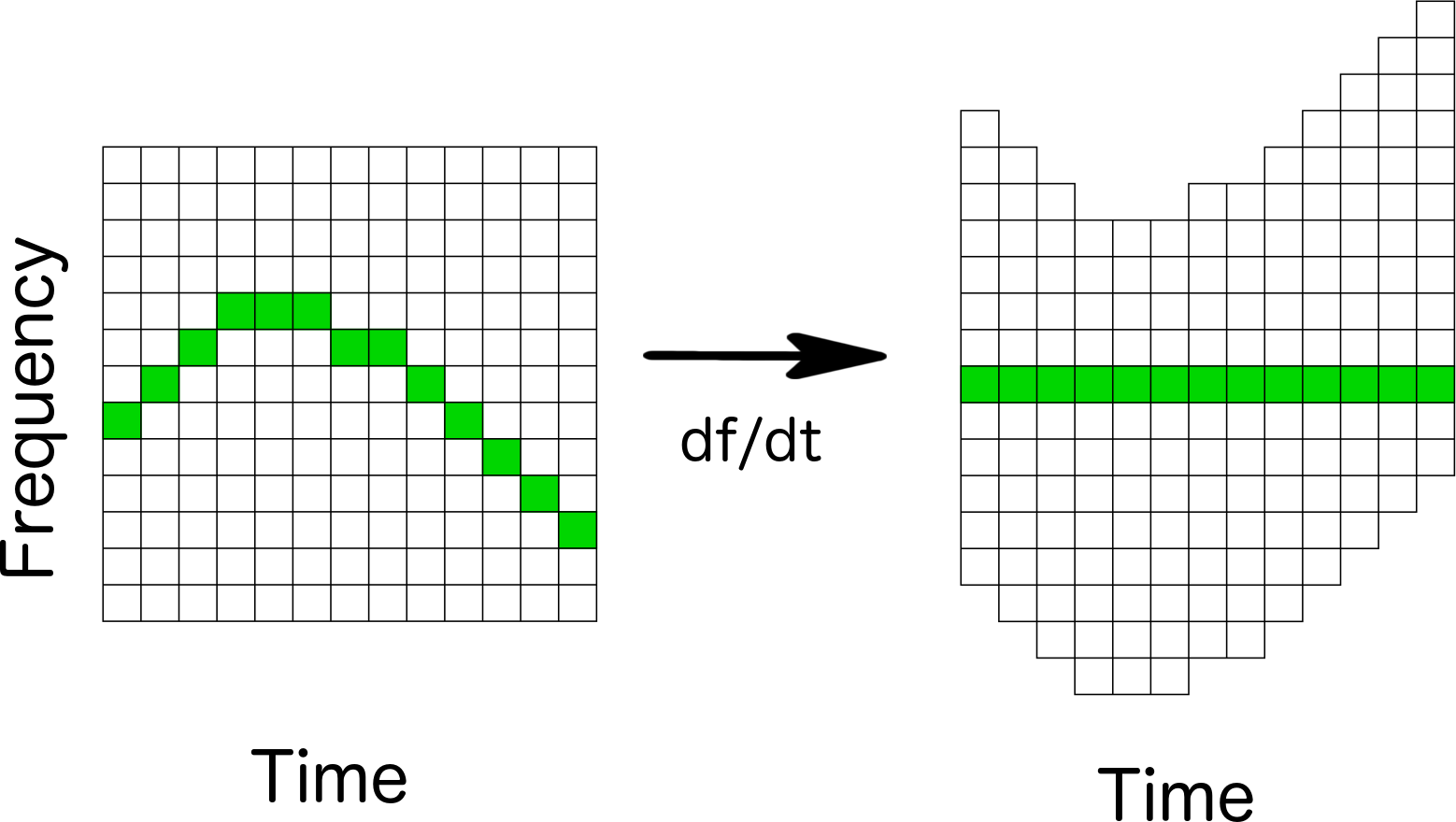}
\caption{Conceptual illustration of the ``stack-slide'' method in which rows of a spectrogram are shifted up or down in frequency to account
  for Doppler modulations.}
\label{fig:stackslidegraphic}
\end{center}
\end{figure}

For short-enough $\Tcoh$, no frequency demodulation need be applied within a single segment.
One can simply sum the strain power from each bin in an DFT containing the
frequency of the signal for that time interval. Figure~\ref{fig:stackslidegraphic} illustrates the simplest version of this approach,
known as ``stack-slide''~\citep{bib:stackslide1,bib:stackslideimplementation}.
In a spectrogram where each column represents DFT powers for a given $\Tcoh$, the
bin containing the signal frequency (indicated by the green square) varies in frequency from one column to the next. Correcting for
the frequency modulation by shifting columns up or down leads to the signal's power being contained
in a horizontal track in the demodulated spectrogram. For a relatively
narrow frequency band, the amount of vertical shift for a given column is nearly the same for all frequencies in the band, for a given
set of source parameters, including sky location. Hence by stacking powers across rows in the demodulated spectrogram, one can look for
an outlier indicating a signal.

To be concrete, define the power $\Pki$ to be the strain power spectral density measured in bin $i$ of DFT $k$, where the bin $i$ is the appropriate bin
after ``sliding'':
\begin{equation}
  \label{eqn:powerdef}
  \Pki = {2|\Dkidemod|^2\over\Tcoh}.
\end{equation}
Following~\citep{bib:cwallskyS4,bib:stackslideimplementation},
this power is renormalized to form a dimensionless quantity $\etaki$
\begin{equation}
  \etaki = {\Pki\over\Shki},
\end{equation}
\noindent where $\Shki$ is the one-sided power spectral density expected in the absence of signal.
This quantity differs from the $\rhoi^2$ defined in Eqn.~\ref{eqn:rhodefinition}, both
in the implicit demodulation associated with bin $i$ and in a factor of 2. Here $\etaki$ has an expectation
value of 1 in the absence of signal. 

The stack-slide detection statistic $\Pssi$ then is the average value of $\etaki$ over the $\NDFT$ DFT's used in the analysis:
\begin{equation}
  \Pssi = {1\over\NDFT} \sum_{k=1}^{\NDFT} \etaki.
\end{equation}
\noindent This quantity has an expectation value of 1 in the absence of signal and a variance of $1/\NDFT$.
Signal candidates are chosen based on exceeding a threshold corresponding to a false alarm probability $\alpha$,
from which detection sensitivity is determined from a desired false dismissal probability $\beta$.
Appendix B of~\citep{bib:cwallskyS4} details the statistical behavior. In brief, the quantity (similar to $\Ri$ of Eqn.~\ref{eqn:Ridef} above)
\begin{equation}
  \Pss = 2\NDFT\etaki
\end{equation}
has the probability density distribution of a non-central $\chi^2$ with $2\NDFT$ degrees of freedom and a non-centrality parameter
$2\NDFT<\!\!d^2\!\!>$ which is the expectation value of the estimator in Eqn.~\ref{eqn:lambdadefinition} when evaluated over a single DFT.
Hence the
probability density distribution for $\Pss$ follows:
\begin{eqnarray}
  p_{\rm signal+noise}(\Pss;\NDFT,d) & = & {I_{\NDFT-1}\left(\sqrt{\Pss\NDFT <\!d^2\!>}\right)\over\left(\NDFT<\!d^2\!>\right)^{\NDFT-1}} \nonumber \\
                                  & \times  & \Pss^{{\NDFT-1\over2}} e^{-\left(\NDFT+<\!{d^2\over2}\!>\right)}.
\end{eqnarray}
\noindent Numerical evaluation~\citep{bib:cwallskyS4} for $\alpha=0.01$ and $\beta=0.10$ leads (in the large $\NDFT$ limit)
to a sensitivity $<\!d^2\!>^{(90)} \approx 7.385 / \sqrt{\NDFT}$ and to a strain sensitivity {\it for a single template search}
of $h_0^{(90)} \approx 7.7\sqrt{\Sh}/\left(\Tcoh\Tobs\right)^{1/4}$, where $\Tobs$ refers here to the total observing time analyzed and where
stationary data is implicitly assumed. In practice, however, this method is applied to wide-parameter searches for which trials factors lead
to much worse strain sensitivities~\citep{bib:TenorioEtalLoudestCandidate}. \cite{bib:PrixShaltev} carry out a detailed analysis of maximizing sensitivity at fixed computational cost for
different stack-slide search configurations. 

\subsubsection{The Powerflux method}
\label{sec:PowerFlux}

The PowerFlux method~\citep{bib:cwallskyS4},
in its simplest form, is similar to the stack-slide method, with the following
refinements: 1) an explicit polarization is assumed for each signal template searched,
with an antenna pattern correction applied; 2) detection statistic variance is minimized in
the presence of non-stationary noise; and the detection statistic itself is a direct measure
of strain amplitude.

Using the same notation as above (see Eqn.~\ref{eqn:powerdef}), the PowerFlux detection statistic $\Rpf$ for a given set of orientation
parameters $\iota$ and $\psi$ is written:
\begin{equation}
  \label{eqn:PFdef}
  \Rpf = {2\over\Tcoh} { \sum_{i=1}^{\NDFT} \Wi \Pki / (\Fi)^2 \over \sum_{i=1}^{\NDFT} \Wi },
\end{equation}
\noindent where the weights are defined as
\begin{equation}
   \Wi = [(\Fi)^2]^2/\Shi^2,
\end{equation}
\noindent and where $\Fi$ is the antenna pattern weight calculated for the midpoint of the
time segment $i$ for the assumed polarization such that the detector amplitude response can be
written as $h_{{\rm det},i} = h_0\Fi$. In practice, searches have been carried out for
circular polarization ($\iota=0$ or $\pi$) and for particular linear polarization angles $\psi$
(with $\iota=\pi/2$) to define ``best-case'' and ``worst-case'' orientations, respectively.

The choice of weight definition comes from minimizing the variance of the strain amplitude
estimator $\Pki/(\Fi)^2$, where the noise (in the weak signal regime) is assumed to be dominated
in each time segment $i$  by a power spectral density $\Shi$ with underlying Gaussian distributions for
real and imaginary DFT components. Under that assumption, the variance of the noise is
proportional to $(\Shi)^2$. As a result, each term in the numerator of Eqn.~\ref{eqn:PFdef}
is proportional to $(\Fi)^2\Pki/\Shi^2$, which gives higher weight to segments with higher $\Fi$ magnitude
and lower noise $\Shi$, as one would wish. For a given polarization choice defined by $(\iota,\psi)$
the detection statistic $\Rpf$ is a direct measure of total strain power such that subtracting
the expectation value based on neighboring bin yields a direct estimator for signal power.

\subsubsection{Hough transform methods}
\label{sec:hough}

Hough transform methods refer, in practice, to an application of a pattern recognition algorithm
first developed for use in the 1960's by high energy particle physicists~\citep{bib:houghibm1,bib:houghibm2}
to reconstruct a charged particle's trajectory from discrete
positions (``hits''), measured by a tracking detector.
The method is best suited to
data that is ``sparse'' and for which a simple transformation from the raw measurements to
the signal parameter space can amplify the detection statistic. In the original application
to particle tracking, the hits were two-dimensional projections for which looking
for straight lines built out of all hit combinations
was computationally intensive (especially in the 1960's!).
To represent a straight line, instead of offset and slope,
the vector of its minimum distance to the origin, in polar coordinates
$(r,\theta)$, is used.
A point $(x,y)$ belonging to that line sets the relation
$r=x\cos\theta+y\sin\theta$ which is a sinusoidal curve in the $\theta$-$r$ plane.
Cells in that plane count how many curves pass within their boundaries,
and the most occupied cell identifies $(r,\theta)$ of the original track.

In the case of CW searches, two different Hough transform methods (``Sky Hough'' and ``Frequency Hough'')
have been used in recent years,
both of which accumulate excess power from frequency-demodulated DFTs. In the
Sky Hough method~\citep{bib:houghmethod}, the transformation is from a narrow frequency band and
frequency derivative to right ascension and declination, where broad patches of sky
are searched collectively. In the Frequency Hough method~\citep{bib:freqhough1,bib:freqhough2}, the transform
is from a time-frequency plane to a plane of frequency and frequency derivative. In each case, one searches
for a statistically significant excess among the pixels and applies a thresholding
to individual accumulated powers, in order to reduce computational cost in the accumulation.

The Hough number count is defined as a weighted sum of binary counts $\numi$:
\begin{equation}
n = \sum_{i=1}^{\NDFT} \wi \numi,
\end{equation}
\noindent where $\numi$ = 1 if the normalized segment power $\etaki$ exceeds a threshold $\etastar$
and zero otherwise,
and where the weights favor low-noise times and are optimized for circular polarization~\citep{bib:freqhough1,bib:cwallskyS4}:
\begin{equation}
\wi \propto {1\over\Shi} \left[(\Fplusi)^2+(\Fcrossi)^2\right], 
\end{equation}
\noindent with a normalization chosen to satisfy:
\begin{equation}
  \sum_{i=1}^{\NDFT} \wi = \NDFT.
\end{equation}

In the Sky Hough method~\citep{bib:houghmethod}, so-called ``Hough maps'' in right ascension and declination are created for each assumed frequency and frequency derivative,
where signal outliers produce ``hot'' pixels in the sky patch for which the map applies.
In the Frequency Hough method~\citep{bib:freqhough1,bib:freqhough2}, the Hough map is
created instead in the plane of frequency and frequency derivative for each
localized sky point. The primary motivations for this alternative mapping to parameter space are
reduction of inaccuracies arising from approximations and non-linearities in the mapping to the sky;
avoidance of artifact ``pileup'' in which certain regions of the sky are contaminated over subbands by particular narrowband artifacts;
and the possibility to use over-resolution in frequency, at negligible additional computational cost~\citep{bib:freqhough1}.

Regardless of the choice of parameter space mapping, the statistical character of the Hough number counts
is governed by the value of the threshold used to define the binary counts $\numi$.
The mean number count in the absence of a signal is $\nbar = \NDFT p$, where $p$ is the probability that the normalized power
$\etaki$ exceeds a threshold value $\etastar$. For unity weighting, the standard deviation is
$\sigma_{\nbar} = \NDFT p(1-p)$. For the more general weighting, this becomes:
\begin{equation}
  \sigma = \left[p(1-p)\sum_{i=1}^{\NDFT}\wi^2\right]^{1/2}.
\end{equation}
\noindent For $\NDFT\gg1$, the underlying distribution can be approximated as Gaussian,
in which case a threshold $\nth(\alpha)$ corresponding to a false alarm rate $\alpha$ is
given by~\citep{bib:houghmethod}
\begin{equation}
  \nth = \NDFT\,p+\sqrt{2}\,\sigma\, \erfc^{-1}(2\alpha),
\end{equation}
\noindent where it is natural to regard the significance of a given measured $n$ to
be
\begin{equation}
  s = {n-\nbar\over\sigma}.
\end{equation}
In \citep{bib:houghmethod,bib:cwallskyS2,bib:cwallskyS4} an optimal choice of the normalized power threshold parameter
is found to be $\etastar\approx 1.6$, for which $p=e^{-\etastar} \approx0.2$. 

One can compute~\citep{bib:cwallskyS4} a sensitivity $h_0^{1-\beta}(\alpha)$ for
a false dismissal probability $\beta$ and false alarm probability $\alpha$:
\begin{equation}
  h_0^{1-\beta}(\alpha) \approx 3.38\mathcal(S)^{1/2} \left({||{\bf w}||\over{\bf w\cdot X}}\right)^{1/2}\,\sqrt{1\over\Tcoh},
\end{equation}
\noindent where $||{\bf w}|| = \sum_{i=1}^{\NDFT}\wi^2$ and
\begin{eqnarray}
  \mathcal{S} & = & \erfc^{-1}(2\alpha)+\erfc^{-1}(2\beta), \\
  X_i & =  & {1\over \Shi} \left[\left(F_+^i\right)^2+\left(F_\times^i\right)^2\right],
\end{eqnarray}
\noindent and where $F_{+/\times}^i$ refer to the antenna pattern functions for the $+$ and $\times$ polarizations evaluated
at the midpoint of time segment $i$.

Other improvements to the Sky Hough method have included incorporating a hierarchical approach~\citep{bib:SkyHoughHierarchical},
adaptation to a search for stars in binary systems~\citep{bib:binaryskyhough1} (see section~\ref{sec:allskybinary}),
clustering of outliers~\citep{bib:skyhoughclustering} and systematic outlier
follow-up~\citep{bib:TenorioEtal}.

\subsubsection{The Stacked \fstatistic\ method}
\label{sec:stackedfstatistic}

The semi-coherent approach used above (in various approaches) with DFT coefficients can also be applied to
longer segments of time for each of which the coherent \fstatistic\ is computed. This approach permits deeper
sensitivity since the \fstatistic\ can be computed without degradation of signal coherence for arbitrarily long
periods of time. The disadvantage is that the much finer resolution in parameter space associated
with such sensitivity leads to much greater computational cost, coming from the fine stepping needed within each
segment and from the mapping with negligible signal loss from one segment to the next. A variety of \fstatistic\ ``stacking''
methods\footnote{``Stacking'' the \fstatistic\ values is more subtle than in the stacking used in
the stack-slide and other semi-coherent methods based on summing DFT powers because the demodulations to obtain
the \fstatistic\ values differ across time segments.}
have been implemented over the years, both inside and outside of the framework of the Einstein@Home
distributed computing system (see section~\ref{sec:allsky}).
When computing the \fstatistic\ over short time segments, a modified variation, the \fabstatistic, which avoids
degeneracy due to minimal antenna pattern modulation
can be more effective~\citep{bib:CovasPrixModFstat}.

Many of the considerations discussed in semi-coherent summing of DFT power have analogs in \fstatistic\ summing,
including the use of thresholding and the use of Hough transform mapping. Particular implementations will be discussed
below in sections~\ref{sec:templatesdirected} and \ref{sec:allsky}. One critical issue in these computationally costly searches is the optimum
placement of signal templates in parameter space, to be discussed next, more generally.
Another important consideration is clustering of initial outlier candidates~\citep{bib:SteltnerEtalClustering}
to reduce computational cost in hierarchical searches
prior to follow up with deeper search algorithms.

\subsection{Template placement}
\label{sec:templates}

Computationally demanding searches must choose step sizes in signal parameter space, with finer spacing leading
to greater cost, in general. The choices are typically governed by what is considered an acceptable maximum ``mismatch'',
normally parametrized by the fractional decrease in detection statistic for a given offset in parameter space.

\begin{figure}[t!]
\begin{center}
\includegraphics[width=13.cm]{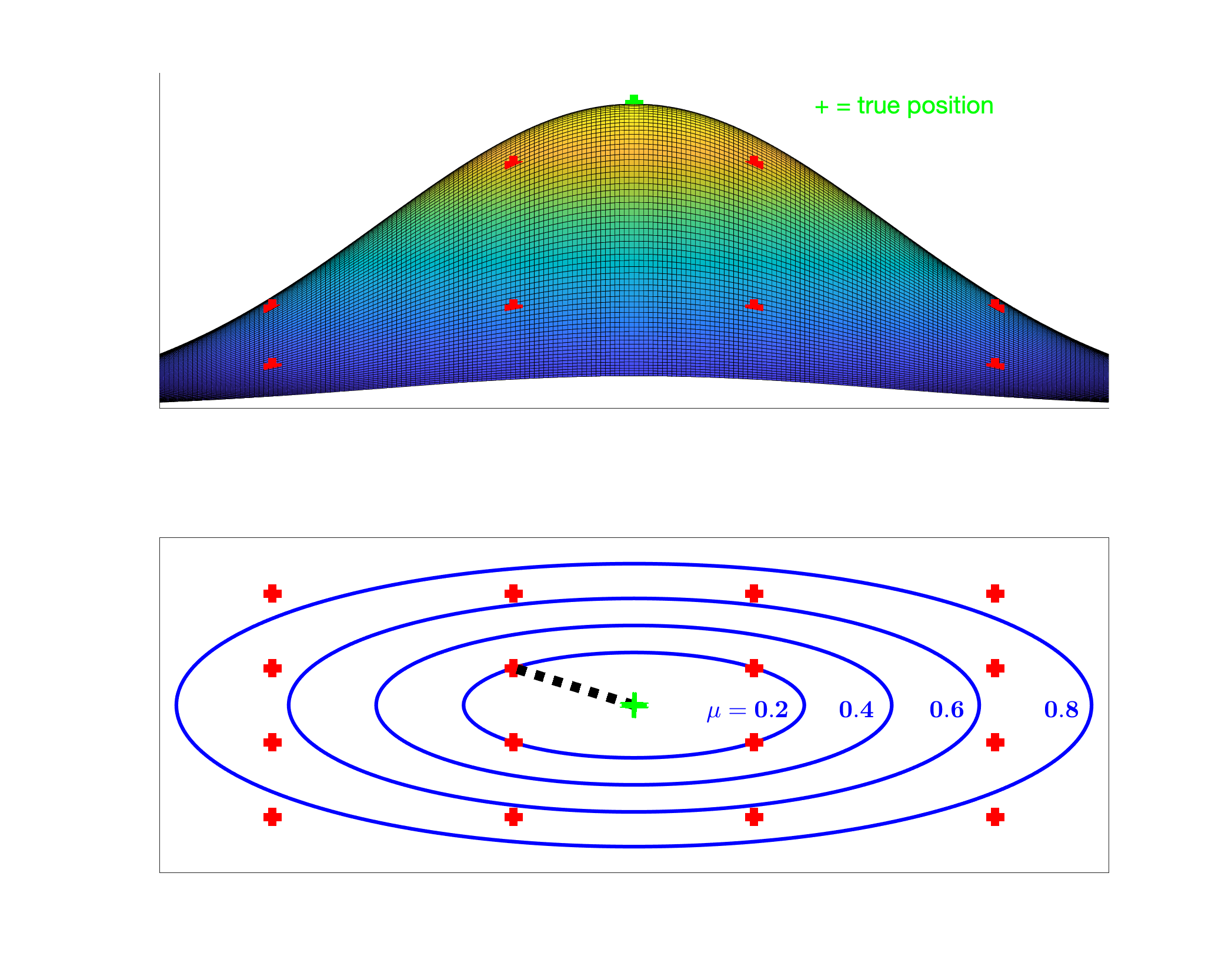}
\caption{Illustration of mismatch for a generic detection statistic. The upper panel shows
  a ``surface'' of height equal to the detection statistic for a pure signal above a plane
  defined by two signal-defining parameters (with zero covariance for simplicity). The green cross marks the true location for
  the two parameters and the maximum possible detection statistic. The lower panel shows detection statistic
  contours in the two-parameter space, where the contours correspond to mismatch
  values of 20\%, 40\%, 60\%\ and 80\%.
  The red crosses define a search template grid chosen to be least optimal for this signal location in that the true
  signal location is centered in a 2-dimensional cell, which maximizes the possible minimum mismatch (20\%) between
  the detection statistics for the true signal and the closest template. The dashed diagonal line defines
  the ``distance'' in the 2-dimensional parameter space between the true signal location and
  the closest search template.
}
\label{fig:templatediagram}
\end{center}
\end{figure}

For an $n$-dimensional, hypercubic grid defined by $n$ search parameters, one can regard the mismatch parameter $\mm$ as governing
the maximum half-length of the diagonal of the $n$-dimensional cell containing the correct signal parameters.
Conceptually, we imagine having made the least optimum choice of grid offset such that the true parameters lie at
the center of the cell, and no matter which of the $2^n$ corners of the cell is sampled, the value of the detection statistic
is no smaller than $1-\mm$ of the value obtained, had the center of the cell been sampled.
Figure~\ref{fig:templatediagram} illustrates the concept with a detection statistic ``surface'' above a plane
in two signal parameters, where the contours correspond to mismatch
values of 20\%, 40\%, 60\%\ and 80\%.

In the following,
general considerations of template placement are considered, first for directed searches for particular points on
the sky, for which placement is relatively straightforward, and then for all-sky searches, where template placement
is quite subtle and remains an active research front.

\subsubsection{Template placement in directed searches}
\label{sec:templatesdirected}

For coherent directed searches, the phase evolution Eqn.~\ref{eqn:phaseevolution} governs template placement,
where for multi-day analyses, the effects of amplitude modulation can be safely neglected in choosing
template spacing~\citep{bib:PrixTemplates1,bib:PrixTemplatesEfficient}.
Consider for a moment a highly simplified detection statistic based on multiplying in the time domain
an assumed sinusoidal signal template having
a particular phase constant $\phi_0$ and frequency $f_0$ against the raw data $x(t)$, assumed to be a sum of
random Gaussian noise $n(t)$ and a sinusoid signal having amplitude $h_0$, phase constant $\phi_0'$ and frequency $f_0'$:
\begin{eqnarray}
  \label{eqn:templateFbasic}
  F(\phi_0',f_0') & = & \left|{2\over T}\int_0^T e^{-i(\phi_0+2\pi f_0t)}\>x(t)dt\right|^2 \\
    & = & \left|{2\over T}\int_0^T e^{-i(\phi_0+2\pi f_0t)}\>[n(t)+h_0\cos(\phi_0'+2\pi f_0't)]dt\right|^2.
\end{eqnarray}

In the limit of large $T$ and strong signal (neglecting $n(t)$), the expectation value of $F$ when maximized over possible template values for
$f_0$ is simply $h_0^2$, independent of $\phi_0$, $\phi_0'$ and $f_0'$, where $F$ is maximized for
$\Delta f\equiv f_0'-f_0 = 0$. To understand how rapidly $F$ decreases as  $|\Delta f|$ departs from zero, it's helpful
to rewrite $\cos(\phi_0'+2\pi f_0't) = {1\over2}(e^{i(\phi_0'+2\pi f_0't)}+e^{-i(\phi_0'+2\pi f_0't)})$, where in the strong-signal limit of large $T$
and for small $|\Delta f|$ such that the second term of the cosine expansion can be neglected,
$F$ approaches
\begin{eqnarray}
  F & \approx & \left|{h_0\over T}\int_0^T e^{i[\Delta\phi+2\pi \Delta ft]}\>dt\right|^2 \\
  & = & h_0^2 \left| \sinc(\pi\Delta fT)\right|^2 \\
  & \approx & h_0^2\left[1-{1\over3}\left(\pi\Delta fT\right)^2\right],
\end{eqnarray}
\noindent where $\Delta\phi\equiv\phi_0'-\phi_0$ drops out and where the convention $\sinc(x) \equiv {\sin(x)\over x}$ is chosen. If we rewrite this last result
as $F\approx h_0^2\cos^2(\Delta\phi_{\rm mismatch})$, then the tolerance in $\Delta f$ for a phase mismatch value
$\Delta\phi_{\rm mismatch}$ is
\begin{equation}
\Delta f_{\rm mismatch} \approx {\sqrt{3}\over\pi T}\Delta\phi_{\rm mismatch},
\end{equation}
\noindent which is $2\sqrt{3}$ larger than the naive underestimate of 
Eqn.~\ref{eqn:ftolerance}. Consequently, in a search that automatically maximizes $F$ over the unknown
phase constant, one need not search as finely in frequency as suggested by Eqn.~\ref{eqn:ftolerance},
which implies reduced computational costs in large-scale searches.

Given the importance of template placement to those costs, in fact, a systematic approach is merited. Following methodology developed
originally for template placement in compact binary merger searches~\citep{bib:SathyaFilterbank,bib:OwenTemplates,bib:SathyaTemplates}, one can
rewrite and generalize the simplified detection statistic in Eqn.~\ref{eqn:templateFbasic}, replacing the data with another template and
address the reduction in $F$'s value due to mismatch of template parameters
\begin{eqnarray}
  \label{eqn:templateFgeneral1}
  F(\vlam,\vlamp) & = & \left|{1\over T}\int_0^T e^{-i\Phi(t;\vlam)}e^{i\Phi(t;\vlamp)}\>dt\right|^2 \\
  \label{eqn:templateFgeneral3}
                & = & \left|{1\over T}\int_0^T e^{i\Delta\Phi(t;\vlam,\dvlam)}\>dt\right|^2,
\end{eqnarray}
\noindent where $\vlam$ and  $\vlamp$ refer to a set of $N$ parameters, such as phase and frequency derivatives,
and where $\dvlam \equiv \vlamp-\vlam$ is taken small enough that 2nd-order $\dvlam$ corrections in
$\Delta\Phi\equiv\Phi(t;\vlam+\dvlam)-\Phi(t;\vlam)$ can be neglected.
Clearly, for $\dvlam = 0$, $F = 1$ and is maximum, with vanishing first
partial derivatives. Hence we expect $F$ to have the following form in the vicinity of $\dvlam=0$:
\begin{equation}
  F \approx 1 + {1\over2}\sum_{k,\ell=1}^N {\partial^2 F\over\partial\dlam_k\partial\dlam_\ell}\biggr|_{\dvlam=0}\dlam_k\dlam_\ell,
\end{equation}
where the diagonal 2nd-partial derivatives are negative and which leads to the definition of a {\it metric}:
\begin{equation}
  \gkell = - {1\over2}{\partial^2 F\over\partial\dlam_k\partial\dlam_\ell}\biggr|_{\dvlam=0},
\end{equation}
such that the {\it mismatch} $\mm$ of a template deviation is $\mm = \sum_{k,\ell}\gkell\dlam_k\dlam_\ell$. Hence the
appropriate spacing of templates in parameter space to avoid excessive mismatch is governed by the form of $\gkell$.

A general treatment of finding $\gkell$~\citep{bib:OwenTemplates} can be approached by Taylor-expanding the exponential
in equation~\ref{eqn:templateFgeneral3}: $e^{i\Delta\Phi} \approx 1 + i\Delta\Phi - {1\over2}\Delta\Phi^2$ and evaluating
the second derivatives of $F$ with respect to $\dlam_k$ and $\dlam_\ell$. In the limit $\dvlam\rightarrow0$, one
finds:
\begin{equation}
  \label{eqn:metricpartials}
       -{1\over2} {\partial^2 F\over\partial\dlam_k\partial\dlam_\ell}\biggr|_{\dvlam=0}
        = \left[\left<\!{\partial\Delta\Phi\over\partial\dlam_k}{\partial\Delta\Phi\over\partial\dlam_\ell}\!\right>
    - \left<\!{\partial\Delta\Phi\over\partial\dlam_k}\!\right>\left<\!{\partial\Delta\Phi\over\partial\dlam_\ell}\!\right>\right]_{\dvlam=0},
\end{equation}
\noindent where
\begin{equation}
  \left<f(t)\right> \equiv {1\over T}\int_0^T f(t)\>dt.
\end{equation}
More specifically, in the context of the Taylor $N$th-order expansion of the phase function
(henceforth omitting $\vlam$ dependence in $\Delta\Phi$):
\begin{equation}
  \Delta\Phi(t;\dvlam) \approx \Delta\phi_0 + 2\pi\sum_{m=0}^N  {\Delta f^{(m)}t^{m+1}\over(m+1)!}, 
\end{equation}
\noindent where $f^{(m)} = {d^mf\over dt^m}\bigr|_{t=0}$, and the set of frequency derivatives can be treated as a
parameter vector $\mathbf{f} \equiv [f^{(0)},f^{(1)},...,f^{(N)}]$. The detection statistic $F$ can be expanded:
\begin{eqnarray}
  F & \approx & \left|{1\over T}\int_0^T e^{i\left(\Delta\phi_0 + 2\pi\sum_{m=0}^N  {\Delta f^{(m)}t^{m+1}\over(m+1)!}\right) }\>dt\right|^2  \\
  & \approx & |e^{i\Delta\phi_0} |^2\Biggl|{1\over T}\int_0^T \biggl[1 + i\,2\pi\sum_{m=0}^N{\Delta f^{(m)}t^{m+1}\over(m+1)!}  \nonumber \\
  & &  - {1\over2} (2\pi)^2\sum_{m,n=0}^N{\Delta f^{(m)}\Delta f^{(n)}t^{m+n+2}\over(m+1)!(n+1)!}\biggr] \>dt \Biggr|^2 \\
  & \approx & \Biggl|{1\over T} \biggl[T + i\,2\pi\sum_{m=0}^N{\Delta f^{(m)}T^{m+2}\over(m+2)!}  \nonumber \\
  & &  - {1\over2} (2\pi)^2\sum_{m,n=0}^N{\Delta f^{(m)}\Delta f^{(n)}T^{m+n+3}\over(m+1)!(n+1)!(m+n+3)}\biggr] \Biggr|^2  \\
  & \approx & \biggl[1 + (2\pi)^2\sum_{m,n=0}^N{\Delta f^{(m)}\Delta f^{(n)}T^{m+n+2}\over(m+2)!(n+2)!} \nonumber \\
  & &  -  (2\pi)^2\sum_{m,n=0}^N{\Delta f^{(m)}\Delta f^{(n)}T^{m+n+2}\over(m+1)!(n+1)!(m+n+3)}\biggr] \\
  & = & 1 - (2\pi)^2\sum_{m,n=0}^N{\Delta f^{(m)}\Delta f^{(n)}T^{m+n+2}(m+1)(n+1)\over(m+2)!(n+2)!(m+n+3)}.
\end{eqnarray}
Terms higher in order than $\Delta f^{(m)}\Delta f^{(n)}$ have been neglected in the above.
From this last expression, we conclude that the metric $\gkell$ can be written:
\begin{equation}
  \gkell = (2\pi)^2{T^{k+\ell+2}(k+1)(\ell+1)\over(k+2)!(\ell+2)!(k+\ell+3)}.
\end{equation}
\noindent See~\citep{bib:cwcasamethod} for the same expression for the
metric for the \fstatistic~\citep{bib:JKS} in a directed search.

As examples, consider the 2-parameter metric with respect to frequency $f_0$ and its first derivative $f_1$:
\begin{eqnarray}
  g_{00} & = & {1\over3}\left(\pi T\right)^2, \\
  g_{01} & = & {1\over6}\left(\pi T^{3/2}\right)^2, \\
  g_{11} & = & {4\over45}\left(\pi T^2\right)^2. 
\end{eqnarray}
For a given desired mismatch $\Delta M$, define nominal offsets $\Delta f_0^*$ and $\Delta f_1^*$, using only the
diagonal metric elements: ($\Delta f_k^* \equiv \sqrt{\Delta M}/g_{kk}$)  
\begin{eqnarray}
  \label{eqn:fstar0def}
  \Delta f_0^* & = & {\sqrt{3\Delta M}\over\pi T}, \\
  \label{eqn:fstar1def}
  \Delta f_1^* & = & {3\sqrt{5\Delta M}\over\pi T^2}.
\end{eqnarray}

\begin{figure}[t!]
\begin{center}
\includegraphics[width=13.cm]{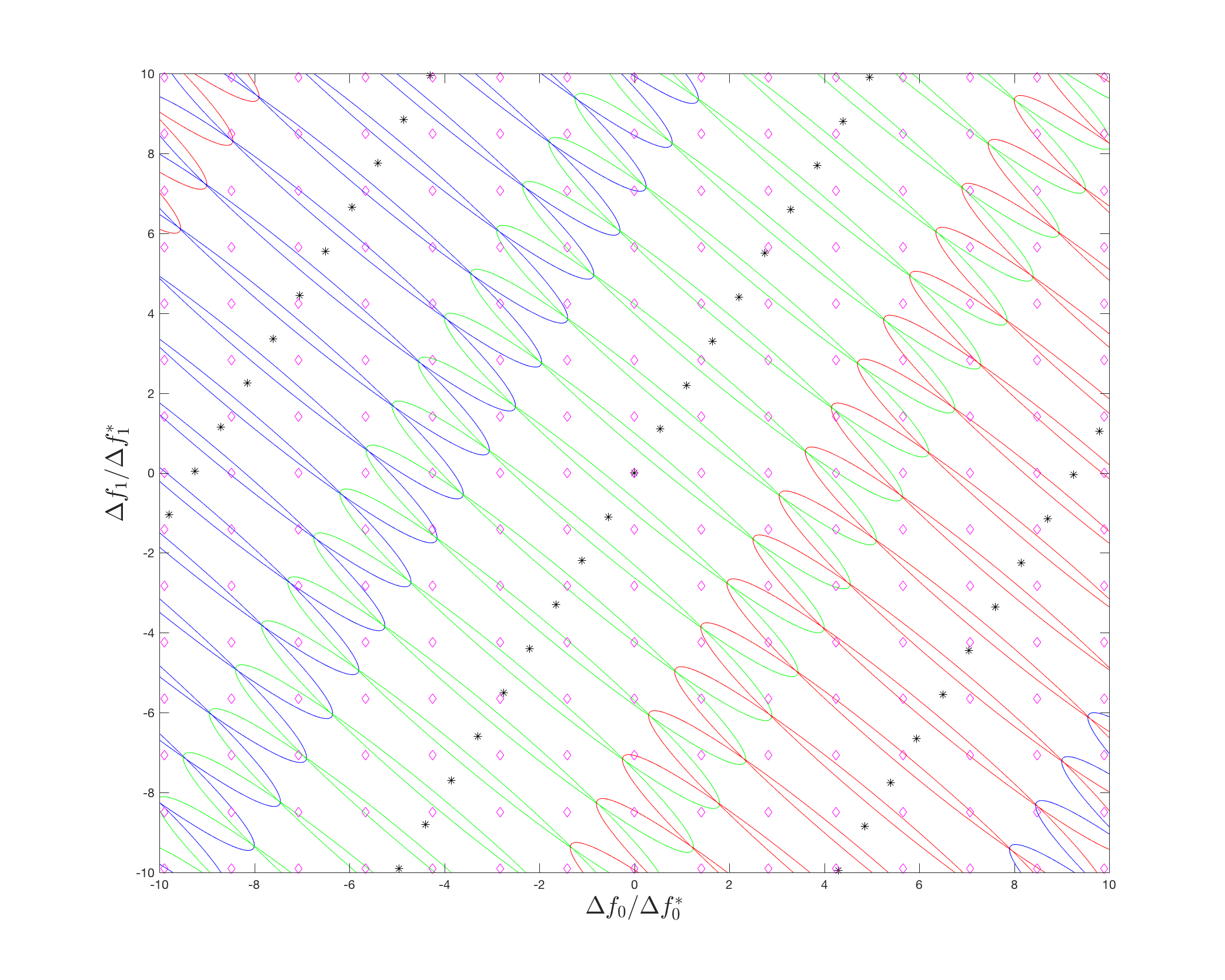}
\caption{Illustration of a ($\Delta f_0$, $\Delta f_1$) template grid (black stars) and constant-mismatch elliptical contours for which the grid point placement gives complete coverage. The values of the frequency and frequency derivative are given in
  normalized units of $\Delta f_0^*$ and $\Delta f_1^*$ defined in Eqns.~\ref{eqn:fstar0def}-\ref{eqn:fstar1def}.
The magenta diamonds indicate a rectangular grid with full coverage when the off-diagonal metric term is ignored.}
\label{fig:templatepoints}
\end{center}
\end{figure}

Since off-diagonal terms in the metric are non-zero, a rectangular grid using only diagonal terms will, in general,
be inefficient. Figure~\ref{fig:templatepoints} illustrates for a 2-dimensional slice of $\Delta f_0$ \vs\ $\Delta f_1$ (=$\Delta\fgwdot$) a template grid that accounts for these correlations in mismatch. A grid placement based on only
the diagonal metric elements would lead to inefficient coverage, as shown.
\cite{bib:PrixTemplatesEfficient} and \cite{bib:WetteTemplates1} discuss more generally and in more
detail template grid placement for CW searches, with special focus on searches over
the three-dimensional parameter space ($\fgw$,$\fgwdot$,$\fgwddot$).
As noted above, however, for short coherence
times, the range of $\fgwddot$ searches may be smaller than $\Delta\fgwddot^*$ in regions of parameter space,
depending on braking-index assumptions and the value of $\fgwdot$.

\subsubsection{Template placement in all-sky searches}
\label{sec:templatesallsky}

Template placement in all-sky searches is relatively straightforward for semi-coherent searches using short coherence times $\Tcoh$
of $\sim$hours or less, but is quite subtle for coherent searches using much longer coherence times (days) and for semi-coherent
searches using long coherence times for each data segment.

Short-$\Tcoh$ template grids can be factorized over sky location ($\alpha$, $\delta$) and over ($\fgw$, $\fgwdot$), using
isotropic grid point placement, \eg, density proportional to $\cos(\delta$) and uniform in $\alpha$, with a rectangular
grid in ($\fgw$, $\fgwdot$), with spacings determined empirically or semi-analytically for a given data run. For example,
the rule of thumb given in Eqn.~\ref{eqn:angres} overestimates the density needed for short observation times of
$\sim$few months because of correlations~\citep{bib:PrixItoh}
in the dependence of a semi-coherent power sum on sky location and frequency
parameters. For a data set collected over 1-2 months of the Earth's orbit, the average acceleration of the
detector toward the Sun creates an apparent offset in the spin-down of a putative source. Hence a search over a band of
frequencies and 1st derivatives
may detect a signal with nearly as high an SNR as the nominal maximum, but with correlated offsets in
the four parameters ($\alpha$, $\delta$, $\fgw$, $\fgwdot$). For longer observation times, these near degeneracies
in parameter space become less helpful; signal templates must be placed more densely.
At additional computational cost, these semi-coherent
searches may also search explicitly over source polarizations, or may choose to apply a circular polarization weighting
and sacrifice some sensitivity to near-linear polarizations~\citep{bib:cwallskyS4}.

Template placement for much longer coherence times is more challenging
because analytic 
approximations break down for long coherence times and because naive grid spacings depend on the specific region
of the Earth's orbit covered by a particular coherence time, making the systematic matching of signal candidates
across different time segments non-trivial in semi-coherent searches. Template placement for the \fstatistic\
has received much attention in the last decade and a half~\citep{bib:Whitbeck,bib:PrixTemplates1,bib:PrixTemplatesEfficient,bib:WettePrix,bib:WetteTemplates1},
in part because of its use in the Einstein@Home (see section~\ref{sec:allsky}) distributed computing platform.
From Eqn.~\ref{eqn:lambdadefinition}, one can define a mismatch analogous to that of
section~\ref{sec:templatesdirected}: 
\begin{equation}
  1 - \mu \equiv {(h_{\vlam+\dvlam}|h_{\vlam+\dvlam})\over (h_{\vlam}|h_{\vlam})},
\end{equation}
\noindent where we expect an approximately quadratic falloff from unity for small $|\dvlam|$ (but
see~\citep{bib:AllenTemplates} for a discussion of template placement for larger $|\dvlam|$ and
see~\citep{bib:AllenTemplates2} which distinguishes between optimality for setting rigorous
upper limits and optimality for signal detection).

The complexity of the definition of $(h|h)$ (see Eqns.~\ref{eqn:aoftdef}-\ref{eqn:boftdef}
and \ref{eqn:hoftdef}-\ref{eqn:hidefinitions2})
do not yield a definition of $(h|h)$ in the convenient form of Eqn.~\ref{eqn:templateFgeneral1}.
In particular, the sidereal antenna pattern modulations due to the Earth's rotation are not accommodated
by the phase-only dependence of the simplified form.
For long observation times, however, amplitude modulation effects can be averaged with sufficient
accuracy~\citep{bib:PrixTemplates1}. Phase modulation from the Earth's motion is captured
by Eqn.~\ref{eqn:templateFgeneral1}, allowing use of Eqn.~\ref{eqn:metricpartials} to
determine the \fstatistic\ metric with respect to frequency parameters and sky location. 

Following the treatment of \citep{bib:PrixTemplates1}, a more explicit phase evolution can be written:
\begin{equation}
  \label{eqn:phaseexplicit}
  \Phi(t) = \phi_0 + 2\pi\sum_{m=0}^N{f^{(m)}(\tau_{\rm ref})\,\tau(t)^{m+1}\over(m+1)!},
\end{equation}
\noindent where $\tau(t)$ is the SSB arrival time of the signal. Ignoring the Shapiro and Einstein
delays in Eqn.~\ref{eqn:phasedefinition} for metric definition, one can write:
\begin{equation}
  \label{eqn:ttotau}
\tau(t) = t + {\vec r(t)\cdot\hat n\over c} - \tau_{\rm ref},
\end{equation}
\noindent where $\vec r(t)$ is the position of the detector at time $t$, $\hat n$ is the unit
vector pointing from the detector to the source, and $\tau_{\rm ref}$ is the reference time
in the SSB frame at which the frequency and its derivatives are defined.

The phase derivatives entering Eqn.~\ref{eqn:metricpartials}
can then be written:
\begin{eqnarray}
  {\partial \Phi\over \partial f^{(k)}} & = & 2\pi{\tau(t)^{k+1}\over(k+1)!}, \\
  {\partial \Phi\over \partial n_i} & = & 2\pi {r_i(t)\over c}\sum_{m=0}^N{f^{(m)}(\tau_{\rm ref})\,\tau(t)^m\over m!},
\end{eqnarray}
\noindent where $\mathbf{r}(t)\cdot\hat{n} = \sum_i r_i(t) n_i$,
from which the metric terms for a particular point in parameter space ($\vec f$, $\hat n$)
can be computed via numerical integration of Eqn.~\ref{eqn:metricpartials}
over the observation span, with precise description
of $\mathbf{r}(t)$, accounting for the non-zero eccentricity of the Earth's orbit. It is convenient in some studies, though,
to make the ``Ptolemaic'' approximation~\citep{bib:ptolemymetric,bib:Whitbeck} in which the Earth's orbit
is treated as circular, for which analytic but quite lengthy trigonometric expressions can be obtained~\citep{bib:Whitbeck}.

As shown in \cite{bib:JKS}, the GW phase described by Eqns.~\ref{eqn:phaseexplicit}-\ref{eqn:ttotau} can be
well approximated (setting the reference time $\tau_{\rm ref}=0$ for convenience) by
\begin{equation}
  \Phi(t) = \phi_0 + 2\pi\sum_{m=0}^N{f^{(m)}t^{m+1}\over(m+1)!} +  2\pi {\vec r(t)\cdot \hat n\over c} \left(\sum_{k=0}^N {f^{(k)} t^k\over k!}\right).
  \label{eqn:SSBexpansion}
\end{equation}

The last term in Eqn~\ref{eqn:SSBexpansion} can be usefully decomposed into the orbital motion
of the Earth's center and the spin of the detector about the Earth's center with 
orbital and spin phases ($\vec r(t) = \vec r_{\rm orb}(t) + \vec r_{\rm spin}(t)$):
\begin{eqnarray}
  \label{eqn:phaseorb}
  \Phi_{\rm orb}(t) & = & 2\pi {\vec r_{\rm orb}(t)\cdot \hat n\over c} \left(\sum_{k=0}^N {f^{(k)} t^k\over k!}\right), \\
  \Phi_{\rm spin}(t) & = & 2\pi {\vec r_{\rm spin}(t)\cdot \hat n\over c} \left(\sum_{k=0}^N {f^{(k)} t^k\over k!}\right).
\end{eqnarray}

An inconvenient property of the metric defined above using the parameters ($\vec f$, $\hat n$) is that
converting the 3-D Cartesian $\hat n$ components to the 2-D sky coordinates $\alpha$ and $\delta$ leads to
a sky spacing that depends on the parameter themselves.
A metric more convenient for large-scale CW searches over the entire sky can be obtained by
using {\it global correlations} in parameter space~\citep{bib:Pletsch,bib:PletschAllen}.
Some searches exploiting these correlations are known as ``GCT'' searches for ``Global Correlation Transform.''
For multi-day coherence times short compared to one orbital year, one can Taylor-expand
the rescaled position of the Earth's center $\vec\xi(t) \equiv \mathbf{r}_{\rm orb}(t)/c$ in Eqn.~\ref{eqn:ttotau} about the
midpoint $t_0$ of the coherence time span $\Tcoh$:
\begin{equation}
  \label{eqn:xidef}
  \vec \xi(t) = \vec \xi(t_0) + \sum_{n=1}^\infty {\vec\xi^{(n)}(t_0)(t-t_0)^n\over n!}.
\end{equation}

The Earth's orbital motion contribution to signal phase (Eqn.~\ref{eqn:phaseorb}) can then be rewritten:
\begin{eqnarray}
  \label{eqn:OrbitExpansion}
  \Phi_{\rm orb}(t) & = & 2\pi \left(\sum_{k=0}^N {f^{(k)} (t-t_0)^k\over k!}\right)\left(\sum_{\ell=0}^\infty{(t-t_0)^\ell\over\ell!}\vec\xi^{(\ell)}\cdot\hat n\right) \\
  & = & 2\pi \sum_{m=0}^\infty (t-t_0)^m\left(\sum_{n=0}^{m'} {f^{(n)}\vec\xi^{(m'-n)}\over n!(m'-n)!}\cdot\hat n\right),
\end{eqnarray}
where $m'$ = min($m$,$N$).

It is also convenient to define new sky coordinates that capture the vector difference in signal
phase (radians) between
the source direction ($\alpha$, $\delta$) and the detector's direction from the Earth's center ($\alpha_D(t_0)$, $\delta_D$) at
time $t_0$~\citep{bib:Pletschmetric}:
\begin{eqnarray}
  \label{eqn:nxdef}
  n_x(t_0) & \equiv & 2\pi f(t_0) \tau_{\rm E} \cos(\delta)\cos(\delta_D)\cos[\alpha-\alpha_D(t_0)], \\
  \label{eqn:nydef}
  n_y(t_0) & \equiv & 2\pi f(t_0) \tau_{\rm E} \cos(\delta)\cos(\delta_D)\sin[\alpha-\alpha_D(t_0)],
\end{eqnarray}
\noindent where $\tau_{\rm E} = R_{\rm E}/c$ is the light travel time from the Earth's center to the detector.
(See~\citep{bib:JK1999} for a similar sky coordinate definition.)

Using Eqns.~\ref{eqn:phaseexplicit}-\ref{eqn:ttotau} and \ref{eqn:xidef}-\ref{eqn:nydef}, including
the approximation in Eqn.~\ref{eqn:SSBexpansion}, and absorbing
phase constants into a single term $\phi_0'$, one obtains~\citep{bib:Pletschmetric}:
\begin{eqnarray}
  \label{eqn:phiandnu}
  \Phi(t) & = & \phi_0' + \sum_{k=0}^N\nu^{(k)}(t_0)\left({t-t_0\over\Tcoh}\right)^{k+1}2^{k+1} \nonumber \\ 
          &   & +\>n_x(t_0)\cos(\Omega t) + n_y(t_0)\sin(\Omega t),
\end{eqnarray}
\noindent where $\nu^{(k)}$ are new coordinates, serving the role of effective frequencies and effective
frequency derivatives and $\Omega$ is the Earth's rotational angular velocity (sidereal time):
\begin{eqnarray}
  \nu^{(k)}(t_0) & = & 2\pi\left({\Tcoh\over2}\right)^{k+1}\biggl[{f^{(k)}(t_0)\over(k+1)!} + \nonumber \\
           &   & +\sum_{\ell=0}^{k+1}{f^{(\ell)}(t_0)\over\ell!(k-\ell+1)!}\vec\xi^{(k-\ell+1)}(t_0)\cdot\hat n\biggr],
\end{eqnarray}
\noindent where the insertion of powers of $\Tcoh$ is to make the coordinates dimensionless. Since large-parameter-space
all-sky searches to date
have used only up to 1st-order frequency derivatives in first-stage analysis, it is useful to express $\nu(t_0)$ and
$\dot\nu(t_0) \equiv \nu^{(1)}(t_0)$ explicitly (neglecting higher-order derivatives and setting $N=1$ in
Eqn.~\ref{eqn:OrbitExpansion}). One obtains~\citep{bib:Pletschmetric}:
\begin{eqnarray}
  \nu(t_0) & = & 2\pi{\Tcoh\over2}\left[f(t_0)+f(t_0)\dot{\vec\xi}(t_0)\cdot\hat n+\dot f(t_0)\vec\xi(t_0)\cdot\hat n\right], \\
  \dot\nu(t_0) & = & 2\pi\left({\Tcoh\over2}\right)^2\left[{\dot f(t_0)\over2}+{f(t_0)\over2}\ddot{\vec\xi}(t_0)\cdot\hat n+\dot f(t_0)\dot{\vec\xi}(t_0)\cdot\hat n\right]. 
\end{eqnarray}

The form of
Eqn.~\ref{eqn:phiandnu} indicates the phase is linear with respect to the coordinates $\nu^{(k)}$, $n_x$ and $n_y$,
which permits an analytic evaluation of the metric components~\citep{bib:Pletschmetric} for a coherent search.
Expressions appropriate for searching over a 2nd-order frequency derivative can be found in~\citep{bib:Pletschmetric}

Further, in the context of a semi-coherent search constructed from $\Ncoh$ coherently analyzed segments, one
can systematically apply a refined metric in summing \fstatistic\ values over the segments. In practice, a
``coarse grid'' for each segment $j$ is defined by evaluating Eqn.~\ref{eqn:metricpartials} to obtain
the $g_{\alpha\beta}^{[j]}$. In summing the \fstatistic\ values, one must use a ``fine grid'' to avoid needless
loss of SNR from signal evolution over the full observation period. As shown in \citep{bib:Pletschmetric},
for the global correlation parameters, one can obtain the following approximation to the fine-grid metric from
\begin{equation}
  \bar g_{\alpha\beta} = {1\over\Ncoh}\sum_{j=1}^{\Ncoh} g_{\alpha\beta}^{[j]}.
  \label{eqn:GCTsemi}
\end{equation}
Explicit evaluation of $\bar g_{\alpha\beta}$ over many sidereal days leads to a fine grid that scales as ${1\over\Ncoh}$ for
only the $\dot\nu$ coordinate~\citep{bib:Pletschmetric}, which is unsurprising,
since the frequency derivative is the parameter driving
the evolution of the frequency over time. Explicit expressions for $g_{\alpha\beta}^{[j]}$ and $\bar g_{\alpha\beta}$ may be
found in \citep{bib:Pletschmetric}. One criticism~\citep{bib:WetteTemplates2} of this fine-grid metric approximation, however, is that
it does not explicitly take into account the changes in reference time implicit in each Taylor expansion for each segment.
Nonetheless, one finds empirically~\citep{bib:WetteTemplates3} that for semi-coherent searches
the effective \fstatistic\ mismatch grows much more slowly than implied by the Taylor expansion in Eqn.~\ref{eqn:xidef},
allowing Eqn.~\ref{eqn:GCTsemi} to be used successfully in Einstein@Home searches with large nominal metric mismatches.

The \weave\ software infrastructure provides a more systematic approach to covering the parameter space volume in
a templated search to ensure acceptable loss of SNR for true signals lying between
template points~\citep{bib:WetteTemplates4}.
The \weave\ program combines together recent developments 
in template placement to use an optimal parameter-space metric~\citep{bib:WettePrix,bib:WetteTemplates2} and
optimal template lattices~\citep{bib:WetteTemplates1}. The package is versatile enough to be used in all-sky searches for unknown
sources and in directed searches for particular sources, such as the
\casa\ and \vela\ supernova remnants~\citep{bib:cwdirectedO3aCasAVelaJr}.

In brief, \weave\ creates a template grid in the parameter space for each time segment,
a grid that is appropriate to computing the \fstatistic\ for a
coherence time $\Tcoh$ equal to the total observation period $\Tobs$ divided by $\Nseg$. The spacing of the
grid points in parameter space is set according to a metric~\citep{bib:WettePrix,bib:WetteTemplates2} that ensures a worst-case
maximum mismatch $\mcoh$ defined by the fractional loss in summed \fstatistic\ value due to a true signal not coinciding with a
search template.

Separately, a much finer grid is defined for the full observation period with respect to the midpoint of the
observation period, one with its own mismatch parameter $\msemicoh$, analogous to $\mcoh$, where
the semi-coherent metric is the average of all the coherent metrics, which (unlike in the (GCT) approximation)
use a common reference time. The choice of the $\msemicoh$ value
is set empirically in a tradeoff between sensitivity and computational cost. The \weave\
package creates at initialization a mapping between each point in the semi-coherent template grid and a
nearest corresponding point in each of the separate, coarser segment grids, accounting for frequency evolution.

The discussion above has implicitly assumed analysis of data from 
a single detector. One may wonder if detection statistics based
on two or more detectors require a finer template spacing, given the potential for better discrimination of signals
by requiring coherent signal phase consistency among the detectors. For short coherence times there is indeed
a finer discrimination from coherent summing when phase consistency is enforced
and hence a need for finer sampling of frequency and sky location~\citep{bib:GoetzRilessftsumming}.
For much longer coherence times, however, this statement no longer holds. For example, the
multi-detector \fstatistic~\citep{bib:CutlerSchutzmultifstat} has a coherent parameter space metric that is essentially
unchanged from that of a single-detector \fstatistic~\citep{bib:PrixTemplates1}.

This perhaps surprising
result can be understood from considering the intrinsic motions of the detectors on the surface of an Earth in
orbit. In order to maintain phase coherence for a single detector over the course of one day, one
must track the detector's relative motion around the Earth's center a distance
of order the diameter of the Earth ($\sim$13,000 km),
larger than any detector pair separation. In addition, the Earth's center travels a distance in its orbit
of about 2.6 million km in one day, and more important to template spacing, deviates from a straight line
by approximately 22,000 km. Given the phase fidelity needed to account for these Earth-induced motions over
coherence times much longer than this, the incorporation of additional detectors on the face of the Earth does not
impose an extra burden on template placement. Note, though, that combining data coherently from
$N_{\rm det}$ detectors of equal sensitivity and similar livetime fractions does improve SNR by the nominal desired
$\sqrt{N_{\rm det}}$~\citep{bib:PrixTemplates1}. 

Finally, although the above approach of defining the template spacing according to a metric computed
in the strong-signal regime is widespread in the CW literature, an important alternative instead
places templates according to isoheights of the autocovariance function of the {\it signal-free}
detection statistic~\citep{bib:cwexplorer2,bib:JaranowskiKrolaktext,bib:tdfstatistic,bib:PisarskiEtalTemplateBank,
  bib:PisarskiJaranowski}.
See Appendix A of \cite{bib:PisarskiJaranowski} for a comparison of these two approaches for
searches based on the \fstatistic. 

\subsubsection{Viterbi methods and machine learning}

\label{sec:viterbiother}

All of the search methods described so far use signal templates, explicitly or implicitly via favored frequency evolution.
When searching a large parameter space volume with fine resolution, computational cost becomes formidable and
often determinative of achievable sensitivity. Alternative approaches receiving increased attention rely upon more
generic pattern recognition.

The generic approach that has received most attention in recent years is based on Viterbi dynamical programming~\citep{bib:ViterbiOriginal}.
To illustrate with a simplified example, consider finding a signal ``trajectory'' in a spectrogram, such as shown in Figure~\ref{fig:samplesignalspectrogram}.
A templated search might sum up the power for every possible trajectory allowed by the signal model and
declare one or more candidate outliers based on a summed power of spectrogram pixels exceeding a pre-determined threshold.
The Viterbi method (in its simplest form) dispenses with templates, seeking instead for the loudest trajectory
that ``moves'' in time from left to right, where the degree of contiguity from one vertical column to the next
is tunable. For example, a trajectory traveling from a pixel in column $n$ and row $m_n$ to column $n+1$ may be constrained
to change by no more than one row: $|m_{n+1}-m_n|\le 1$. For a trajectory that begins in row $m_1$ in column 1 and travels
to row $m_N$ in the last column ($N$), the number of possible trajectories is $3^{N-1}$. Maximizing the power over all possible
such trajectories does not, however, require explicitly evaluating each power. The Viterbi algorithm  leads to the insight that the trajectory with
the highest summed power (for a strong enough signal) is also locally maximum, which allows rapid elimination of the vast majority of non-optimum trajectory segments
and a remarkably fast evaluation of the detection statistic.

The Viterbi method was first demonstrated in CW searches via a ``spectrogram'' with
each pixel representing a Bessel-weighted \fstatistic\ evaluated over a 10-day period for Scorpius X-1~\citep{bib:sidebandviterbi} over the course
of the initial LIGO S6 run (part of a Sco X-1 mock data challenge~\citep{bib:ScoX1MDC1}). Follow-up analyses with additional refinements
have been applied to~\citep{bib:ViterbiPaperII} or proposed~\citep{bib:ViterbiPaperIII} for searches from the
Advanced LIGO and Advanced Virgo O1, O2 and O3 data~\citep{bib:ViterbiO1,bib:cwdirectedO2ScoX1Viterbi,bib:ViterbiCygX1}
(see section~\ref{sec:directedbinary}). Simultaneous tracking of stellar rotational phase and orbital phase~\citep{bib:ViterbiPaperIII}
offers a significant improvement in strain sensitivity relative to tracking of orbital phase alone~\citep{bib:ViterbiPaperII}.
In addition, the Viterbi method has also been applied to searches
for accreting millisecond pulsars~\citep{bib:ViterbiFiveLMXBsO2,bib:cwAXMPO3},
isolated neutron stars~\citep{bib:ViterbiSNRmethod,bib:ViterbiSNRO2,bib:cwdirectedO3aSNRs}
and for a post-merger remnant from the BNS merger GW170817~\citep{bib:Postmerger2}. The Viterbi method may see its largest gain in computation cost,
though, from application to all-sky searches~\citep{bib:ViterbiGlasgow,bib:cwallskyO3FourPipelines}.

Although the hidden Markov Viterbi method has dramatic potential for reducing computational cost, it also has another
important virtue; it is robust with respect to unknown and potentially stochastic frequency evolution that deviates
from templated models. That flexibility makes the methodology especially important for accreting systems like LMXBs
(see section~\ref{sec:targets}) and for extremely young sources, such as newborn neutron stars and post-merger hypermassive neutron stars 
(see section~\ref{sec:transients}).

Machine learning techniques, such as convolutional neural networks, have received less attention, but offer similar gains in computational cost.
One trains an algorithm on noise samples and signal+noise samples, for which machine learning detects an underlying pattern, producing an
opaque but potentially effective algorithm for quickly yielding high detection statistic values for true signals. An early study~\citep{bib:DreissigackerEtalCNN} of single-detector data
confirms the enormous gain in computing cost possible, but does not suggest such automated algorithms achieve greater sensitivity.
A follow-up study~\citep{bib:DreissigackerPrix} examined machine learning on multi-detector data sets with realistic data gaps and
non-Gaussian noise. Another study~\citep{bib:BeheshtipourPapa1} found that a convolutional neural network proved efficient in clustering
Einstein@Home search outliers, to reduce computational cost in follow-up, with
a different tuning found effective for identifying weak signals~\citep{bib:BeheshtipourPapa2}.
Another recent study examined the potential for combining convolutional neural network analysis with
Doppler demodulation for the Earth's diurnal rotation in an all-sky search~\citep{bib:YamamotoTanaka}.

These generic methods are powerful in yielding rapid results, but require some care in use. For example, when searching a narrow band with
instrumental artifacts, the Viterbi method may seize upon the artifact and miss a nearby signal, although imposing consistency between different detectors
can mitigate this problem~\citep{bib:ViterbiGlasgow}. An area of active research is understanding better the statistics of the loudest outlier
in a Viterbi search, specifically, to understand the effective trials factor, a large value of which degrades strain sensitivity.
In the event of a first detection via non-templated methods, there remain, of course, fully templated methods available to assess
more quantitatively a candidate signal's credibility and to estimate source parameters. 

\subsection{Coping with non-Gaussian instrumental artifacts}
\label{sec:lines}

Non-Gaussian instrumental artifacts, especially spectral line artifacts, degrade CW searches.
The degradation depends on the nature of the search. Stationary, narrow line artifacts generally do not significantly degrade
targeted searches for known pulsars, for which long observation times permit extremely fine frequency resolution
and known ephemerides permit that resolution to be exploited. Periods during which a frequency-modulated signal
overlaps with a known artifact can be vetoed or deweighted. On the other hand, an all-sky search is prone to
contamination, especially in short data runs for which frequency modulation from certain sky directions may
be limited, making a stationary instrumental line resemble a signal template, at least in the first stage
of a hierarchical search.

For low assumed source spin-down (and no binary source modulation), the templates most prone to contamination
lie near the ecliptic poles, where signal frequency modulation due to the Earth's orbital motion would be small. At larger
spin-down magnitudes, a stationary line can also lead to contamination of signal templates for which the
frequency shift due to the Earth's average acceleration toward the Sun largely cancels the assumed source spin-down.
The associated templates tend to lie in a circular band concentric with the Sun's average direction during the run
with a radius and skyband thickness depending on the assumed frequency, spin-down and on the coherence time of the search~\citep{bib:cwallskyS4}.
Such contamination is most pronounced for data runs short relative to a year.

In principle, even a stationary line near an ecliptic pole should not be mistaken for a true signal once a fully coherent
algorithm is applied to assess that discrimination. The chance of an instrumental line displaying the residual frequency
modulation (including that due to the Earth's daily rotation) and associated phase modulation of a true signal is
quite small. Moreover, the chance that two different detectors would display the same line artifact with precisely the
right time-dependent offset in phase to account for the daily change in relative positions of the detectors is quite small.
For example, one veto method~\citep{bib:ZhuEtalDoppler} is based on turning off demodulation in the vicinity of an outlier template to determine if
an even louder candidate is found.
Another veto method, specific to the Frequency Hough search pipeline (see section~\ref{sec:hough}),
exploits characteristic patterns in the detection statistic variation across search template parameter space
created by stationary lines~\citep{bib:IntiniEtalDopplerVeto}. Similar considerations can be applied to
following up outliers from Viterbi-based searches~\citep{bib:JonesEtalDoppler} (see section~\ref{sec:viterbiother}).
Nonetheless, lines are a major problem in CW searches because at initial stages of hierarchical searches, such discrimination
is not available with tractable computational cost. Strong lines can trigger apparent loud signal outliers over regions of
parameter space, making outlier follow-up challenging. Simply vetoing such a region because of a known contamination risks
overlooking a true signal that would be recoverable in a deep search.

Several methods have been developed for coping with these line-induced problems in early search
stages~\citep{bib:freqhough2,bib:LeaciFilteringDisturbances,bib:TenorioKeitelSintesreview},
to reduce the burden of needless outlier followup while maintaining satisfactory detection efficiency for true signals. The simplest
method is to veto outliers known to be contaminated by a known line. This approach is effective in reducing computational cost,
but does risk throwing away real and detectable signals. A more refined approach, one that need not rely upon prior knowledge
of particular lines is imposing consistency in signal strength seen in two or more detectors. For example, for two detectors
of similar sensitivity one can require that individual detection statistic strengths in both detectors exceed a threshold and
that the combined detection strength exceed both individual-detector strengths. Similarly, in a Bayesian approach one can
impose consistency in the definition of the combined detections statistic~\citep{bib:lineveto1,bib:lineveto2,bib:lineveto3}.
An empirical background estimation to account for non-Gaussian contribution can be obtained~\citep{bib:IsiEtalSkyShift} via ``sky-shifting,'' that is,
by evaluating template recovery strengths for identical source parameters except for offsets in sky location.
One can also require consistency in SNR across different data subsets for a putative outlier template, such as
via a $\chi^2$ test for the separate contributions to the detection statistic~\citep{bib:SanchodelaJordanaSintes}.

Another approach is ``cleaning'' of data prior to searching for CW signal templates. Time-domain data cleaning has been used
for general-purpose
analysis~\citep{bib:OttewillAllencleaning,bib:MeadorsEtalcleaning,bib:TiwariEtalRegression,bib:DriggersEtalcleaning,bib:DavisEtalcleaning,bib:VajenteEtalcleaning,bib:DavisEtalDetchar,bib:VietsWadeNoiseSubtraction}
where an auxiliary witness channel permits
regression of known noise. Such cleaning can remove both broadband and narrow contamination~\citep{bib:DriggersEtalcleaning}.
A more CW-specific procedure can be carried out in the frequency domain in the absence of a witness channel -- if a non-astrophysical
source is clear. After creating DFTs one can replace bins known to be contaminated with randomly generated DFT coefficients
consistent in magnitude with noise in neighboring bins~\citep{bib:cwallskyEatHS4}. This approach potentially renders particular true signals less detectable
or undetectable, particularly for sky locations near the ecliptic poles; hence injection simulations are needed to assess
efficiency loss when setting upper limits in the absence of a signal.

Many spectral lines in a detector's gravitational wave strain channel can be identified via correlation / coherence with
lines observed in auxiliary channels, such as for magnetometers or accelerometers, that monitor the environment and
that have no sensitivity to true astrophysical systems~\citep{bib:detcharS6,bib:linesO1O2}. Others may not have a reliable
witness channel, but come in ``combs'' of many lines with equal frequency spacings between adjacent lines, inconsistent
with a plausible astrophysical source, allowing safe veto or cleaning~\citep{bib:GoetzEtalO3linelists}. Efficient tracking of known lines is an active
area of investigation, including tracking of lines that wander slightly in frequency~\citep{bib:IWAVE}.

Traditionally, transient instrumental glitches in LIGO data that create nuisances (sometimes severe) in searches for transient gravitational wave
signal have not troubled CW searches much because their effect on overall noise level integrated over long time periods
has been small. In the LIGO O3 data, however, a new class of extremely loud glitches with spectra peaking at low frequencies but
visible as high as $\sim$500 Hz appeared. These glitches of uncertain origin plagued both LIGO interferometers and occurred
loudly and frequently enough to degrade sensitivity to CW signals in the low-frequency band. To cope with this new artifact,
an {\it ad hoc} ``self-gating'' algorithm~\citep{bib:SelfGating1} was developed to taper the data in the time domain to
zero during the affected intevals of $\sim$seconds before creating DFTs for Fourier analysis. A more sophisticated, adaptive
self-gating method~\citep{bib:SelfGating2} achieved transient suppression with reduced deadtime.
An earlier gating algorithm~\citep{bib:SFTdatabase}
was developed to cope with loud gltiches in initial Virgo data and later refined~\citep{bib:freqhough2}.

\subsection{Sensitivity depth}
 
\label{sec:depth}

A rough rule of thumb is convenient when assessing the detectability of a prospective CW signal for a given data set.
Such a figure of merit is the {\it sensitivity depth}~\citep{bib:SensitivityDepth}. Its use arose in part because
of the large variations in 1) methodologies with cost / sensitivity dependence on parameter space volume searched;
2) durations $\Tobs$ of data runs (or subsets) used in analyses; and 3) intrinsic detector sensitivity vs frequency.
In part too, it avoids sometimes unwarranted assumptions based on idealized scaling with observation time.
For example, a semi-coherent search with sensitivity improvements proportional to $\Tobs^{1/4}$ may require
more computational resources than are available if $\Tobs$ becomes too large, especially since increasing $\Tobs$
usually requires stepping more finely in parameter space.

Instead, the sensitivity depth~\citep{bib:DreissigackerPrixWette} addresses the ``bottom line'' with respect to a given intrinsic detector strain
amplitude spectral noise density (square root of power spectral noise density $S_h$):
\begin{equation}
  \D \equiv {\sqrt{S_h}\over h_0},
\end{equation}
\noindent where $h_0$ is the quantity of interest, typically the 90\%\ or 95\%\ upper limit on a strain
amplitude. By design the depth does not include a parametrized scaling with observation time. Hence
the values for a given algorithm do depend on the particular data set.
\cite{bib:DreissigackerPrixWette} examines in detail the sensitivity depths achieved in searches of LIGO and Virgo
data from the early initial LIGO S2 run to the first Advanced LIGO / Virgo run O1.
Values range for templated searches from $\sim$1000 for
targeted searches of $\sim$2 years down to $\sim$20 for the most sensitive all-sky search for CW signals in unknown binary systems.
 
\subsection{Upper limits and Sensitivities}
\label{sec:upperlimits}

The CW search literature is rife with different conventions on how negative results (non-discoveries) are reported.
This section gives a brief guide to the reader in understanding those variations and the reasons for them.

Most analyses have produced frequentist upper limits at 95\%\ (or 90\%) confidence level, meaning that in a hypothetical
ensemble of repeated experiments with the same underlying random noise contributions (but the same, non-random instrumental artifacts),
a signal at the nominal upper limit value would have yielded a higher detection statistic 95\%\ (90\%) of the time.
These upper limits are derived from or at least validated by simulated signals (injections) and are quoted over narrow bands
in frequency (usually 1 Hz or less), where wider bands necessarily have somewhat higher upper limits than most of the
narrower bands from which they are composed.

Deriving rigorous upper limits with extensive simulations in each individual band is computationally
expensive (particularly for 95\% C.L.),
so it has become common in recent years to derive instead ``sensitivities'' at, say, 95\%\ efficiency after following up
and ruling out every outlier in each search band that lies above a nominal threshold (where the choice of threshold depends
on a target false alarm probability that varies considerably across different searches).
These sensitivities are calibrated by deriving upper limits in a sparse sampling
of narrow bands over the full search spectrum and finding an empirical scale factor between upper limits and average strain amplitude spectral densities
for the data set, using a weighted average appropriate to the search. These sensitivities are not rigorous upper limits, particularly in
disturbed bands, but give a useful interpretation of a non-detection.

In highly disturbed bands with one or more strong instrumental lines,
it is sometimes impractical to derive rigorous upper limits for some search methods or even to derive useful sensitivities. Such bands are vetoed and
no upper limit quoted. As noted in  section~\ref{sec:lines}, when SFT cleaning of instrumental lines is used, one must take into account the
resulting loss in detection efficiency in quoting upper limits. When strong lines are not vetoed or cleaned, upper limits in affected nearby bands
typically suffer and may not apply at all to regions very near the ecliptic poles.

Most quoted frequentist upper limits are population-averaged over the parameter space searched, assuming random orientation of the stellar
spin axis, and in the case of all-sky searches, random position on the sky. Detection efficiency varies substantially for different
angles of stellar inclination $\iota$ (best efficiency for $|\cos(\iota)|$ near one,
corresponding to circular polarization), and to a lesser extent over different regions of the sky.
Because of this variation in sensitivity, the PowerFlux pipeline (see section~\ref{sec:PowerFlux})
derives separate upper limits for circular polarization and linear polarization,
where in each case the 95\%\ C.L. upper limits are strict in the sense that 95\%\ coverage is maintained separately for every position on the sky.
Approximate population-averaged upper limits can then be derived from the strict circular-polarization limits via multiplying by a scale factor (typically $\sim$2.3) empirically determined from simulations in a given data set, including its non-stationarity and non-Gaussian
contaminations~\citep{bib:cwallskyEatHO1,bib:cwallskyO3aPowerFlux}.

As described in section~\ref{sec:timedomainPE}, an alternative Bayesian analysis technique has been applied to targeted searches
for known pulsars. In that approach a 95\%\ credible Bayesian upper limit on strain amplitude is obtained, which is
interpreted as the analyst's confidence that the true amplitude of a signal lies below that value, given the observed data
and (conservative) prior beliefs in the parameter values.
Bayesian notions of prior expectation have also influenced the construction of frequentist detection statistics.

\subsection{Transient CW sources}
\label{sec:transients}

In recent years, and particularly since the discovery of the binary neutron star merger GW170817, attention
has turned to signal models that deviate from the canonical CW source of near-constant amplitude and very
low intrinsic frequency evolution. Searches for two distinct classes of ``near-CW'' signals have been developed,
one for sources of stable intrinsic frequency, but of large amplitude variations, and one for sources
of rapid spin-down and concomitant amplitude decrease. The primary target motivating the first type of search
is a neutron star glitch, in which a sudden stellar deformation appears, such as a ruptured crust, causing a
sudden increase in the strength of gravitational waves emitted at twice the spin frequency of the
star~\citep{bib:PrixGiampanisMessenger,bib:YimJones}.
The resulting stellar spin-down would be modest, leading to only small relative changes in frequency during
the time required for the deformation to heal. Hence the search methods differ from ``standard'' CW methods
primarily in allowing for a time-dependent strength.

The danger in using the standard methods on a ``transient CW''
signal is that the data used prior to the glitch tends to reduce the integrated SNR, as does amplitude decay. To avoid this problem,
an \fstatistic-based method segments the data and look separately for signals within individual segments and coherently or semi-coherently
across different combinations of
segments~\citep{bib:PrixGiampanisMessenger,bib:lineveto3,bib:KeitelEtal,bib:TransientFstatGPU,bib:cwnarrowbandO3,bib:ModafferiEtal}.
See~\cite{bib:MoraguesEtalGlitchDetectionProspects} for a recent, detailed assessment of the prospects (near-term and long-term)
for detecting quasi-monochromatic
gravitational emission in the aftermath of glitches, based on a study of 726 previously observed electromagnetic glitches.

Another class of near-CW source is a
post-merger remnant, in which two neutron stars form a hypermassive neutron star (2-3 solar masses).
Although one naively expects such a star to collapse promptly into a black hole, rapid rotation (rigid-body or differential)
can delay the collapse for certain equations of state~\citep{bib:BaiottiRezzolla,bib:PiroEtal,bib:RaviLasky}. In extreme equations of state,
the collapse may be delayed until the star's rotation frequency has decreased dramatically~\citep{bib:RaviLasky}.
Given the enormous initial quadrupole asymmetry as two neutron stars begin to merge, one might hope for a
substantial residual asymmetry in the minutes, hours or even days during which a post-merger remnant persists.
That asymmetry might well lead to a rapid spin-down, one for which the truncated Taylor expansion in
Eqn.~\ref{eqn:phaseevolution} is a poor approximation.

A recent search in LIGO data for a post-GW170817
remnant~\citep{bib:Postmerger2} used instead a model (for sensitivity determination) in which the frequency has an evolution similar to
that of Eqn.~\ref{eqn:spindownpowerlaw}, but with a different normalization convention:
\begin{equation}
  {d\Omega \over dt} = -k\Omega^n,
\end{equation}
\noindent where $\Omega(t)$ is the angular frequency of rotation, $n$ is the braking index and $k$ is a positive real constant.
This equation leads to an explicit form for $\fgw(t)$~\citep{bib:PostmergerWaveforms,bib:SarinEtal}:
\begin{equation}
  \fgw(t) = {\fgw(0)\over\left(1+{t\over\tauconst}\right)^{1\over n-1}},
\end{equation}
\noindent where $\tauconst$ is a characteristic time scale for spin-down:
\begin{equation}
  \tauconst = {1\over k(n-1)\Omega_0^{n-1}},
\end{equation}
\noindent and where $\Omega_0 = \Omega(t=0)$.

Since the amplitude depends on frequency for fixed ellipticity (see Eqn.~\ref{eqn:hexpected}), one expects the
amplitude to decrease monotonically too:
\begin{equation}
h_0 = {4\,\pi^2G\epsilon\Izz\fgw^2(0)\over c^4r}{1\over\left(1+{t\over\tauconst}\right)^{2\over n-1}}.
\end{equation}
\noindent In addition, the product $\epsilon\Izz$ is likely to decrease as the post-merger remnant spins down.

A more significant hurdle to detection than fidelity of the signal model, however, is the typical distance at which
binary neutron star mergers occur.
GW170817 lay approximately 40 Mpc away,
several orders of magnitude farther than the neutron stars sought in our own galaxy. The necessary ellipticity to generate
a detectable signal is hence enormous; at the same time, such an ellipticity ensures a rapid-enough spin-down that no appreciable SNR could be
achieved through integration over the signal's duration at current detector sensitivities.
Based on the total number of definitive BNS detections (two)~\citep{bib:GW170817,bib:GW190425,bib:GWTC3} during the O1 through O3 data runs and
on the volume-time sampled in those runs,
it appears that GW170817 was closer than the bulk of the BNS mergers expected in future runs. Detecting a CW signal from
a post-merger remnant may require significantly more sensitive detectors than those that detected GW170817.
Applicable search methods for such a rapidly evolving signal have been developed both well before~\citep{bib:ThraneEtalStamp} and
especially after~\citep{bib:ThraneEtalStamp,bib:MillerEtalPostmerger,bib:SunMelatos,bib:OliverKeitelSintes,bib:LongTransientsHMM,bib:MytidisEtalNewborn,bib:MillerEtalLongTransient} the discovery of GW170817.

\section{Results of continuous wave searches}
\label{sec:results}

Searches have been carried out for continuous gravitational waves for five decades,
starting with data from early detector prototypes~\citep{bib:LevineStebbins,bib:EarlyBarLimits,bib:LivasArticle,bib:Suzuki}.
Although transient gravitational wave discoveries to date
have relied upon coincident signal detections in two or more detectors, 
a definitive continuous-wave source discovery can be
accomplished, at least in principle, with a single gravitational wave detector.
By definition, the source remains on, allowing follow-up verification of the signal strength and of
the distinctive Doppler 
modulations of signal frequency due to the Earth's motion.
In the event of an all-sky discovery, for which intrinsic sensitivity is necessarily limited
by computational realities (see section~\ref{sec:challenges}), it is likely that a stable continuous signal
could then {\it a posteriori} be detected in prior data sets via targeted searches.
Hence a relatively large number of CW searches were carried out with both bar
detectors and interferometer prototypes in the decades before the major
1st-generation interferometers began collecting data, 
as summarized in~\citep{bib:cwtargetedS1}.

The most sensitive of the resulting early upper limits~\citep{bib:EarlyBarLimits,bib:Suzuki,bib:cwexplorer1} came from bar detectors
in their narrow bands of sensitivity.
The Explorer detector reported~\citep{bib:cwexplorer1} an upper limit on 
spin-downless CW signals from the
galactic center of \sci{2.9}{-24} in a 0.06 Hz band near 921 Hz, based on
96 days of observation.
A broader-band ($\sim$1 Hz) upper limit of \sci{2.8}{-23}
was also reported~\citep{bib:cwexplorer2} from the Explorer detector
based on a coherent 2-day search that allowing for stellar spin-down.
In addition, searches for spin-downless CW waves from the galactic center and from
the pulsar-rich globular cluster 47 Tucanae in two 1 Hz bands near 900 Hz
were carried out in Allegro detector data, yielding upper limits~\citep{bib:cwallegro} of \sci{8}{-24}. 
Finally, a narrowband (0.05 Hz) 
search~\citep{bib:cwtama} was carried out with the TAMA interferometer near
935 Hz for continuous waves from the direction of Supernova 1987A, with
an upper limit of \sci{5}{-23} reported.

When the initial LIGO interferometers and later the initial Virgo interferometer
began collecting data in the 2000's, CW searches became more sensitive, both from
improved detector sensitivity, and to a lesser extent, because search algorithms
improved. In the following, brief summaries of the results from those searches
will be given, with emphasis on results from searches in advanced detector
data. As of this writing, numerous results from the third LIGO-Virgo observing
run (O3) have appeared and will be featured where available, along with many results from the O1 and O2
runs, to illustrate the progression of sensitivities and algorithms during the Advanced LIGO and Virgo era to date.
In recent years, research groups outside of the LIGO Scientific Collaboration, Virgo Collaboration and KAGRA
Collaboration (LVK) have also carried out analyses of the public GW data, which is released approximately 18 months
after collection. Because of that delay, many additional results from the O3 data, beyond those described here,
can be expected in the coming months and years, perhaps in parallel with LVK results from the upcoming O4 data
run~\citep{bib:obsscenario}.

\subsection{Targeted and narrowband searches for known pulsars}
\label{sec:targeted}

In {\it targeted} searches for known pulsars using measured ephemerides from radio, optical, X-ray or $\gamma$-ray
observations valid over the gravitational wave observation time, one can apply
precise, well known corrections for phase of the signal, including modulations, because one knows
the source phase evolution, its location and motion, the Earth's location and motion, and the
detector's position and orientation on the Earth.

Various approaches have been used in targeted searches in LIGO and Virgo data to date:
1) A time-domain heterodyne method~\citep{bib:DupuisWoan} in which Bayesian posteriors are determined on
the signal parameters that govern absolute phase, amplitude and
amplitude modulations (see section~\ref{sec:timedomainPE}); 
2) a Fourier-domain determination of a ``carrier'' strength along with the strengths
of two pairs of sidebands created by amplitude modulation from the Earth's sidereal
rotation of each detector's antenna pattern (``5-Vector'' method)~\citep{bib:fivevector,bib:fivevectorupdates} (see section~\ref{sec:fivevectormethod});
and 3) a matched-filter method in which marginalization
is carried out over unknown orientation parameters (the ``\fstatistic'')~\citep{bib:JKS,bib:gstatisticmethod} (see section~\ref{sec:fstatistic}).

The first application of the heterodyne Bayesian method~\citep{bib:cwtargetedS1} 
in LIGO and GEO 600 S1 data (separately to each interferometer) led to upper limits
on $h_0$ of a few times 10$^{-22}$ for PSR J1939$+$2134 ($\frot$ = 642 Hz).
Comparable upper limits were obtained from an implementation of the (frequentist) \fstatistic~\citep{bib:cwtargetedS1}.
Later applications of the heterodyne Bayesian method incorporated a variety
of improvements, including coherent treatment of multiple interferometers, marginalization
over noise parameters, a Markov Chain Monte Carlo search method for parameter estimation and
joint searching over one and two times the stellar rotation frequency.
At the same time the number of stars searched in each data run increased, along with
closer partnership with radio and X-ray astronomers who provided ephemerides.
In the S2 data, limits were placed on 28 pulsars, with a lowest strain limit
of \sci{1.7}{-24}~\citep{bib:cwtargetedS2}. In the S3 and S4 data (analyzed jointly), limits were
placed on 78 pulsars, with a lowest strain limit of \sci{2.6}{-25}~\citep{bib:cwtargetedS3S4}.
In the S5 data, limits were placed on 116 pulsars, with a lowest strain
limit of \sci{2.3}{-26} (PSR J1603$-$7202)~\citep{bib:cwtargetedS5}. The lowest limit placed on ellipticity
from the S5 search was \sci{7.0}{-8} (PSR J2124$-$3358).

The final targeted-search results from initial LIGO
and Virgo presented joint results from the LIGO S5 and S6 runs, and for the two low-frequency Crab and Vela
pulsars, results from the Virgo VSR2 and VSR4 runs~\citep{bib:cwtargetedS5S6VSR24}. This synoptic paper
presented results for 195 pulsars in total, where the lowest obtained strain limit was only slightly
better than obtained from the S5 data alone:
\sci{2.1}{-26} (PSR J1910$-$5959D), with a lowest ellipticity upper limit of \sci{6.7}{-8} (PSR J2124$-$3358).
The use of Virgo VSR2 and VSR4 data in this last analysis did, however, open up a new low-frequency
spectrum, giving sensitivities approaching the spin-down limits for several pulsars other than the Crab,
most notably the Vela pulsar, for which the spin-down limit was beaten~\citep{bib:cwtargetedvela,bib:cwtargetedS5S6VSR24}.
The S5, S6, VSR2 and VSR4 analyses also included searches using the \fstatistic\ and 5-vector algorithms
applied to ``high value'' isolated pulsars for which the spin-down limits were approached or beaten.
As expected, sensitivities obtained were comparable to those found in the Bayesian analysis.
All three methods typically obtain somewhat better sensitivities when exploiting the
inclination and polarization angles $\iota$ and $\psi$ inferred from pulsar wind nebulae observations for
known pulsars, such as Crab and Vela (for example, the \fstatistic\ is refined to a more specific
\gstatistic~\citep{bib:gstatisticmethod}), although an unfavorable orientation can also lead to worse $h_0$ sensitivity.

\begin{figure}[t!]
\begin{center}
\includegraphics[width=11.cm]{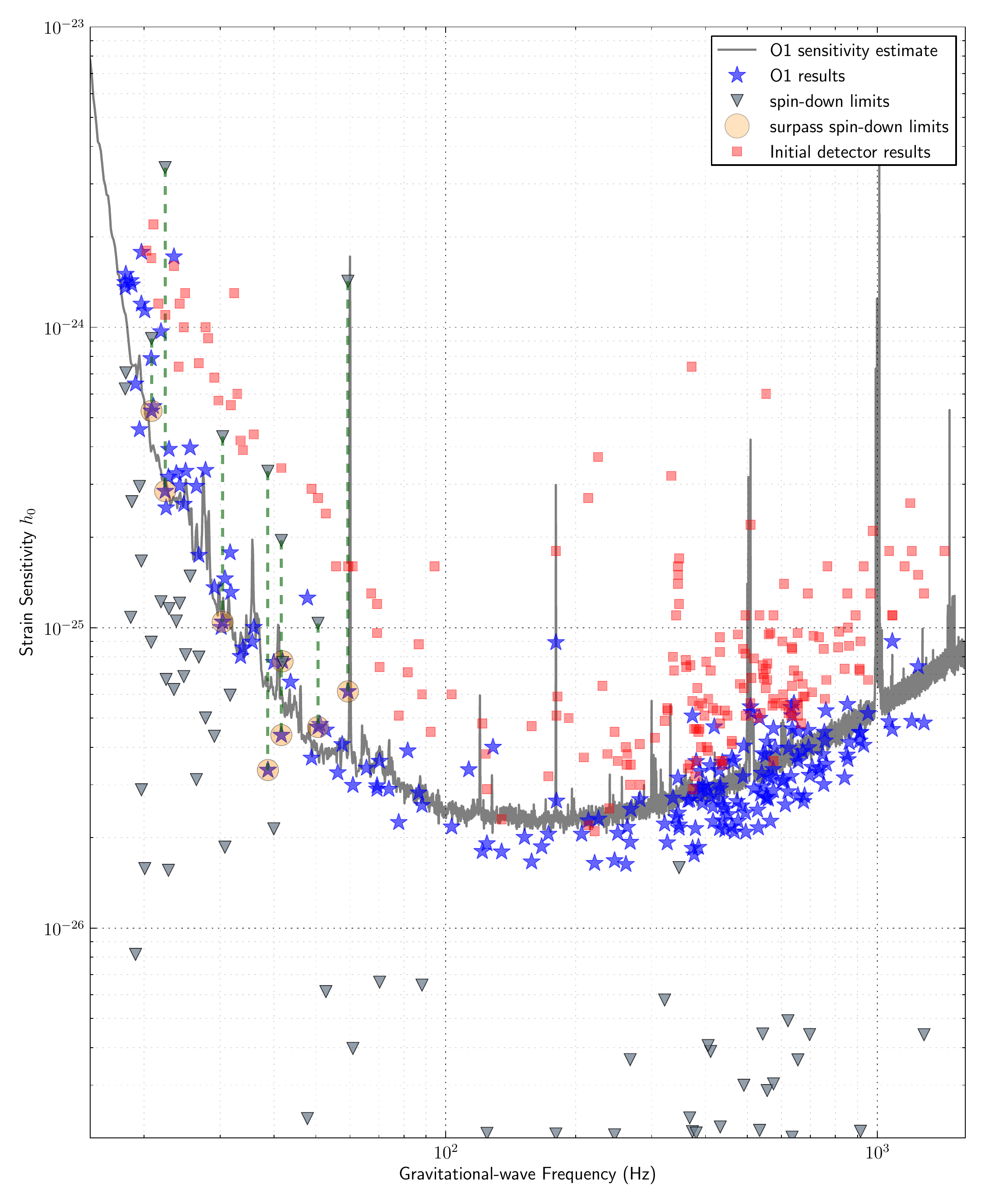}
\caption{Upper limits (95\%\ CL) on $h_0$ for known pulsars from targeted searches in the 
LIGO O1 data~\citep{bib:cwtargetedO1} (closed stars). The gray band shows the {\it a priori} estimated 
sensitivity range of the search. Also plotted (closed squares)
are the lowest upper limits from searches in initial LIGO and Virgo data and spin-down limits (closed triangles).
Upper limits that lie below spin-down limits are outlined with a circle.
\figpermission{\cite{bib:cwtargetedO1}}{AAS}}
\label{fig:cwtargetedO1}
\end{center}
\end{figure}

When Advanced LIGO data collection began in fall 2015 there was a significant improvement in broadband sensitivity
and a dramatic improvement at the lowest frequency, thanks to improved seismic isolation~\citep{bib:aligodetector1,bib:aligodetector2}.
The low-frequency improvements were, of course, helpful to the first binary black hole merger detection~\citep{bib:GW150914},
but they also made a large number of known young pulsars accessible with respect to spin-down limit (energy conservation).
Targeted searches were carried out in the O1 data using each search
program~\citep{bib:cwtargetedO1}, where
method 1) was applied to 200 stars, and methods 2) and 3) were applied to 11 and 10 stars,
respectively, for which the spin-down limit (Eqn.~\ref{eqn:spindownlimit}) was likely
to be beaten or approached, given the detector sensitivity. Results are shown in
Figure~\ref{fig:cwtargetedO1}, along with those from initial LIGO and Virgo searches.
Highlights of these O1 searches included
setting a lowest upper limit on strain amplitude of \sci{1.6}{-26}
(PSR J1918$-$0642),
setting a lowest upper limit on ellipticity of \sci{1.3}{-8} (PSR J0636$+$5129) and beating
the spin-down limit on 8 stars (PSR J0205$+$6449, J0534$+$2200, J0835$-$4510, J1302$-$6350, J1813$-$1246, J1952$+$3252,
J2043$+$2740, J2229$+$6114).
Perhaps the most notable result was setting
an upper limit on the Crab pulsar's (PSR J0534$+$2200) energy loss to gravitational radiation at a level
of 0.2\% of the star's total rotational enegy loss inferred from measured rotational spin-down.

\begin{figure}[t!]
\begin{center}
\includegraphics[width=11.cm]{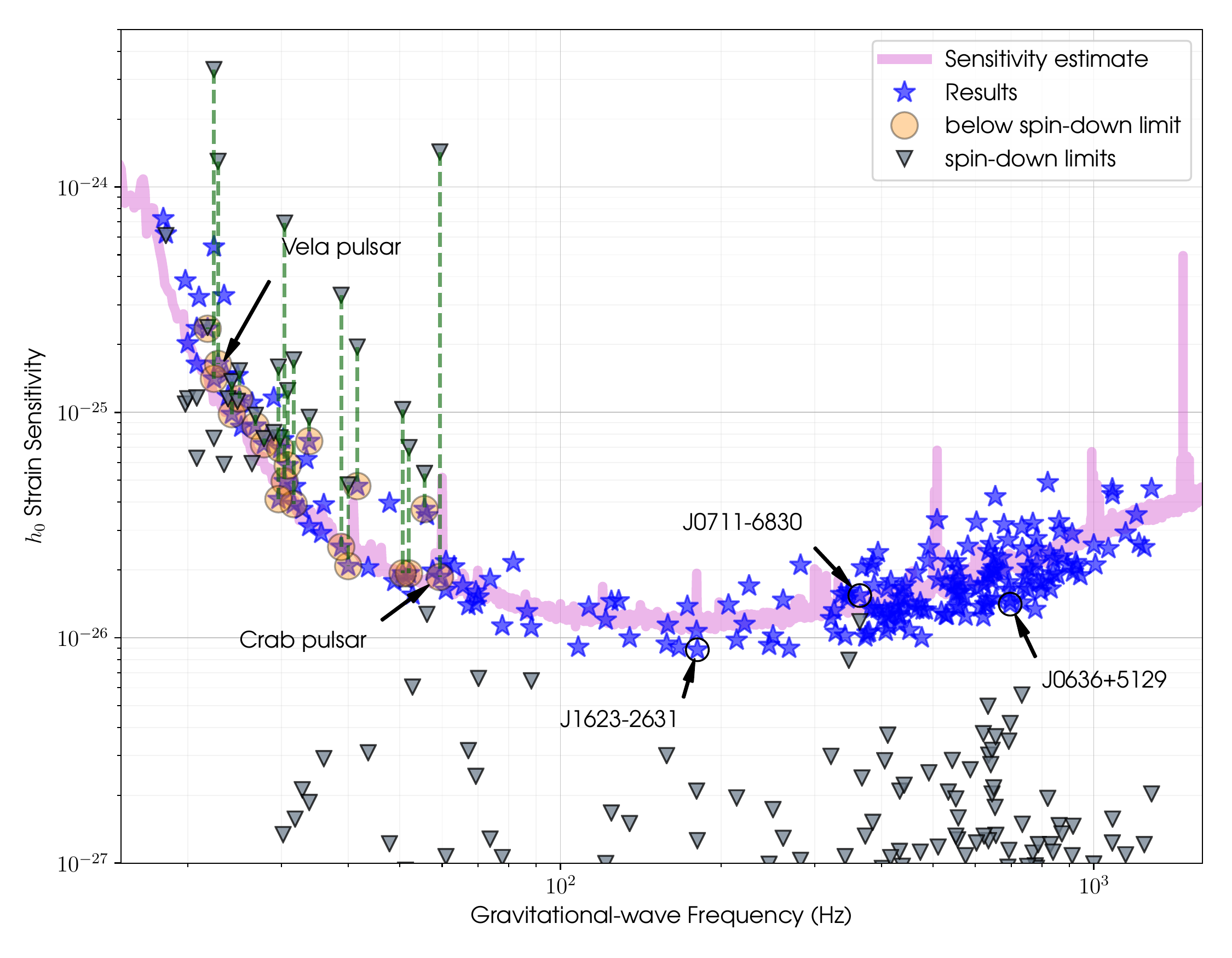}
\caption{Upper limits (95\%\ CL) on $h_0$ for 221 known pulsars from targeted searches in the 
LIGO O1 and/or O2 data~\citep{bib:cwtargetedO2} (closed stars). The pink band shows the {\it a priori} estimated 
sensitivity range of the search. Also plotted are spin-down limits (closed triangles).
Upper limits that lie below spin-down limits are outlined with a circle. \figpermission{\cite{bib:cwtargetedO2}}{AAS}}
\label{fig:cwtargetedO2}
\end{center}
\end{figure}

Similar searches were carried out for 221 known pulsars in the O1 and/or O2 data, with results summarized
in Figure~\ref{fig:cwtargetedO2}~\citep{bib:cwtargetedO2}.
Highlights included beating the
spin-down limit on 20 pulsars, 
a lowest upper limit on strain of \sci{8.9}{-27} (PSR J1623$-$2631), a lowest upper limit on ellipticity
of \sci{5.8}{-9} (PSR J0636$+$5129) and an upper limit on the Crab pulsar's fractional energy loss to gravitational
radiation of 0.02\%. In addition, the upper limit on strain amplitude (\sci{1.5}{-26})
for the MSP PSR J0711$-$6830 ($\frot=182$ Hz) was
only 30\%\ above the star's spin-down limit, corresponding to an ellipticity upper limit of \sci{1.2}{-8}.
Upper limits are also presented in~\citep{bib:cwtargetedO2} on signals at the stellar rotation frequencies,
along with upper limits on the mass quadrupole moment
$Q_{22} \equiv \epsilon \Izz\sqrt{15\over8\pi}$~\citep{bib:UshomirskyEtal}.
Initial analysis of the first six months of the LIGO and Virgo O3 data set
reduced further the upper limits on the Crab and Vela pulsar, along with those of three
recycled pulsars, for which the spin-down limit has now been beaten~\citep{bib:cwtargetedO3a}.
Similarly, a targeted search for the young, highly energetic star PSR J0537$-$6910~\citep{bib:cwtargetedO3J0537} dived below the
spin-down limit to an upper limit (95\%\ CL) of \sci{1}{-26} for a GW frequency (123.8 Hz) of twice
the rotation frequency.

A separate analysis~\citep{bib:nonlvctargetedO2} of the Advanced LIGO O1 and O2 data for a newly discovered
gamma-ray pulsar (PSR J0952$-$0607) also set an upper limit on emission amplitude of \sci{6.6}{-26}.
Upper limits on GW emission of amplitude \sci{3.0}{-26} were also set on the black widow $\gamma$-ray pulsar
PSR J1653$-$0158~\citep{bib:NiederEtalBlackWidow} discovered in an Einstein@Home search.

\begin{figure}[t!]
\begin{center}
\includegraphics[width=11.cm]{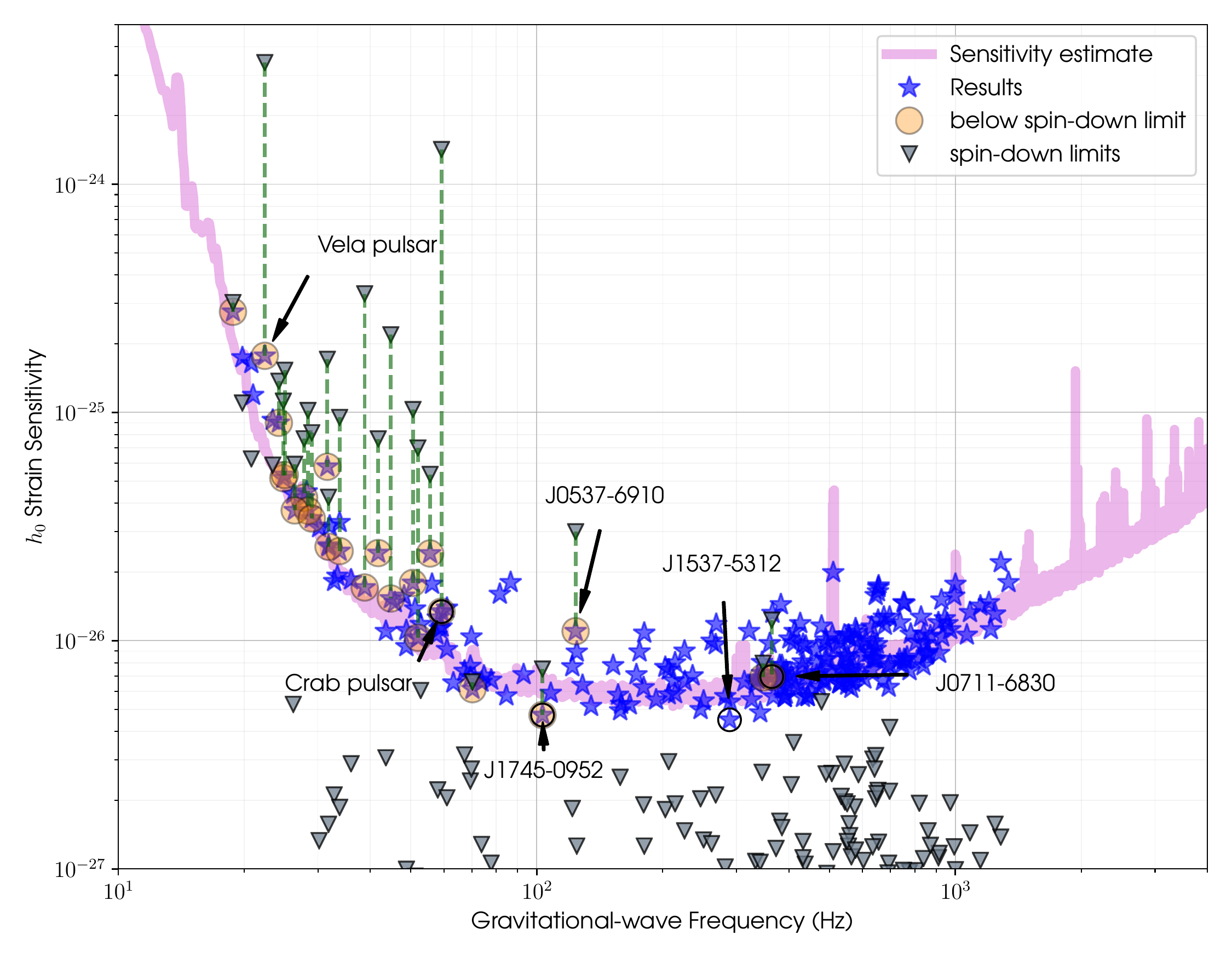}
\caption{Upper limits (95\%\ CL) on $h_0$ for 237 known pulsars from targeted searches in the 
  cumulative LIGO and Virgo O1--O3 data~\citep{bib:cwtargetedO3,bib:cwtargetedO3J0537}.
  The stars show 95\% credible upper limits on the amplitudes of $h_0$, while gray triangles represent the spin-down
  limits for each pulsar.  For those pulsars which surpass their spin-down limits, their results are
  plotted within shaded circles. The pink curve gives an estimate of the expected strain sensitivity of all
  three detectors combined during the course of O3. The highlighted pulsars are those with the best $h_0$,
  $Q_{22}$ and spin-down ratio out of the pulsars which surpassed their spin-down limit,
  as well as the best $h_{0}$ limit out of the whole sample. 
  \figpermission{\cite{bib:cwtargetedO3}}{AAS}}
\label{fig:cwtargetedO3}
\end{center}
\end{figure}

Recent cumulative results from targeted searches from the O1, O2 and O3 data runs~\citep{bib:cwtargetedO3}
for 236 known pulsars in total are shown in
Figure~\ref{fig:cwtargetedO3}. Highlights include beating the spin-down limit on 23 pulsars,
a lowest upper limit on strain of \sci{4.7}{-27} (PSR J1745$-$0952), a lowest upper limit on ellipticity of
\sci{5.26}{-9} (PSR J0711$-$6830), and an upper limit on the  Crab pulsar's fractional energy loss to gravitational
radiation of 0.009\%. In addition, the spin-down limit was beaten for two millisecond pulsars: 
PSR J0711$-$6830 ($h_0 <$ \sci{7.0}{-27}, $\epsilon<$ \sci{5.3}{-9} for $\fgw\approx$364 Hz) and
PSR J0437$-$4715 ($h_0 <$ \sci{6.9}{-27}, $\epsilon<$ \sci{8.5}{-9} for $\fgw\approx$347 Hz).
These results also include a more general analysis searching simultaneously for a signal at
one and two times the rotation frequency~\citep{bib:PitkinEtalTwotones,bib:cwtargetedO3}.
Results from cumulative O1--O3a searches for seven additional pulsars were presented in
\citep{bib:AshokEtal}.

The progressive improvement in noise level for the LIGO and Virgo detectors over the O1, O2 and O3 runs is reflected
in Figures~\ref{fig:cwtargetedO1}-\ref{fig:cwtargetedO3}. Although more refined analyses have been brought to bear
in parallel, the gains in astrophysical sensitivity come primarily from improving the instruments, for these targeted
searches which already approach optimality.

These upper limits assume the correctness of General Relativity in that antenna pattern calculations
used in the searches assume two tensor polarizations in strain. Alternative theories of gravity can,
in principle, support four additional polarizations (two scalar and two vector modes), which would
lead to different antenna pattern sensitivities~\citep{bib:TGRmethod}. Searches have been carried out for evidence of signals
from the 200 targeted pulsars in the O1 data exhibiting these other polarizations, using the heterodyned data products.
In no case was significant evidence of a non-standard signal seen, and upper limits were placed~\citep{bib:TGRO1}.

The targeted-search upper limits in Figure~\ref{fig:cwtargetedO1}
assume a fixed phase relation between stellar rotation (measured by
electromagnetic pulses) and gravitational wave emission ($f_s=\frot$). To allow for a
more general scenario, such as slight differential rotation of EM- and GW-emitting regions,
searches have also been carried out for signals very near in parameter space to those expected
from an ideal phase relation. These so-called ``narrowband'' searches allow a relative frequency deviation
of $\sim$$10^{-3}$. The first such search, using the \fstatistic,
was for the Crab pulsar in the Inital LIGO S5 data set~\citep{bib:cwtargetedcrabS5},
which set a limit slightly below the Crab spin-down limit, despite a large trials factor of \sci{3}{7}, when
the orientation of the assumed signal was aligned with observed Crab pulsar wind nebula X-ray jet axes~\citep{bib:NgRomani},
a limit five times higher than achieved in the same data set using a targeted search.
A similar narrowband search was later carried out in initial Virgo VSR4 data for the Crab and Vela pulsars, using the
5-vector program, a search which yielded a Crab upper limit about two times below the spin-down limit and
a Vela upper limit slightly higher than its spin-down limit~\citep{bib:cwnarrowbandVSR4}.

\begin{figure}[t!]
\begin{center}
\includegraphics[width=11.cm]{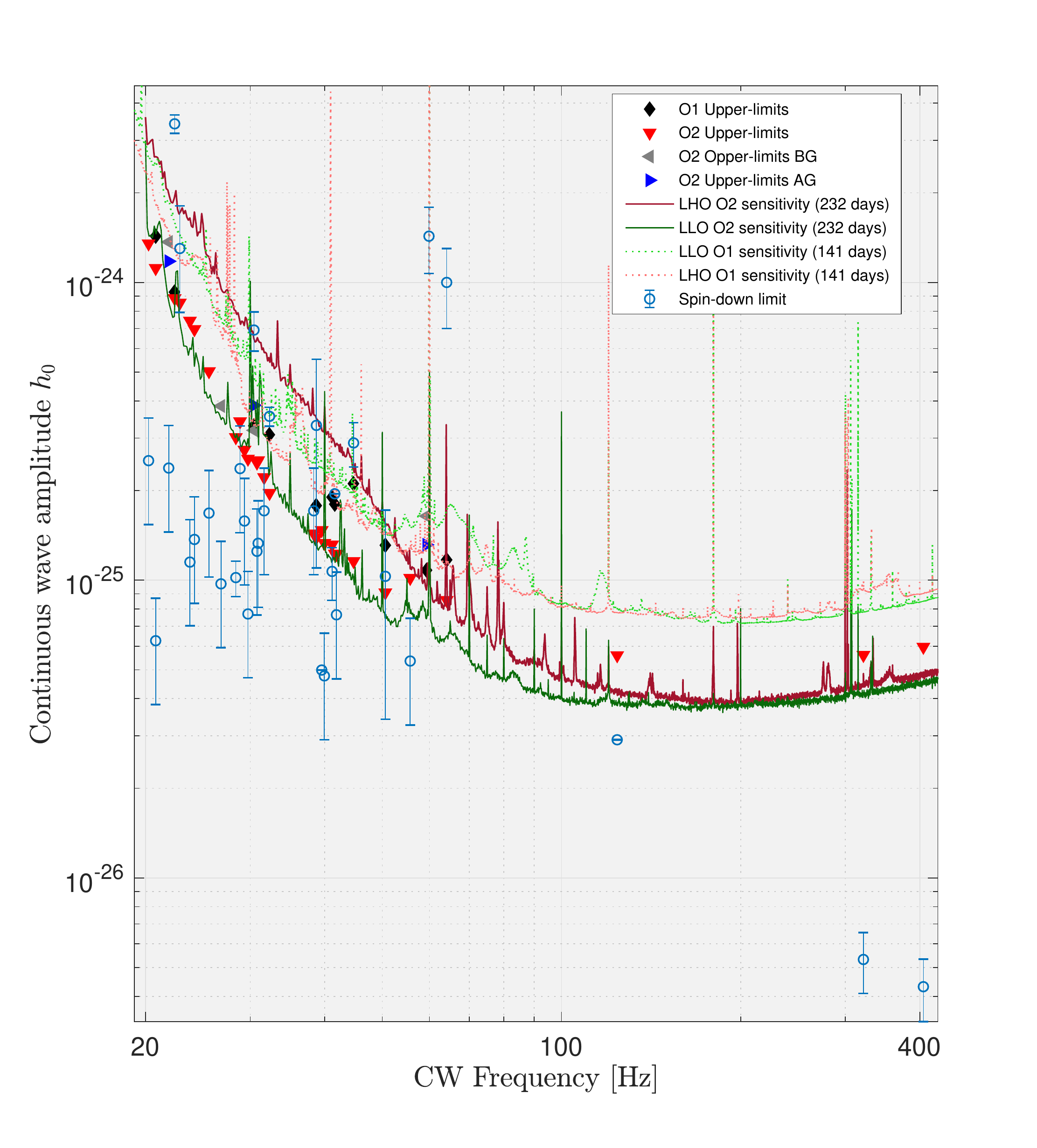}
\caption{Upper limits (95\%\ CL) on $h_0$ for (11) 33 known pulsars from narrowband searches in the 
  LIGO (O1) O2 data~\citep{bib:cwnarrowbandO2} (closed diamonds and triangles), where the GW frequency and derivative are
  allowed to vary by $\sim$$10^{-3}$ with respect to the expectation from electromagnetic observations.
  For those pulsars known to have glitched in the O2 run, separate upper limits are shown for the epochs
  before the glitch (BG) and afterward (AG). Spin-down limits are shon as open circles, where error bars denote
  the uncertainties due to pulsar distances. Curves denote nominal sensitivities for the O1 and O2 runs for
  the individual LIGO Hanford  (LHO) and Livingston (LLO) interferometers.
  \figpermission{\cite{bib:cwnarrowbandO2}}{APS}}
\label{fig:cwnarrowbandO1O2}
\end{center}
\end{figure}

The 5-vector program was applied again to the Advanced LIGO O1 data set.
Results from searches for 11 stars with expected sensitivities near
the spin-down limits have been obtained from O1 data~\citep{bib:cwnarrowbandO1}. In general, these limits are expected
and found to be higher than the corresponding upper limits from targeted searches above because the
increased parameter space search implies an additional trials factor. Nonetheless, this first advanced detector narrowband
search beat the spin-down limit on the Crab (PSR J0534$+$2200), Vela (PSR J0835$-$4510) and PSR J2229$+$6114.
Later, a 5-vector search of LIGO O2 data~\citep{bib:cwnarrowbandO2} for 33 known pulsars yielded the upper limits shown in
Figure~\ref{fig:cwnarrowbandO1O2}, along with the 11 (higher) O1 upper limits. In this analysis, the spin-down
limit was beaten for 8 known pulsars, despite trials factors ranging from $\sim$$10^6$ to $\sim$$10^9$.
For the Crab pulsar, the strain upper limit was an order of magnitude lower than the spin-down limit,
leading to a limit on fractional energy loss to gravitational waves of $\sim$1\%.

\begin{figure}[t!]
\begin{center}
\includegraphics[width=11.cm]{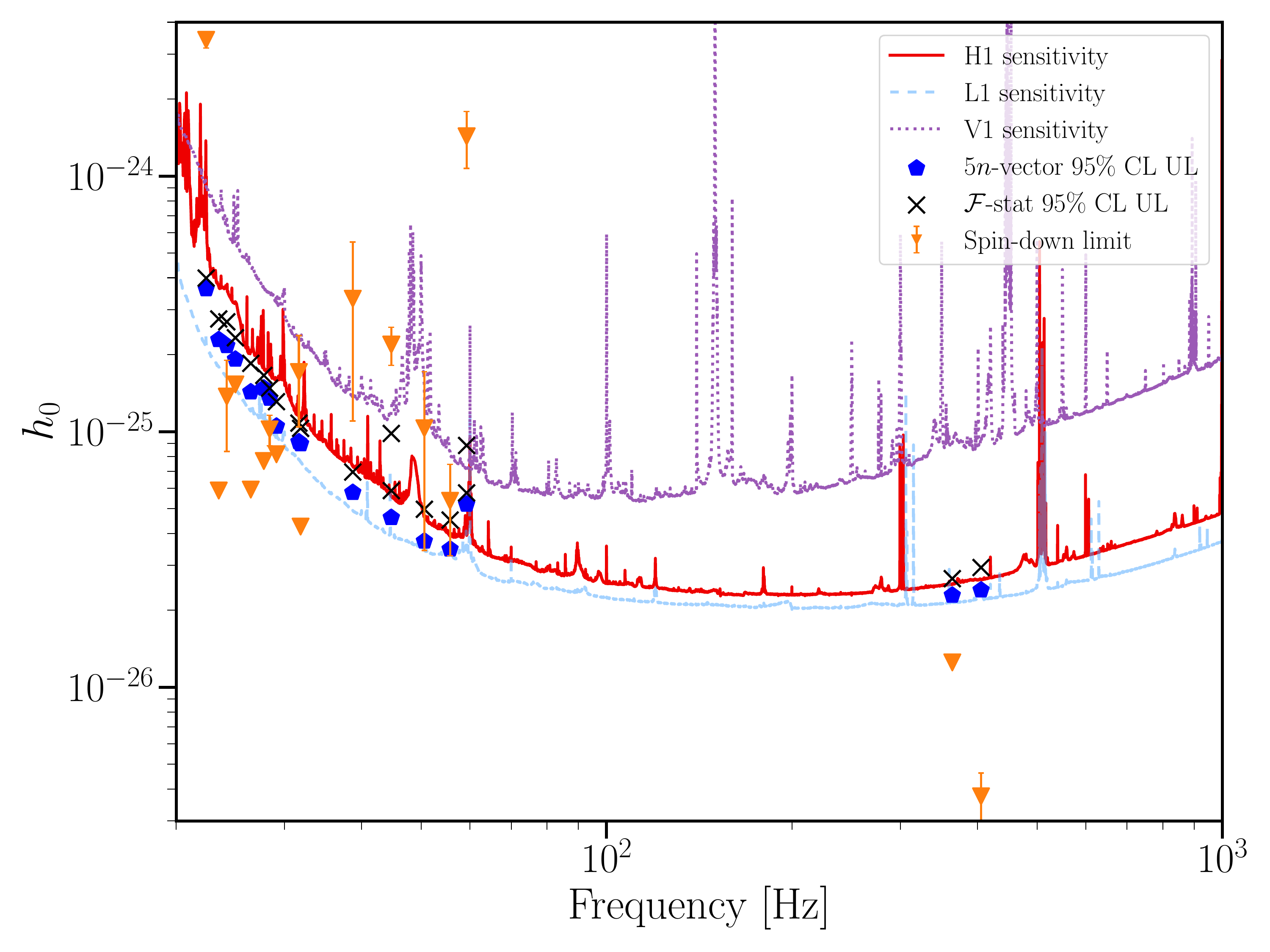}
\caption{Upper limits (95\%\ CL) on $h_0$ for known pulsars from {\it narrowband} searches in the 
  LIGO O3 data~\citep{bib:cwnarrowbandO3}.
  The red solid, blue dashed, and purple dotted curves show the expected sensitivities for H1, L1, and V1, respectively.
  The blue pentagons indicate the median 95\%\ CL ULs from the 5$n$-vector search across all $10^{-4}$~Hz sub-bands for each source.
  The black crosses indicate 95\% CL ULs from the \fstatistic\ search, which are set across the full search range for each target.
  The orange triangles indicate the spin-down limit, $\hsd$, with error bars that reflect uncertainty in the distance to each source.
  In a few cases the error bars are smaller than the size of the markers.
  \figpermission{\cite{bib:cwnarrowbandO3}}{the author(s)}}
\label{fig:cwnarrowbandO3}
\end{center}
\end{figure}

Most recently, further sensitivity improvement was seen in O3 narrowband results~\citep{bib:cwnarrowbandO3}, as shown
in Fig.~\ref{fig:cwnarrowbandO3} for 18 known pulsars with spin-down limits within a factor of 3 of the expected sensitivity
for which the spin-down limit is beaten for six pulsars. A separate analysis of O1--O3a data for seven other
pulsars was carried out in \citep{bib:AshokEtal}.

Searches for accreting X-ray millisecond pulsars (AXMPs) (see section~\ref{sec:targets}) require a modified
narrowband approach in that nominal rotation frequencies are known, but with poor precision compared to that available
for pulsars for which sustained monitoring is feasible. Their frequencies can vary rapidly (and likely with
significant stochasticity) during active (accreting) phases and during quiescent phases can generally only be inferred.
Given these uncertainties, including unknown stochastic contributions, a hidden Markov Viterbi method based on the
\jstatistic\footnote{The \jstatistic\ is a weighted sum of powers from a large number of orbital sidebands generated by
evaluating the \fstatistic\ for a binary source, using a weighting governed by a set of Bessel functions $J_n$ arising from the frequency
modulation and incorporating the orbital phase of the binary system~\citep{bib:ViterbiPaperII}.}
has been applied to searches for five AXMPs in the O2 LIGO data~\citep{bib:ViterbiFiveLMXBsO2}
and to 20 AXMPs in the O3 LIGO data~\citep{bib:cwAXMPO3}. The O3 search yielded strain amplitude sensitivities
in the range (5-24)$\times$10$^{-26}$, where estimated spin-down limits based on measured frequency derivatives lie in the range
$10^{-28}$--$10^{-27}$ with comparable to somewhat larger estimates based on torque balance~\citep{bib:cwAXMPO3} (see section~\ref{sec:spindown}).

\subsection{Directed searches for isolated stars}
\label{sec:directedisolated}

Directed searches are those for which the source location is precisely known, but for which
the signal's gravitational wave phase evolution is unknown or poorly known.
As discussed in section~\ref{sec:templatesdirected}, the 
implied parameter space volume of a truly broadband search will
then depend sensitively upon the assumed age of the star. For a very young
pulsar, one must search over not only the frequency and first frequency derivative
(spin-down), but also over the second and possibly higher derivatives.

Directed-search methods are also appropriate when searching for \rmodes\ from known pulsars.
The search band lies nominally near $4/3$ the star's rotation frequency, but has large systematic
uncertainties of order 10\%\ that depend on the unknown equation of state governing the modes~\citep{bib:IdrisyOwenJones,bib:CarideIntaOwenRajbhandari}.
Hence, while the search
band is much smaller than that for, say, a young neutron star with unknown frequency, the band is
also much larger than that used in narrowband searches (see section~\ref{sec:targeted}),
arguing for careful balancing of computational cost against sensitivity~\citep{bib:CarideIntaOwenRajbhandari}.

The computational cost of fully coherent directed searches can be understood qualitatively from
the scalings with coherence time implied by Eqns.~\ref{eqn:ftolerance}-\ref{eqn:fddottolerance},
with more quantitative estimates based on the template placement considerations discussed in
section~\ref{sec:templatesdirected}. Semi-coherent searches have more complex scalings, but for long
observation spans, generally achieve improved strain sensitivity with respect to fully coherent searches
carried out over shorter subsets of the data set (which is typically necessary). 

The first such analysis in initial LIGO data used the \fstatistic\ algorithm~\citep{bib:cwcasa} to 
search for the central compact object (X-ray source) at the center
of the Cassiopeia A supernova remnant. 
As discussed in section~\ref{sec:sources}, given the $\sim$300-year presumed
age of the star, one can derive a frequency-dependent upper limit on
its strain emission of $\sim$\sci{1.3}{-24}, assuming its rotational energy loss 
has been dominated by gravitational wave emission. A coherent search was carried out
in a 12-day period of LIGO S5 data over the band 100-300 Hz, for which it was
expected that the age-based limit could be tested with that data set~\citep{bib:cwcasamethod}.
The resulting upper limits did indeed beat the age-based limit over that
band, reaching a minimum upper limit of \sci{7}{-25} at 150 Hz.
That the limits were more than an order of magnitude higher than found
in the full-S5 targeted searches for known pulsars in that band reflected
not only the much shorter observation time used (12 days \vs\ 23 months), 
but also the higher SNR threshold
necessary to apply when searching over $\sim$$10^{12}$ templates in $f_s$, $\dot f_s$ and
$\ddot f_s$ for a 300-year old star.

This coherent approach over tractable intervals~\citep{bib:cwcasamethod} was later applied to searches in the data from
the last initial LIGO data run (S6) for nine young
supernova remnants~\citep{bib:S6NineSNRs} and to a possible source at the core of the
globular cluster NGC 6544~\citep{bib:S6NGC6544}, achieving upper limits on strain comparable
to those found in the S5 data, with lowest values ranging over $\sim$4--7$\times10^{-25}$,
depending on source age (lower limits for older sources with lower trials factors from
searching over frequency derivatives).

\begin{figure}[t!]
\begin{center}
\includegraphics[width=0.49\textwidth]{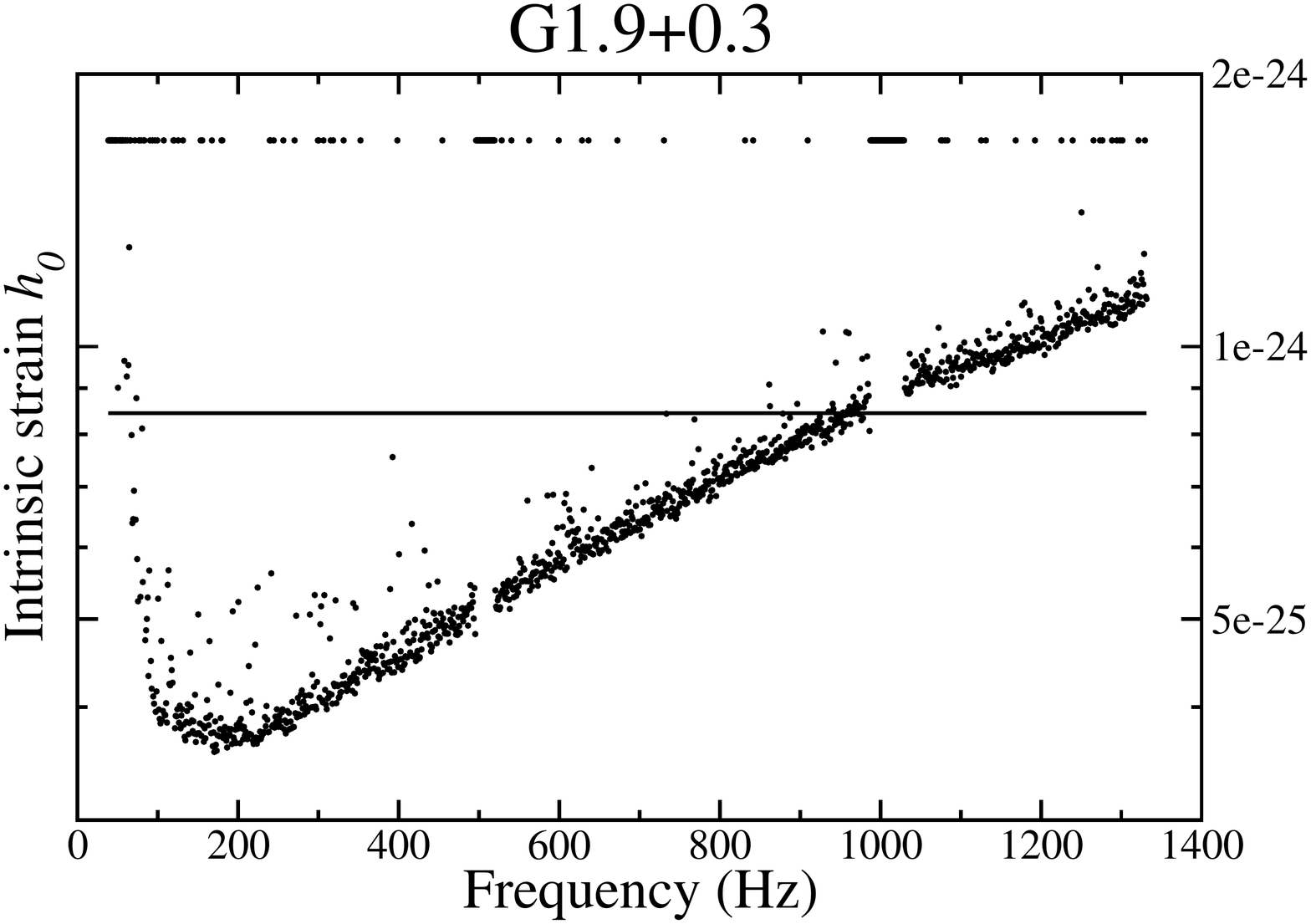}
\includegraphics[width=0.49\textwidth]{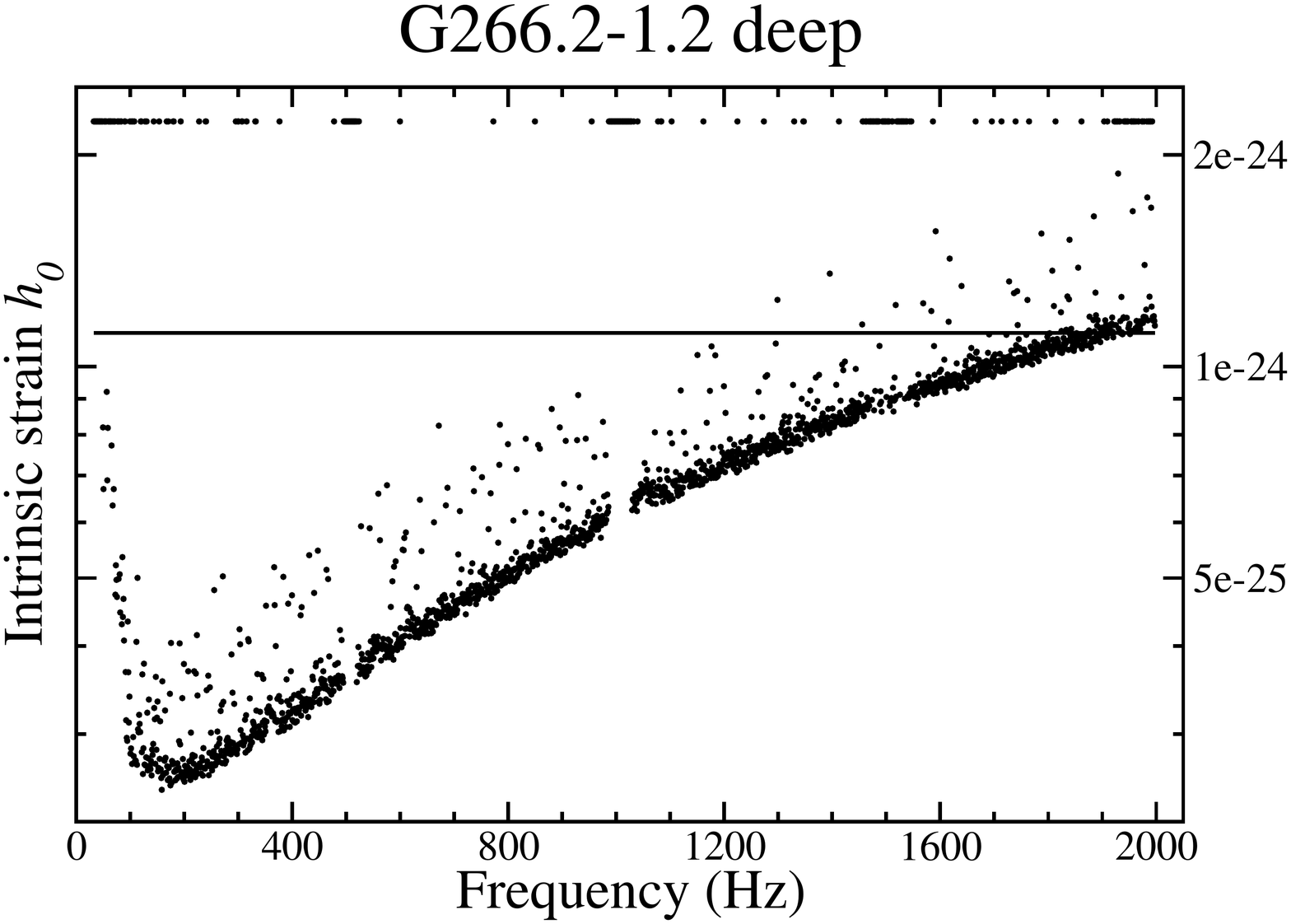}
\includegraphics[width=0.49\textwidth]{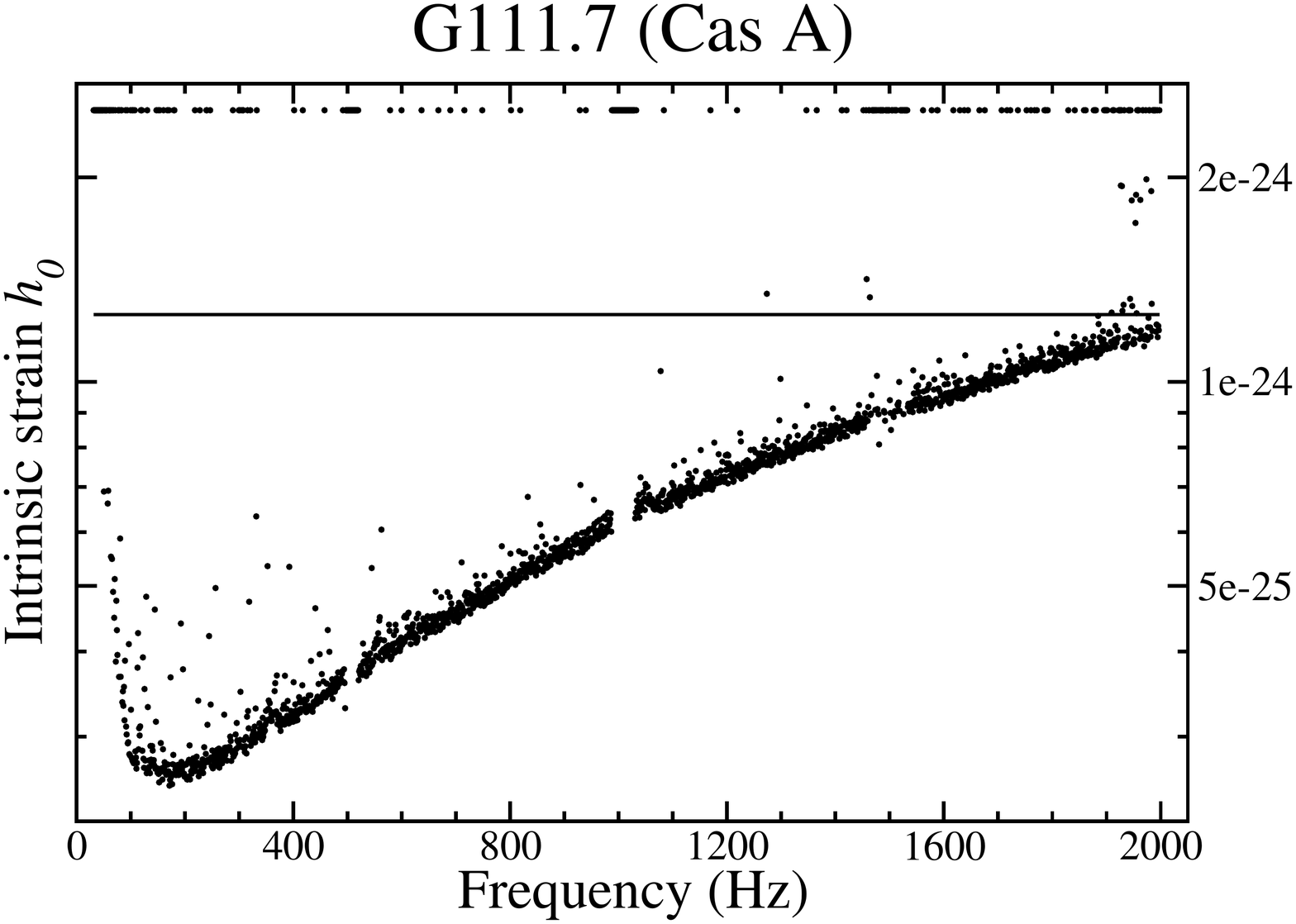}
\includegraphics[width=0.49\textwidth]{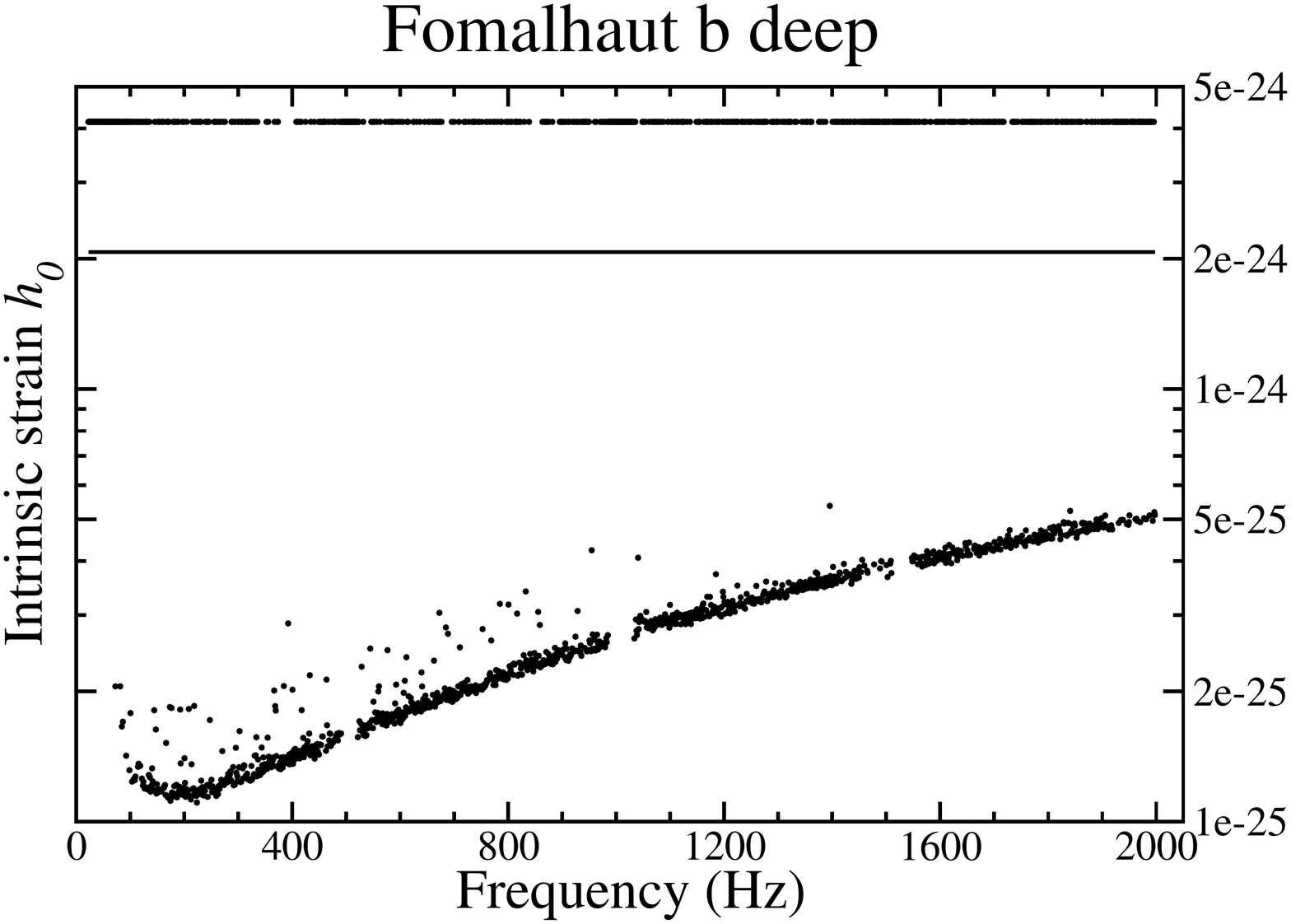}
\caption{Upper limits (95\%\ CL) on $h_0$ (dots) for 3 supernova remnant cores and
  nominal exoplanet (but possible neutron star) Fomalhaut b, using
  coherent \fstatistic\ searches of O1 data~\citep{bib:cwdirectedSNRO1}.
  Upper left: G1.9+0.3; upper right: Vela~Jr.; lower left: Cassiopeia A; lower right: Fomalhaut b.
  The horizontal lines indicate nominal age-based limits (Eqn.~\ref{eqn:agebasedlimit}).
  \figpermission{\cite{bib:cwdirectedSNRO1}}{AAS}}
\label{fig:O1CoherentSNRexamples}
\end{center}
\end{figure}

The coherent \fstatistic\ approach was applied to Advanced LIGO O1 data~\citep{bib:cwdirectedSNRO1}
in a search for 15 supernova remnants and one nominal exoplanet with an unusual apparent orbit,
which has been suggested to be a very
nearby neutron star~\citep{bib:NeuhauserEtalFomalhautb} (see section~\ref{sec:targets}). Best upper limits obtained ranged over $\sim$1-4$\times10^{-25}$,
depending on assumed source range. Figure~\ref{fig:O1CoherentSNRexamples} shows sample results
for three of the supernova remnants, including Cas~A, along with that for Fomalhaut b. In this analysis,
separate ``deep'' and ``wide'' analyses were applied to three of the supernova remnants, including Vela~Jr.,
to account for large uncertainties in source age, where deep searches could be carried out for older sources,
requiring a smaller range in frequency derivatives.
A similar approach was used to probe the O2 data for 12 supernova remnants, restricting attention to
frequencies below 150 Hz, applying coherence times ranging from 12 to 55.9 days~\citep{bib:LindblomOwen}.
A recent coherent search of 8.7-day and 12.8-day subsets O2 LIGO data~\citep{bib:OwenEtalSn1987A} for CW radiation from a Supernova 1987A remnant
beat the age-based indirect limit.
A Viterbi-based search for Fomalhaut b was applied to O2 data~\citep{bib:ViterbiFomalontbO2}.

As one might imagine, a semi-coherent approach has the potential to improve upon a single coherent directed search.
One demonstration of the method in initial LIGO S5 data searched for a source at the galactic center~\citep{bib:cwdirectedgalacticcenterS5},
using 630 segments of 11.5 hours each, where \fstatistic\ values averaged over the segments were computed,
where the global correlation transform template mapping was used in combining the \fstatistic\ values
over the segments.
A similar but more sensitive semi-coherent approach was applied in a
computationally intensive Einstein@Home (see section~\ref{sec:allsky})
S6 search for a CW signal from Cas~A~\citep{bib:directedEatHS6}. This search used
44 segments of 140 hours, again applying the global correlation transform template gridding
and summing.

\begin{figure}[t!]
\begin{center}
\includegraphics[width=12.cm]{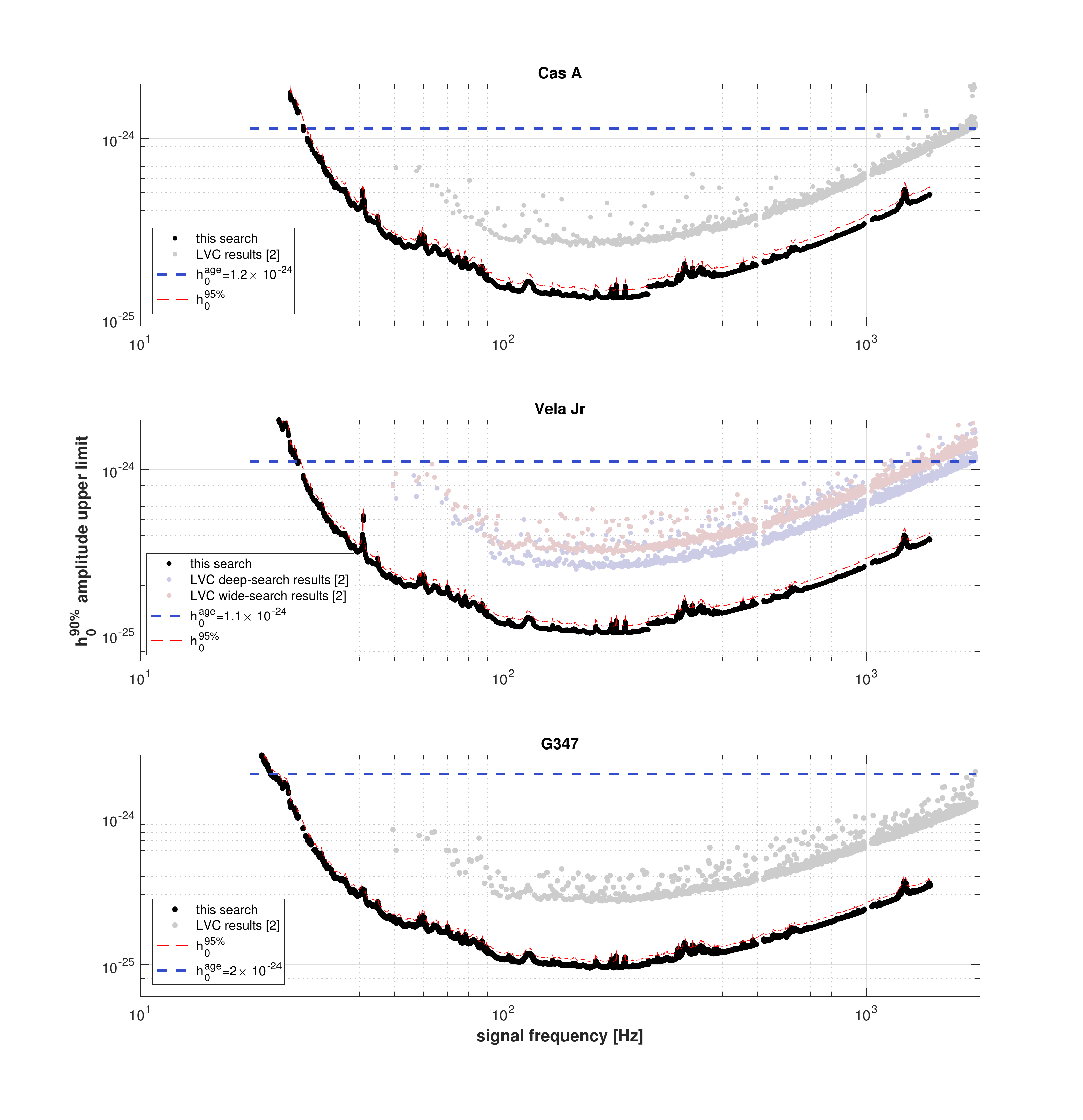}
\caption{Upper limits (90\%\ CL) on $h_0$ for Cassiopeia A, Vela~Jr. and G347.3 (dots)
  from (``this search'') semi-coherent Einstein@Home directed \fstatistic\ searches in Advanced LIGO O1
  data,~\citep{bib:cwdirectedSNREatHO1,bib:cwdirectedSNREatHO1subthreshold}, shown with
  previous (higher) coherent-search limits~\citep{bib:cwdirectedSNRO1} (``LVC results'') using subsets of the O1 data. The dashed curves denote estimated 95\%\ CL upper limits
  based on the 90\%\ CL values. \figpermission{\cite{bib:cwdirectedSNREatHO1}}{APS}}
\label{fig:cwdirectedSNRO1EatH}
\end{center}
\end{figure}

The same method was applied to an Einstein@Home search in Advanced LIGO O1 data
for three supernova remnants: Cas~A, Vela~Jr. and G347.3~\citep{bib:cwdirectedSNREatHO1}.
Figure~\ref{fig:cwdirectedSNRO1EatH} shows the results of the three searches, together with results
from the coherent search of a subset of the same data set~\citep{bib:cwdirectedSNRO1}.
The semi-coherent search 
which exploits the full data set, typically achieves a factor
of two improvement in strain sensitivity over the coherent search over a data subset.
This search also applied a search optimization method~\citep{bib:MingEtalOptimization}
to choose coherence times and segmentations for
each source, where the optimization attempts to take into account relative probabilities for
detection, given available astronomical information. In this instance the segmentations chosen
were 12 245-hour segments (Cas~A), 8 369-hour segments (Vela~Jr.) and 6 489-hour segments (G347.3).
As seen in Figure~\ref{fig:cwdirectedSNRO1EatH}, best upper limits obtained were $\sim$$\scimm{1}{-25}$.
Another semi-coherent directed search for the galactic center, based on the Frequency Hough method~\citep{bib:freqhough1},
accelerated by the Band-Sampled Data (BSD) use of DFTs~\citep{bib:BSD}, was carried out
using the Advanced LIGO O2 data~\citep{bib:PiccinniEtalGalacticCenter}.
Another O2 analysis~\citep{bib:MingEtAlG347} searched for a signal from
the G347.3 using a semi-coherent \fstatistic\ implementation in Einstein@Home.

\begin{figure*}
	\centering
		\subfigure[][G65.7+1.2]
	{
		\label{fig:ulBSD_G65_7}
		\scalebox{0.25}{\includegraphics{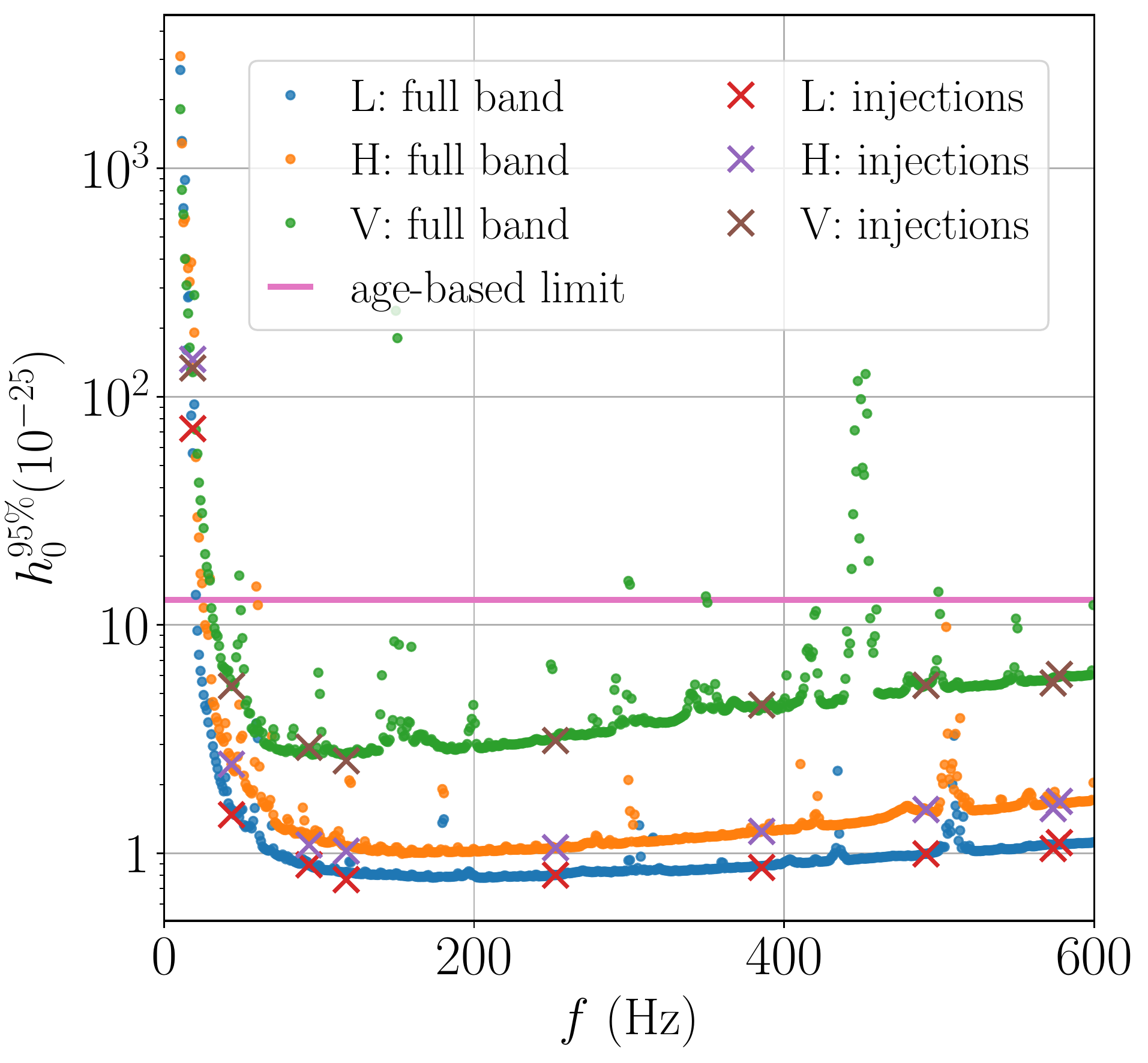}}
	}
	\subfigure[][G189.1+3.0]
	{
		\label{fig:ulBSD_G189_1}
		\scalebox{0.25}{\includegraphics{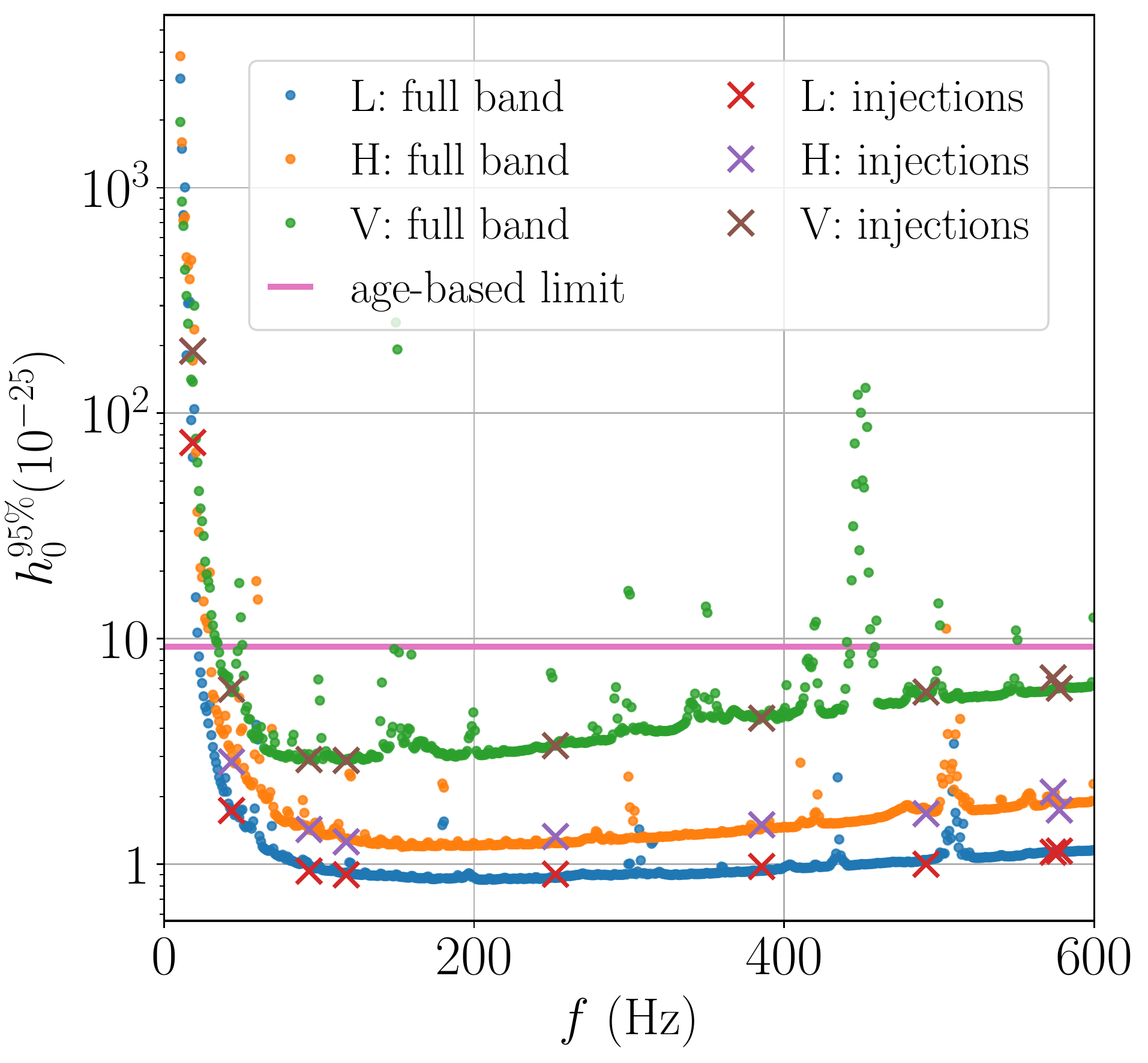}}
	}
 	\subfigure[][G266.2--1.2/Vela Jr.]
	{
		\label{fig:ulBSD_G266_2}
		\scalebox{0.25}{\includegraphics{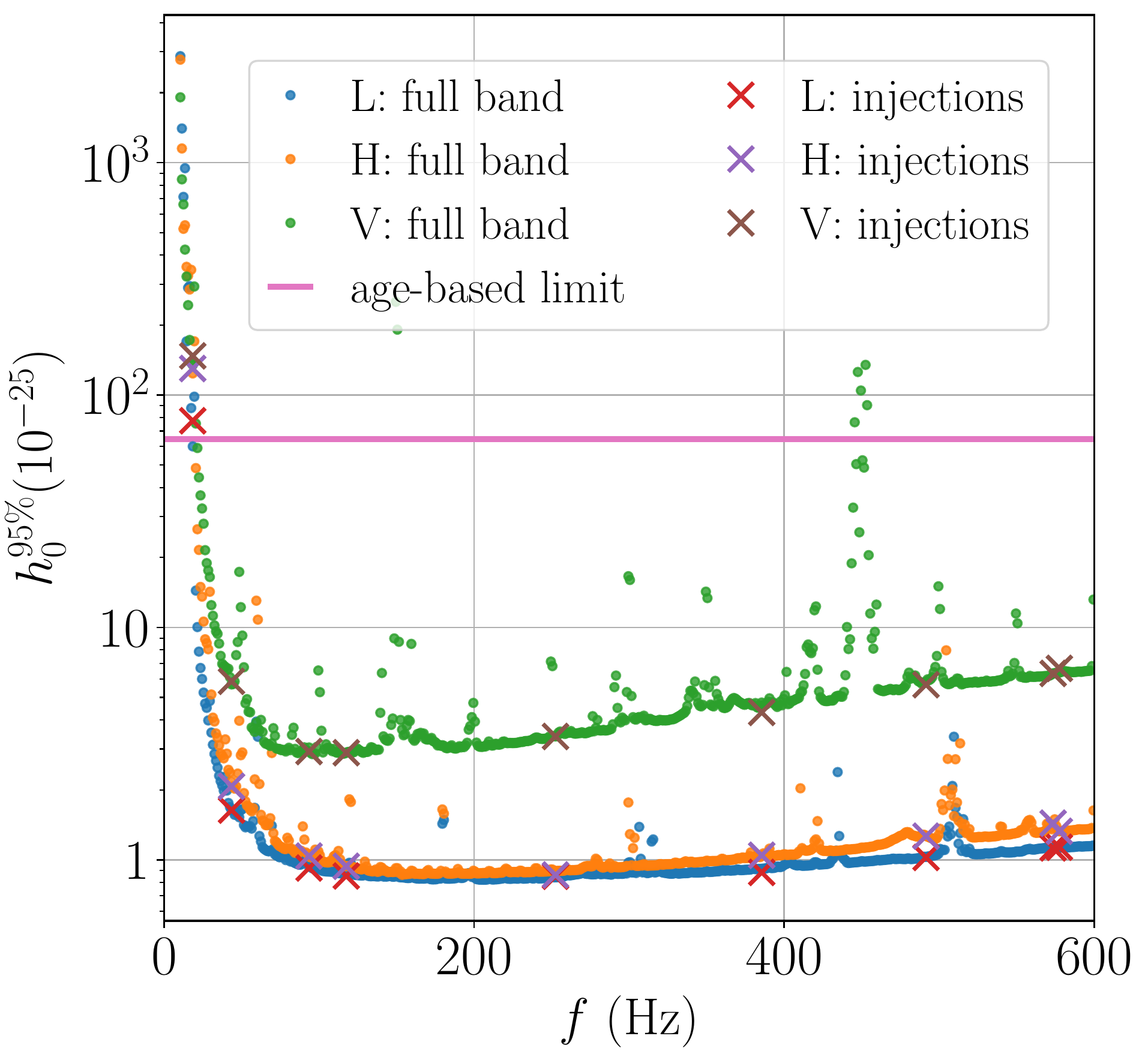}}
	}
	\subfigure[][G39.2--0.3]
	{
		\label{fig:ulBSD_G39_2}
		\scalebox{0.25}{\includegraphics{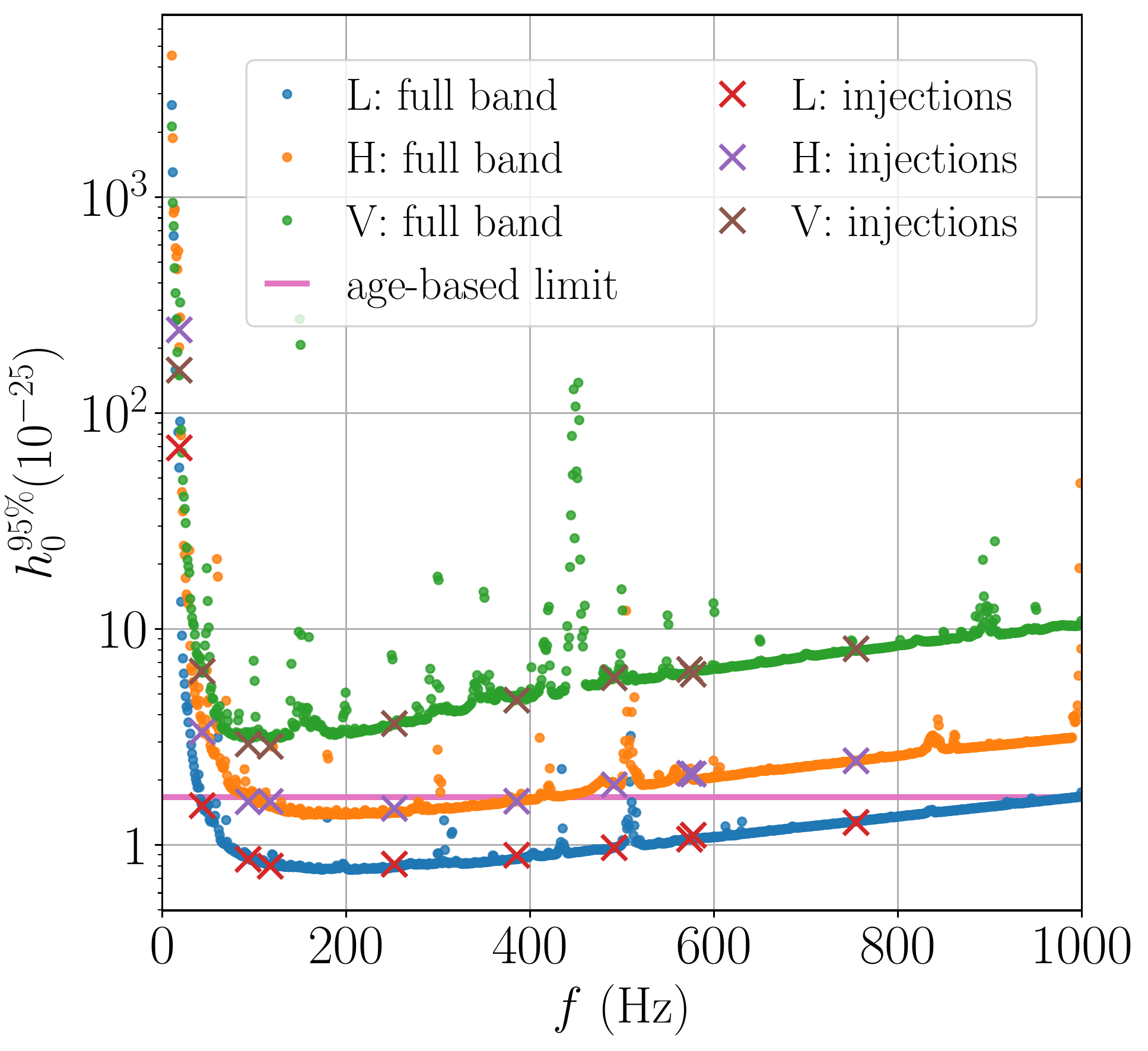}}
	}
	\subfigure[][G93.3+6.9]
	{
		\label{fig:ulBSD_G93_3}
		\scalebox{0.25}{\includegraphics{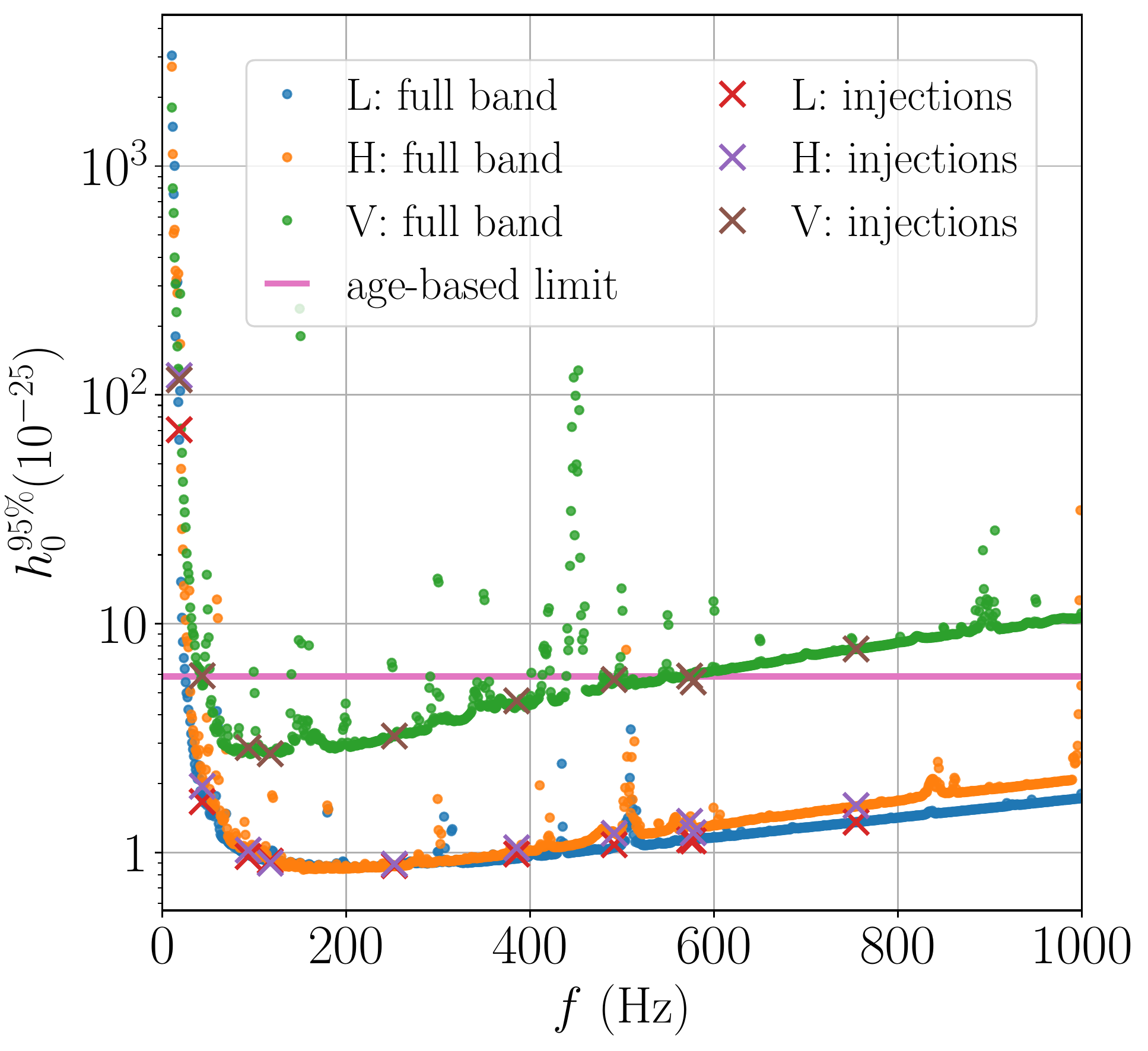}}
	}
	\subfigure[][G18.9--1.1]
	{
		\label{fig:ulBSD_G18_9}
		\scalebox{0.25}{\includegraphics{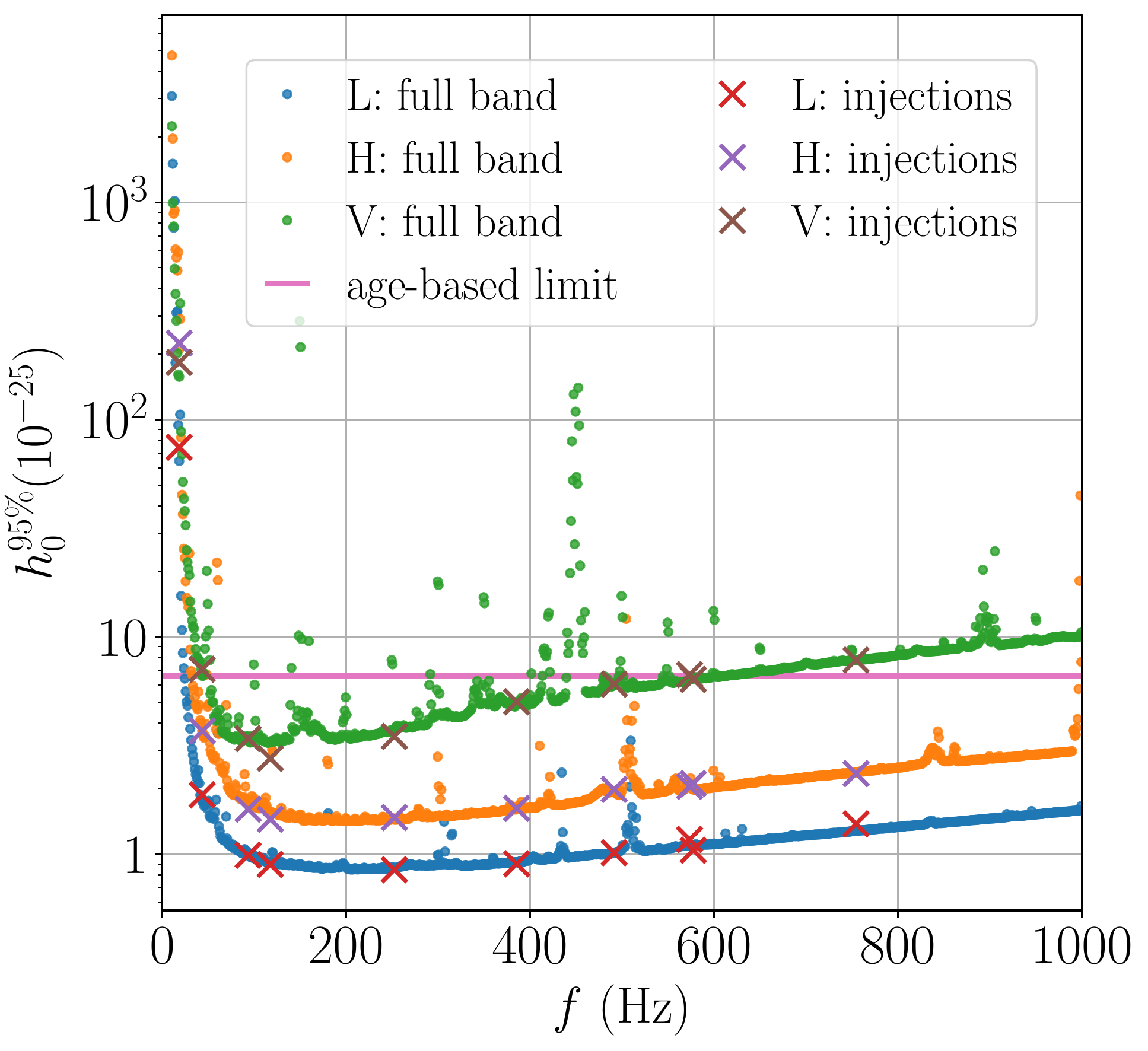}}
	}
	\subfigure[][G353.6--0.7]
	{
		\label{fig:ulBSD_G353_6}
		\scalebox{0.25}{\includegraphics{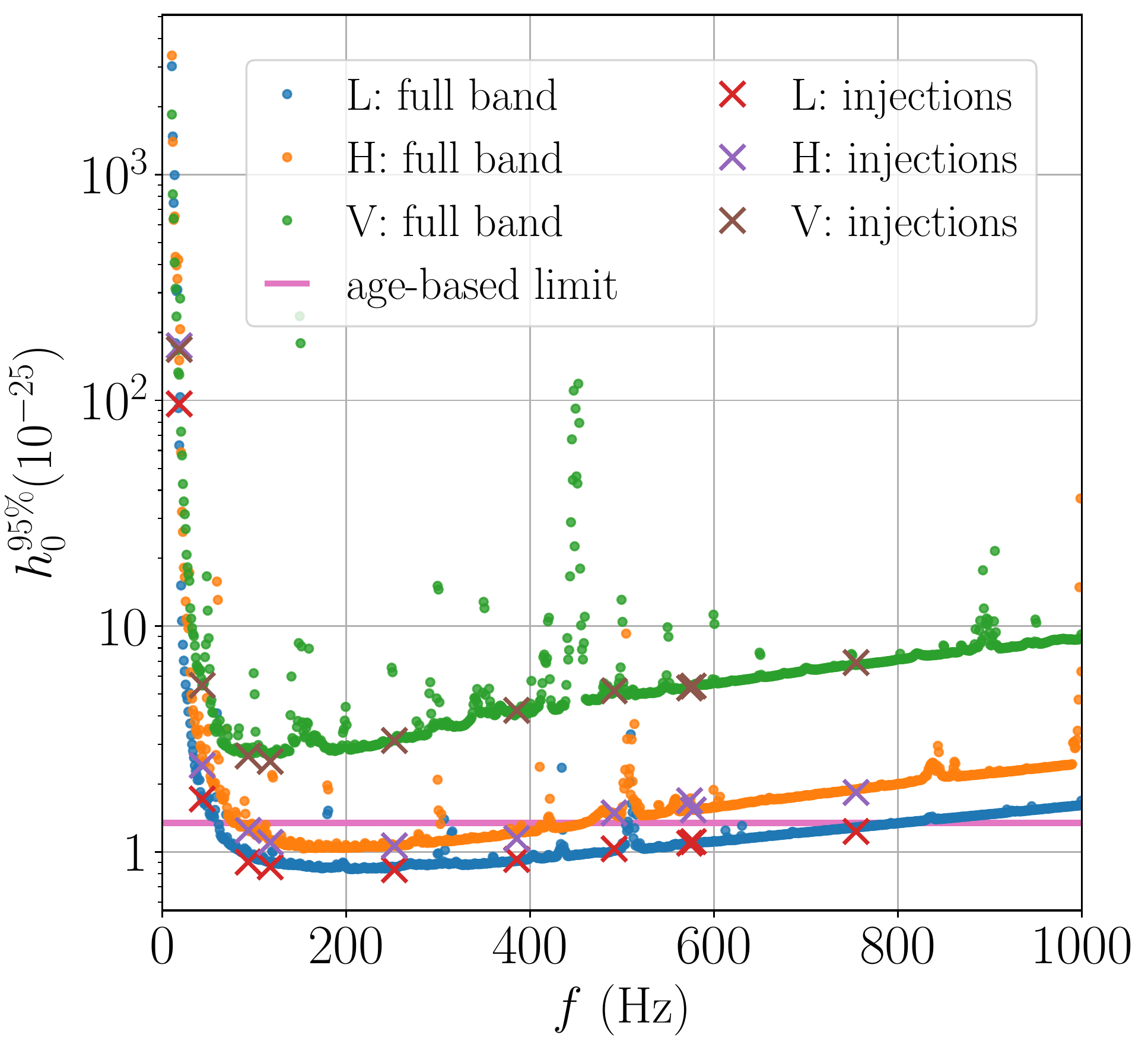}}
	}
	\caption{Sensitivity estimates (95\%\ efficiency) $h_0^{95\%}$ obtained from an O3a semi-coherent Frequency Hough (BSD-accelerated) search~\citep{bib:cwdirectedO3aSNRs}.
          The dotted curves represent the estimated $h_0^{95\%}$ in the full band of H, L and V detectors searched by the pipeline.
          The crosses represent the frequentist strain upper limits at 95\% confidence level obtained empirically in sample sub-bands of 1~Hz.
          Horizontal lines are the indirect age-based limit (Eqn.~\ref{eqn:agebasedlimit}). The limit is beaten across the full band also using
          Virgo data, except for the most disturbed regions, for G65.7+1.2, G189.1+3.0 and G266.2--1.2/Vela Jr.
          The remaining curves beat the limit on a limited parameter space and/or not for every detector. \figpermission{\cite{bib:cwdirectedO3aSNRs}}{AAS}}
	\label{fig:cwdirectedSNRO3a}
\end{figure*}

Several distinct semi-coherent directed searches have been carried out to date using O3 data.
First came results from three methods applied to 15 supernova remnants using the O3a data~\citep{bib:cwdirectedO3aSNRs},
one method being a semi-coherent, BSD-accelerated Frequency Hough search 
and the other two methods being less sensitive but more robust Viterbi~\citep{bib:ViterbiSNRmethod} searches.
The two Viterbi methods searched for either signal at only a single frequency~\citep{bib:ViterbiSNRmethod} (assumed to be twice
the unknown rotation frequency) or at both one frequency and its doubled value~\citep{bib:ViterbiTwoFrequencies}. 
All three methods were applied to seven stars, with only the single-frequency Viterbi method applied
to another eight stars. The results from the Frequency Hough search are shown in Fig.~\ref{fig:cwdirectedSNRO3a}.
A separate publication~\citep{bib:BeniwalEtal} described a Viterbi search for a potential
unidentified pulsar powering HESS~J1427-608, a spatially unresolved TeV gamma-ray point source~\citep{bib:HessUnidentifiedSources}.
Another publication described a coherent \fstatistic\ search (2 days from O3b data) for CW emission
from the center of the 840-year-old supernova remnant G4.8+6.2~\citep{bib:LiuZhouSNR}.

\begin{figure*}[t!]
\begin{center}
  \includegraphics[width=4.75in]{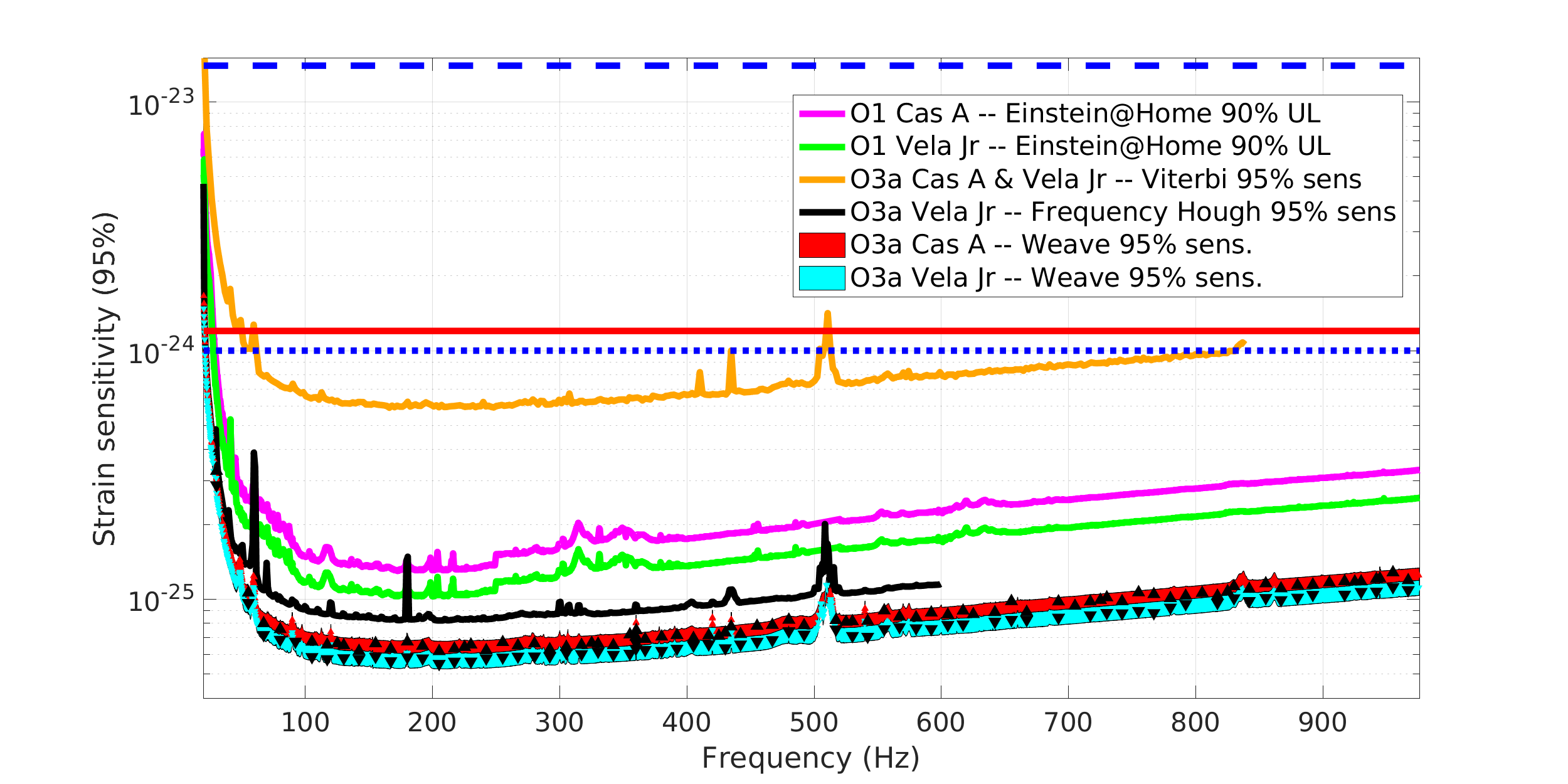}
  \includegraphics[width=4.75in]{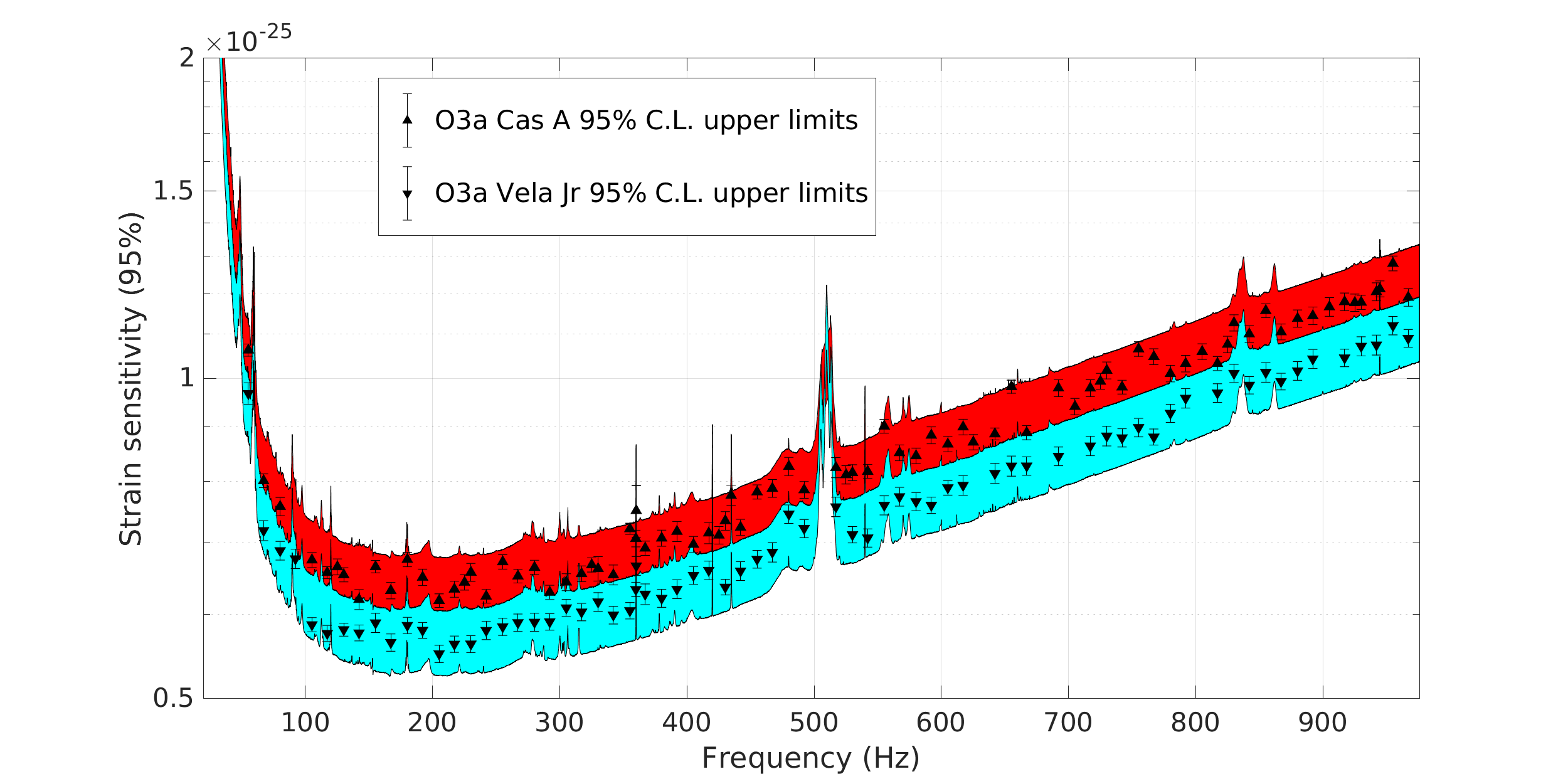}
  \caption{{\it Top panel:} Estimated gravitational wave strain amplitude sensitivities (95\%\ efficiency) in each 0.1 Hz sub-band for the
    O3a \casa\ (red band) and \vela\ (cyan band) searches~\citep{bib:cwdirectedO3aCasAVelaJr}. Conservative uncertainty bands of $\pm$7\% are indicated,
    to account for statistical and systematic uncertainties in estimating sensitivity depths, including calibration uncertainties.
    Black triangles (upright -- \casa, inverted -- \vela) denote 0.1 Hz bands for which rigorous upper limits are used to determine estimated
    sensitivity \vs\ frequency. 
    Additional results from prior searches for \casa\ and \vela\ are also shown:
    O1 Einstein@Home 90\%\ C.L. upper limits for \casa\ (magenta curve) and for \vela\ (green curve)~\citep{bib:cwdirectedSNREatHO1};
    O3a \casa\ and \vela\ 95\%\ C.L. upper limits using a model-robust Viterbi method (orange curve)~\citep{bib:cwdirectedO3aSNRs};
    O3a \vela\ 95\%\ C.L. upper limits using the template-based Frequency Hough method (black curve)~\citep{bib:cwdirectedO3aSNRs}.
    The solid red horizontal line indicates the age-based upper limit on \casa\ strain amplitude.
    The dashed (dotted) horizonal blue lines indicate the optimistic (pessimistic) age-based upper limit on \vela\ strain amplitude,
    assuming an age and distance of 700 yr and 0.2 kpc (5100 yr and 1.0 kpc).
    {\it Bottom panel:} Magnification of the sensitivity bands from the O3a Weave search over most of the search band ($\sim$40--976 Hz), with 1-$\sigma$ statistical uncertainties shown for
    the individual sparsely sampled upper limits used to estimate the depth.
    \figpermission{\cite{bib:cwdirectedO3aCasAVelaJr}}{APS}}
  \label{fig:cwdirectedO3aCasaVelaJr}
\end{center}
\end{figure*}

A separate O3a analysis~\citep{bib:cwdirectedO3aCasAVelaJr} used the \weave\ implementation (see section~\ref{sec:templatesallsky})
of a semi-coherent \fstatistic\ search.
While the package is versatile enough to be used in all-sky searches for unknown
sources, a simpler configuration, applicable to well localized sources, was used to search in the O3a data for the \casa\ and \vela\
supernova remnants. Figure~\ref{fig:cwdirectedO3aCasaVelaJr} shows the results in comparison with earlier searches
for these two sources in O1, O2 and O3a data. 

\begin{figure*}[htbp]
  \includegraphics[width=4.9in]{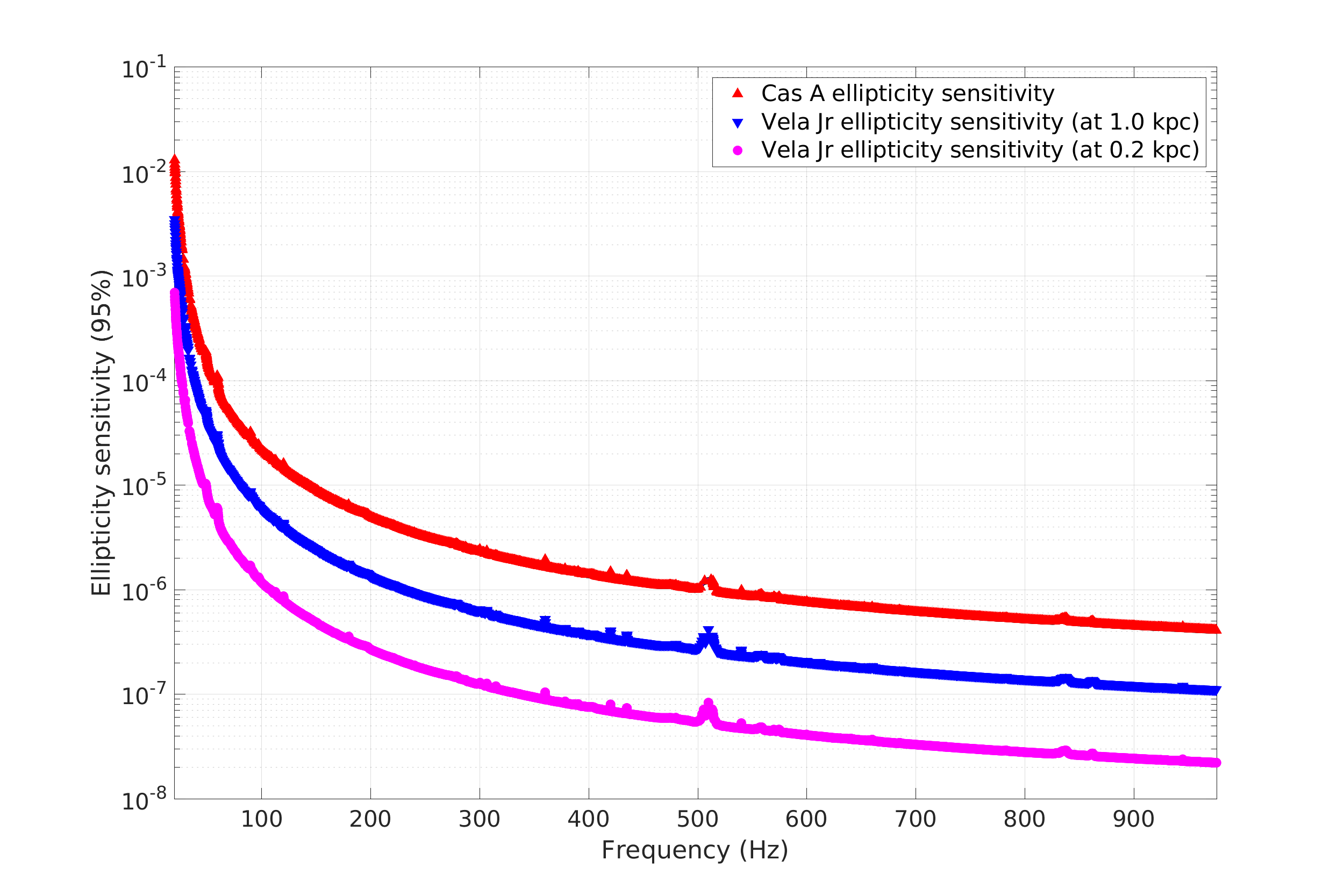}
   \caption{Estimated ellipticity sensitivities (95\%\ efficiency) in each 0.1 Hz sub-band for the O3a (Weave-based) \casa\ (red) and \vela\ (blue, magenta) searches,
    derived from the strain amplitude sensitivities shown in Fig.~\ref{fig:cwdirectedO3aCasaVelaJr} assuming a source distance of
    3.3 kpc for \casa, and assuming source distances of 1.0 kpc and 0.2 kpc for \vela\
    (color online).}
  \label{fig:ellipticitysensitivitiesO3aCasAVelaJr}
\end{figure*}

These 95\%-efficiency sensitivities to \casa\ and \vela\ can be translated into sensitivities to ellipticity,
as shown in Fig.~\ref{fig:ellipticitysensitivitiesO3aCasAVelaJr}. The quadratic dependence of strain on frequency for
fixed ellipticity (see Eqn.~\ref{eqn:hexpected}) leads to dramatically better sensitivity to ellipticity at
higher frequencies, reaching as low as $\epsilon\approx\scimm{2}{-8}$ near 1000 Hz for the more optimistic assumption
of \vela\ distance (0.2 kpc).

Another approach~\citep{bib:xcorrmethod1} for directed searches is based on 
cross correlation of independent data streams. The most straightforward method
defines bins in detector-frame frequency and uses short coherence times, as 
in directional searches for stochastic gravitational radiation,~\citep{bib:radiometermethod,bib:DirectedStochasticO1}
which can be used to search for both isolated and binary sources, albeit with limited sensitivity.
One can use finer frequency binning, however, when correcting explicitly for Doppler modulation of the signal.
Cross-correlation methods are especially robust against wrong assumptions about phase
evolution and are attractive in searching for a very young object, such as
a hypothetical neutron star remaining from Supernova 1987A (see~\citep{bib:AshtonPrixJones} for
a discussion of potential degradation of coherent searches from neutron star glitches,
\citep{bib:PageEtalSNR1987A} for evidence of a hidden star from an excess of infrared emission
and \citep{bib:GrecoEtalSN1987A} for evidence of pulsar wind nebula).
A cross-correlation search for SN 1987A, including demodulation for effects from the motion
of the Earth,~\citep{bib:xcorrmethod2}
was carried out in initial LIGO data~\citep{bib:xcorrS5}. Recent application of cross-correlation methods to
directed searches for binary sources will be discussed in the next section (\ref{sec:directedbinary}).

Directed searches for particular sources require making choices, that is, to prioritize among a
wide set of potential targets in deciding how best to apply computational resources and analyst time.
Recent work~\citep{bib:MingEtalOptimization,bib:targetchoice2} has taken a probabilistic approach to address this
problem, based on source age and distance information (including sometimes large uncertainties) along with detector sensitivity,
an approach that may be generalized to parameter choices in both directed and all-sky searches.

Searches for \rmodes\ radiation from known pulsars are less challenging computationally than truly broadband
directed searches, because the range of expected frequencies is better known. Nonetheless there is substantial theoretical
uncertainty in the ratio between GW emission frequency and rotation frequency. Although the nominal ratio is 4/3
in the slow-spinning, non-relativistic regime, there are substantial corrections for fast-spinning stars and
for stellar compactness that depend on the equation of
state~\citep{bib:YoshidaYoshidaEriguchi,bib:JasiulekChirenti,bib:IdrisyOwenJones,bib:CarideIntaOwenRajbhandari},
leading to a significant range in possible ratios. Following~\citep{bib:CarideIntaOwenRajbhandari}, the ratio can
be written:
\begin{equation}
  {\fgw\over\frot} \approx A - B\left({\frot\over\fkepler}\right)^2,
\end{equation}
\noindent where $\fkepler$ is the Kepler frequency of the star (rotation frequency at which
centrifugal forces destroy the star), $A$ is a parameter dependent on the equation of state,
with an estimated allowed range of 1.39-1.57~\citep{bib:IdrisyOwenJones} and $B$ is
a correction term for high spin with an estimated maximum value of $B_{\rm max}$ = 0.195~\citep{bib:CarideIntaOwenRajbhandari}.

\begin{figure}[htb]
    \centering
    \includegraphics[width=\columnwidth]{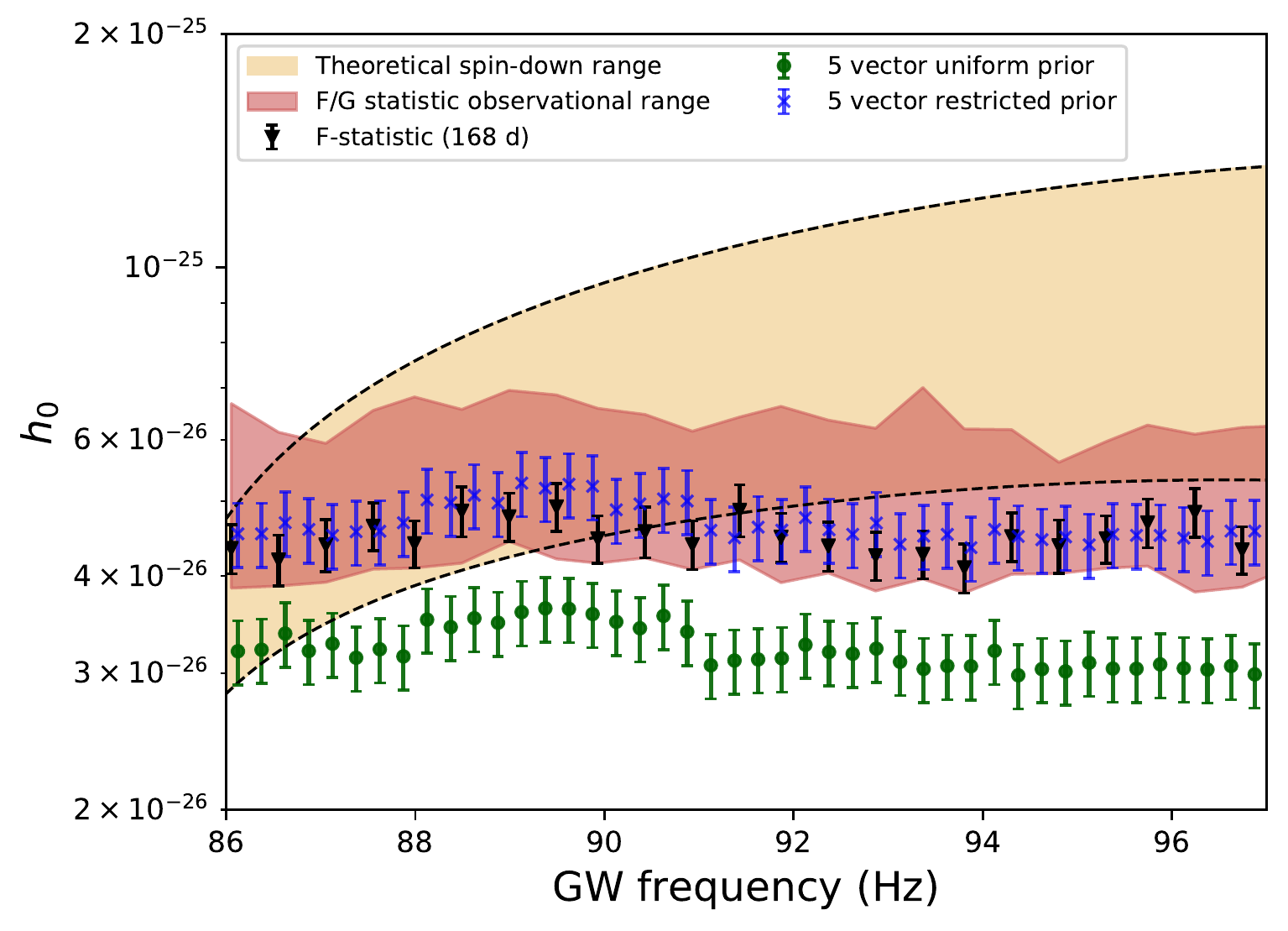}
    \caption{
      Upper limits on GW amplitude $h_0$ obtained from searches for \rmodes\ emission from PSR J0537-6910 in the O3 LIGO data using the \fstatistic/\gstatistic\ and
      5-vector methods~\citep{bib:cwdirectedO3J0537rmodes}. The shaded band indicates the full range of results of the \fstatistic/\gstatistic\ pipeline.
      The dashed lines are defined by the stiffest and softest equations of state considered in the analysis and enclose a range of theoretical $h_0$.
    }
    \label{fig:cwdirectedO3J0537rmodes}
\end{figure}

Using these assumptions, several searches have been carried out explicitly for such \rmodes: 1) an analysis of O1 and O2 data for emission from
the young, energetic pulsar PSR J0537$-$6910~\citep{bib:FesikPapa} (see section~\ref{sec:targets}),
which reached to within an order of magnitude of the strain spin-down limit; 2) an analysis of O1 and O2 data for
emission from the younger, comparably energetic and much closer Crab pulsar, for which the spin-down limit was surpassed by an
order of magnitude~\citep{bib:RajbhandariOwenCarideInta}; and a recent search in the O3 LIGO and Virgo data~\citep{bib:cwdirectedO3J0537rmodes}
which placed stringent constraints on theoretical models for \rmode-driven spin-down in J0537–6910,
especially for higher frequencies for which upper limits reach below the spin-down limit. These latter results
which attempt to address directly the evidence for \rmodes\ in inter-glitch J0537-6910 spin-down are shown for
the frequency band of the search (86-97 Hz) in Fig.~\ref{fig:cwdirectedO3J0537rmodes}.

\subsection{Directed searches for binary stars}
\label{sec:directedbinary}

For known binary pulsars with measured timing ephemerides,
targeted searches work well, and upper limits have been reported for many 
stars, as described in section~\ref{sec:targeted}.
But searching for known (possibly accreting) neutron stars in binary systems
not exhibiting pulsations  or for entirely unknown stars in binary systems
once again significantly increases the parameter space,
relative to the corresponding isolated star searches,
posing new algorithmic challenges and computing costs. 

\begin{figure}[t!]
\begin{center}
\includegraphics[width=12.cm]{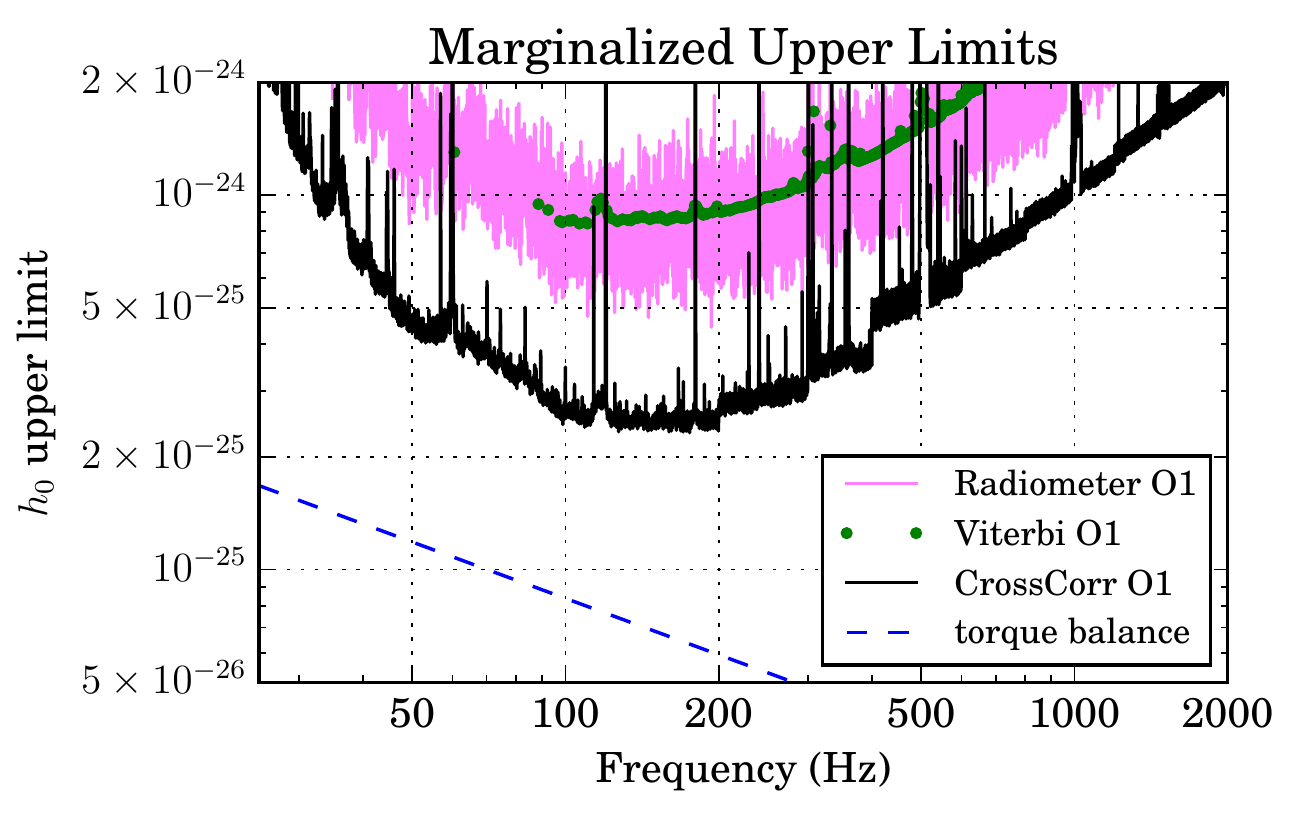}
\caption{Upper limits (95\%\ CL) on $h_0$ for Scorpius X-1 from Advanced LIGO O1 data, using several different search methods:
  a ``radiometer'' search using stochastic analysis methods~\citep{bib:DirectedStochasticO1} and fine frequency binning, a Viterbi method based on
  a Bessel-weighted \fstatistic~\citep{bib:ViterbiO1} and a templated cross-correlation method~\citep{bib:O1CrossCorr}.
  The dashed line indicates the torque-balance benchmark defined in Eqn.~\ref{eqn:torquebalance2} for accretion at the stellar radius. \figpermission{\cite{bib:O1CrossCorr}}{AAS}}
\label{fig:O1CrossCorrScoX1}
\end{center}
\end{figure}

Searches for Sco X-1 in O1 data were carried out with several
methods: 1) a ``Sideband'' method~\citep{bib:sidebandmethod1,bib:sidebandmethod2,bib:sidebandviterbi,bib:ViterbiO1,bib:SunThesis}
based on summing power in orbital sideband frequencies;
2) a non-demodulated cross-correlation method~\citep{bib:BallmerRadiometer,bib:DirectedStochasticO1}
and 3) a demodulated cross-correlation method~\citep{bib:xcorrmethod3,bib:O1CrossCorr,bib:CrossCorrResampling}.
The demodulated cross-correlation method has proven to be the most sensitive method to date
in such searches on a fixed data set for templated signal models without stochasticity,
as expected from a previous mock data challenge~\citep{bib:ScoX1MDC1} including these methods
and others~\citep{bib:twospectmethod,bib:GoetzRilessftsumming,bib:twospectdirectedmethod,bib:twospectS6,bib:polynomial},
and as shown in Figure~\ref{fig:O1CrossCorrScoX1}. Computationally intensive methods using the
\fstatistic, however, may eventually improve upon it~\citep{bib:StackedFstatScoX1Method}.
Follow-up Sco X-1 searches of the O2 data were based on the Viteri method using the
\jstatistic~\citep{bib:cwdirectedO2ScoX1Viterbi,bib:cwdirectedO3ScoX1Viterbi}.

\begin{figure}[t!]
\begin{center}
\includegraphics[width=12.cm]{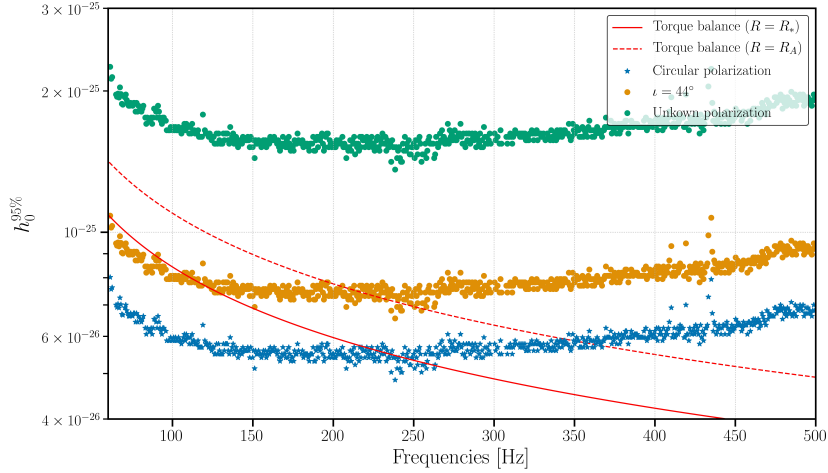}
\caption{Upper limits on strain amplitude $h_0$ from a hidden-Markov model search for Scorpius X-1 at $95\%$ confidence from LIGO O3 data~\citep{bib:cwdirectedO3ScoX1Viterbi} as a function of sub-band frequency, for three scenarios: circular polarization with $\iota=0$ (blue stars), $\iota \approx 44^{\circ}$ based on radio observations (see~\citep{bib:FomalontEtalScoX1}; orange dots), and a flat prior on $\rm cos\,\iota$ (green dots). Indirect torque-balance upper limits (see Section \ref{sec:spindown}) for two torque lever arms are also shown: the stellar radius (red solid line) and the \alfven\ radius (dashed red line).
  \figpermission{\cite{bib:cwdirectedO3ScoX1Viterbi}}{APS}}
\label{fig:O3ScoX1polarizations}
\end{center}
\end{figure}

One complication in Sco X-1 searches is potential spin wandering due to fluctuations in accretion
from its companion~\citep{bib:spinwandering}, which limits the length of a coherence time that can
be assumed safe for a signal template.
One previous fully coherent search~\citep{bib:sidebandS5} restricted its coherence length to 10 days,
to be conservative. Semi-coherent and cross-correlation
methods~\citep{bib:sidebandviterbi,bib:twospectdirectedmethod,bib:radiometermethod,bib:xcorrmethod3}
should be more robust against wandering.
Figure~\ref{fig:O3ScoX1polarizations} shows results from the recent Viterbi-based Sco X-1 search~\citep{bib:cwdirectedO3ScoX1Viterbi} in the O3 data
using the \jstatistic~\citep{bib:ViterbiPaperII}, in which results for different assumptions about Sco X-1 orientation are made.
The implied limits on intrinsic strain amplitude $h_0$ are lowest in the most favorable case of circular polarization,
less favorable for an inclination angle $\iota$ = 44$^\circ$ consistent with observations of
its radio lobes~\citep{bib:FomalontEtalScoX1}, and least favorable for a strain amplitude marginalized over unknown inclination.
Also shown are torque-balance limits assuming both a stellar radius and \alfven\ radius for the accretion lever arm (see section~\ref{sec:spindown}).

\begin{figure}[t!]
\begin{center}
\includegraphics[width=12.cm]{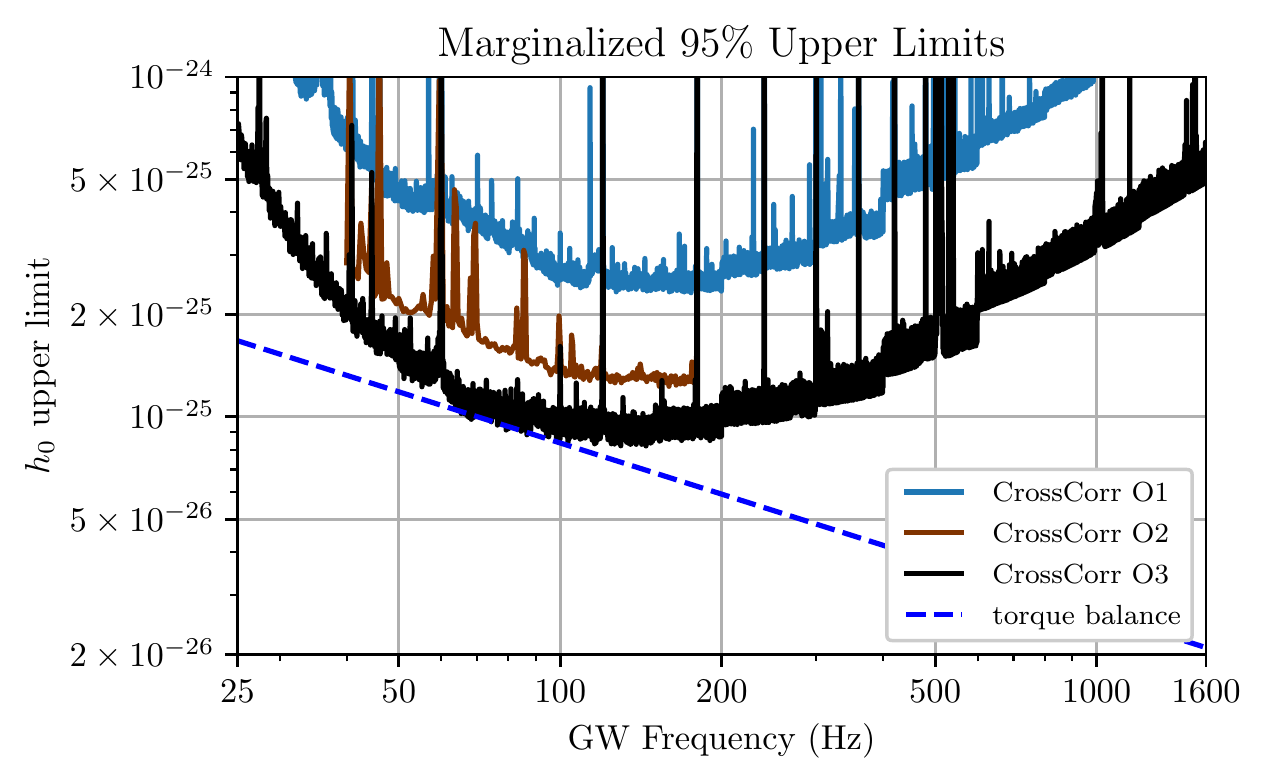}
\caption{Comparison of 95\%\ CL upper limits on $h_0$ due to Sco X-1 emission from searches using the CrossCorr method carried out
  in O1, O2 and O3 data:
  blue solid -- O1 CrossCorr search~\citep{bib:O1CrossCorr},
  brown solid -- O2 CrossCorr search~\citep{bib:O2CrossCorrAEI},
  black solid -- O3 CrossCorr search~\cite{bib:O3CrossCorr}.
  The indirect torque-balance upper limits (see Section V C), using the stellar radius 
  are also plotted (blue dashed line), marginalized over stellar inclination angle.
  \figpermission{\cite{bib:O3CrossCorr}}{the author(s)}}
\label{fig:O2O3ScoX1Comparison}
\end{center}
\end{figure}

\begin{figure}[t!]
\begin{center}
\includegraphics[width=12.cm]{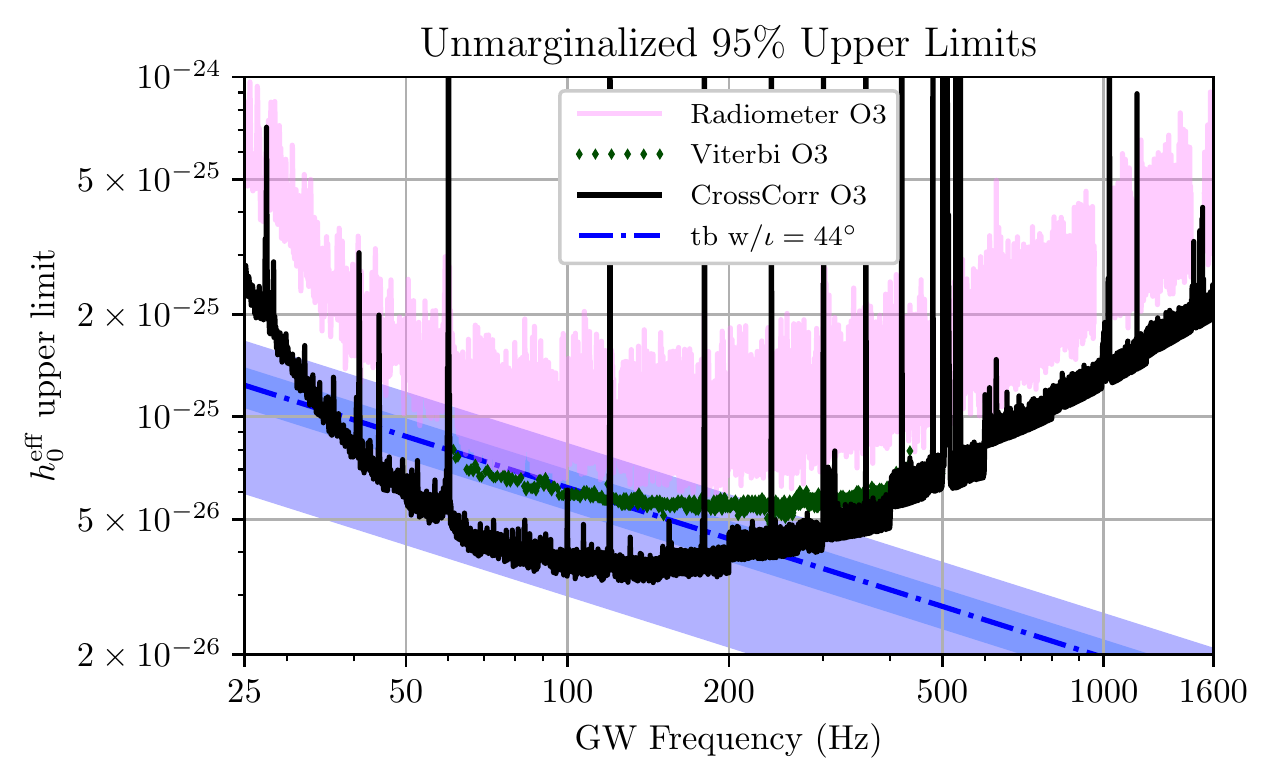}
\caption{Upper limits on {\it effective} strain amplitude $h_0^{\rm eff}$ (defined in text) at $95\%$ confidence from three different searches
  for Scorpius X-1 emission from LIGO O3 data. The CrossCorr limits (black)~\citep{bib:O3CrossCorr} probe the torque-balance limit expectation
    over a broad frequency band and range of assumed inclinations (light purple band). The Viterbi limits (green)~\citep{bib:cwdirectedO3ScoX1Viterbi}
    and Radiometer limits (pink)~\citep{bib:O3Radiometer} shown assume the
    most favorable inclination ($\iota=0$ or $\pi$, producing circular polarization).
    The dashed-dotted blue line and darkened blue band show the torque-balance limits and uncertainty,
    assuming the inclination favored by radio lobe observations~\citep{bib:FomalontEtalScoX1}.
    \figpermission{\cite{bib:O3CrossCorr}}{the author(s)}}

\label{fig:O2O3ScoX1Comparison2}
\end{center}
\end{figure}

Figure~\ref{fig:O2O3ScoX1Comparison} shows a comparison of Sco X-1 upper limits (marginalized over the unknown stellar inclination angle)
obtained from the CrossCorr method applied to the O1, O2 and O3 LIGO data,
with comparisons to the torque balance limit for a stellar radius lever arm.
The O3 inclination-averaged strain upper limits~\citep{bib:cwdirectedO3ScoX1Viterbi}
shown in Figure~\ref{fig:O2O3ScoX1Comparison} now reach as low as the torque-balance benchmark in
Eqn.~(\ref{eqn:torquebalance2}) for a narrow frequency band below 100 Hz.
Less conservatively, Fig.~\ref{fig:O2O3ScoX1Comparison2} shows the results of three search methods applied to the O3 data
in terms of upper limits on ``effective strain'' amplitude, which takes into account the
inclination angle $\iota$ of the star:
\begin{eqnarray}
 ( h_0^{\rm eff})^2 & \equiv & (h_+)^2+(h_\times)^2 \\
                  & = & h_0^2 \> { (1+\cos^2(\iota))/2]^2 + \cos^2(\iota) \over 2},
\end{eqnarray}
\noindent where $h_0^{\rm eff} = h_0$ for circular polarization and $h_0^{\rm eff} = {1\over\sqrt{8}}h_0$ for linear polarization.
The torque-balance upper limit is shown as a band, depending on assumed inclination, with the value favored by radio lobes
observations highlighted by the dashed-dotted line. In this comparison, one can see that recent CW searches 
probe the torque-balance hypothesis over a much broader band below several hundred Hz, depending on inclination assumptions.
As advanced detector sensitivities continue to improve
and with longer data runs, future searches should progressively probe to higher frequencies along this benchmark.

Possessing more definitive information on
the rotation frequency of Sco X-1 could potentially make the
difference between missing and detecting its gravitational
waves in advanced detector data, by both permitting longer coherence-time searches and reducing the statistical 
trials factor and thereby the threshold needed to identify
an interesting outlier. More intensive measurements and
  analysis of Sco X-1 X-ray emission could yield a dramatic scientific payoff~\citep{bib:GalaudageEtal}.

Until recently, CW searches for known LXMB systems focused almost exclusively on
Scorpius X-1, although~\citep{bib:twospectS6} did also include limits from narrowband searches
around three particular frequencies of interest for XTE J1751$-$305, given
X-ray observations of a potential \rmode\ excitation~\citep{bib:strohmayermahmoodifar1}.
More attention is turning now to other accreting systems, such as Cygnus X-2~\citep{bib:GallowayCygX2Source,bib:GalaudageEtal}.
In addition, recent searches were carried out in Advanced LIGO O2 data for five systems~\citep{bib:ViterbiFiveLMXBsO2}
and in O3 data for 20 accreting millisecond pulsars~\citep{bib:cwAXMPO3},
both using a Viterbi hidden Markov method~\citep{bib:ViterbiPaperII} and both exploiting
the relatively good precision with which the stellar rotation frequencies are known (see section~\ref{sec:targeted}).

\subsection{All-sky searches for isolated stars}
\label{sec:allsky}

\subsubsection{Overview of search pipelines in use}

Various semi-coherent algorithmic approaches have been tried, many based in some way on the
``Stack Slide'' algorithm~\citep{bib:stackslide1,bib:stackslide2,bib:stackslide3,bib:stackslideimplementation}
in which the strain powers from Fourier
transforms computed over each coherently analyzed segment are stacked on each other after
sliding each transform some number of bins to account for Doppler modulation of
the source frequency (see section~\ref{sec:stackslide}).
One algorithm is a direct implementation of this idea called StackSlide~\citep{bib:stackslideimplementation}.

Other implementations~\citep{bib:houghmethod,bib:freqhough1} are based on the Hough transform
approach,~\citep{bib:houghibm1,bib:houghibm2} 
in which for each segment a detection statistic is compared to a threshold and given
a value of 0 or 1. The unity values were later refined to be adaptive non-unity weights, to account for
variations in noise and detector antenna pattern~\citep{bib:adaptivefreqhough,bib:adaptiveskyhough}.
The sums of those weights are accumulated in parameter space ``maps,''
with high counts warranting follow-up. The Hough approach offers
greater computational efficiency from reducing floating point operations, along with
robustness against non-Gaussian artifacts~\citep{bib:cwallskyS4} (see section~\ref{sec:lines}).
The Hough approach has been implemented in two distinct search pipelines, the ``Sky Hough''~\citep{bib:houghmethod,bib:cwallskyS2} and
``Frequency Hough''~\citep{bib:freqhough1,bib:freqhough2,bib:cwallskyfreqhoughVSR2VSR4} programs, named after the different parameter spaces chosen in which to
accumulate weight sums.

Another implementation, known as PowerFlux,~\citep{bib:cwallskyS4,bib:PowerFlux1,bib:PowerFluxPol,bib:PowerFlux2,bib:loosecoherence,bib:universalstatistic}
improves upon the StackSlide method by weighting segments by the inverse variance
of the estimated (usually non-stationary) noise and by searching explicitly over
different assumed polarizations while including the antenna pattern correction factors
in the noise weighting (see section~\ref{sec:PowerFlux}).

Yet another method uses coincidences among \fstatistic\ outliers (see section~\ref{sec:fstatistic}) in
multiple time segments typically longer than those used in the semi-coherent
approaches~\citep{bib:tdfstatistic,bib:cwallskyfstatVSR1}, where the implementation is
carried out in the time domain (hereafter denoted as \tdf), with systematic follow-up of
outliers carried out through progressive increase of coherence time~\citep{bib:SieniawskaBejgerKrolak}.

The deepest wideband searches (including wide in frequency derivative range) achieved to date in given fixed data sets
have stacked \fstatistic\ values over time segments
semi-coherently (see section~\ref{sec:stackedfstatistic}) and have used the resources of the distributed computing
project Einstein@Home~\citep{bib:cwallskyEatHS4} based on the same software infrastructure (BOINC)~\citep{bib:boinc}
developed for the Seti@Home project~\citep{bib:seti@home}.
Einstein@Home encourages
volunteers to download narrow-band segments of LIGO data and carry out a semi-coherent
\fstatistic\ search over a small patch of sky. Results are automatically returned to 
an Einstein@Home server and recorded, with every set of templates analyzed independently
by host computers owned by at least two different volunteers. Einstein@Home Scientists then carry out post-processing
to follow up on promising outliers found. This project has been remarkably successful
in engaging the public (hundreds of thousands of volunteers and 750,000 host computers to date) 
in forefront science  while making
good use of idle computer cycles to carry out searches that would otherwise exceed
the capacity of dedicated gravitational wave computing clusters.

The availability of the Einstein@Home platform has driven the evolution of
semi-coherent stacked \fstatistic\ techniques. This evolution has led to 
imcreased search sophistication and sensitivity over the last decade and a half, in
general, including for related pipelines outside of that distributed computing framework, such as \weave\ (which has
a memory footprint incompatible with Einstein@Home).
Particular improvements have included search setup optimization~\citep{bib:stackslide3,bib:PrixShaltev,bib:shaltev},
more efficient semi-coherent stacking and template placement,~\citep{bib:PrixTemplates1,bib:PrixTemplatesEfficient,bib:Pletsch,bib:PletschAllen,bib:WettePrix,bib:WetteTemplates1,bib:WetteTemplates2,bib:WetteTemplates3,bib:WetteTemplates4,bib:WalshEtaltemplates}
automated vetoing of instrumental lines,~\citep{bib:lineveto1,bib:lineveto2,bib:lineveto3} and 
hierarchical outlier followup and
veto~\citep{bib:shaltevetal,bib:papafollowup,bib:singh,bib:ZhuEtalDoppler,bib:AshtonPrix,bib:IntiniEtalDoppler}.

Technical challenges in distributed computing include efficient data transfer to/from
host computers and running on many computing platforms of greatly varying
CPU. GPU and memory capabilities. The large computing resources available via distributed
computing can be used to enlarge the parameter space searched or to probe more deeply
in the noise than is feasible on current computing clusters, but optimization must
account for scaling of computing cost with the target range of frequency and
frequency derivative and weigh the benefit of longer coherence time for sensitivity
against the incurred cost (see section~\ref{sec:challenges}).

The \fstatistic-stacking techniques can also be used, of course, in
less powerful computing environments, with different tunings, \eg, shorter
coherence times per segment. These techniques can also be used for systematic
follow-up of outliers found in first-stage semi-coherent \fstatistic\ searches or in
searches using other semi-coherent methods~\citep{bib:WalshEtaltemplates}, including
both all-sky and directed searches.
One general-purpose, multi-stage approach uses the python wrapper PyFstat for \fstatistic\ summing~\citep{bib:AshtonPrix,bib:pyFstat}
and a Markov Chain Monte Carlo search through parameter space to ``zero in'' on signals~\citep{bib:TenorioEtal}.
This method systematically lengthens segment coherence times (hence reducing segment counts per observational run)
simultaneously with narrowing of the parameter space volume, while guided by the parameters of
the loudest survivors from each stage.

A comparison of many of these all-sky search methods was carried out via a mock
data challenge using initial LIGO data,~\citep{bib:allskymdc}
and these methods have been applied to 
searches of the Advanced LIGO O1--O3 data
sets~\citep{bib:cwallskyO1paper1,bib:cwallskyEatHO1,bib:cwallskyO1paper2,bib:cwallskyO2,bib:cwallskyO2EatH,bib:cwallskyO3aPowerFlux,bib:cwallskyO3FourPipelines}.
Unsurprisingly, the all-sky search  enabled by Einstein@Home computing resources
displayed consistently better sensitivity than the other methods in the mock data challenge, given
the longer coherence times made possible by those resources.

A newcomer all-sky search pipeline, known as the SOAP pipeline~\citep{bib:ViterbiGlasgow}, uses a Viterbi approach to seek
trajectories in spectrograms for which each time segment is represented by the average spectrum over a 24-hour period
using a 30-minute coherence time. Although not as sensitive as the pipelines described above, the technique is
blazingly fast, in comparison, offering the potential of rapid discovery for observing runs with much improved
detector noise. Perhaps more important, because the algorithm is untemplated, it has the additional potential of detecting
new (strong) signals that do not follow the models sought by other isolated-star pipelines, including long-period
binary systems.

\subsubsection{Results from all-sky, isolated-star searches of LIGO and Virgo data}

The Sky Hough algorithm was used to produce all-sky upper limits in the 200-400 Hz band
of the LIGO S2 data~\citep{bib:cwallskyS2}, based on a total of 3800 30-minute segments of data
from the three LIGO interferometers. The StackSlide, Sky Hough and PowerFlux methods 
were used to produce all-sky upper limits in 
the 50-1000 band of the LIGO S4 data~\citep{bib:cwallskyS4}.
The first Einstein@Home all-sky search was carried out too
on the S4 data~\citep{bib:cwallskyEatHS4}.

The PowerFlux algorithm was
used to produce all-sky upper limits in the 50-1100 Hz band
of the first eight months of LIGO S5 data~\citep{bib:cwallskyearlyS5}. The sheer length
of data for the full 23-month S5 run required substantial upgrade of the program which
was then used to produce all-sky upper limits in the 50-800 Hz band of the full data set,
based on a total of more than 80,000 (50\%-overlapped) 
30-minute segments from the H1 and L1 data.
This PowerFlux result~\citep{bib:cwallskyS5} included a three-stage hierarchical search
with a follow-up procedure of loud candidates based on {\it loose coherence}
(see section~\ref{sec:longlagloose}).
A Sky Hough search of the S5 data consisted of a coincidence analysis of data sets
from two separate approximately 1-year subsets of the data over the 50-1000 Hz band.
Einstein@Home too was applied in sequential analyses to the early S5~\citep{bib:cwallskyEatHearlyS5}
and to the full S5~\citep{bib:cwallskyEatHS5}.
A final all-sky initial LIGO PowerFlux analysis of the S6 data set~\citep{bib:cwallskyS6}
included a 5-stage hierachical search with longer and longer effective coherence times
over 100-1500 Hz within the loose coherence framework.
The S6 Einstein@Home search~\citep{bib:cwallskyEatHS6} achieved the most sensitive all-sky results from any
of the initial LIGO data sets, reaching upper limit values as low as \sci{5.5}{-25}. 

When initial Virgo VSR1 data became available, a direct time-domain implementation of
the \fstatistic~\citep{bib:tdfstatistic}  was applied to a search of it for the 100-1000 Hz
band~\citep{bib:cwallskyfstatVSR1}.
Later, the Frequency Hough method was applied to data from the initial Virgo
VSR2 and VSR4 runs over the 20-128 Hz band, the first time an all-sky search was applied
to frequencies below 50 Hz~\citep{bib:cwallskyfreqhoughVSR2VSR4}.

Since Advanced LIGO observing has begun, multiple all-sky search programs have been applied
to data from the first three observing runs, O1, O2 and O3.
The first publications based on O1 data focused on lower frequencies. Four pipelines
(PowerFlux, Sky Hough, Frequency Hough and \tdf) covered the band 20-475 Hz and a spin-down range
  $[\scimm{-1.0}{-8},\scimm{+1.0}{-9}]$ Hz/s~\citep{bib:cwallskyO1paper1}.
  A separate Einstein@Home search using
  the \gctf\ method drilled deeper in the 20-100 Hz band in a narrower
  spin-down range $[\scimm{-2.65}{-9},\scimm{+2.64}{-10}]$ Hz/s~\citep{bib:cwallskyEatHO1}.
  A follow-up publication using three of the first four pipelines (PowerFlux, Sky Hough and \tdf)
  covered the broader band 475-2000 Hz~\citep{bib:cwallskyO1paper2}.
  
\begin{figure}[htb]
\begin{center}
  \includegraphics[width=5in]{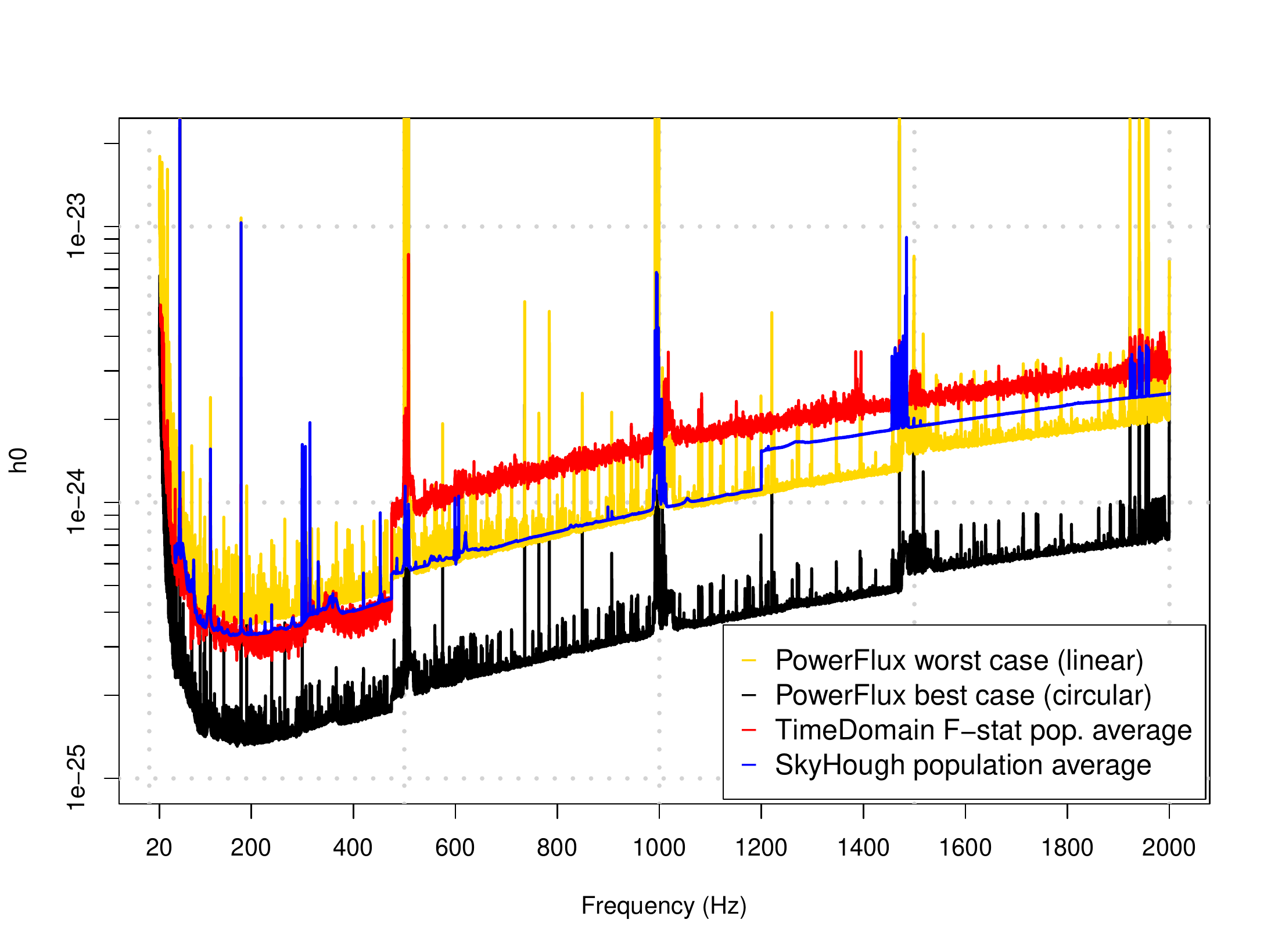}
\vspace{-\abovedisplayskip}
\caption{O1 all-sky upper limits (95\%\ CL) on $h_0$ for isolated stars
  from three semi-coherent search pipelines over the band 20-2000 Hz~\citep{bib:cwallskyO1paper2}.
  The limits shown for the PowerFlux method correspond to best-case (circular polarization) and worst-case (linear polarization)
  over the entire sky,  while the limits shown for the time-domain \fstatistic\ and SkyHough methods correspond to population
  averages over the sky and source orientations. The steps in sensitivty apparent in the limits correspond to reductions
  in FFT coherence time as frequencies increase.
  \figpermission{\cite{bib:cwallskyO1paper2}}{the author(s)}}
\label{fig:cwallskyO1fullband}
\end{center}
\end{figure}

\begin{figure}[htb]
  \begin{center}
  \includegraphics[width=5in]{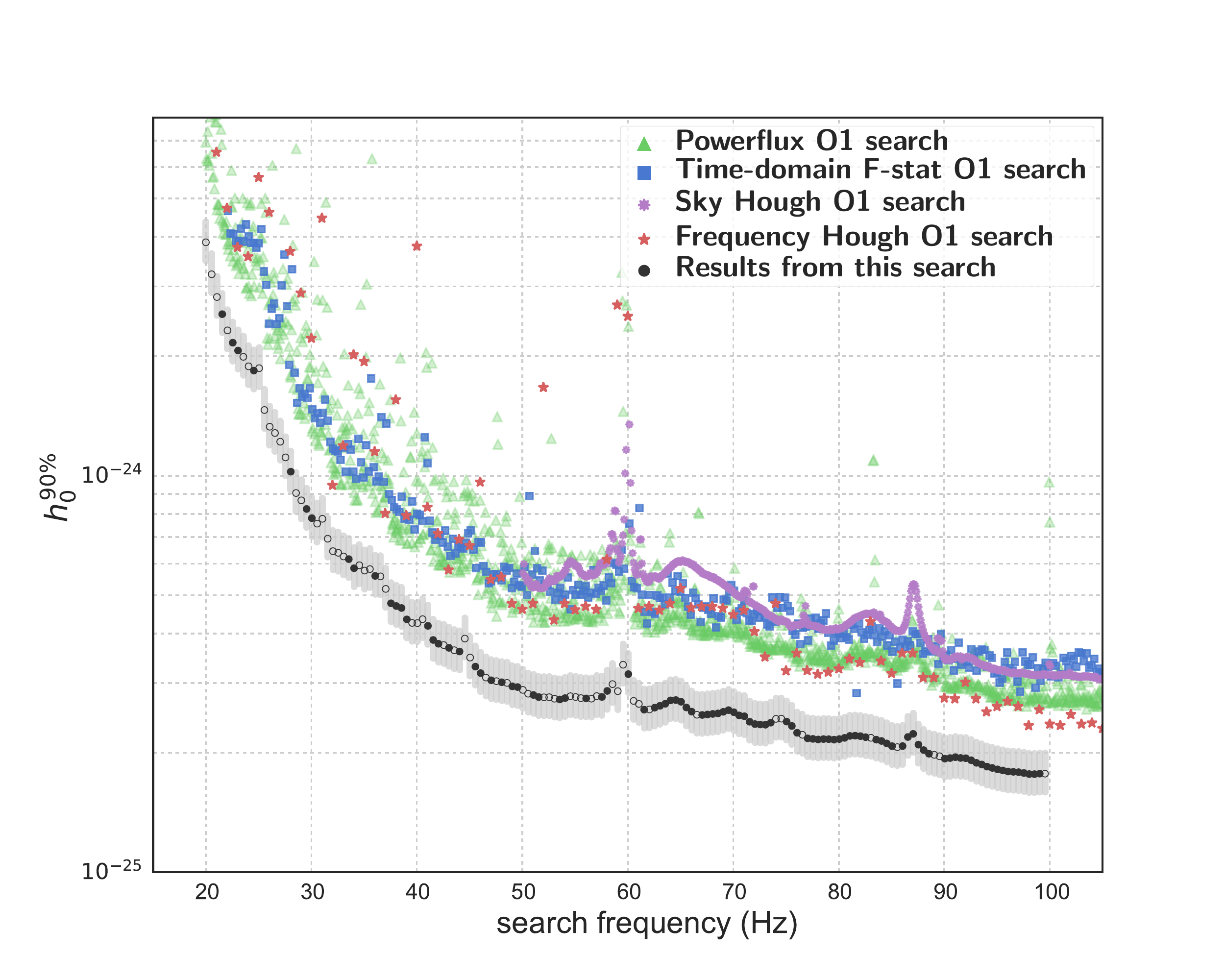}
\vspace{-\abovedisplayskip}
\caption{O1 all-sky upper limits (95\%\ CL) on $h_0$ for isolated stars in the low-frequency band (20-100 Hz) for five semi-coherent
  pipelines~\citep{bib:cwallskyEatHO1}, including an Einstein@Home \gctf\ search (``this search''). The PowerFlux limits here
  are population-averaged, unlike those shown in Figure~\ref{fig:cwallskyO1fullband}.
  \figpermission{\cite{bib:cwallskyEatHO1}}{the author(s)}}
\label{fig:cwallskyEatHO1}
\end{center}
\end{figure}

  Figure~\ref{fig:cwallskyO1fullband} shows the full-band O1 results from~\citep{bib:cwallskyO1paper2}
  for these three pipelines. The PowerFlux results shown are defined differently from those
  shown for the other searches. PowerFlux upper limits are derived as strict frequentist over
  the full sky, that is, a 95\%\ CL limit provides at least 95\%\ coverage, regardless of sky
  position, making it quite conservative. At the same time, however, limits are shown for
  an optimistic polarization assumption (circular polarization corresponding to $|\cos(\iota)| =1$)
  and for a pessimistic assumption (linear polarization corresponding to $\cos(\iota)=0$ for
  the least favorable choice of polarization angle $\psi$). These limits are derived directly
  from the corresponding detection statistics (see section~\ref{sec:PowerFlux}).
  The other limits shown are conventional frequentist population-based values, averaged over
  source orientation and sky position.
  Figure~\ref{fig:cwallskyEatHO1} shows the low-frequency band up to 100 Hz, comparing the
  limits obtained in~\citep{bib:cwallskyO1paper1} with those from the \gctf\ search on
  Einstein@Home~\citep{bib:cwallskyEatHO1} (which use a smaller spin-down range),
  where the PowerFlux limits have been reevaluated via explicit simulation
  to produce population-averaged values for comparison.

\begin{figure}[htb]
\begin{center}
  \includegraphics[width=5in]{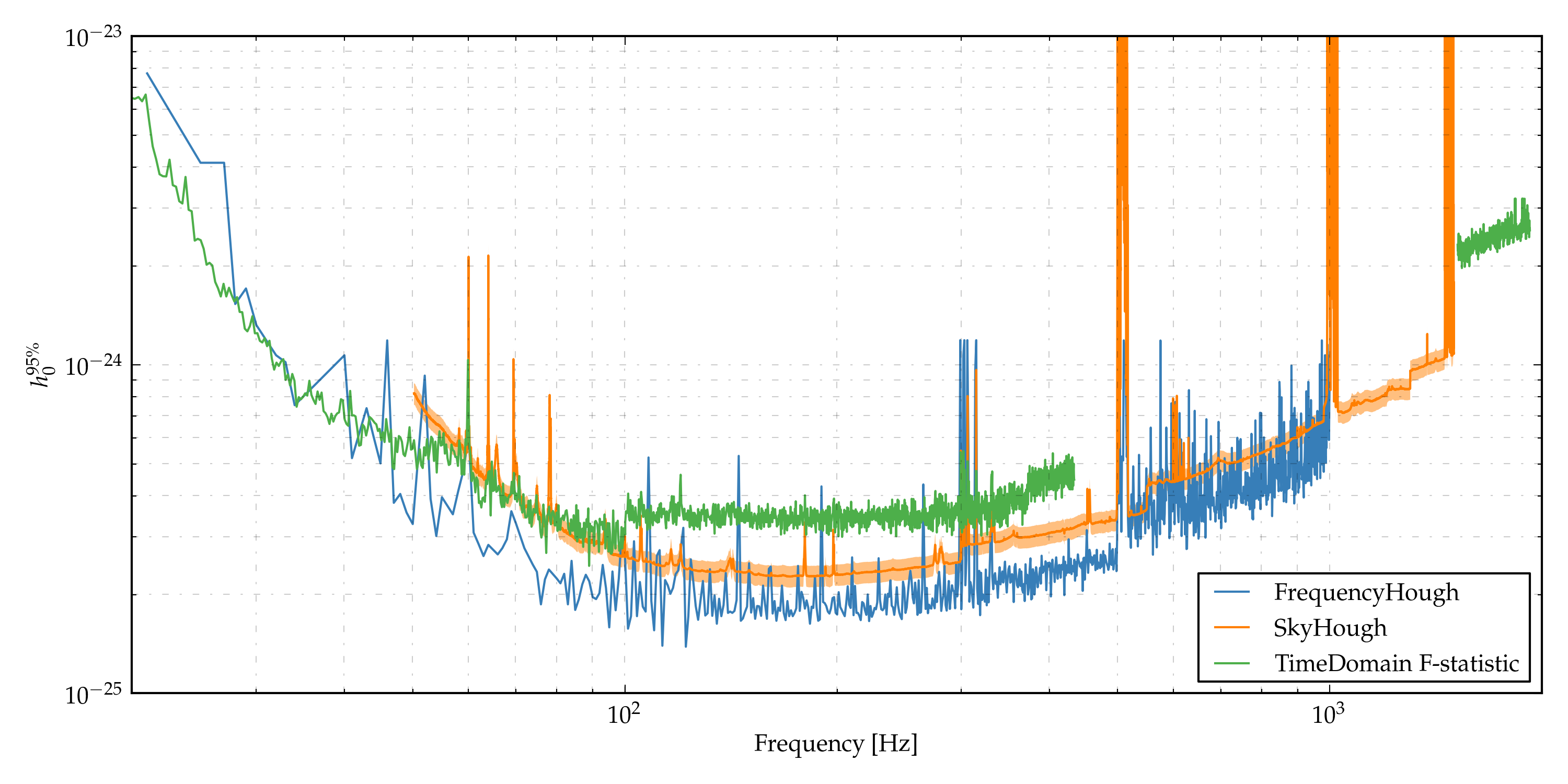}
\vspace{-\abovedisplayskip}
\caption{O2 all-sky upper limits (95\%\ CL) on $h_0$ for isolated stars from three semi-coherent search pipelines over the band 20-1922 Hz.
  As in Figure~\ref{fig:cwallskyO1fullband}, step changes in sensitivity correspond to reductions in FFT coherence time with
  increasing frequency.
  \figpermission{\cite{bib:cwallskyO2}}{the author(s)}}
\label{fig:cwallskyO2}
\end{center}
\end{figure}

\begin{figure}[htb]
\begin{center}
  \includegraphics[width=5in]{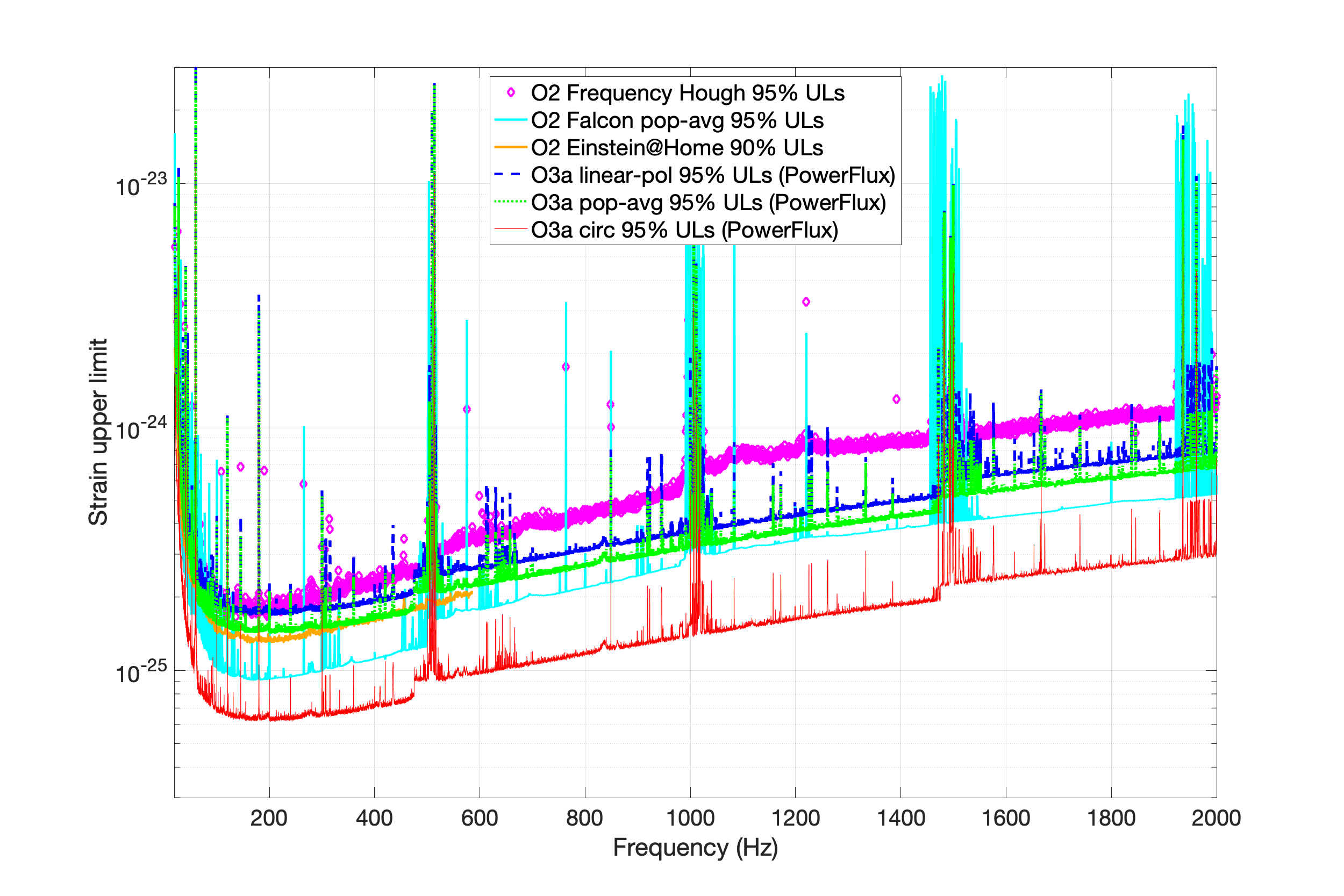}
\vspace{-\abovedisplayskip}
\caption{O3a all-sky upper limits (95\%\ CL) on $h_0$ for isolated stars from the O3a PowerFlux search in comparison with earlier
  O2 searches. The corresponding parameter space areas in $\fgw$--$\fgwdot$ are shown in Fig.~\ref{fig:O3aPFvsO2paramspace}.
  \figpermission{\cite{bib:cwallskyO3aPowerFlux}}{APS}}
\label{fig:O3aPFvsO2limits}
\end{center}
\end{figure}

  All-sky results from three pipelines (Sky Hough, Frequency Hough and \tdf)
  were applied to the O2 data set~\citep{bib:cwallskyO2,bib:PalombaEtalAxion} over the 20-1922 Hz band and
  a spin-down range $[\scimm{-1.0}{-8},\scimm{+2.0}{-9}]$,
  where frequency coverage varied by pipeline. Resulting upper limits
  are shown in Figure~\ref{fig:cwallskyO2}. The Frequency Hough search was later extended up to 2024 Hz~\citep{bib:PalombaEtalAxion}
  (see Fig.~\ref{fig:O3aPFvsO2limits}).
  A dedicated Einstein@Home search of the O2 data~\citep{bib:cwallskyO2EatH} over
  the 20-585 Hz band achieved significantly lower upper limits in the overlapping frequency band (see Fig.~\ref{fig:O3aPFvsO2limits}) for
  a spin-down range about four times smaller.

  An intriguing set of O1 and O2 all-sky searches using the Falcon pipeline
  (derived from PowerFlux, but implemented with approximations and exploiting additional
  symmetries~\citep{bib:FalconPaper,bib:cwallskyFalconO1,bib:cwallskyFalconO2MidFreq,bib:cwallskyFalconO2HighFreq,bib:cwallskyFalconO2LowFreq}),
  focused on deeper searches. The O1 search~\citep{bib:FalconPaper,bib:cwallskyFalconO1}
  doubled the first-stage effective coherence time from that used in the O1 PowerFlux
  search~\citep{bib:cwallskyO1paper1,bib:cwallskyO1paper2} while covering the same spin-down range over the 100-600 Hz band.
  The O2 searches, on the other hand, targeted
  low-ellipticity pulsars~\citep{bib:WoanEtalMSP} 
  by severely restricting the spin-down range (\eg, $|\fgwdot|<\scimm{3}{-12}$ Hz/s) in the 500-1500 Hz band.
  This vast reduction in parameter space permits using loose coherence
  with an effective coherence time of 12 hours in its initial search stage, albeit with a necessarily reduced
  astrophysical range because of the spin-down restriction.
  
  Another deep O2 search~\citep{bib:WetteEtalDeep} focused on the narrow 171-172 Hz band while restricting spin-down
  magnitudes below $\sim$\sci{3}{-13} Hz/s. This search used a semi-coherent \fstatistic\ technique
  with Graphics Processing Unit acceleration in the \fstatistic\ computation, where the frequency band chosen
  was meant to optimize probability density of detection in a narrow band based on detector sensitivity and the known pulsar
  population.

\begin{figure}[htb]
\begin{center}
  \includegraphics[width=5in]{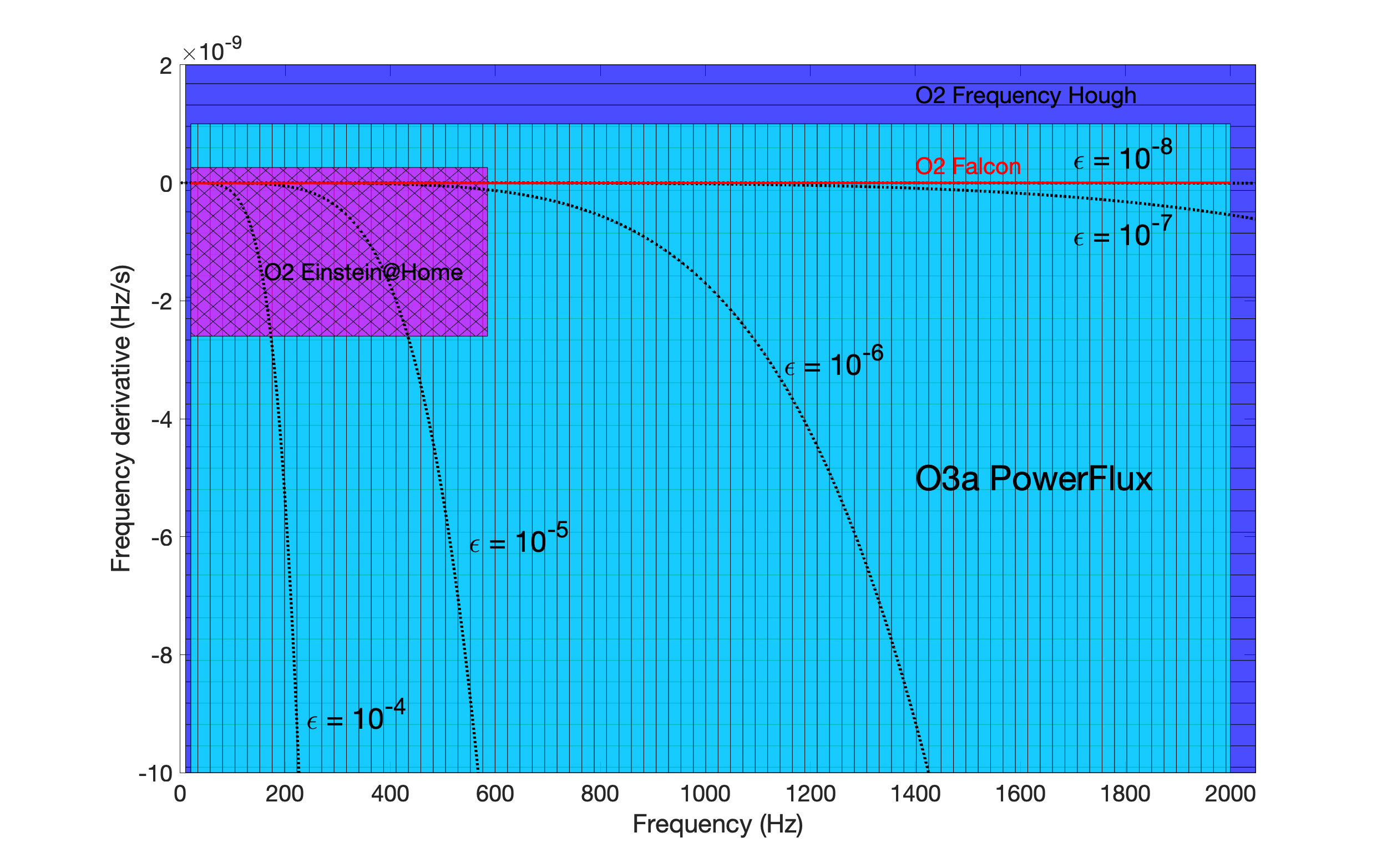}
\vspace{-\abovedisplayskip}
\caption{Comparison of parameter space areas in O2 all-sky searches vs the O3a PowerFlux search.
      The shaded rectangle with vertical bars shows the 20--2000 Hz and
      $-10^{-8}$--$10^{-9}$ Hz/s range for the O3a search~\citep{bib:cwallskyO3aPowerFlux}. The slightly larger rectangle with horizontal bars shows the
      region searched in the O2 data with the Frequency Hough method~\citep{bib:cwallskyO2,bib:PalombaEtalAxion}. The smaller rectangle
      with crossed diagonal bars shows the region searched by the distributed-computing project Einstein@Home~\citep{bib:cwallskyO2EatH}.
      The solid line at zero spin-down depicts the specialized O2 search for low-ellipticity millisecond pulsars using the
      Falcon method~\citep{bib:cwallskyFalconO2MidFreq,bib:cwallskyFalconO2HighFreq,bib:cwallskyFalconO2LowFreq} (the thickness of the line overstates the coverage in spin-down range).
      The dotted curves indicate contours of constant equatorial ellipticity $\epsilon$ = ($10^{-8}$, $10^{-7}$, $10^{-6}$, $10^{-5}$ and $10^{-4}$) for a star with stellar spin-down dominated by gravitational wave emission.
  \figpermission{\cite{bib:cwallskyO3aPowerFlux}}{APS}}
\label{fig:O3aPFvsO2paramspace}
\end{center}
\end{figure}

  The first all-sky search of O3 data for isolated CW sources~\citep{bib:cwallskyO3aPowerFlux} used the PowerFlux pipeline to examine the O3a data for
  the same broad parameter space in frequency and spin-down as used in the O1 search. A comparison of upper limits obtained
  from several O2 searches with those obtained from the O3a search are shown in Fig.~\ref{fig:O3aPFvsO2limits}.
  Figure~\ref{fig:O3aPFvsO2paramspace} shows the corresponding parameter space coverages.
  
\begin{figure}[htb]
\begin{center}
  \includegraphics[width=5in]{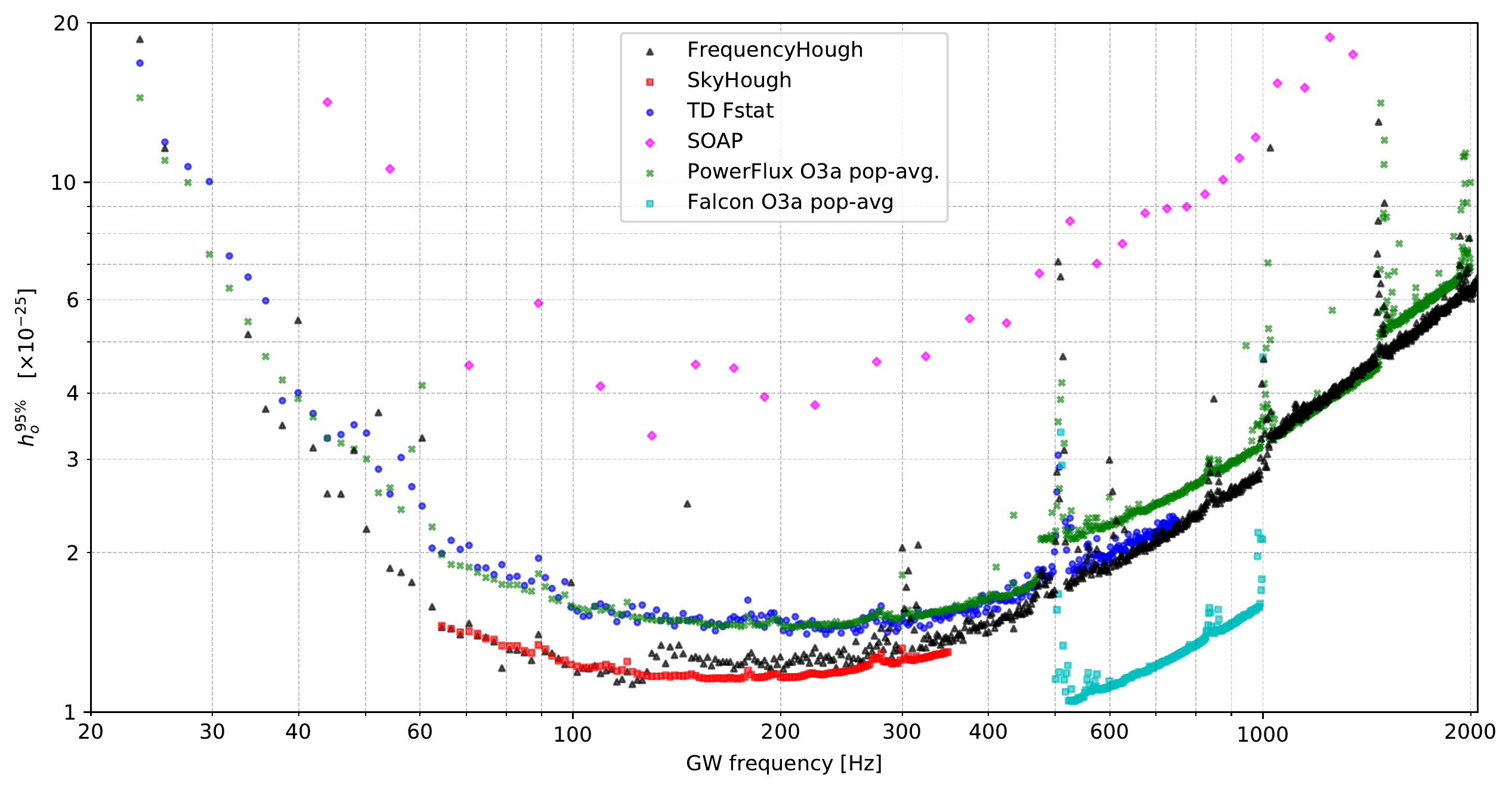}
\vspace{-\abovedisplayskip}
\caption{O3 all-sky upper limits (95\%\ CL) on $h_0$ for isolated stars from four pipelines~\citep{bib:cwallskyO3FourPipelines},
  in comparison with the O3a PowerFlux~\citep{bib:cwallskyO3aPowerFlux} and Falcon results~\citep{bib:cwallskyFalconO3aMidFreq}.
  The corresponding parameter space areas in $\fgw$--$\fgwdot$ are shown in Fig.~\ref{fig:O3FourPipelinesparamspace}.}
\label{fig:O3FourPipelineslimits}
\end{center}
\end{figure}

\begin{figure}[htb]
\begin{center}
  \includegraphics[width=5in]{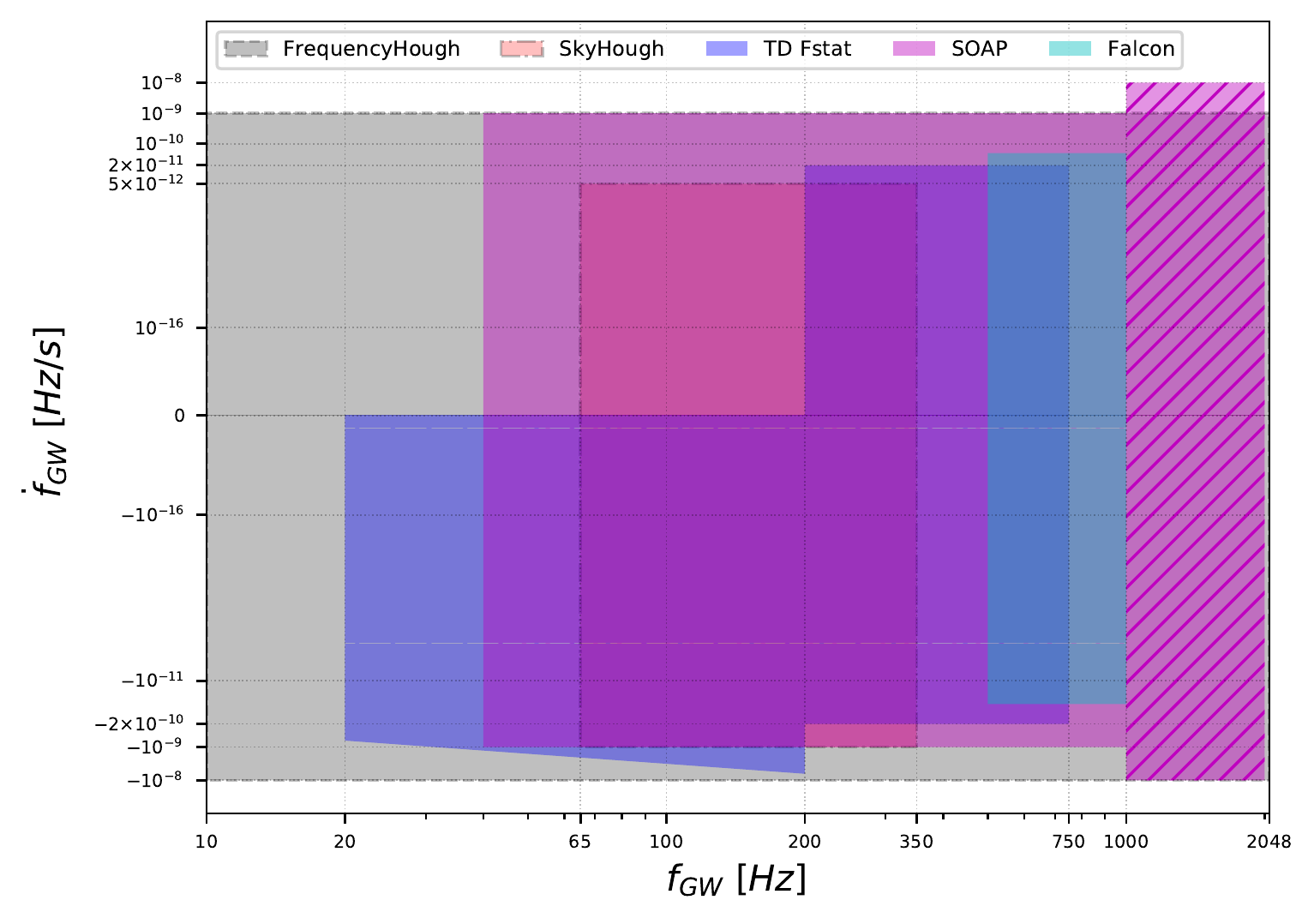}
\vspace{-\abovedisplayskip}
\caption{Comparison of  $\fgw$--$\fgwdot$ parameter space coverage for the four search pipelines used in the full-O3 all-sky
  searches~\citep{bib:cwallskyO3FourPipelines} and for the restricted-spindown O3a Falcon search.~\citep{bib:cwallskyO3FourPipelines}.}
\label{fig:O3FourPipelinesparamspace}
\end{center}
\end{figure}

  The most sensitive all-sky results to date for broad coverage of both frequency and spin-down were obtained recently
  from the full O3 data from three pipelines (Sky Hough, Frequency Hough and \tdf)~\citep{bib:cwallskyO3FourPipelines} and are shown in
  Fig.~\ref{fig:O3FourPipelineslimits}, in comparison with the O3a PowerFlux results~\citep{bib:cwallskyO3aPowerFlux} and with the results from
  the new Viterbi-based, less sensitive but blazing-fast, SOAP pipeline. Also shown are recent O3a Falcon results~\citep{bib:cwallskyFalconO3aMidFreq} over
  a restricted spin-down range. Figure~\ref{fig:O3FourPipelinesparamspace} shows a comparison of the parameter space coverages
  of these different searches.

\subsection{All-sky searches for binary stars}
\label{sec:allskybinary}

Several methods have been proposed and implemented for carrying out a CW all-sky binary search.
The first method, which was
used in a published search of initial LIGO S6 data~\citep{bib:twospectresultsS6} is
known as TwoSpect~\citep{bib:twospectmethod}. The program carries out a
semi-coherent search over an observation time long compared to
the maximum orbital period considered, while using coherence times short with
respect to the orbital period. Fourier transforms are carried
out over each row (fixed frequency bin) in a $\sim$year-long
spectrogram, and the resulting frequency-frequency plot is searched
for characteristic harmonic patterns. 

Another developed pipeline, known
as Polynomial,~\citep{bib:polynomial} searches coherently using
matched filters over an observation time short compared to the minimum orbital period
considered. A bank of frequency polynomials in time is used for
creating the matched filters, where for a small segment of an orbit,
the frequency should vary as a low-order polynomial.
Other proposed methods, which
offer potentially substantial computational savings at a cost in sensitivity, include 
autocorrelations in the time-frequency plane~\citep{bib:vicereautocorr} and
stochastic-background techniques~\citep{bib:BallmerRadiometer}, with
computational costs gains achieved by using skymaps with sidereal-day folding~\citep{bib:stochfolding,bib:stochfolding2,bib:ASAFO3}.

\begin{figure}[htb]
\begin{center}
  \includegraphics[width=5in]{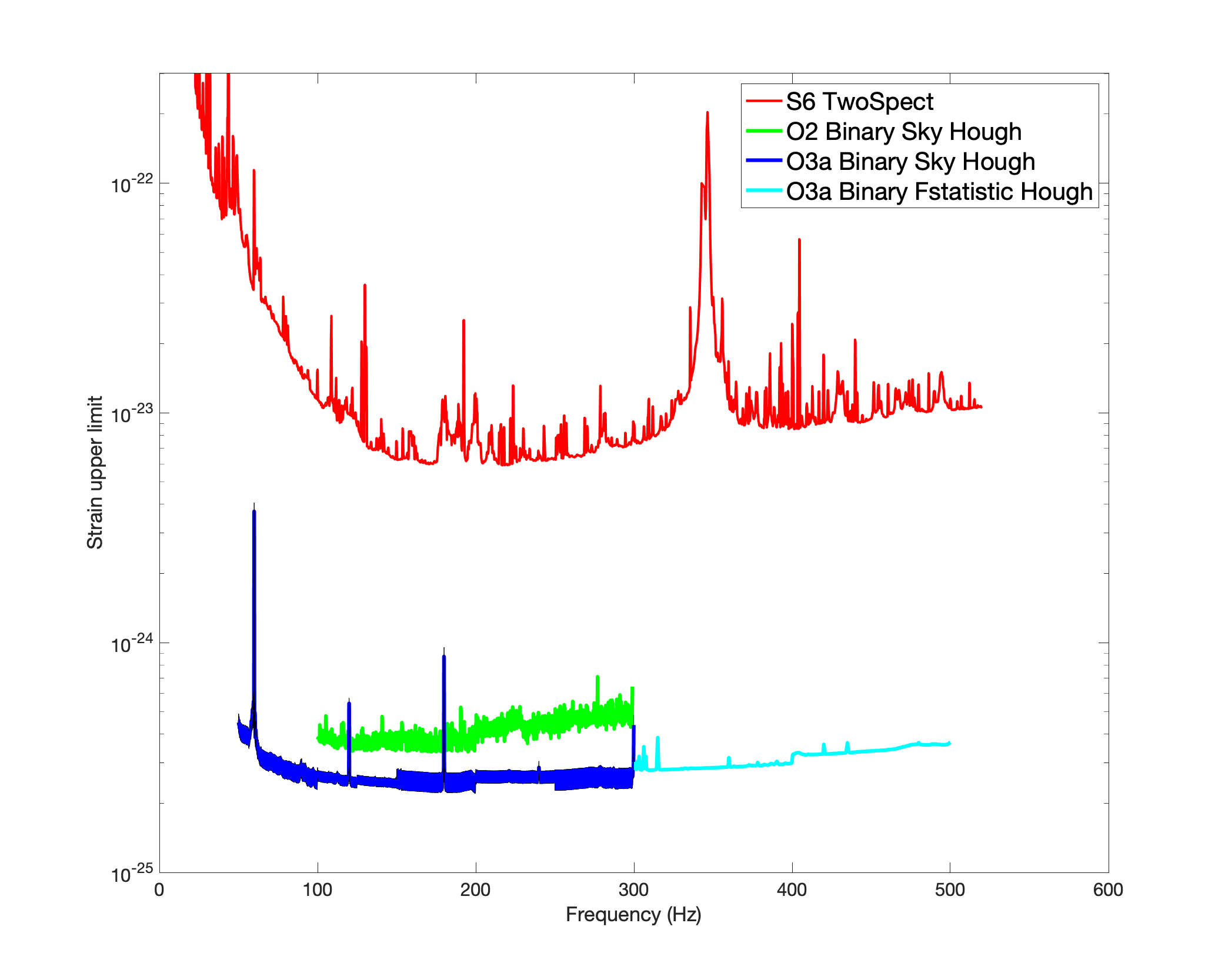}
\vspace{-\abovedisplayskip}
\caption{All-sky upper limits (95\%\ CL) on $h_0$ for stars in binary systems.
  Upper limits are shown from the inital LIGO S6 TwoSpect search~\citep{bib:twospectresultsS6},
  from the GPU-enhanced O2 Binary Sky Hough search~\citep{bib:binaryskyhough2} (100-300 Hz), from the
  O3a Binary Sky Hough search~\citep{bib:cwallskybinaryO3aBinarySkyHough} (50-300 Hz) and from the O3a Binary \fstatistic\
  search~\citep{bib:cwallskybinaryO3aBinaryFstat} (300-500 Hz). See references for details on orbital parameter space regions covered
  by the different analyses. The O2 and O3a Binary Sky Hough values shown are 95\%\ sensitivities with bands to indicate uncertainties.}
\label{fig:cwallskybinaryskyhoughO2}
\end{center}
\end{figure}

More recently, the implementation of graphics processor units software in the framework of the
Sky Hough all-sky program has led to a breakthrough in all-sky binary search
sensitivity~\citep{bib:binaryskyhough1}.
Upper limits were initially obtained over 100-300 Hz and over a broad range of binary orbital parameters
from the LIGO O2 data~\citep{bib:binaryskyhough2}. Although this approach does not yet cover the full orbital parameter space possible with
the TwoSpect program, the intrinsic sensitivity is dramatically better, with extension of the method to shorter
orbital periods a natural future improvement. A follow-up analysis in the O3a data~\citep{bib:cwallskybinaryO3aBinarySkyHough,bib:TenorioProceedings}
expanded the search band slightly (50-300 Hz), and a parallel development using a similar Hough transform framework but with a
\fstatistic~\citep{bib:CovasPrixModFstat,bib:BinaryHoughFstat} tailored to multi-hour segments,
has been applied to the O3a data in the 300-500 Hz band~\citep{bib:cwallskybinaryO3aBinaryFstat}. All of these results
are shown in Figure~\ref{fig:cwallskybinaryskyhoughO2}. See references for details on orbital parameter space regions covered
by the different analyses, which vary considerably.

In addition, searches for isolated stars retain some sensitivity to long-period binaries,
as shown in a recent studies~\citep{bib:aeibinarystudy,bib:SinghPapaLongPeriodBinaries}.

\subsection{Searches for CW transients and other CW-like signals}

The first dedicated search for CW transients following a known pulsar's glitch
addressed glitches detected by radio astronomers during the Advanced LIGO / Virgo
O2 data run. The search used the transient \fstatistic\ method~\citep{bib:PrixGiampanisMessenger}
and focused on periods following glitches in the Crab and Vela pulsars~\citep{bib:KeitelEtal}.
A recent O3 analysis~\citep{bib:cwnarrowbandO3,bib:ModafferiEtal} searched for CW transients following nine glitches
across six pulsars (one glitch each from five stars: PSR J0534+2200, J0908–4913, J1105–6107, J1813–1749 and J1826–1334; and
four glitches from the intriguing source, PSR J0537-6910 -- see sections~\ref{sec:spindown}, \ref{sec:targets} and \ref{sec:directedisolated}).
No significant candidates were observed, although two marginal outliers were seen after one PSR J0537-6910 glitch, albeit with
implied strengths well above those consistent with the inferred glitch energies.
In fact, the upper limits obtained for post-glitch energy emission from all glitches examined lay
above the maximum expected in a simple two-fluid model~\citep{bib:PrixGiampanisMessenger}, with
strain limits for PSR J1105-6107 approaching most closely to that benchmark (within a factor of $\sim$1.6).

The first dedicated search for long-lived, CW-like signals from a post-merger remnant looked
for a signal from the post-GW170817 remnant, but as expected, given the $\sim$40 Mpc distance
to the merger, no signal was detected in the immediate aftermath~\citep{bib:Postmerger1} ($\sim$500 s)
or in a multi-hour to multi-day period afterward~\citep{bib:Postmerger2}. Should another opportunity arise
(from a nearby binary neutron star merger or a galactic supernova), search methods are available for
use~\citep{bib:ThraneEtalStamp,bib:MillerEtalPostmerger,bib:SunMelatos,bib:OliverKeitelSintes,bib:LongTransientsHMM,bib:MytidisEtalNewborn,bib:MillerEtalLongTransient}.

More exotic recent analyses seeking CW or CW-like signals include:
\begin{itemize}
\item Searches
for CW signals from Bose-Einstein clouds~\citep{bib:DantonioEtalBosoncloud,bib:cwallskyO3BosonCloud} (see section~\ref{sec:axions}).
\item A search for non-black-hole weakly interacting compact dark objects
with mass below $10^{-7}$ $\msolar$ orbiting within the Sun about its center~\citep{bib:HorowitzPapaReddy}.
\item Searches for ultralight dark photon or scalar boson dark matter creating an extremely narrowband ($\Delta f/f\sim10^{-6}$) spectral
excess with stochastic phase~\citep{bib:DPDM_PRZ,bib:DPDM_GRYZ,bib:MillerEtalDPDM,bib:DPDMO3,bib:GroteStadnik,bib:GEOscalar}.
\item A search for binary systems of planetary-scale / asteroid-scale primordial
  black holes near to the Earth~\citep{bib:MillerEtalPBH,bib:planetaryBBH,bib:cwallskyO3FourPipelines}.
\item There has also been a proposal to apply CW search techniques to suspected Thorne-\.Zytkow objects~\citep{bib:TZO} (TZOs)
for which a neutron star orbiting inside of a giant star (slow inspiral decay) could produce signals in the band of ground-based
gravitational wave detectors~\citep{bib:TZOsearchpaper}.
\end{itemize}

\section{Outlook}
\label{sec:outlook}

\subsection{Prospects for discovery}

Over the next several years, the Advanced LIGO~\citep{bib:aligodetector1} and Virgo~\citep{bib:avirgodetector} detectors are expected
to approach and eventually surpass their original design sensitivities in strain, increasing the range within the galaxy which
CW searches can access, thereby increasing detection likelihood.
As the sensitive ranges of different search methods approach the dense galactic core, detection chances may rise more
rapidly. In parallel to detector improvement, algorithms continue to improve, as researchers
find more effective tradeoffs between computational cost and detection efficiency,
while Moore's Law, including Graphics Processing Unit (GPU)
exploitation~\citep{bib:TransientFstatGPU,bib:binaryskyhough1,bib:WetteEtalDeep,bib:FstatGPU,bib:LaRosaEtal}, ensures
increased computing resources for searches. All of these trends are encouraging
for successful CW detection.

At the same time, theoretical uncertainties in what sensitivity is needed for the
first CW detection are very large. While the spin-down limits based on gravitar assumptions
and on either energy conservation or known age have been beaten for a handful
of sources and will be beaten for more sources in the coming years, the gravitar
model is surely optimistic -- most stellar spin-downs are likely dominated by
electromagnetic interactions. Whether the first detection is imminent or still many
years distant remains unclear. A recent phenomenological population synthesis
study~\citep{bib:CieslarEtalPopSynth}, based on an exponentially decaying ellipticity
that starts at its allowed maximum $\sim$$10^{-5}$ with a supernova rate of
once per century concluded that the expected number of detectable, young and isolated neutron stars
for Advanced LIGO sensitivity is less than one and is $\sim$10 for Einstein Telescope.  

Electromagnetic astronomers could prove pivotal in hastening detection by identifying
new nearby or young neutron stars, or discovering pulsations from known stars, perhaps most usefully
from the accreting Sco X-1 system~\citep{bib:GalaudageEtal}. Given the computational challenges of most CW searches,
narrowing the parameter space of a search exploiting electromagnetic observations could make
the difference between a gravitational wave miss and a discovery. 

\subsection{Confirming and exploiting a discovery}

There are several aspects of confirming a nominal CW discovery, including establishing
the statistical significance of the outlier, verifying consistency of the signal
with the CW model, excluding an environmental or instrumental cause, and (optionally
but ideally) confirming consistency with prior or follow-up electromagnetic observations.
Once that discovery is established, exploiting it to understand neutron star astrophysics
(or fundamental particle physics should a superradiant boson cloud be observed) will be a rich endeavor.

The degree of statistical confidence with which a putative CW signal detection can be confirmed
depends on the type of search that leads to the candidate. The statistical significance
of an outlier depends on a trials factor that may range from $\sim$1 for targeted
searches for known pulsars using electromagnetically derived ephemerides to
$\sim$$10^{15}$ for all-sky searches. Hence the SNR threshold for, say, a ``5-sigma''
discovery varies too. In practice, though, for a candidate emerging from a hierarchical
search with multiple stages of ``zooming in'', the SNR for a surviving outlier may
be so much higher than the sensitivity-defining SNR threshold used in the first stage that
the initial trials factor is irrelevant. Establishing the statistical confidence of
a targeted-search candidate, on the other hand, may simply require steady accumulation of
additional data while fully exploiting all data in hand from all detectors with appreciable
sensitivity in that band. Empirically assessing the significance of loudest outliers is
discussed in detail in~\citep{bib:TenorioEtalLoudestCandidate}.

If known detector artifacts are
degrading sensitivity in the frequency band of the candidate, it may be feasible to focus
detector commissioning to mitigate the artifact prior to the next observing run. Another
possible approach, although potentially detrimental to other GW observations, is ``narrowbanding.''
In narrowbanding the detector sensitivity is improved in a narrow band at the expense of broadband
sensitivity by adjusting the position of the ``signal recycling'' mirror at the
output port of the interferometer~\citep{bib:aligodetector1}; with the advent of
quantum squeezing in advanced detectors~\citep{bib:squeezingO3}, however,
the potential gains from narrowbanding are less pronounced.

Even with a promising outlier, a discovery claim would need more than statistical inconsistency with
detector noise. One would seek consistency with the signal model, particularly for candidates originating
in hierarchical searches where early stages look primarily for excess power that is only roughly consistent
with a particular template and where the spacing between templates is relatively coarse.
The expected Doppler modulations due to the Earth's motion should be present~\citep{bib:ZhuEtalDoppler,bib:IntiniEtalDoppler}.
One wants to see a signal for which a fully coherent search over all data yields an SNR consistent with expectation from
the putative source. Ideally, a residual spectrum from subtracting the reconstructed signal would be
consistent with random background noise. 

In the case of a targeted or directed search for which the source location is {\it a priori} known,
one would want to verify that the highest-SNR template observed in that region of the sky and near the template's frequency parameters
is indeed consistent with the correct sky location. Although one could impose a similar constraint on frequency and frequency derivatives
for a targeted search, narrowband searches do allow those parameters to differ slightly from the nominal ones, to accommodate
differential stellar rotation. Hence seeing the SNR peak at precisely the right location in parameter space for a known pulsar
would lend credence to the signal, but seeing the SNR peak at a nearby point in parameter space can still mean discovery, albeit with
a trials factor appropriate to a narrowband search.

The possibility of a rotational glitch during an observational period
presents additional complications. One can no longer safely apply a fully coherent
search over the time span and expect a monotonically increasing SNR.
In the case of a targeted search with ephemerides in hand indicating a glitch,
breaking the observation time into two (or more) segments is
straightforward~\citep{bib:cwtargetedS5},
but in the case of a source without independent timing information, one may
have a true signal detection but lack the confidence to declare discovery without
additional data taking because of apparent phase inconsistency in the available data.

In any gravitational wave analysis using interferometers that push the frontier of
technology (and which are routinely operated at maximum achievable sensitivity), one
must consider whether or not instrumental or environmental contamination leads to
a false signal. As discussed in section~\ref{sec:lines}, narrow lines can contribute
to accumulated power in a templated search. As part of confirming a discovery, one
would need to quantify that contamination for a putative signal lying near a known
instrumental spectral line. More challenging and more realistic for a signal candidate
surviving multiple hierarchical search stages are spectral lines that are {\it not} immediately
apparent in the strain channel spectrum, especially lines that are non-stationary with
respect to time or frequency (``wandering''). To address that possibility, one would
look comprehensively at auxiliary data channels, such as readouts from magnetometers,
accelerometers, seismometers, microphones and from any servo control channels that could
impose tiny actuations on the gravitational wave strain channel. Those investigations
would include examination of averaged spectra for peaks coincident with the strain
signal frequency and more probing searches for cross-coherence between the auxiliary
channels and the strain channel that is inconsistent with statistical fluctuation.

The fact that the Virgo and KAGRA interferometers have made different technological choices
means that instrumental contaminations differ, in general, from those of the LIGO
interferometers, allowing better discrimination of astrophysical from instrumental sources.
For example, a seemingly trivial but important difference is that the U.S. power mains used
by LIGO, and the western Japanese power mains used by KAGRA provide alternating
electrical currents at 60 Hz, but Virgo depends on Italian
power mains which operate at 50 Hz. Since nearly all observatory electronics derive
power from the mains, low-lying line contamination from the fundamental oscillation, from its higher harmonics
and from sidebands due to non-linear interactions with mechanical and other electrical oscillations, are
difficult to mitigate completely. Such technical differences reduce the
chance of coincident false CW signals between the nominally identical LIGO detectors and
the other two detectors. 

Consistency in SNR detected by different interferometers offers another important astrophysical reality test. Unlike
with transient GW detections, for which antenna pattern variations among detectors are critical
to localization~\citep{bib:obsscenario}, any single interferometer can strongly localize a long-lived continous waves signal,
largely removing antenna pattern / polarization ambiguity and permitting signal strength consistency comparisons across
independent data sets in different interferometers (assuming the correctness of General Relativity GW polarization). 

Finally, in confirming a continuous gravitational wave signal one would, ideally, want
confirmation via electromagnetic observations. For targeted or narrowband searches of known pulsars,
the observations already exist, and the primary task is to establish statistical confidence
of their consistency with gravitational data.
For other known sources, however, gravitational wave measurements may provide the necessary clues
to allow detection of previously undetected pulsations. For example, detection of a CW signal
from Scorpius X-1 could permit discrimination of X-ray pulsations from a stochastic
background dominated by accretion emission.
For a previously unknown source found in an all-sky search, determining the source location
from coherent integration over months of data (with sub-arcsecond resolution possible from the
aperture formed by the Earth's orbit) may suffice for radio, X-ray, gamma-ray (and perhaps
even optical) astronomers to find the counterpart. If electromagnetic pulsations were detected
and agreed with expectation, the confirmation of the gravitational wave signal would be ironclad.

An interesting challenge to confirmation would be continuous gravitational radiation due to
boson cloud superradiance for an isolated black hole (see section~\ref{sec:axions}). If there
were no accretion disk or companion to induce an electromagnetic signal,
one would have to rely heavily upon the evolution of
the gravitational wave signal itself to infer the nature of the source. In particular, the source frequency
governed by the boson's apparent mass in the potential of the black hole
could spin {\it up} instead of down as the black hole loses mass energy to gravitational radiation, thereby reducing the magnitude of
the negative binding
energy correction to the unbound boson mass~\citep{bib:axionArvanitaki}.

Once a continuous gravitational wave detection has been confirmed electromagnetically,
one will want to exploit the correlations to understand the source. Below are a sampling
of potential measurements possible, along with questions they help to address:
\begin{itemize}
\item Relation between rotational and gravitational wave frequencies, determining the fundamental
  mechanism of emission (see section~\ref{sec:sources} and see \cite{bib:JonesFrequency} for
  a detailed discussion).
\item Correlation of the gravitational and electromagnetic phase constants in
  the event of consistent frequency (phase) evolution.
  If an equatorial mass ``bulge'' explains the GW signal, how well does the implied
  quadrupolar axis align with a pulsar's inferred magnetic dipole projection? In an
  accreting system, for example, does added mass accumulate near the magnetic poles?
\item Differential frequency (phase) evolution. Is there differential rotation between
  the stellar crust and its interior? If electromagnetic frequency glitches are observed,
  what is seen gravitationally before, during and after the glitch?
\item If there is evidence for {\rmode}s from, say, an approximate 4/3 ratio of GW
  signal frequency to stellar rotation frequency, how does the GW frequency evolve with time
  and how does the ratio evolve? Is there evidence of amplitude growth from instability?
  Decay from viscosity? 
\item Inferred quadrupole moment. Although ellipticity is a convenient dimensionless parameter,
  it is approximately the product of the ellipticity times the stellar moment of inertia about
  its spin axis that determines the signal strength for a mass-quadrupole radiator. Given
  the uncertainties in neutron star equation of state, there are large uncertainties in the
  moment of inertia and hence ambiguity in extracting ellipticity. Ambiguity at the level of near-degeneracy
  would arise in the absence of an independent determination of
  source distance from electromagnetic observations~\citep{bib:SieniawskaJones}.
  A CW detection in a binary
  system would offer an opportunity for determination of the stellar mass. 
  Other stellar properties potentially accessible include the stellar radius (inferred from luminosity
  and temperature, if measurable). A precessing star with detectable electromagnetic pulsations
  offers additional opportunity for understanding internal structure~\citep{bib:GaoEtal}.
\item Even in the absence of detected electromagnetic pulsations, one can use the known source of an
  electromagnetic counterpart to infer a star's moment of inertia, equatorial ellipticity,
  and the component of the magnetic dipole moment perpendicular to its rotation axis~\citep{bib:LuEtalInferringProperties}.
  For a close enough neutron star without an electromagnetic counterpart,
  parallax inference from the GW signal alone could resolve the degeneracy among source distance, moment of
  inertia and equatorial ellipticity~\citep{bib:SieniawskaEtalParallax}
  (see Eqn.~\ref{eqn:hexpected}).
\item Boson properties from superradiance. In the event of detecting superradiance from a
  boson cloud around a black hole, determining the boson's mass will be immediate from the
  signal frequency (at least for the annihilation channel expected to dominate) with the
  boson intrinsic spin determination more model dependent, based on signal strength and frequency
  evolution with some knowledge of the black hole source needed. 
\end{itemize}

In addition to exploiting CW detection to understand the source, one can also carry out precise
tests of General Relativity by measuring the polarization of the propagating gravitational wave.
In Einstein's theory there should be two independent transverse, quadrupolar polarizations for which the relative strengths
depend on the source orientation relative to the line of sight. In non-standard theories of
gravity other polarization modes, including scalar, vector and longitudinal polarizations,
may be present~\citep{bib:TGRmethod}. Testing for these additional polarizations with
transient gravitational wave detections to date has been challenging because nearly all of the
signal-to-noise ratio has come from the two nearly aligned LIGO detectors such that they mainly
detect the same polarization projection. In contrast, even for a single detector, a CW signal permits
disentangling multiple polarization contributions as the sidereal rotation of the Earth changes
the detector's (polarimeter's) orientation with respect to the source direction deterministically~\citep{bib:TGRmethod,bib:KuwaharaAsada}.
In fact, even in the absence of a CW signal, one can set upper limits on the non-standard polarizations~\citep{bib:TGRO1},
just as is possible for standard polarizations.

Nature has blessed the gravitational wave community with a bounty of compact star mergers, including the remarkable first
detected BBH merger, GW150914, and the even more remarkable and informative multi-messenger
detection of the GW170817 BNS event.
Should such kindness continue, one may hope soon for a multi-messenger detection of a CW source that not only could be observed into
the foreseeable future, but could mark the first of a large collection to come, as GW150914 proved to be.

\begin{acknowledgements}
The author is deeply grateful to current and former colleagues in the LIGO Scientific Collaboration
and Virgo collaboration Continuous Waves Search Group for close collaboration
from which he has benefited in preparing this article. The author also thanks 
Julian Carlin, Pep Covas, Vladimir Dergachev, Francesco Fidecaro, Bryn Haskell, Wynn Ho, Ian Jones,
David Keitel, Andrew Melatos, Ben Owen, Maria Alessandra Papa, Lilli Sun, Rodrigo Tenorio,
Karl Wette and Graham Woan for helpful suggestions concerning the manuscript.
Special thanks to the two anonymous referees for their thorough readings and for many
constructive suggestions (and corrections). 
Thanks too to LIGO, Virgo and the Max Planck Institute f. Gravitational Physics
for the use of figures. This work was supported in part
by National Science Foundation Awards PHY-1505932 and PHY-1806577.
The author thanks the Institute of Nuclear Theory at the University of Washington
and the Perimeter Institute of Waterloo for hosting workshops
that proved helpful in composing this review.
The author is grateful for computational resources provided by the LIGO Laboratory and is supported by the National
Science Foundation.
This material is based in part upon work supported by NSF's LIGO Laboratory which is a major facility
fully funded by the National Science Foundation.
\end{acknowledgements}

\def\Journal#1#2#3#4{{#1} {\bf #2} #3 (#4)}
\def\RNC{{\em Riv. Nuovo Cim.}}
\def\NCA{{\em Nuovo Cimento} A}
\def\PHYS{{\em Physica}}
\def\NPA{{\em Nucl. Phys.} A}
\def\MATH{{\em J. Math. Phys.}}
\def\PRO{{\em Prog. Theor. Phys.}}
\def\NPB{{\em Nucl. Phys.} B}
\def\PLA{{\em Phys. Lett.} A}
\def\PLB{{\em Phys. Lett.} B}
\def\PLD{{\em Phys. Lett.} D}
\def\PL{{\em Phys. Lett.}}
\def\PRL{\em Phys. Rev. Lett.}
\def\PREV{\em Phys. Rev.}
\def\PREP{\em Phys. Rep.}
\def\PRA{{\em Phys. Rev.} A}
\def\PRD{{\em Phys. Rev.} D}
\def\PRC{{\em Phys. Rev.} C}
\def\PRB{{\em Phys. Rev.} B}
\def\PPNP{\em Prog. Part. Nuc. Phys.}
\def\IAUC{\em Intl. Astron. Union Circ.}
\def\ZPC{{\em Z. Phys.} C}
\def\ZPA{{\em Z. Phys.} A}
\def\ANNP{\em Ann. Phys. (N.Y.)}
\def\RMP{{\em Rev. Mod. Phys.}}
\def\CHEM{{\em J. Chem. Phys.}}
\def\INT{{\em Int. J. Mod. Phys.} E}
\def\IJMPD{{\em Int. J. Mod. Phys.} D}
\def\APJ{\em Astroph. J.}
\def\ASJ{\em Astron. J.}
\def\APJL{\em Astroph. J. Lett.}
\def\APJS{\em Astroph. J. Supp.}
\def\ASTAST{\em Astron. \&\ Astroph.}
\def\JCAP{\em J. Cosm. Astropart. Phys.}
\def\LRRE{\em Liv. Rev. Rel.}
\def\CQG{\em Class. Quant. Grav.}
\def\NAR{\em New Astron. Rev.}
\def\NAT{\em Nature}
\def\nat{\em Nature}
\def\NATP{\em Nature Phys.}
\def\NIMA{{\em Nuc. Inst. Meth.} A}
\def\SOVA{\em Sov. Astron.}
\def\MNRAS{\em Mon. Not. Roy. Astron. Soc.}
\def\ASTSPACE{\em Astrophys. Space Sci.}
\def\SCI{\em Science}
\def\SSR{\em Space Sci. Rev.}
\def\PTRSLA{{\em Phil. Trans. R. Soc. Lon.} A}
\def\RPP{\em Rep. Prog. Phys.}
\def\PNAS{\em Proc. Nat. Acad. Sci.}
\def\PCPS{\em  Proc. Cam. Phil. Soc.}
\def\ARAA{\em  Ann. Rev. Astron. Astroph.}
\def\NJP{\em New J. Phys.}
\def\PT{\em Phys. Today}
\def\OL{\em Opt. Lett.}
\def\PASP{\em Pubs. Astron. Soc. Pacif.}
\def\PASA{\em Pubs. Astron. Soc. Australia}
\def\PSPIE{\em Proc. SPIE}
\def\IEEE{\em IEEE Proc.}
\def\IEEEIT{\em IEEE Trans. Inform. Theory}
\def\CJAAS{\em Chin. J. Astron. Astroph. Sup.}
\def\JPCS{\em J. Phys. Conf. Ser.}
\def\MPLA{{\em Mod. Phys. Lett.} A}
\def\AJP{\em Amer. J. Phys.}
\def\RSI{\em Rev. Sci. Inst.}
\def\APB{{\em Appl. Phys.} B}
\def\APP{\em Astropart. Phys.}
\def\JMO{\em J. Mod. Opt.}
\def\AO{\em Appl. Opt.}
\def\LP{\em Las. Phys.}
\def\JAP{\em J. Appl. Phys.}
\def\SJETP{\em Sov. J. Exp. \&\ Theor. Phys.}
\def\JETPL{\em J. Exp. \&\ Theor. Phys. Lett.}
\def\JGR{\em J. Geophys. Res.}
\def\PTEP{\em Prog. Theor. Exp. Phys.}
\def\LRR{\em Liv. Rev. Rel.}
\def\CACM{\em Commun. A.C.M.}
\def\etal{{\it et al.}}

\def\aas{J.~Aasi \etal}
\def\aba{J.~Abadie \etal}
\def\bab{B.~Abbott \etal}



\bibliography{lrre}{}

\begin{thebibliography}{646}
\providecommand{\natexlab}[1]{#1}
\providecommand{\url}[1]{{#1}}
\providecommand{\urlprefix}{URL }
\expandafter\ifx\csname urlstyle\endcsname\relax
  \providecommand{\doi}[1]{DOI~\discretionary{}{}{}#1}\else
  \providecommand{\doi}{DOI~\discretionary{}{}{}\begingroup
  \urlstyle{rm}\Url}\fi
\providecommand{\eprint}[2][]{\url{#2}}

\bibitem[{Aasi et~al.(2013{\natexlab{a}})}]{bib:cwdirectedgalacticcenterS5}
Aasi J, et~al. (2013{\natexlab{a}}) {Directed search for continuous
  gravitational waves from the Galactic center}. Physical Review D 88(10),
  \doi{10.1103/physrevd.88.102002}

\bibitem[{Aasi et~al.(2013{\natexlab{b}})}]{bib:cwallskyEatHS5}
Aasi J, et~al. (2013{\natexlab{b}}) {Einstein@Home all-sky search for periodic
  gravitational waves in LIGO S5 data}. Phys Rev D 87(4):042001,
  \doi{10.1103/PhysRevD.87.042001}, \eprint{1207.7176}

\bibitem[{Aasi et~al.(2014{\natexlab{a}})}]{bib:twospectresultsS6}
Aasi J, et~al. (2014{\natexlab{a}}) {First all-sky search for continuous
  gravitational waves from unknown sources in binary systems}. Phys Rev D
  90(6):062010, \doi{10.1103/PhysRevD.90.062010}, \eprint{1405.7904}

\bibitem[{Aasi et~al.(2014{\natexlab{b}})}]{bib:cwtargetedS5S6VSR24}
Aasi J, et~al. (2014{\natexlab{b}}) {Gravitational Waves from Known Pulsars:
  Results from the Initial Detector Era}. The Astrophysical Journal 785(2):119,
  \doi{10.1088/0004-637x/785/2/119}

\bibitem[{Aasi et~al.(2014{\natexlab{c}})}]{bib:cwallskyfstatVSR1}
Aasi J, et~al. (2014{\natexlab{c}}) {Implementation of an
  $\mathcal{F}$-statistic all-sky search for continuous gravitational waves in
  Virgo {VSR}1 data}. Classical and Quantum Gravity 31(16):165014,
  \doi{10.1088/0264-9381/31/16/165014}

\bibitem[{Aasi et~al.(2015{\natexlab{a}})}]{bib:aligodetector1}
Aasi J, et~al. (2015{\natexlab{a}}) {Advanced {LIGO}}. Classical and Quantum
  Gravity 32(7):074001, \doi{10.1088/0264-9381/32/7/074001}

\bibitem[{Aasi et~al.(2015{\natexlab{b}})}]{bib:detcharS6}
Aasi J, et~al. (2015{\natexlab{b}}) {Characterization of the LIGO detectors
  during their sixth science run}. Classical and Quantum Gravity 32(11):115012,
  \doi{10.1088/0264-9381/32/11/115012}

\bibitem[{Aasi et~al.(2015{\natexlab{c}})}]{bib:sidebandS5}
Aasi J, et~al. (2015{\natexlab{c}}) {Directed search for gravitational waves
  from Scorpius X-1 with initial LIGO data}. Phys Rev D 91:062008,
  \doi{10.1103/PhysRevD.91.062008}

\bibitem[{Aasi et~al.(2015{\natexlab{d}})}]{bib:cwnarrowbandVSR4}
Aasi J, et~al. (2015{\natexlab{d}}) {Narrow-band search of continuous
  gravitational-wave signals from Crab and Vela pulsars in Virgo VSR4 data}.
  Physical Review D 91(2), \doi{10.1103/physrevd.91.022004}

\bibitem[{Aasi et~al.(2015{\natexlab{e}})}]{bib:S6NineSNRs}
Aasi J, et~al. (2015{\natexlab{e}}) {{Searches} {for} {continuous}
  {gravitational} {waves} {from} {nine} {young} {supernova} {remnants}}. The
  Astrophysical Journal 813(1):39, \doi{10.1088/0004-637x/813/1/39}

\bibitem[{Aasi et~al.(2016{\natexlab{a}})}]{bib:cwallskyfreqhoughVSR2VSR4}
Aasi J, et~al. (2016{\natexlab{a}}) {First low frequency all-sky search for
  continuous gravitational wave signals}. Phys Rev D 93:042007,
  \doi{10.1103/PhysRevD.93.042007}

\bibitem[{Aasi et~al.(2016{\natexlab{b}})}]{bib:CWOrionSpur}
Aasi J, et~al. (2016{\natexlab{b}}) {Search of the Orion spur for continuous
  gravitational waves using a loosely coherent algorithm on data from LIGO
  interferometers}. Phys Rev D 93:042006, \doi{10.1103/PhysRevD.93.042006}

\bibitem[{Abadie et~al.(2010)}]{bib:cwcasa}
Abadie J, et~al. (2010) {First search for gravitational waves from the youngest
  known neutron star}. The Astrophysical Journal 722(2):1504–1513,
  \doi{10.1088/0004-637x/722/2/1504}

\bibitem[{Abadie et~al.(2011)}]{bib:cwtargetedvela}
Abadie J, et~al. (2011) {Beating the Spin-down Limit on Gravitational Wave
  Emission from the Vela pulsar}. The Astrophysical Journal 737(2):93,
  \doi{10.1088/0004-637x/737/2/93}

\bibitem[{Abadie et~al.(2012)}]{bib:cwallskyS5}
Abadie J, et~al. (2012) {All-sky search for periodic gravitational waves in the
  full S5 LIGO data}. Physical Review D 85(2):022001,
  \doi{10.1103/physrevd.85.022001}

\bibitem[{Abazajian(2011)}]{bib:Abazajian}
Abazajian KN (2011) {The Consistency of Fermi-LAT Observations of the Galactic
  Center with a Millisecond Pulsar Population in the Central Stellar Cluster}.
  JCAP 03:010, \doi{10.1088/1475-7516/2011/03/010}, \eprint{1011.4275}

\bibitem[{Abbott et~al.(2004)}]{bib:cwtargetedS1}
Abbott BP, et~al. (2004) {Setting upper limits on the strength of periodic
  gravitational waves from PSR $\mathrm{J}1939+2134$ using the first science
  data from the GEO 600 and LIGO detectors}. Phys Rev D 69:082004,
  \doi{10.1103/PhysRevD.69.082004}

\bibitem[{Abbott et~al.(2005{\natexlab{a}})}]{bib:cwallskyS2}
Abbott BP, et~al. (2005{\natexlab{a}}) {First all-sky upper limits from LIGO on
  the strength of periodic gravitational waves using the Hough transform}. Phys
  Rev D 72:102004, \doi{10.1103/PhysRevD.72.102004}, \eprint{gr-qc/0508065}

\bibitem[{Abbott et~al.(2005{\natexlab{b}})}]{bib:cwtargetedS2}
Abbott BP, et~al. (2005{\natexlab{b}}) {Limits on Gravitational-Wave Emission
  from Selected Pulsars Using LIGO Data}. Phys Rev Lett 94:181103,
  \doi{10.1103/PhysRevLett.94.181103}

\bibitem[{Abbott et~al.(2007{\natexlab{a}})}]{bib:cwfstatS2}
Abbott BP, et~al. (2007{\natexlab{a}}) {Searches for periodic gravitational
  waves from unknown isolated sources and Scorpius X-1: Results from the second
  LIGO science run}. Phys Rev D 76:082001, \doi{10.1103/PhysRevD.76.082001}

\bibitem[{Abbott et~al.(2007{\natexlab{b}})}]{bib:cwtargetedS3S4}
Abbott BP, et~al. (2007{\natexlab{b}}) {Upper limits on gravitational wave
  emission from 78 radio pulsars}. Phys Rev D 76:042001,
  \doi{10.1103/PhysRevD.76.042001}

\bibitem[{Abbott et~al.(2008{\natexlab{a}})}]{bib:cwallskyS4}
Abbott BP, et~al. (2008{\natexlab{a}}) {All-sky search for periodic
  gravitational waves in LIGO S4 data}. Phys Rev D 77:022001,
  \doi{10.1103/PhysRevD.77.022001}, [Erratum: Phys.Rev.D 80, 129904 (2009)],
  \eprint{0708.3818}

\bibitem[{Abbott et~al.(2008{\natexlab{b}})}]{bib:cwtargetedcrabS5}
Abbott BP, et~al. (2008{\natexlab{b}}) {Beating the Spin-Down Limit on
  Gravitational Wave Emission from the Crab Pulsar}. Astrophys J Lett
  683(1):L45, \doi{10.1086/591526}, \eprint{0805.4758}

\bibitem[{Abbott et~al.(2009{\natexlab{a}})}]{bib:cwallskyearlyS5}
Abbott BP, et~al. (2009{\natexlab{a}}) {All-sky LIGO Search for Periodic
  Gravitational Waves in the Early S5 Data}. Phys Rev Lett 102:111102,
  \doi{10.1103/PhysRevLett.102.111102}, \eprint{0810.0283}

\bibitem[{Abbott et~al.(2009{\natexlab{b}})}]{bib:cwallskyEatHearlyS5}
Abbott BP, et~al. (2009{\natexlab{b}}) {Einstein@Home search for periodic
  gravitational waves in early S5 LIGO data}. Phys Rev D 80:042003,
  \doi{10.1103/PhysRevD.80.042003}, \eprint{0905.1705}

\bibitem[{Abbott et~al.(2009{\natexlab{c}})}]{bib:cwallskyEatHS4}
Abbott BP, et~al. (2009{\natexlab{c}}) {The Einstein@Home search for periodic
  gravitational waves in LIGO S4 data}. Phys Rev D 79:022001,
  \doi{10.1103/PhysRevD.79.022001}, \eprint{0804.1747}

\bibitem[{Abbott et~al.(2010)}]{bib:cwtargetedS5}
Abbott BP, et~al. (2010) {Searches for Gravitational Waves from Known Pulsars
  with Science Run 5 LIGO Data}. The Astrophysical Journal 713(1):671--685,
  \doi{10.1088/0004-637x/713/1/671}

\bibitem[{Abbott et~al.(2016{\natexlab{a}})}]{bib:cwallskyS6}
Abbott BP, et~al. (2016{\natexlab{a}}) {Comprehensive all-sky search for
  periodic gravitational waves in the sixth science run LIGO data}. Phys Rev D
  94(4):042002, \doi{10.1103/PhysRevD.94.042002}, \eprint{1605.03233}

\bibitem[{Abbott et~al.(2016{\natexlab{b}})}]{bib:GW150914}
Abbott BP, et~al. (2016{\natexlab{b}}) {GW150914: The Advanced LIGO Detectors
  in the Era of First Discoveries}. Phys Rev Lett 116(13):131103,
  \doi{10.1103/PhysRevLett.116.131103}, \eprint{1602.03838}

\bibitem[{Abbott et~al.(2016{\natexlab{c}})}]{bib:aligodetector2}
Abbott BP, et~al. (2016{\natexlab{c}}) {GW150914: The Advanced LIGO Detectors
  in the Era of First Discoveries}. Phys Rev Lett 116:131103,
  \doi{10.1103/PhysRevLett.116.131103}

\bibitem[{Abbott et~al.(2016{\natexlab{d}})}]{bib:GW151226}
Abbott BP, et~al. (2016{\natexlab{d}}) {GW151226: Observation of Gravitational
  Waves from a 22-Solar-Mass Binary Black Hole Coalescence}. Phys Rev Lett
  116:241103, \doi{10.1103/PhysRevLett.116.241103}

\bibitem[{Abbott et~al.(2016{\natexlab{e}})}]{bib:cwallskyEatHS6}
Abbott BP, et~al. (2016{\natexlab{e}}) {Results of the deepest all-sky survey
  for continuous gravitational waves on LIGO S6 data running on the
  Einstein@Home volunteer distributed computing project}. Phys Rev D
  94(10):102002, \doi{10.1103/PhysRevD.94.102002}, \eprint{1606.09619}

\bibitem[{Abbott et~al.(2017{\natexlab{a}})}]{bib:cwallskyO1paper1}
Abbott BP, et~al. (2017{\natexlab{a}}) {All-sky Search for Periodic
  Gravitational Waves in the O1 LIGO Data}. Phys Rev D 96(6):062002,
  \doi{10.1103/PhysRevD.96.062002}, \eprint{1707.02667}

\bibitem[{Abbott et~al.(2017{\natexlab{b}})}]{bib:DirectedStochasticO1}
Abbott BP, et~al. (2017{\natexlab{b}}) {Directional Limits on Persistent
  Gravitational Waves from Advanced LIGO's First Observing Run}. Phys Rev Lett
  118:121102, \doi{10.1103/PhysRevLett.118.121102}

\bibitem[{Abbott et~al.(2017{\natexlab{c}})}]{bib:CosmicExplorer}
Abbott BP, et~al. (2017{\natexlab{c}}) {Exploring the sensitivity of next
  generation gravitational wave detectors}. Classical and Quantum Gravity
  34(4):044001, \doi{10.1088/1361-6382/aa51f4}

\bibitem[{Abbott et~al.(2017{\natexlab{d}})}]{bib:cwallskyEatHO1}
Abbott BP, et~al. (2017{\natexlab{d}}) {First low-frequency Einstein@Home
  all-sky search for continuous gravitational waves in Advanced LIGO data}.
  Phys Rev D 96(12):122004, \doi{10.1103/PhysRevD.96.122004},
  \eprint{1707.02669}

\bibitem[{Abbott et~al.(2017{\natexlab{e}})}]{bib:cwnarrowbandO1}
Abbott BP, et~al. (2017{\natexlab{e}}) {First narrow-band search for continuous
  gravitational waves from known pulsars in advanced detector data}. Phys Rev D
  96:122006, \doi{10.1103/PhysRevD.96.122006}

\bibitem[{Abbott et~al.(2017{\natexlab{f}})}]{bib:cwtargetedO1}
Abbott BP, et~al. (2017{\natexlab{f}}) {First search for gravitational waves
  from known pulsars with Advanced LIGO}. Astrophys J 839(1):12,
  \doi{10.3847/1538-4357/aa677f}, [Erratum: Astrophys.J. 851, 71 (2017)],
  \eprint{1701.07709}

\bibitem[{Abbott et~al.(2017{\natexlab{g}})}]{bib:GW170817GRB}
Abbott BP, et~al. (2017{\natexlab{g}}) {Gravitational Waves and Gamma-Rays from
  a Binary Neutron Star Merger: {GW}170817 and {GRB} 170817A}. The
  Astrophysical Journal 848(2):L13, \doi{10.3847/2041-8213/aa920c}

\bibitem[{Abbott et~al.(2017{\natexlab{h}})}]{bib:GW170104}
Abbott BP, et~al. (2017{\natexlab{h}}) {GW170104: Observation of a
  50-Solar-Mass Binary Black Hole Coalescence at Redshift 0.2}. Phys Rev Lett
  118:221101, \doi{10.1103/PhysRevLett.118.221101}

\bibitem[{Abbott et~al.(2017{\natexlab{i}})}]{bib:GW170608}
Abbott BP, et~al. (2017{\natexlab{i}}) {GW170608: Observation of a
  19-solar-mass Binary Black Hole Coalescence}. Astrophys J 851(2):L35,
  \doi{10.3847/2041-8213/aa9f0c}, \eprint{1711.05578}

\bibitem[{Abbott et~al.(2017{\natexlab{j}})}]{bib:GW170814}
Abbott BP, et~al. (2017{\natexlab{j}}) {GW170814: A Three-Detector Observation
  of Gravitational Waves from a Binary Black Hole Coalescence}. Phys Rev Lett
  119:141101, \doi{10.1103/PhysRevLett.119.141101}

\bibitem[{Abbott et~al.(2017{\natexlab{k}})}]{bib:GW170817}
Abbott BP, et~al. (2017{\natexlab{k}}) {GW170817: Observation of Gravitational
  Waves from a Binary Neutron Star Inspiral}. Phys Rev Lett 119:161101,
  \doi{10.1103/PhysRevLett.119.161101}

\bibitem[{Abbott et~al.(2017{\natexlab{l}})}]{bib:GW170817MMA}
Abbott BP, et~al. (2017{\natexlab{l}}) {Multi-messenger Observations of a
  Binary Neutron Star Merger}. Astrophys J Lett 848(2):L12,
  \doi{10.3847/2041-8213/aa91c9}, \eprint{1710.05833}

\bibitem[{Abbott et~al.(2017{\natexlab{m}})}]{bib:S6NGC6544}
Abbott BP, et~al. (2017{\natexlab{m}}) {Search for continuous gravitational
  waves from neutron stars in globular cluster NGC 6544}. Phys Rev D 95:082005,
  \doi{10.1103/PhysRevD.95.082005}

\bibitem[{Abbott et~al.(2017{\natexlab{n}})}]{bib:ViterbiO1}
Abbott BP, et~al. (2017{\natexlab{n}}) {Search for gravitational waves from
  Scorpius X-1 in the first Advanced LIGO observing run with a hidden Markov
  model}. Phys Rev D 95:122003, \doi{10.1103/PhysRevD.95.122003}

\bibitem[{Abbott et~al.(2017{\natexlab{o}})}]{bib:Postmerger1}
Abbott BP, et~al. (2017{\natexlab{o}}) {Search for Post-merger Gravitational
  Waves from the Remnant of the Binary Neutron Star Merger GW170817}. Astrophys
  J Lett 851(1):L16, \doi{10.3847/2041-8213/aa9a35}, \eprint{1710.09320}

\bibitem[{Abbott et~al.(2017{\natexlab{p}})}]{bib:O1CrossCorr}
Abbott BP, et~al. (2017{\natexlab{p}}) {Upper Limits on Gravitational Waves
  from Scorpius X-1 from a Model-based Cross-correlation Search in Advanced
  {LIGO} Data}. The Astrophysical Journal 847(1):47,
  \doi{10.3847/1538-4357/aa86f0}

\bibitem[{Abbott et~al.(2018{\natexlab{a}})}]{bib:TGRO1}
Abbott BP, et~al. (2018{\natexlab{a}}) {First search for nontensorial
  gravitational waves from known pulsars}. Phys Rev Lett 120(3):031104,
  \doi{10.1103/PhysRevLett.120.031104}, \eprint{1709.09203}

\bibitem[{Abbott et~al.(2018{\natexlab{b}})}]{bib:cwallskyO1paper2}
Abbott BP, et~al. (2018{\natexlab{b}}) {Full Band All-sky Search for Periodic
  Gravitational Waves in the O1 LIGO Data}. Phys Rev D 97(10):102003,
  \doi{10.1103/PhysRevD.97.102003}, \eprint{1802.05241}

\bibitem[{Abbott et~al.(2018{\natexlab{c}})}]{bib:GW170817EOS}
Abbott BP, et~al. (2018{\natexlab{c}}) {GW170817: Measurements of Neutron Star
  Radii and Equation of State}. Phys Rev Lett 121:161101,
  \doi{10.1103/PhysRevLett.121.161101}

\bibitem[{Abbott et~al.(2019{\natexlab{a}})}]{bib:cwallskyO2}
Abbott BP, et~al. (2019{\natexlab{a}}) {All-sky search for continuous
  gravitational waves from isolated neutron stars using Advanced LIGO O2 data}.
  Phys Rev D 100(2):024004, \doi{10.1103/PhysRevD.100.024004},
  \eprint{1903.01901}

\bibitem[{Abbott et~al.(2019{\natexlab{b}})}]{bib:GWTC1}
Abbott BP, et~al. (2019{\natexlab{b}}) {GWTC-1: A Gravitational-Wave Transient
  Catalog of Compact Binary Mergers Observed by LIGO and Virgo during the First
  and Second Observing Runs}. Phys Rev X 9:031040,
  \doi{10.1103/PhysRevX.9.031040}

\bibitem[{Abbott et~al.(2019{\natexlab{c}})}]{bib:cwnarrowbandO2}
Abbott BP, et~al. (2019{\natexlab{c}}) {Narrow-band search for gravitational
  waves from known pulsars using the second LIGO observing run}. Physical
  Review D 99(12), \doi{10.1103/physrevd.99.122002}

\bibitem[{Abbott et~al.(2019{\natexlab{d}})}]{bib:Postmerger2}
Abbott BP, et~al. (2019{\natexlab{d}}) {Search for gravitational waves from a
  long-lived remnant of the binary neutron star merger GW170817}. Astrophys J
  875(2):160, \doi{10.3847/1538-4357/ab0f3d}, \eprint{1810.02581}

\bibitem[{Abbott et~al.(2019{\natexlab{e}})}]{bib:cwdirectedO2ScoX1Viterbi}
Abbott BP, et~al. (2019{\natexlab{e}}) {Search for gravitational waves from
  Scorpius X-1 in the second Advanced LIGO observing run with an improved
  hidden Markov model}. Physical Review D 100(12),
  \doi{10.1103/physrevd.100.122002}

\bibitem[{{Abbott} et~al.(2019{\natexlab{a}})}]{bib:cwdirectedSNRO1}
{Abbott} BP, et~al. (2019{\natexlab{a}}) {Searches for Continuous Gravitational
  Waves from 15 Supernova Remnants and Fomalhaut b with Advanced LIGO}. The
  Astrophysical Journal 875(2):122, \doi{10.3847/1538-4357/ab113b},
  \eprint{1812.11656}

\bibitem[{{Abbott} et~al.(2019{\natexlab{b}})}]{bib:cwtargetedO2}
{Abbott} BP, et~al. (2019{\natexlab{b}}) {Searches for Gravitational Waves from
  Known Pulsars at Two Harmonics in 2015-2017 LIGO Data}. The Astrophysical
  Journal 879(1):10, \doi{10.3847/1538-4357/ab20cb}, \eprint{1902.08507}

\bibitem[{Abbott et~al.(2020{\natexlab{a}})}]{bib:GW190425}
Abbott BP, et~al. (2020{\natexlab{a}}) {GW190425: Observation of a Compact
  Binary Coalescence with Total Mass $\sim 3.4 M_{\odot}$}. Astrophys J Lett
  892(1):L3, \doi{10.3847/2041-8213/ab75f5}, \eprint{2001.01761}

\bibitem[{Abbott et~al.(2020{\natexlab{b}})}]{bib:obsscenario}
Abbott BP, et~al. (2020{\natexlab{b}}) Prospects for observing and localizing
  gravitational-wave transients with advanced ligo, advanced virgo and kagra.
  Living Reviews in Relativity 23(1):3, \doi{10.1007/s41114-020-00026-9}

\bibitem[{Abbott et~al.(2020{\natexlab{c}})}]{bib:cwtargetedO3a}
Abbott R, et~al. (2020{\natexlab{c}}) {Gravitational-wave Constraints on the
  Equatorial Ellipticity of Millisecond Pulsars}. The Astrophysical Journal
  Letters 902(1):L21, \doi{10.3847/2041-8213/abb655}

\bibitem[{Abbott et~al.(2020{\natexlab{d}})}]{bib:GW190521}
Abbott R, et~al. (2020{\natexlab{d}}) {GW190521: A Binary Black Hole Merger
  with a Total Mass of $150 ~ M_{\odot}$}. Phys Rev Lett 125(10):101102,
  \doi{10.1103/PhysRevLett.125.101102}, \eprint{2009.01075}

\bibitem[{Abbott et~al.(2020{\natexlab{e}})}]{bib:GW190814}
Abbott R, et~al. (2020{\natexlab{e}}) {{GW}190814: Gravitational Waves from the
  Coalescence of a 23 Solar Mass Black Hole with a 2.6 Solar Mass Compact
  Object}. The Astrophysical Journal 896(2):L44, \doi{10.3847/2041-8213/ab960f}

\bibitem[{Abbott et~al.(2021{\natexlab{a}})}]{bib:cwallskyO3aPowerFlux}
Abbott R, et~al. (2021{\natexlab{a}}) {All-sky search for continuous
  gravitational waves from isolated neutron stars in the early O3 LIGO data}.
  Physical Review D 104(8), \doi{10.1103/physrevd.104.082004}

\bibitem[{Abbott
  et~al.(2021{\natexlab{b}})}]{bib:cwallskybinaryO3aBinarySkyHough}
Abbott R, et~al. (2021{\natexlab{b}}) {All-sky search in early O3 LIGO data for
  continuous gravitational-wave signals from unknown neutron stars in binary
  systems}. Phys Rev D 103:064017, \doi{10.1103/PhysRevD.103.064017}

\bibitem[{Abbott et~al.(2021{\natexlab{c}})}]{bib:cwdirectedO3J0537rmodes}
Abbott R, et~al. (2021{\natexlab{c}}) {Constraints from {LIGO} O3 Data on
  Gravitational-wave Emission Due to R-modes in the Glitching Pulsar {PSR}
  J0537{\textendash}6910}. The Astrophysical Journal 922(1):71,
  \doi{10.3847/1538-4357/ac0d52}

\bibitem[{Abbott et~al.(2021{\natexlab{d}})}]{bib:cwtargetedO3J0537}
Abbott R, et~al. (2021{\natexlab{d}}) {Diving below the Spin-down Limit:
  Constraints on Gravitational Waves from the Energetic Young Pulsar PSR
  J0537-6910}. The Astrophysical Journal Letters 913(2):L27,
  \doi{10.3847/2041-8213/abffcd}

\bibitem[{Abbott et~al.(2021{\natexlab{e}})}]{bib:GWTC2}
Abbott R, et~al. (2021{\natexlab{e}}) {GWTC-2: Compact Binary Coalescences
  Observed by LIGO and Virgo during the First Half of the Third Observing Run}.
  Phys Rev X 11:021053, \doi{10.1103/PhysRevX.11.021053}

\bibitem[{Abbott et~al.(2021{\natexlab{f}})}]{bib:GWTC3}
Abbott R, et~al. (2021{\natexlab{f}}) {GWTC-3: Compact Binary Coalescences
  Observed by LIGO and Virgo During the Second Part of the Third Observing
  Run}. \eprint{2111.03606}

\bibitem[{Abbott et~al.(2021{\natexlab{g}})}]{bib:NSBH}
Abbott R, et~al. (2021{\natexlab{g}}) {Observation of Gravitational Waves from
  Two Neutron Star\textendash{}Black Hole Coalescences}. Astrophys J Lett
  915(1):L5, \doi{10.3847/2041-8213/ac082e}, \eprint{2106.15163}

\bibitem[{Abbott et~al.(2021{\natexlab{h}})}]{bib:O3Radiometer}
Abbott R, et~al. (2021{\natexlab{h}}) {Search for anisotropic
  gravitational-wave backgrounds using data from Advanced LIGO and Advanced
  Virgo\textquoteright{}s first three observing runs}. Phys Rev D
  104(2):022005, \doi{10.1103/PhysRevD.104.022005}, \eprint{2103.08520}

\bibitem[{Abbott et~al.(2021{\natexlab{i}})}]{bib:cwdirectedO3aSNRs}
Abbott R, et~al. (2021{\natexlab{i}}) {Searches for Continuous Gravitational
  Waves from Young Supernova Remnants in the Early Third Observing Run of
  Advanced LIGO and Virgo}. The Astrophysical Journal 921(1):80,
  \doi{10.3847/1538-4357/ac17ea}

\bibitem[{Abbott et~al.(2022{\natexlab{a}})}]{bib:ASAFO3}
Abbott R, et~al. (2022{\natexlab{a}}) All-sky, all-frequency directional search
  for persistent gravitational waves from advanced ligo's and advanced virgo's
  first three observing runs. Phys Rev D 105:122001,
  \doi{10.1103/PhysRevD.105.122001}, \eprint{2110.09834}

\bibitem[{Abbott et~al.(2022{\natexlab{b}})}]{bib:cwallskyO3FourPipelines}
Abbott R, et~al. (2022{\natexlab{b}}) All-sky search for continuous
  gravitational waves from isolated neutron stars using advanced ligo and
  advanced virgo o3 data. Phys Rev D 106:102008,
  \doi{10.1103/PhysRevD.106.102008}, \eprint{2201.00697}

\bibitem[{Abbott et~al.(2022{\natexlab{c}})}]{bib:cwallskyO3BosonCloud}
Abbott R, et~al. (2022{\natexlab{c}}) {All-sky search for gravitational wave
  emission from scalar boson clouds around spinning black holes in LIGO O3
  data}. Phys Rev D 105:102001, \doi{10.1103/PhysRevD.105.102001},
  \eprint{2111.15507}

\bibitem[{Abbott et~al.(2022{\natexlab{d}})}]{bib:DPDMO3}
Abbott R, et~al. (2022{\natexlab{d}}) {Constraints on dark photon dark matter
  using data from LIGO's and Virgo's third observing run}. Phys Rev D
  105:063030, \doi{10.1103/PhysRevD.105.063030}, \eprint{2105.13085}

\bibitem[{Abbott et~al.(2022{\natexlab{e}})}]{bib:O3CrossCorr}
Abbott R, et~al. (2022{\natexlab{e}}) {Model-based Cross-correlation Search for
  Gravitational Waves from the Low-mass X-Ray Binary Scorpius X-1 in LIGO O3
  Data}. The Astrophysical Journal Letters 941(2):L30,
  \doi{10.3847/2041-8213/aca1b0}, \eprint{2209.02863}

\bibitem[{Abbott et~al.(2022{\natexlab{f}})}]{bib:cwnarrowbandO3}
Abbott R, et~al. (2022{\natexlab{f}}) Narrowband searches for continuous and
  long-duration transient gravitational waves from known pulsars in the
  ligo-virgo third observing run. The Astrophysical Journal 932(2):133,
  \doi{10.3847/1538-4357/ac6ad0},
  \urlprefix\url{https://dx.doi.org/10.3847/1538-4357/ac6ad0},
  \eprint{2112.10990}

\bibitem[{Abbott et~al.(2022{\natexlab{g}})}]{bib:cwAXMPO3}
Abbott R, et~al. (2022{\natexlab{g}}) {Search for continuous gravitational
  waves from 20 accreting millisecond x-ray pulsars in O3 LIGO data}. Phys Rev
  D 105:022002, \doi{10.1103/PhysRevD.105.022002}

\bibitem[{Abbott et~al.(2022{\natexlab{h}})}]{bib:cwdirectedO3ScoX1Viterbi}
Abbott R, et~al. (2022{\natexlab{h}}) {Search for gravitational waves from
  Scorpius X-1 with a hidden Markov model in O3 LIGO data}. Phys Rev D
  106(6):062002, \doi{10.1103/PhysRevD.106.062002}, \eprint{2201.10104}

\bibitem[{Abbott et~al.(2022{\natexlab{i}})}]{bib:cwdirectedO3aCasAVelaJr}
Abbott R, et~al. (2022{\natexlab{i}}) {Search of the Early O3 LIGO Data for
  Continuous Gravitational Waves from the Cassiopeia A and Vela Jr. Supernova
  Remnants}. Phys Rev D 105:082005, \doi{10.1103/PhysRevD.105.082005},
  \eprint{2111.15116}

\bibitem[{Abbott et~al.(2022{\natexlab{j}})}]{bib:cwtargetedO3}
Abbott R, et~al. (2022{\natexlab{j}}) Searches for gravitational waves from
  known pulsars at two harmonics in the second and third ligo-virgo observing
  runs. The Astrophysical Journal 935(1):1, \doi{10.3847/1538-4357/ac6acf},
  \urlprefix\url{https://dx.doi.org/10.3847/1538-4357/ac6acf},
  \eprint{2111.13106}

\bibitem[{Acernese et~al.(2014)}]{bib:avirgodetector}
Acernese F, et~al. (2014) {Advanced Virgo: a second-generation interferometric
  gravitational wave detector}. Classical and Quantum Gravity 32(2):024001,
  \doi{10.1088/0264-9381/32/2/024001}

\bibitem[{Ackermann et~al.(2017)}]{bib:AckermannEtal}
Ackermann M, et~al. (2017) {The Fermi Galactic Center GeV Excess and
  Implications for Dark Matter}. Astrophys J 840(1):43,
  \doi{10.3847/1538-4357/aa6cab}, \eprint{1704.03910}

\bibitem[{{Aharonian} et~al.(2005)}]{bib:HESSJ1813}
{Aharonian} F, et~al. (2005) {A New Population of Very High Energy Gamma-Ray
  Sources in the Milky Way}. Science 307(5717):1938--1942,
  \doi{10.1126/science.1108643}, \eprint{astro-ph/0504380}

\bibitem[{{Aharonian} et~al.(2008)}]{bib:HessUnidentifiedSources}
{Aharonian} F, et~al. (2008) {HESS very-high-energy gamma-ray sources without
  identified counterparts}. Astron Astrophys 477(1):353--363,
  \doi{10.1051/0004-6361:20078516}, \eprint{0712.1173}

\bibitem[{Alarie et~al.(2014)Alarie, Bilodeau, and
  Drissen}]{bib:AlarieEtalCasA}
Alarie A, Bilodeau A, Drissen L (2014) {A hyperspectral view of Cassiopeia A}.
  Monthly Notices of the Royal Astronomical Society 441:2996--3008

\bibitem[{Alford and Schwenzer(2014)}]{bib:alfordschwenzeryoungpulsar}
Alford MG, Schwenzer K (2014) {Gravitational wave emission and spindown of
  young pulsars}. Astrophys J 781:26, \doi{10.1088/0004-637X/781/1/26},
  \eprint{1210.6091}

\bibitem[{Alford and Schwenzer(2015)}]{bib:alfordschwenzerMSP}
Alford MG, Schwenzer K (2015) {Gravitational wave emission from oscillating
  millisecond pulsars}. Mon Not Roy Astron Soc 446(4):3631--3641,
  \doi{10.1093/mnras/stu2361}, \eprint{1403.7500}

\bibitem[{Allen(2019)}]{bib:AllenTemplates}
Allen B (2019) {Spherical ansatz for parameter-space metrics}. Phys Rev D
  100(12):124004, \doi{10.1103/PhysRevD.100.124004}, \eprint{1906.01352}

\bibitem[{Allen(2021)}]{bib:AllenTemplates2}
Allen B (2021) {Optimal template banks}. Phys Rev D 104(4):042005,
  \doi{10.1103/PhysRevD.104.042005}, \eprint{2102.11254}

\bibitem[{Allen et~al.(1999)Allen, Hua, and
  Ottewill}]{bib:OttewillAllencleaning}
Allen B, Hua W, Ottewill A (1999) {Automatic cross-talk removal from
  multi-channel data}. \eprint{gr-qc/9909083}

\bibitem[{Allen et~al.(2002)Allen, Papa, and Schutz}]{bib:AllenPapaSchutz}
Allen B, Papa MA, Schutz BF (2002) {Optimal strategies for sinusoidal signal
  detection}. Phys Rev D 66:102003, \doi{10.1103/PhysRevD.66.102003}

\bibitem[{Allen et~al.(2015)Allen, Chow, DeLaney, Filipovic, Houck, Pannuti,
  and Stage}]{bib:AllenEtalVelaJr}
Allen GE, Chow K, DeLaney T, Filipovic MD, Houck JC, Pannuti TG, Stage MD
  (2015) {On the Expansion Rate, Age, and Distance of the Supernova Remnant
  G266.2-1.2 (Vela Jr.)}. Astrophys J 798(2):82,
  \doi{10.1088/0004-637X/798/2/82}, \eprint{1410.7435}

\bibitem[{{Alpar} et~al.(1982){Alpar}, {Cheng}, {Ruderman}, and
  {Shaham}}]{bib:AlparEtalRecycling}
{Alpar} MA, {Cheng} AF, {Ruderman} MA, {Shaham} J (1982) {A new class of radio
  pulsars}. Nature 300(5894):728--730, \doi{10.1038/300728a0}

\bibitem[{Althouse et~al.(1998)Althouse, Jones, and
  Lazzarini}]{bib:AlthouseJonesLazzarini}
Althouse B, Jones L, Lazzarini A (1998) {Determination of Global and Local
  Coordinate Axes for the LIGO Sites}. LIGO Report T980044,
  \urlprefix\url{https://dcc.ligo.org/T980044}

\bibitem[{Anderson et~al.(2002)Anderson, Cobb, Korpela, Lebofsky, and
  Werthimer}]{bib:seti@home}
Anderson DP, Cobb J, Korpela E, Lebofsky M, Werthimer D (2002) {SETI@home: An
  Experiment in Public-Resource Computing}. Commun ACM 45(11):56–61,
  \doi{10.1145/581571.581573}

\bibitem[{Andersson(1998)}]{bib:rmodes1}
Andersson N (1998) {A New class of unstable modes of rotating relativistic
  stars}. Astrophys J 502:708--713, \doi{10.1086/305919},
  \eprint{gr-qc/9706075}

\bibitem[{Andersson(2019)}]{bib:AnderssonText}
Andersson N (2019) {Gravitational-Wave Astronomy: Exploring the Dark Side of
  the Universe}. Oxford University Press,
  \doi{10.1093/oso/9780198568032.001.0001}

\bibitem[{Andersson et~al.(2014)Andersson, Jones, and
  Ho}]{bib:AnderssonJonesHo}
Andersson N, Jones DI, Ho WCG (2014) {Implications of an r-mode in XTE
  J1751-305: Mass, radius and spin evolution}. Mon Not Roy Astron Soc
  442(2):1786--1793, \doi{10.1093/mnras/stu870}, \eprint{1403.0860}

\bibitem[{Andersson et~al.(2018)Andersson, Antonopoulou, Espinoza, Haskell, and
  Ho}]{bib:rmodeJ0537-6910}
Andersson N, Antonopoulou D, Espinoza C, Haskell B, Ho W (2018) {The Enigmatic
  Spin Evolution of PSR J0537--6910: r-modes, Gravitational Waves, and the Case
  for Continued Timing}. Astrophys J 864(2):137,
  \doi{10.3847/1538-4357/aad6eb}, \eprint{1711.05550}

\bibitem[{{Antoniadis} et~al.(2013){Antoniadis}, {Freire}, {Wex}, {Tauris},
  {Lynch}, {van Kerkwijk}, {Kramer}, {Bassa}, {Dhillon}, {Driebe}, {Hessels},
  {Kaspi}, {Kondratiev}, {Langer}, {Marsh}, {McLaughlin}, {Pennucci}, {Ransom},
  {Stairs}, {van Leeuwen}, {Verbiest}, and {Whelan}}]{bib:AntoniadisJ0348}
{Antoniadis} J, {Freire} PCC, {Wex} N, {Tauris} TM, {Lynch} RS, {van Kerkwijk}
  MH, {Kramer} M, {Bassa} C, {Dhillon} VS, {Driebe} T, {Hessels} JWT, {Kaspi}
  VM, {Kondratiev} VI, {Langer} N, {Marsh} TR, {McLaughlin} MA, {Pennucci} TT,
  {Ransom} SM, {Stairs} IH, {van Leeuwen} J, {Verbiest} JPW, {Whelan} DG (2013)
  {A Massive Pulsar in a Compact Relativistic Binary}. Science 340(6131):448,
  \doi{10.1126/science.1233232}, \eprint{1304.6875}

\bibitem[{Antonopoulou et~al.(2018)Antonopoulou, Espinoza, Kuiper, and
  Andersson}]{bib:Antonopoulouetal}
Antonopoulou D, Espinoza CM, Kuiper L, Andersson N (2018) {Pulsar spin-down:
  the glitch-dominated rotation of PSR J0537$-$6910}. Mon Not Roy Astron Soc
  473(2):1644--1655, \doi{10.1093/mnras/stx2429}, \eprint{1708.09459}

\bibitem[{Antonucci et~al.(2008)Antonucci, Astone, D'Antonio, Frasca, and
  Palomba}]{bib:freqhough1}
Antonucci F, Astone P, D'Antonio S, Frasca S, Palomba C (2008) {Detection of
  periodic gravitational wave sources by Hough transform in the f versus
  {\textbackslash}skew6{\textbackslash}dot f plane}. Classical and Quantum
  Gravity 25(18):184015, \doi{10.1088/0264-9381/25/18/184015}

\bibitem[{de~Araujo et~al.(2016)de~Araujo, Coelho, and
  Costa}]{bib:DeAraujoCoelhoCosta1}
de~Araujo JCN, Coelho JG, Costa CA (2016) {Gravitational Waves from Pulsars and
  their Braking Indices: The Role of a Time Dependent Magnetic Ellipticity}.
  The Astrophysical Journal 831(1):35, \doi{10.3847/0004-637x/831/1/35}

\bibitem[{de~Araujo et~al.(2017)de~Araujo, Coelho, and
  Costa}]{bib:DeAraujoCoelhoCosta2}
de~Araujo JCN, Coelho JG, Costa CA (2017) {Gravitational waves from pulsars in
  the context of magnetic ellipticity}. The European Physical Journal C 77(5),
  \doi{10.1140/epjc/s10052-017-4925-3}

\bibitem[{Archibald et~al.(2009)Archibald, Stairs, Ransom, Kaspi, Kondratiev,
  Lorimer, McLaughlin, Boyles, Hessels, Lynch, van Leeuwen, Roberts, Jenet,
  Champion, Rosen, Barlow, Dunlap, and Remillard}]{bib:tMSPfirst3}
Archibald AM, Stairs IH, Ransom SM, Kaspi VM, Kondratiev VI, Lorimer DR,
  McLaughlin MA, Boyles J, Hessels JWT, Lynch R, van Leeuwen J, Roberts MSE,
  Jenet F, Champion DJ, Rosen R, Barlow BN, Dunlap BH, Remillard RA (2009) {A
  Radio Pulsar/X-ray Binary Link}. Science 324(5933):1411--1414,
  \doi{10.1126/science.1172740}

\bibitem[{Archibald et~al.(2016)Archibald, Gotthelf, Ferdman, Kaspi, Guillot,
  Harrison, Keane, Pivovaroff, Stern, Tendulkar, and
  et~al.}]{bib:ArchibaldEtal_2016}
Archibald RF, Gotthelf EV, Ferdman RD, Kaspi VM, Guillot S, Harrison FA, Keane
  EF, Pivovaroff MJ, Stern D, Tendulkar SP, et~al (2016) {A High Braking Index
  for a Pulsar}. The Astrophysical Journal 819(1):L16,
  \doi{10.3847/2041-8205/819/1/l16}

\bibitem[{Arras et~al.(2003)Arras, Flanagan, Morsink, Schenk, Teukolsky, and
  Wasserman}]{bib:rmodesdoubts}
Arras P, Flanagan EE, Morsink SM, Schenk A, Teukolsky SA, Wasserman I (2003)
  {Saturation of the R mode instability}. Astrophys J 591:1129--1151,
  \doi{10.1086/374657}, \eprint{astro-ph/0202345}

\bibitem[{Arvanitaki and Dubovsky(2011)}]{bib:axiverseArvanitaki2}
Arvanitaki A, Dubovsky S (2011) {Exploring the String Axiverse with Precision
  Black Hole Physics}. Phys Rev D 83:044026, \doi{10.1103/PhysRevD.83.044026},
  \eprint{1004.3558}

\bibitem[{Arvanitaki et~al.(2010)Arvanitaki, Dimopoulos, Dubovsky, Kaloper, and
  March-Russell}]{bib:axiverseArvanitaki1}
Arvanitaki A, Dimopoulos S, Dubovsky S, Kaloper N, March-Russell J (2010)
  {String axiverse}. Phys Rev D 81:123530, \doi{10.1103/PhysRevD.81.123530}

\bibitem[{Arvanitaki et~al.(2015)Arvanitaki, Baryakhtar, and
  Huang}]{bib:axionArvanitaki}
Arvanitaki A, Baryakhtar M, Huang X (2015) {Discovering the QCD Axion with
  Black Holes and Gravitational Waves}. Phys Rev D 91(8):084011,
  \doi{10.1103/PhysRevD.91.084011}, \eprint{1411.2263}

\bibitem[{Arvanitaki et~al.(2017)Arvanitaki, Baryakhtar, Dimopoulos, Dubovsky,
  and Lasenby}]{bib:axionBBHmerger}
Arvanitaki A, Baryakhtar M, Dimopoulos S, Dubovsky S, Lasenby R (2017) {Black
  hole mergers and the QCD axion at Advanced LIGO}. Phys Rev D 95:043001,
  \doi{10.1103/PhysRevD.95.043001}

\bibitem[{{Arzoumanian} et~al.(2018){Arzoumanian}, {Brazier}, {Burke-Spolaor},
  {Chamberlin}, {Chatterjee}, {Christy}, {Cordes}, {Cornish}, {Crawford},
  {Thankful Cromartie}, {Crowter}, {DeCesar}, {Demorest}, {Dolch}, {Ellis},
  {Ferdman}, {Ferrara}, {Fonseca}, {Garver-Daniels}, {Gentile}, {Halmrast},
  {Huerta}, {Jenet}, {Jessup}, {Jones}, {Jones}, {Kaplan}, {Lam}, {Lazio},
  {Levin}, {Lommen}, {Lorimer}, {Luo}, {Lynch}, {Madison}, {Matthews},
  {McLaughlin}, {McWilliams}, {Mingarelli}, {Ng}, {Nice}, {Pennucci}, {Ransom},
  {Ray}, {Siemens}, {Simon}, {Spiewak}, {Stairs}, {Stinebring}, {Stovall},
  {Swiggum}, {Taylor}, {Vallisneri}, {van Haasteren}, {Vigeland}, {Zhu}, and
  {NANOGrav Collaboration}}]{bib:ArzoumanianEtalJ1614}
{Arzoumanian} Z, {Brazier} A, {Burke-Spolaor} S, {Chamberlin} S, {Chatterjee}
  S, {Christy} B, {Cordes} JM, {Cornish} NJ, {Crawford} F, {Thankful Cromartie}
  H, {Crowter} K, {DeCesar} ME, {Demorest} PB, {Dolch} T, {Ellis} JA, {Ferdman}
  RD, {Ferrara} EC, {Fonseca} E, {Garver-Daniels} N, {Gentile} PA, {Halmrast}
  D, {Huerta} EA, {Jenet} FA, {Jessup} C, {Jones} G, {Jones} ML, {Kaplan} DL,
  {Lam} MT, {Lazio} TJW, {Levin} L, {Lommen} A, {Lorimer} DR, {Luo} J, {Lynch}
  RS, {Madison} D, {Matthews} AM, {McLaughlin} MA, {McWilliams} ST,
  {Mingarelli} C, {Ng} C, {Nice} DJ, {Pennucci} TT, {Ransom} SM, {Ray} PS,
  {Siemens} X, {Simon} J, {Spiewak} R, {Stairs} IH, {Stinebring} DR, {Stovall}
  K, {Swiggum} JK, {Taylor} SR, {Vallisneri} M, {van Haasteren} R, {Vigeland}
  SJ, {Zhu} W, {NANOGrav Collaboration} (2018) {The NANOGrav 11-year Data Set:
  High-precision Timing of 45 Millisecond Pulsars}. Astrophys J Suppl
  235(2):37, \doi{10.3847/1538-4365/aab5b0}, \eprint{1801.01837}

\bibitem[{Ashok et~al.(2021)Ashok, Beheshtipour, Papa, Freire, Steltner,
  Machenschalk, Behnke, Allen, and Prix}]{bib:AshokEtal}
Ashok A, Beheshtipour B, Papa MA, Freire PCC, Steltner B, Machenschalk B,
  Behnke O, Allen B, Prix R (2021) {New Searches for Continuous Gravitational
  Waves from Seven Fast Pulsars}. The Astrophysical Journal 923(1):85,
  \doi{10.3847/1538-4357/ac2582}

\bibitem[{Ashton and Prix(2018)}]{bib:AshtonPrix}
Ashton G, Prix R (2018) {Hierarchical multistage MCMC follow-up of continuous
  gravitational wave candidates}. Phys Rev D 97(10):103020,
  \doi{10.1103/PhysRevD.97.103020}, \eprint{1802.05450}

\bibitem[{Ashton et~al.(2017)Ashton, Prix, and Jones}]{bib:AshtonPrixJones}
Ashton G, Prix R, Jones DI (2017) {Statistical characterization of pulsar
  glitches and their potential impact on searches for continuous gravitational
  waves}. Phys Rev D 96(6):063004, \doi{10.1103/PhysRevD.96.063004},
  \eprint{1704.00742}

\bibitem[{Ashton et~al.(2019)Ashton, Lasky, Graber, and
  Palfreyman}]{bib:AshtonLaskyGraberPalfreyman}
Ashton G, Lasky PD, Graber V, Palfreyman J (2019) {Rotational evolution of the
  Vela pulsar during the 2016 glitch}. Nature Astronomy 3(12):1143--1148,
  \doi{10.1038/s41550-019-0844-6}

\bibitem[{Astone et~al.(2002{\natexlab{a}})Astone, Borkowski, Jaranowski, and
  Kr\'olak}]{bib:cwexplorer2}
Astone P, Borkowski KM, Jaranowski P, Kr\'olak A (2002{\natexlab{a}}) {Data
  analysis of gravitational-wave signals from spinning neutron stars. IV. An
  all-sky search}. Phys Rev D 65:042003, \doi{10.1103/PhysRevD.65.042003}

\bibitem[{Astone et~al.(2005)Astone, Frasca, and Palomba}]{bib:SFTdatabase}
Astone P, Frasca S, Palomba C (2005) {The short {FFT} database and the peak map
  for the hierarchical search of periodic sources}. Classical and Quantum
  Gravity 22(18):S1197--S1210, \doi{10.1088/0264-9381/22/18/s34}

\bibitem[{Astone et~al.(2008)Astone, Bassan, Bonifazi, Borkowski, Budzyński,
  Chincarini, Coccia, D’Antonio, Emilio, Fafone, and
  et~al.}]{bib:VirgoBarycentering}
Astone P, Bassan M, Bonifazi P, Borkowski KM, Budzyński RJ, Chincarini A,
  Coccia E, D’Antonio S, Emilio MDP, Fafone V, et~al (2008) {All-sky search
  of NAUTILUS data}. Classical and Quantum Gravity 25(18):184012,
  \doi{10.1088/0264-9381/25/18/184012}

\bibitem[{Astone et~al.(2010{\natexlab{a}})Astone, Borkowski, Jaranowski,
  Pietka, and Kr\'olak}]{bib:tdfstatistic}
Astone P, Borkowski KM, Jaranowski P, Pietka M, Kr\'olak A (2010{\natexlab{a}})
  {Data analysis of gravitational-wave signals from spinning neutron stars. V.
  A narrow-band all-sky search}. Phys Rev D 82:022005,
  \doi{10.1103/PhysRevD.82.022005}

\bibitem[{Astone et~al.(2010{\natexlab{b}})Astone, D'Antonio, Frasca, and
  Palomba}]{bib:fivevector}
Astone P, D'Antonio S, Frasca S, Palomba C (2010{\natexlab{b}}) {A method for
  detection of known sources of continuous gravitational wave signals in
  non-stationary data}. Class Quant Grav 27:194016,
  \doi{10.1088/0264-9381/27/19/194016}

\bibitem[{Astone et~al.(2012)Astone, Colla, D'Antonio, Frasca, and
  Palomba}]{bib:fivevectorupdates}
Astone P, Colla A, D'Antonio S, Frasca S, Palomba C (2012) {Coherent search of
  continuous gravitational wave signals: extension of the 5-vectors method to a
  network of detectors}. J Phys Conf Ser 363:012038,
  \doi{10.1088/1742-6596/363/1/012038}, \eprint{1203.6733}

\bibitem[{Astone et~al.(2014{\natexlab{a}})Astone, Colla, D'Antonio, Frasca,
  and Palomba}]{bib:freqhough2}
Astone P, Colla A, D'Antonio S, Frasca S, Palomba C (2014{\natexlab{a}})
  {Method for all-sky searches of continuous gravitational wave signals using
  the frequency-Hough transform}. Phys Rev D 90:042002,
  \doi{10.1103/PhysRevD.90.042002}

\bibitem[{Astone et~al.(2014{\natexlab{b}})Astone, Colla, D'Antonio, Frasca,
  Palomba, and Serafinelli}]{bib:VirgoResampling}
Astone P, Colla A, D'Antonio S, Frasca S, Palomba C, Serafinelli R
  (2014{\natexlab{b}}) {Method for narrow-band search of continuous
  gravitational wave signals}. Phys Rev D 89:062008,
  \doi{10.1103/PhysRevD.89.062008}

\bibitem[{Astone et~al.(2002{\natexlab{b}})}]{bib:cwexplorer1}
Astone P, et~al. (2002{\natexlab{b}}) {Search for periodic gravitational wave
  sources with the explorer detector}. Phys Rev D 65:022001,
  \doi{10.1103/PhysRevD.65.022001}, \eprint{gr-qc/0011072}

\bibitem[{Baade and Zwicky(1934)}]{bib:BaadeZwicky}
Baade W, Zwicky F (1934) {Cosmic Rays from Super-Novae}. Proceedings of the
  National Academy of Sciences 20(5):259--263, \doi{10.1073/pnas.20.5.259},
  \eprint{https://www.pnas.org/content/20/5/259.full.pdf}

\bibitem[{Baiko and Chugunov(2018)}]{bib:BaikoChugunov}
Baiko DA, Chugunov AI (2018) {Breaking properties of neutron star crust}. Mon
  Not Roy Astron Soc 480(4):5511--5516, \doi{10.1093/mnras/sty2259},
  \eprint{1808.06415}

\bibitem[{Baiotti and Rezzolla(2017)}]{bib:BaiottiRezzolla}
Baiotti L, Rezzolla L (2017) {Binary neutron star mergers: a review of
  Einstein’s richest laboratory}. Reports on Progress in Physics
  80(9):096901, \doi{10.1088/1361-6633/aa67bb}

\bibitem[{Balasubramanian et~al.(1996)Balasubramanian, Sathyaprakash, and
  Dhurandhar}]{bib:SathyaTemplates}
Balasubramanian R, Sathyaprakash B, Dhurandhar S (1996) {Gravitational waves
  from coalescing binaries: Detection strategies and Monte Carlo estimation of
  parameters}. Phys Rev D 53:3033--3055, \doi{10.1103/PhysRevD.53.3033},
  [Erratum: Phys.Rev.D 54, 1860 (1996)], \eprint{gr-qc/9508011}

\bibitem[{Ballmer(2006{\natexlab{a}})}]{bib:radiometermethod}
Ballmer SW (2006{\natexlab{a}}) {A radiometer for stochastic gravitational
  waves}. Classical and Quantum Gravity 23(8):S179--S185,
  \doi{10.1088/0264-9381/23/8/s23}

\bibitem[{Ballmer(2006{\natexlab{b}})}]{bib:BallmerRadiometer}
Ballmer SW (2006{\natexlab{b}}) {A radiometer for stochastic gravitational
  waves}. Classical and Quantum Gravity 23(8):S179--S185,
  \doi{10.1088/0264-9381/23/8/s23}

\bibitem[{Banagiri et~al.(2019)Banagiri, Sun, Coughlin, and
  Melatos}]{bib:LongTransientsHMM}
Banagiri S, Sun L, Coughlin MW, Melatos A (2019) {Search strategies for long
  gravitational-wave transients: Hidden Markov model tracking and seedless
  clustering}. Phys Rev D 100:024034, \doi{10.1103/PhysRevD.100.024034}

\bibitem[{{Bassa} et~al.(2017){Bassa}, {Pleunis}, {Hessels}, {Ferrara},
  {Breton}, {Gusinskaia}, {Kondratiev}, {Sanidas}, {Nieder}, {Clark}, {Li},
  {van Amesfoort}, {Burnett}, {Camilo}, {Michelson}, {Ransom}, {Ray}, and
  {Wood}}]{bib:BassaEtal}
{Bassa} CG, {Pleunis} Z, {Hessels} JWT, {Ferrara} EC, {Breton} RP, {Gusinskaia}
  NV, {Kondratiev} VI, {Sanidas} S, {Nieder} L, {Clark} CJ, {Li} T, {van
  Amesfoort} AS, {Burnett} TH, {Camilo} F, {Michelson} PF, {Ransom} SM, {Ray}
  PS, {Wood} K (2017) {LOFAR Discovery of the Fastest-spinning Millisecond
  Pulsar in the Galactic Field}. The Astrophysical Journal Letters 846(2):L20,
  \doi{10.3847/2041-8213/aa8400}, \eprint{1709.01453}

\bibitem[{Baumann et~al.(2019)Baumann, Chia, Stout, and ter
  Haar}]{bib:BaumannChiaStoutTerHaar}
Baumann D, Chia HS, Stout J, ter Haar L (2019) {The Spectra of Gravitational
  Atoms}. JCAP 12:006, \doi{10.1088/1475-7516/2019/12/006}, \eprint{1908.10370}

\bibitem[{Bayley et~al.(2020)Bayley, Messenger, and Woan}]{bib:ViterbiGlasgow}
Bayley J, Messenger C, Woan G (2020) Robust machine learning algorithm to
  search for continuous gravitational waves. Physical Review D 102(8),
  \doi{10.1103/physrevd.102.083024}, \eprint{2007.08207}

\bibitem[{Baym et~al.(1969)Baym, Pethick, Pines, and Ruderman}]{bib:BaymEtal}
Baym G, Pethick C, Pines D, Ruderman M (1969) {Spin Up in Neutron Stars : The
  Future of the Vela Pulsar}. Nature 224(5222):872--874, \doi{10.1038/224872a0}

\bibitem[{Becker(2009)}]{bib:NSandPulsars-Becker}
Becker W (2009) {Neutron Stars and Pulsars}. Astrophysics and Space Science
  Library, vol 357. Springer, Berlin, Heidelberg.,
  \doi{10.1007/978-3-540-76965-1}

\bibitem[{{Becker} et~al.(2006){Becker}, {Hui}, {Aschenbach}, and
  {Iyudin}}]{bib:BeckerEtalVelaJr}
{Becker} W, {Hui} CY, {Aschenbach} B, {Iyudin} A (2006) {Exploring the Central
  Compact Object in the RX J0852.0-4622 Supernova Remnant with XMM-Newton}.
  arXiv e-prints astro-ph/0607081, \eprint{astro-ph/0607081}

\bibitem[{Beheshtipour and Papa(2020)}]{bib:BeheshtipourPapa1}
Beheshtipour B, Papa M (2020) {Deep learning for clustering of continuous
  gravitational wave candidates}. Physical Review D 101(6),
  \doi{10.1103/physrevd.101.064009}

\bibitem[{Beheshtipour and Papa(2021)}]{bib:BeheshtipourPapa2}
Beheshtipour B, Papa MA (2021) {Deep learning for clustering of continuous
  gravitational wave candidates. II. Identification of low-SNR candidates}.
  Phys Rev D 103:064027, \doi{10.1103/PhysRevD.103.064027}

\bibitem[{Behnke et~al.(2015)Behnke, Papa, and Prix}]{bib:SensitivityDepth}
Behnke B, Papa MA, Prix R (2015) {Postprocessing methods used in the search for
  continuous gravitational-wave signals from the Galactic Center}. Phys Rev D
  91:064007, \doi{10.1103/PhysRevD.91.064007}

\bibitem[{Bender et~al.(1996)}]{bib:LISA}
Bender P, et~al. (1996) {MPQ Reports, MPQ-208}. Tech. rep., Max-Planck-Institut
  fur Quantenoptik, Garching, {URL: https://www.elisascience.org.}

\bibitem[{Beniwal et~al.(2022)Beniwal, Clearwater, Dunn, Strang, Rowell,
  Melatos, and Ottaway}]{bib:BeniwalEtal}
Beniwal D, Clearwater P, Dunn L, Strang L, Rowell G, Melatos A, Ottaway D
  (2022) {Search for continuous gravitational waves from HESS J1427-608 with a
  hidden Markov model}. Phys Rev D 106(10):103018,
  \doi{10.1103/PhysRevD.106.103018}, \eprint{2210.09592}

\bibitem[{{Bernal} et~al.(2010){Bernal}, {Lee}, and
  {Page}}]{bib:BernalEtalAccretion}
{Bernal} CG, {Lee} WH, {Page} D (2010) {Hypercritical accretion onto a
  magnetized neutron star surface: a numerical approach}. Rev Mex de Astron y
  Astrofís 46:309--322

\bibitem[{Bero and Whelan(2019)}]{bib:BeroWhelan}
Bero JJ, Whelan JT (2019) {An Analytic Approximation to the Bayesian Detection
  Statistic for Continuous Gravitational Waves}. Class Quant Grav 36(1):015013,
  \doi{10.1088/1361-6382/aaed6a}, [Erratum: Class.Quant.Grav. 36, 049601
  (2019)], \eprint{1808.05453}

\bibitem[{Bertone(2010)}]{bib:darkmatterreview}
Bertone Ge (2010) {Particle Dark Matter: Observations, Models and Searches}.
  Cambridge University Press, \doi{10.1017/CBO9780511770739}

\bibitem[{Beskin and Istomin(2022)}]{bib:BeskinIstominDeathValley}
Beskin VS, Istomin AY (2022) {Pulsar death line revisited \textendash{} II.
  \textquoteleft{}The death valley\textquoteright{}}. Mon Not Roy Astron Soc
  516(4):5084--5091, \doi{10.1093/mnras/stac2423}, \eprint{2207.04723}

\bibitem[{Beskin and Litvinov(2022)}]{bib:BeskinLitvinovDeathLine}
Beskin VS, Litvinov PE (2022) {Pulsar death line revisited \textendash{} I.
  Almost vacuum gap}. Mon Not Roy Astron Soc 510(2):2572--2582,
  \doi{10.1093/mnras/stab3575}, \eprint{2201.02875}

\bibitem[{Bhattacharyya(2020)}]{bib:Bhattacharyya}
Bhattacharyya S (2020) {The permanent ellipticity of the neutron star in PSR
  J1023+0038}. Mon Not Roy Astron Soc 498(1):728--736,
  \doi{10.1093/mnras/staa2304}, \eprint{2008.01716}

\bibitem[{Bhattacharyya(2021)}]{bib:BhattacharyyaTwoModes}
Bhattacharyya S (2021) {Spin evolution of neutron stars in two modes:
  implication for millisecond pulsars}. Monthly Notices of the Royal
  Astronomical Society: Letters 502(1):L45–L49, \doi{10.1093/mnrasl/slab001}

\bibitem[{Bildsten(1998)}]{bib:rmodes2}
Bildsten L (1998) {Gravitational Radiation and Rotation of Accreting Neutron
  Stars}. The Astrophysical Journal 501(1):L89--L93, \doi{10.1086/311440}

\bibitem[{Biwer et~al.(2017)Biwer, Barker, Batch, Betzwieser, Fisher, Goetz,
  Kandhasamy, Karki, Kissel, Lundgren, and et~al.}]{bib:hwinjectionpaper}
Biwer C, Barker D, Batch J, Betzwieser J, Fisher R, Goetz E, Kandhasamy S,
  Karki S, Kissel J, Lundgren A, et~al (2017) {Validating gravitational-wave
  detections: The Advanced LIGO hardware injection system}. Physical Review D
  95(6), \doi{10.1103/physrevd.95.062002}

\bibitem[{{Blaes} and {Madau}(1993)}]{bib:BlaesMadauAccretion}
{Blaes} O, {Madau} P (1993) {Can We Observe Accreting, Isolated Neutron Stars?}
  The Astrophysical Journal 403:690, \doi{10.1086/172240}

\bibitem[{Blair et~al.(1991)}]{bib:Blairbook1}
Blair DG, et~al. (1991) {The Detection of Gravitational Waves}. Cambridge
  University Press, Cambridge

\bibitem[{Blair et~al.(2012)}]{bib:Blairbook2}
Blair DG, et~al. (2012) {Advanced Gravitational Wave Detectors}. Cambridge
  University Press, Cambridge

\bibitem[{{Bogdanov} et~al.(2019{\natexlab{a}}){Bogdanov}, {Guillot}, {Ray},
  {Wolff}, {Chakrabarty}, {Ho}, {Kerr}, {Lamb}, {Lommen}, {Ludlam}, {Milburn},
  {Montano}, {Miller}, {Baub{\"o}ck}, {{\"O}zel}, {Psaltis}, {Remillard},
  {Riley}, {Steiner}, {Strohmayer}, {Watts}, {Wood}, {Zeldes}, {Enoto},
  {Okajima}, {Kellogg}, {Baker}, {Markwardt}, {Arzoumanian}, and
  {Gendreau}}]{bib:NICERI}
{Bogdanov} S, {Guillot} S, {Ray} PS, {Wolff} MT, {Chakrabarty} D, {Ho} WCG,
  {Kerr} M, {Lamb} FK, {Lommen} A, {Ludlam} RM, {Milburn} R, {Montano} S,
  {Miller} MC, {Baub{\"o}ck} M, {{\"O}zel} F, {Psaltis} D, {Remillard} RA,
  {Riley} TE, {Steiner} JF, {Strohmayer} TE, {Watts} AL, {Wood} KS, {Zeldes} J,
  {Enoto} T, {Okajima} T, {Kellogg} JW, {Baker} C, {Markwardt} CB,
  {Arzoumanian} Z, {Gendreau} KC (2019{\natexlab{a}}) {Constraining the Neutron
  Star Mass--Radius Relation and Dense Matter Equation of State with NICER. I.
  The Millisecond Pulsar X-Ray Data Set}. Astrophys J Lett 887(1):L25,
  \doi{10.3847/2041-8213/ab53eb}, \eprint{1912.05706}

\bibitem[{{Bogdanov} et~al.(2019{\natexlab{b}}){Bogdanov}, {Lamb},
  {Mahmoodifar}, {Miller}, {Morsink}, {Riley}, {Strohmayer}, {Tung}, {Watts},
  {Dittmann}, {Chakrabarty}, {Guillot}, {Arzoumanian}, and
  {Gendreau}}]{bib:NICERII}
{Bogdanov} S, {Lamb} FK, {Mahmoodifar} S, {Miller} MC, {Morsink} SM, {Riley}
  TE, {Strohmayer} TE, {Tung} AK, {Watts} AL, {Dittmann} AJ, {Chakrabarty} D,
  {Guillot} S, {Arzoumanian} Z, {Gendreau} KC (2019{\natexlab{b}})
  {Constraining the Neutron Star Mass--Radius Relation and Dense Matter
  Equation of State with NICER. II. Emission from Hot Spots on a Rapidly
  Rotating Neutron Star}. Astrophys J Lett 887(1):L26,
  \doi{10.3847/2041-8213/ab5968}, \eprint{1912.05707}

\bibitem[{{Bogdanov} et~al.(2021)}]{bib:NICEREOSIII}
{Bogdanov} S, et~al. (2021) {Constraining the Neutron Star Mass-Radius Relation
  and Dense Matter Equation of State with NICER. III. Model Description and
  Verification of Parameter Estimation Codes}. The Astrophysical Journal
  Letters 914(1):L15, \doi{10.3847/2041-8213/abfb79}, \eprint{2104.06928}

\bibitem[{Bonazzola and Gourgoulhon(1996)}]{bib:BonazzolaGourgoulhon}
Bonazzola S, Gourgoulhon E (1996) {Gravitational waves from pulsars: Emission
  by the magnetic field induced distortion}. Astron Astrophys 312:675,
  \eprint{astro-ph/9602107}

\bibitem[{Bond et~al.(2002)Bond, White, Becker, and O’Brien}]{bib:tMSPfirst1}
Bond H, White R, Becker R, O’Brien M (2002) {FIRST J102347.6+003841: The
  First Radio‐selected Cataclysmic Variable}. Publications of the
  Astronomical Society of the Pacific 114(802):1359–1363,
  \doi{10.1086/344381}

\bibitem[{Bondi and Hoyle(1944)}]{bib:bondi2}
Bondi H, Hoyle F (1944) {On the mechanism of accretion by stars}. Mon Not Roy
  Astron Soc 104:273

\bibitem[{Boztepe et~al.(2020)Boztepe, Göğüş, Güver, and
  Schwenzer}]{bib:BoztepeGogusGuverSchwenzer}
Boztepe T, Göğüş E, Güver T, Schwenzer K (2020) {Strengthening the bounds
  on the r-mode amplitude with X-ray observations of millisecond pulsars}.
  Monthly Notices of the Royal Astronomical Society 498(2):2734–2749,
  \doi{10.1093/mnras/staa2503}

\bibitem[{Brady and Creighton(2000)}]{bib:stackslide2}
Brady PR, Creighton T (2000) {Searching for periodic sources with LIGO. II.
  Hierarchical searches}. Phys Rev D 61:082001,
  \doi{10.1103/PhysRevD.61.082001}

\bibitem[{Brady et~al.(1998)Brady, Creighton, Cutler, and
  Schutz}]{bib:stackslide1}
Brady PR, Creighton T, Cutler C, Schutz BF (1998) {Searching for periodic
  sources with LIGO}. Phys Rev D 57:2101--2116, \doi{10.1103/PhysRevD.57.2101},
  \eprint{gr-qc/9702050}

\bibitem[{Brito et~al.(2017{\natexlab{a}})Brito, Ghosh, Barausse, Berti,
  Cardoso, Dvorkin, Klein, and Pani}]{bib:brito2}
Brito R, Ghosh S, Barausse E, Berti E, Cardoso V, Dvorkin I, Klein A, Pani P
  (2017{\natexlab{a}}) {Gravitational wave searches for ultralight bosons with
  LIGO and LISA}. Phys Rev D 96:064050, \doi{10.1103/PhysRevD.96.064050}

\bibitem[{Brito et~al.(2017{\natexlab{b}})Brito, Ghosh, Barausse, Berti,
  Cardoso, Dvorkin, Klein, and Pani}]{bib:brito1}
Brito R, Ghosh S, Barausse E, Berti E, Cardoso V, Dvorkin I, Klein A, Pani P
  (2017{\natexlab{b}}) {Stochastic and resolvable gravitational waves from
  ultralight bosons}. Phys Rev Lett 119(13):131101,
  \doi{10.1103/PhysRevLett.119.131101}, \eprint{1706.05097}

\bibitem[{Brito et~al.(2020)Brito, Grillo, and
  Pani}]{bib:BritoEtalTensorBosons}
Brito R, Grillo S, Pani P (2020) {Black Hole Superradiant Instability from
  Ultralight Spin-2 Fields}. Phys Rev Lett 124:211101,
  \doi{10.1103/PhysRevLett.124.211101}

\bibitem[{{Brogan} et~al.(2005){Brogan}, {Gaensler}, {Gelfand}, {Lazendic},
  {Lazio}, {Kassim}, and {McClure-Griffiths}}]{bib:BroganEtalJ1813}
{Brogan} CL, {Gaensler} BM, {Gelfand} JD, {Lazendic} JS, {Lazio} TJW, {Kassim}
  NE, {McClure-Griffiths} NM (2005) {Discovery of a Radio Supernova Remnant and
  Nonthermal X-Rays Coincident with the TeV Source HESS J1813-178}. The
  Astrophysical Journal Letters 629(2):L105--L108, \doi{10.1086/491471},
  \eprint{astro-ph/0505145}

\bibitem[{Buballa et~al.(2014)}]{bib:EOSexclusiontwoSM}
Buballa M, et~al. (2014) {EMMI rapid reaction task force meeting on quark
  matter in compact stars}. J Phys G 41(12):123001,
  \doi{10.1088/0954-3899/41/12/123001}, \eprint{1402.6911}

\bibitem[{Buschauer and Benford(1976)}]{bib:BuschauerBenfordCurvatureRadiation}
Buschauer R, Benford G (1976) {General Theory of Coherent Curvature Radiation}.
  Monthly Notices of the Royal Astronomical Society 177(1):109--136,
  \doi{10.1093/mnras/177.1.109}

\bibitem[{Caleb et~al.(2022)Caleb, Heywood, Rajwade, Malenta, Willem~Stappers,
  Barr, Chen, Morello, Sanidas, van~den Eijnden, Kramer, Buckley, Brink, Motta,
  Woudt, Weltevrede, Jankowski, Surnis, Buchner, Bezuidenhout, Driessen, and
  Fender}]{bib:LongPeriodPulsar}
Caleb M, Heywood I, Rajwade K, Malenta M, Willem~Stappers B, Barr E, Chen W,
  Morello V, Sanidas S, van~den Eijnden J, Kramer M, Buckley D, Brink J, Motta
  SE, Woudt P, Weltevrede P, Jankowski F, Surnis M, Buchner S, Bezuidenhout MC,
  Driessen LN, Fender R (2022) Discovery of a radio-emitting neutron star with
  an ultra-long spin period of 76 s. Nature Astronomy
  \doi{10.1038/s41550-022-01688-x}

\bibitem[{University~of California(2002)}]{bib:boinc}
University~of California B (2002) {The Einstein@Home project is built upon the
  BOINC (Berkeley Open Infrastructure for Network Computing) architecture
  described at http://boinc.berkeley.edu/. }

\bibitem[{{Camilo} et~al.(2021){Camilo}, {Ransom}, {Halpern}, and
  {Roshi}}]{bib:CamiloEtalJ1813}
{Camilo} F, {Ransom} SM, {Halpern} JP, {Roshi} DA (2021) {Radio Detection of
  PSR J1813-1749 in HESS J1813-178: The Most Scattered Pulsar Known}. The
  Astrophysical Journal Letters 917(2):67, \doi{10.3847/1538-4357/ac0720},
  \eprint{2106.00386}

\bibitem[{Caplan and Horowitz(2017)}]{bib:CaplanHorowitzPasta}
Caplan ME, Horowitz CJ (2017) {Colloquium: Astromaterial science and nuclear
  pasta}. Rev Mod Phys 89:041002, \doi{10.1103/RevModPhys.89.041002}

\bibitem[{Cardoso et~al.(2018)Cardoso, Dias, Hartnett, Middleton, Pani, and
  Santos}]{bib:CardosoEtal}
Cardoso V, Dias {\'{O}}J, Hartnett GS, Middleton M, Pani P, Santos JE (2018)
  {Constraining the mass of dark photons and axion-like particles through
  black-hole superradiance}. Journal of Cosmology and Astroparticle Physics
  2018(03):043--043, \doi{10.1088/1475-7516/2018/03/043}

\bibitem[{Cardoso et~al.(2020)Cardoso, Duque, and
  Ikeda}]{bib:CardosoDuqueIkeda}
Cardoso V, Duque F, Ikeda T (2020) {Tidal effects and disruption in
  superradiant clouds: a numerical investigation}. Phys Rev D 101(6):064054,
  \doi{10.1103/PhysRevD.101.064054}, \eprint{2001.01729}

\bibitem[{Caride et~al.(2019)Caride, Inta, Owen, and
  Rajbhandari}]{bib:CarideIntaOwenRajbhandari}
Caride S, Inta R, Owen BJ, Rajbhandari B (2019) {How to search for
  gravitational waves from $r$-modes of known pulsars}. Phys Rev D
  100(6):064013, \doi{10.1103/PhysRevD.100.064013}, \eprint{1907.04946}

\bibitem[{Cerda-Duran and Elias-Rosa(2018)}]{bib:CerdaduranEliasrosa}
Cerda-Duran P, Elias-Rosa N (2018) Neutron Stars Formation and Core Collapse
  Supernovae, Springer International Publishing, Cham, pp 1--56.
  \doi{10.1007/978-3-319-97616-7\_1}

\bibitem[{Chakrabarty et~al.(2003)Chakrabarty, Morgan, Muno, Galloway,
  Wijnands, van~der Klis, and Markwardt}]{bib:speedlimit}
Chakrabarty D, Morgan EH, Muno MP, Galloway DK, Wijnands R, van~der Klis M,
  Markwardt CB (2003) {Nuclear-powered millisecond pulsars and the maximum spin
  frequency of neutron stars}. Nature 424:42--44, \doi{10.1038/nature01732},
  \eprint{astro-ph/0307029}

\bibitem[{Chamel and Haensel(2008)}]{bib:lrre-ChamelHaensel}
Chamel N, Haensel P (2008) {Physics of Neutron Star Crusts}. Living Rev Rel
  11:10, \doi{10.12942/lrr-2008-10}, \eprint{0812.3955}

\bibitem[{Chandrasekhar(1970)}]{bib:cfs1}
Chandrasekhar S (1970) {Solutions of Two Problems in the Theory of
  Gravitational Radiation}. Phys Rev Lett 24:611--615,
  \doi{10.1103/PhysRevLett.24.611}

\bibitem[{{Chen} and {Ruderman}(1993)}]{bib:ChenRuderman}
{Chen} K, {Ruderman} M (1993) {Pulsar Death Lines and Death Valley}. Astrophys
  J 402:264, \doi{10.1086/172129}

\bibitem[{Chen(2020)}]{bib:Chen}
Chen WC (2020) {Constraining the ellipticity of millisecond pulsars with
  observed spin-down rates}. Physical Review D 102(4),
  \doi{10.1103/physrevd.102.043020}

\bibitem[{{Chevalier}(1989)}]{bib:ChevalierAccretion}
{Chevalier} RA (1989) {Neutron Star Accretion in a Supernova}. The
  Astrophysical Journal 346:847, \doi{10.1086/168066}

\bibitem[{Christodoulou(1970)}]{bib:penrose2}
Christodoulou D (1970) {Reversible and Irreversible Transformations in
  Black-Hole Physics}. Phys Rev Lett 25:1596--1597,
  \doi{10.1103/PhysRevLett.25.1596}

\bibitem[{Chung et~al.(2011)Chung, Melatos, Krishnan, and
  Whelan}]{bib:xcorrmethod2}
Chung C, Melatos A, Krishnan B, Whelan JT (2011) {Designing a cross-correlation
  search for continuous-wave gravitational radiation from a neutron star in the
  supernova remnant SNR 1987A}. Mon Not Roy Astron Soc 414:2650,
  \doi{10.1111/j.1365-2966.2011.18585.x}, \eprint{1102.4654}

\bibitem[{Cie{\'s}lar et~al.(2021)Cie{\'s}lar, Bulik, Curylo, Sieniawska,
  Singh, and Bejger}]{bib:CieslarEtalPopSynth}
Cie{\'s}lar M, Bulik T, Curylo M, Sieniawska M, Singh N, Bejger M (2021)
  {Detectability of continuous gravitational waves from isolated neutron stars
  in the Milky Way - The population synthesis approach}. A\&A 649:A92,
  \doi{10.1051/0004-6361/202039503}

\bibitem[{Clark et~al.(2016)}]{bib:ClarkEtal_2016}
Clark CJ, et~al. (2016) {The Braking Index of a Radio-quiet Gamma-ray Pulsar}.
  Astrophys J Lett 832(1):L15, \doi{10.3847/2041-8205/832/1/L15},
  \eprint{1611.01292}

\bibitem[{Clark et~al.(2018)}]{bib:fermipulsarsearchexample}
Clark CJ, et~al. (2018) {Einstein@Home discovers a radio-quiet gamma-ray
  millisecond pulsar}. Science Advances 4(2), \doi{10.1126/sciadv.aao7228}

\bibitem[{Contopoulos et~al.(1999)Contopoulos, Kazanas, and
  Fendt}]{bib:ContopoulosKazanasFendt}
Contopoulos I, Kazanas D, Fendt C (1999) {The axisymmetric pulsar
  magnetosphere}. Astrophys J 511:351, \doi{10.1086/306652},
  \eprint{astro-ph/9903049}

\bibitem[{{Cook} et~al.(1994){Cook}, {Shapiro}, and
  {Teukolsky}}]{bib:breakupspeed}
{Cook} GB, {Shapiro} SL, {Teukolsky} SA (1994) {Rapidly Rotating Neutron Stars
  in General Relativity: Realistic Equations of State}. Astrophys J 424:823,
  \doi{10.1086/173934}

\bibitem[{Covas and Sintes(2020)}]{bib:binaryskyhough2}
Covas P, Sintes AM (2020) {First all-sky search for continuous
  gravitational-wave signals from unknown neutron stars in binary systems using
  Advanced LIGO data}. Phys Rev Lett 124(19):191102,
  \doi{10.1103/PhysRevLett.124.191102}, \eprint{2001.08411}

\bibitem[{Covas et~al.(2018)Covas, Effler, Goetz, Meyers, Neunzert, Oliver,
  Pearlstone, Roma, Schofield, Adya, and et~al.}]{bib:linesO1O2}
Covas P, Effler A, Goetz E, Meyers P, Neunzert A, Oliver M, Pearlstone B, Roma
  V, Schofield R, Adya V, et~al (2018) {Identification and mitigation of narrow
  spectral artifacts that degrade searches for persistent gravitational waves
  in the first two observing runs of Advanced LIGO}. Physical Review D 97(8),
  \doi{10.1103/physrevd.97.082002}

\bibitem[{Covas(2020)}]{bib:CovasProperMotion}
Covas PB (2020) {Effects of proper motion of neutron stars on continuous
  gravitational-wave searches}. Monthly Notices of the Royal Astronomical
  Society 500(4):5167--5176, \doi{10.1093/mnras/staa3624}

\bibitem[{Covas and Prix(2022{\natexlab{a}})}]{bib:BinaryHoughFstat}
Covas PB, Prix R (2022{\natexlab{a}}) {Improved all-sky search method for
  continuous gravitational waves from unknown neutron stars in binary systems}.
  Phys Rev D 106(8):084035, \doi{10.1103/PhysRevD.106.084035},
  \eprint{2208.01543}

\bibitem[{Covas and Prix(2022{\natexlab{b}})}]{bib:CovasPrixModFstat}
Covas PB, Prix R (2022{\natexlab{b}}) Improved short-segment detection
  statistic for continuous gravitational waves. Phys Rev D 105:124007,
  \doi{10.1103/PhysRevD.105.124007}, \eprint{2203.08723}

\bibitem[{Covas and Sintes(2019)}]{bib:binaryskyhough1}
Covas PB, Sintes AM (2019) {New method to search for continuous gravitational
  waves from unknown neutron stars in binary systems}. Phys Rev D
  99(12):124019, \doi{10.1103/PhysRevD.99.124019}, \eprint{1904.04873}

\bibitem[{Covas et~al.(2022)Covas, Papa, Prix, and
  Owen}]{bib:cwallskybinaryO3aBinaryFstat}
Covas PB, Papa MA, Prix R, Owen BJ (2022) {Constraints on r-modes and Mountains
  on Millisecond Neutron Stars in Binary Systems}. The Astrophysical Journal
  Letters 929(2):L19, \doi{10.3847/2041-8213/ac62d7}, \eprint{2203.01773}

\bibitem[{Creighton and Anderson(2011)}]{bib:CreightonAndersontext}
Creighton JDE, Anderson WG (2011) {Gravitational-Wave Physics and Astronomy}.
  Wiley-VCH, Weinheim

\bibitem[{{Cromartie} et~al.(2020){Cromartie}, {Fonseca}, {Ransom}, {Demorest},
  {Arzoumanian}, {Blumer}, {Brook}, {DeCesar}, {Dolch}, {Ellis}, {Ferdman},
  {Ferrara}, {Garver-Daniels}, {Gentile}, {Jones}, {Lam}, {Lorimer}, {Lynch},
  {McLaughlin}, {Ng}, {Nice}, {Pennucci}, {Spiewak}, {Stairs}, {Stovall},
  {Swiggum}, and {Zhu}}]{bib:CromartieEtalJ0740}
{Cromartie} HT, {Fonseca} E, {Ransom} SM, {Demorest} PB, {Arzoumanian} Z,
  {Blumer} H, {Brook} PR, {DeCesar} ME, {Dolch} T, {Ellis} JA, {Ferdman} RD,
  {Ferrara} EC, {Garver-Daniels} N, {Gentile} PA, {Jones} ML, {Lam} MT,
  {Lorimer} DR, {Lynch} RS, {McLaughlin} MA, {Ng} C, {Nice} DJ, {Pennucci} TT,
  {Spiewak} R, {Stairs} IH, {Stovall} K, {Swiggum} JK, {Zhu} WW (2020)
  {Relativistic Shapiro delay measurements of an extremely massive millisecond
  pulsar}. Nature Astronomy 4:72--76, \doi{10.1038/s41550-019-0880-2},
  \eprint{1904.06759}

\bibitem[{Cruces et~al.(2019)Cruces, Reisenegger, and
  Tauris}]{bib:CrucesReiseneggerTauris}
Cruces M, Reisenegger A, Tauris TM (2019) {On the weak magnetic field of
  millisecond pulsars: Does it decay before accretion?} Mon Not Roy Astron Soc
  490(2):2013--2022, \doi{10.1093/mnras/stz2701}, \eprint{1906.06076}

\bibitem[{Cutler(2002)}]{bib:CutlerToroidalBFields}
Cutler C (2002) {Gravitational waves from neutron stars with large toroidal B
  fields}. Physical Review D 66(8), \doi{10.1103/physrevd.66.084025}

\bibitem[{Cutler and Schutz(2005)}]{bib:CutlerSchutzmultifstat}
Cutler C, Schutz BF (2005) {Generalized F-statistic: Multiple detectors and
  multiple gravitational wave pulsars}. Physical Review D 72(6),
  \doi{10.1103/physrevd.72.063006}

\bibitem[{{Cutler} et~al.(2005){Cutler}, {Gholami}, and
  {Krishnan}}]{bib:stackslide3}
{Cutler} C, {Gholami} I, {Krishnan} B (2005) {Improved stack-slide searches for
  gravitational-wave pulsars}. Phys Rev D 72(4):042004,
  \doi{10.1103/PhysRevD.72.042004}, \eprint{gr-qc/0505082}

\bibitem[{Dall'Osso and Stella(2021)}]{bib:DallOssoStellMillisecondMagnetar}
Dall'Osso S, Stella L (2021) {Millisecond Magnetars}. Astrophys Space Sci Libr
  465:245--280, \doi{10.1007/978-3-030-85198-9_8}, \eprint{2103.10878}

\bibitem[{Davis et~al.(2019)Davis, Massinger, Lundgren, Driggers, Urban, and
  Nuttall}]{bib:DavisEtalcleaning}
Davis D, Massinger T, Lundgren A, Driggers JC, Urban AL, Nuttall L (2019)
  {Improving the sensitivity of Advanced LIGO using noise subtraction}.
  Classical and Quantum Gravity 36(5):055011, \doi{10.1088/1361-6382/ab01c5}

\bibitem[{Davis et~al.(2021)}]{bib:DavisEtalDetchar}
Davis D, et~al. (2021) {LIGO detector characterization in the second and third
  observing runs}. Class Quant Grav 38(13):135014,
  \doi{10.1088/1361-6382/abfd85}, \eprint{2101.11673}

\bibitem[{Daw et~al.(2022)Daw, Hollows, Jones, Kennedy, Mistry, Edo, Fays, and
  Sun}]{bib:IWAVE}
Daw EJ, Hollows IJ, Jones EL, Kennedy R, Mistry T, Edo TB, Fays M, Sun L (2022)
  {IWAVE -- An adaptive filter approach to phase lock and the dynamic
  characterization of pseudo-harmonic waves}. Review of Scientific Instruments
  93(4):044502, \doi{10.1063/5.0070394}

\bibitem[{{De Luca}(2008)}]{bib:DelucaCCOs}
{De Luca} A (2008) {Central Compact Objects in Supernova Remnants}.
  \doi{10.1063/1.2900173}, \eprint{0712.2209}

\bibitem[{{Degenaar} and {Suleimanov}(2018)}]{bib:DegenaarSuleimanov}
{Degenaar} N, {Suleimanov} VF (2018) {Testing the Equation of State with
  Electromagnetic Observations}. In: {Rezzolla} L, {Pizzochero} P, {Jones} DI,
  {Rea} N, {Vida{\~n}a} I (eds) Astrophysics and Space Science Library,
  Astrophysics and Space Science Library, vol 457, p 185,
  \doi{10.1007/978-3-319-97616-7\_5}

\bibitem[{DeMarchi et~al.(2021)DeMarchi, Sanders, and
  Levesque}]{bib:TZOsearchpaper}
DeMarchi L, Sanders JR, Levesque EM (2021) {Prospects for Multimessenger
  Observations of Thorne–Żytkow Objects}. The Astrophysical Journal
  911(2):101, \doi{10.3847/1538-4357/abebe1}

\bibitem[{{Demorest} et~al.(2010){Demorest}, {Pennucci}, {Ransom}, {Roberts},
  and {Hessels}}]{bib:DemorestEtalJ1614}
{Demorest} PB, {Pennucci} T, {Ransom} SM, {Roberts} MSE, {Hessels} JWT (2010)
  {A two-solar-mass neutron star measured using Shapiro delay}. Nature
  467(7319):1081--1083, \doi{10.1038/nature09466}, \eprint{1010.5788}

\bibitem[{Deneva et~al.(2009)Deneva, Cordes, and Lazio}]{bib:DenevalEtal}
Deneva JS, Cordes JM, Lazio TJW (2009) {Discovery of Three Pulsars from a
  Galactic Center Pulsar Population}. The Astrophysical Journal
  702(2):L177--L181, \doi{10.1088/0004-637x/702/2/l177}

\bibitem[{Dergachev(2005)}]{bib:PowerFlux1}
Dergachev V (2005) {Description of PowerFlux algorithms and implementation},
  {LIGO Report LIGO-T050186}

\bibitem[{Dergachev(2010{\natexlab{a}})}]{bib:PowerFlux2}
Dergachev V (2010{\natexlab{a}}) {Description of PowerFlux 2 algorithms and
  implementation}. LIGO Report T1000272,
  \urlprefix\url{https://dcc.ligo.org/T1000272}

\bibitem[{Dergachev(2010{\natexlab{b}})}]{bib:loosecoherence}
Dergachev V (2010{\natexlab{b}}) {On blind searches for noise dominated
  signals: a loosely coherent approach}. Classical and Quantum Gravity
  27(20):205017, \doi{10.1088/0264-9381/27/20/205017}

\bibitem[{Dergachev(2012)}]{bib:LooseCoherenceWellModeledSignals}
Dergachev V (2012) Loosely coherent searches for sets of well-modeled signals.
  Phys Rev D 85:062003, \doi{10.1103/PhysRevD.85.062003}

\bibitem[{{Dergachev}(2013)}]{bib:universalstatistic}
{Dergachev} V (2013) {Novel universal statistic for computing upper limits in
  an ill-behaved background}. Phys Rev D 87(6):062001,
  \doi{10.1103/PhysRevD.87.062001}, \eprint{1208.2007}

\bibitem[{Dergachev(2018)}]{bib:LooseCoherenceMediumScale}
Dergachev V (2018) {Loosely coherent searches for medium scale coherence
  lengths}. \eprint{1807.02351}

\bibitem[{Dergachev and Papa(2019)}]{bib:FalconPaper}
Dergachev V, Papa MA (2019) {Sensitivity Improvements in the Search for
  Periodic Gravitational Waves Using O1 LIGO Data}. Phys Rev Lett 123:101101,
  \doi{10.1103/PhysRevLett.123.101101}

\bibitem[{Dergachev and Papa(2020{\natexlab{a}})}]{bib:cwallskyFalconO1}
Dergachev V, Papa MA (2020{\natexlab{a}}) {Results from an Extended Falcon
  All-Sky Survey for Continuous Gravitational Waves}. Phys Rev D 101(2):022001,
  \doi{10.1103/PhysRevD.101.022001}, \eprint{1909.09619}

\bibitem[{Dergachev and Papa(2020{\natexlab{b}})}]{bib:cwallskyFalconO2MidFreq}
Dergachev V, Papa MA (2020{\natexlab{b}}) {Results from the First All-Sky
  Search for Continuous Gravitational Waves from Small-Ellipticity Sources}.
  Physical Review Letters 125(17):171101, \doi{10.1103/physrevlett.125.171101}

\bibitem[{Dergachev and
  Papa(2021{\natexlab{a}})}]{bib:cwallskyFalconO2HighFreq}
Dergachev V, Papa MA (2021{\natexlab{a}}) {Results from high-frequency all-sky
  search for continuous gravitational waves from small-ellipticity sources}.
  Phys Rev D 103:063019, \doi{10.1103/PhysRevD.103.063019}

\bibitem[{Dergachev and Papa(2021{\natexlab{b}})}]{bib:cwallskyFalconO2LowFreq}
Dergachev V, Papa MA (2021{\natexlab{b}}) {Search for continuous gravitational
  waves from small-ellipticity sources at low frequencies}. Physical Review D
  104(4):043003, \doi{10.1103/physrevd.104.043003}

\bibitem[{Dergachev and Papa(2022)}]{bib:cwallskyFalconO3aMidFreq}
Dergachev V, Papa MA (2022) {A frequency resolved atlas of the sky in
  continuous gravitational waves}. \eprint{2202.10598}

\bibitem[{Dergachev and Riles(2005)}]{bib:PowerFluxPol}
Dergachev V, Riles K (2005) {PowerFlux Polarization Analysis}, {LIGO Report
  LIGO-T050187}

\bibitem[{Dergachev et~al.(2019)Dergachev, Papa, Steltner, and
  Eggenstein}]{bib:AEIGCTerzan5}
Dergachev V, Papa MA, Steltner B, Eggenstein HB (2019) {Loosely coherent search
  in LIGO O1 data for continuous gravitational waves from Terzan 5 and the
  Galactic Center}. Phys Rev D 99:084048, \doi{10.1103/PhysRevD.99.084048}

\bibitem[{De Lillo et~al.(2022)De Lillo, Suresh, and
  Miller}]{bib:DeLilloEtal}
De Lillo F, Suresh J, Miller AL (2022) {Stochastic gravitational-wave
  background searches and constraints on neutron-star ellipticity}. Monthly
  Notices of the Royal Astronomical Society 513(1):1105--1114,
  \doi{10.1093/mnras/stac984}, \eprint{2203.03536}

\bibitem[{Dhurandhar et~al.(2008)Dhurandhar, Krishnan, Mukhopadhyay, and
  Whelan}]{bib:xcorrmethod1}
Dhurandhar S, Krishnan B, Mukhopadhyay H, Whelan JT (2008) {Cross-correlation
  search for periodic gravitational waves}. Phys Rev D 77:082001,
  \doi{10.1103/PhysRevD.77.082001}, \eprint{0712.1578}

\bibitem[{Dhurandhar et~al.(2017)Dhurandhar, Krishnan, and
  Willis}]{bib:DhurandharKrishnanWillis}
Dhurandhar S, Krishnan B, Willis JL (2017) {Marginalizing the likelihood
  function for modeled gravitational wave searches}. \eprint{1707.08163}

\bibitem[{Dirichlet(1829)}]{bib:Dirichlet}
Dirichlet GL (1829) Sur la convergence des séries trigonométriques qui
  servent à représenter une fonction arbitraire entre des limites données. J
  f{\"u}r Math 1829:157--169, \doi{doi:10.1515/crll.1829.4.157}

\bibitem[{Doneva et~al.(2015)Doneva, Kokkotas, and Pnigouras}]{bib:donevaetal}
Doneva DD, Kokkotas KD, Pnigouras P (2015) {Gravitational wave afterglow in
  binary neutron star mergers}. Phys Rev D 92:104040,
  \doi{10.1103/PhysRevD.92.104040}

\bibitem[{Dreissigacker and Prix(2020)}]{bib:DreissigackerPrix}
Dreissigacker C, Prix R (2020) {Deep-learning continuous gravitational waves:
  Multiple detectors and realistic noise}. Physical Review D 102(2),
  \doi{10.1103/physrevd.102.022005}

\bibitem[{Dreissigacker et~al.(2018)Dreissigacker, Prix, and
  Wette}]{bib:DreissigackerPrixWette}
Dreissigacker C, Prix R, Wette K (2018) {Fast and Accurate Sensitivity
  Estimation for Continuous-Gravitational-Wave Searches}. Phys Rev D
  98(8):084058, \doi{10.1103/PhysRevD.98.084058}, \eprint{1808.02459}

\bibitem[{Dreissigacker et~al.(2019)Dreissigacker, Sharma, Messenger, Zhao, and
  Prix}]{bib:DreissigackerEtalCNN}
Dreissigacker C, Sharma R, Messenger C, Zhao R, Prix R (2019) {Deep-learning
  continuous gravitational waves}. Phys Rev D 100:044009,
  \doi{10.1103/PhysRevD.100.044009}

\bibitem[{Driggers et~al.(2019)Driggers, Vitale, Lundgren, Evans, Kawabe,
  Dwyer, Izumi, Schofield, Effler, Sigg, and et~al.}]{bib:DriggersEtalcleaning}
Driggers J, Vitale S, Lundgren A, Evans M, Kawabe K, Dwyer S, Izumi K,
  Schofield R, Effler A, Sigg D, et~al (2019) {Improving astrophysical
  parameter estimation via offline noise subtraction for Advanced LIGO}.
  Physical Review D 99(4), \doi{10.1103/physrevd.99.042001}

\bibitem[{Dunn et~al.(2021)Dunn, Clearwater, Melatos, and Wette}]{bib:FstatGPU}
Dunn L, Clearwater P, Melatos A, Wette K (2021) {Graphics processing unit
  implementation of the F-statistic for continuous gravitational wave
  searches}. Classical and Quantum Gravity \doi{10.1088/1361-6382/ac4616}

\bibitem[{Dupuis and Woan(2005)}]{bib:DupuisWoan}
Dupuis RJ, Woan G (2005) {Bayesian estimation of pulsar parameters from
  gravitational wave data}. Phys Rev D 72:102002,
  \doi{10.1103/PhysRevD.72.102002}, \eprint{gr-qc/0508096}

\bibitem[{D’Antonio et~al.(2018)D’Antonio, Palomba, Astone, Frasca, Intini,
  La~Rosa, Leaci, Mastrogiovanni, Miller, Muciaccia, and
  et~al.}]{bib:DantonioEtalBosoncloud}
D’Antonio S, Palomba C, Astone P, Frasca S, Intini G, La~Rosa I, Leaci P,
  Mastrogiovanni S, Miller A, Muciaccia F, et~al (2018) {Semicoherent analysis
  method to search for continuous gravitational waves emitted by ultralight
  boson clouds around spinning black holes}. Physical Review D 98(10),
  \doi{10.1103/physrevd.98.103017}

\bibitem[{Einstein(1916)}]{bib:EinsteinGW1}
Einstein A (1916) {N{\"a}herungsweise Integration der Feldgleichungen der
  Gravitation}. Sitzungber K Preuss Akad Wiss 1:688

\bibitem[{Einstein(1918)}]{bib:EinsteinGW2}
Einstein A (1918) {{\"U}ber Gravitationswellen}. Sitzungber K Preuss Akad Wiss
  1:154

\bibitem[{Ertan and Alpar(2021)}]{bib:ErtanAlpar}
Ertan U, Alpar MA (2021) {The minimum rotation period of millisecond pulsars}.
  Monthly Notices of the Royal Astronomical Society: Letters 505(1):L112--L114,
  \doi{10.1093/mnrasl/slab060}

\bibitem[{Espinoza et~al.(2011)Espinoza, Lyne, Kramer, Manchester, and
  Kaspi}]{bib:BfieldEmergence}
Espinoza CM, Lyne AG, Kramer M, Manchester RN, Kaspi VM (2011) {The Braking
  Index of PSR {J1734–3333} and the Magnetar Population}. The Astrophysical
  Journal 741(1):L13, \doi{10.1088/2041-8205/741/1/l13}

\bibitem[{Espinoza et~al.(2017)Espinoza, Lyne, and
  Stappers}]{bib:EspinozaEtal_2016}
Espinoza CM, Lyne AG, Stappers BW (2017) {New long-term braking index
  measurements for glitching pulsars using a glitch-template method}. Mon Not
  Roy Astron Soc 466(1):147--162, \doi{10.1093/mnras/stw3081},
  \eprint{1611.08314}

\bibitem[{{Essick} et~al.(2020){Essick}, {Landry}, and {Holz}}]{bib:EssickEtal}
{Essick} R, {Landry} P, {Holz} DE (2020) {Nonparametric inference of neutron
  star composition, equation of state, and maximum mass with GW170817}.
  Physical Review D 101(6):063007, \doi{10.1103/PhysRevD.101.063007},
  \eprint{1910.09740}

\bibitem[{Fattoyev et~al.(2018)Fattoyev, Horowitz, and
  Lu}]{bib:FattoyevHorowitzLu}
Fattoyev FJ, Horowitz CJ, Lu H (2018) {Crust breaking and the limiting
  rotational frequency of neutron stars}. \eprint{1804.04952}

\bibitem[{{Ferdman} et~al.(2015){Ferdman}, {Archibald}, and
  {Kaspi}}]{bib:FerdmanEtal_2015}
{Ferdman} RD, {Archibald} RF, {Kaspi} VM (2015) {Long-term Timing and Emission
  Behavior of the Young Crab-like Pulsar PSR B0540-69}. Astrophys J 812(2):95,
  \doi{10.1088/0004-637X/812/2/95}, \eprint{1506.00182}

\bibitem[{Ferdman et~al.(2018)Ferdman, Archibald, Gourgouliatos, and
  Kaspi}]{bib:FerdmanEtal}
Ferdman RD, Archibald RF, Gourgouliatos KN, Kaspi VM (2018) {The glitches and
  rotational history of the highly energetic young pulsar PSR J0537$-$6910}.
  Astrophys J 852(2):123, \doi{10.3847/1538-4357/aaa198}, \eprint{1708.08876}

\bibitem[{Ferrand and Safi-Harb(2012)}]{bib:FerrandSafiHarbSNRCatalog}
Ferrand G, Safi-Harb S (2012) {A census of high-energy observations of Galactic
  supernova remnants}. Advances in Space Research 49(9):1313–1319,
  \doi{10.1016/j.asr.2012.02.004}

\bibitem[{Fesen et~al.(2006)Fesen, Hammell, Morse, Chevalier, Borkowski,
  Dopita, Gerardy, Lawrence, Raymond, and van~den Bergh}]{bib:FesenEtalCasA}
Fesen RA, Hammell MC, Morse J, Chevalier RA, Borkowski KJ, Dopita MA, Gerardy
  CL, Lawrence SS, Raymond JC, van~den Bergh S (2006) {The Expansion Asymmetry
  and Age of the Cassiopeia A Supernova Remnant}. The Astrophysical Journal
  645(1):283–292, \doi{10.1086/504254}

\bibitem[{Fesik and Papa(2020)}]{bib:FesikPapa}
Fesik L, Papa MA (2020) {First Search for r-mode Gravitational Waves from {PSR}
  J0537{\textendash}6910}. The Astrophysical Journal 895(1):11,
  \doi{10.3847/1538-4357/ab8193}

\bibitem[{Fomalont et~al.(2001)Fomalont, Geldzahler, and
  Bradshaw}]{bib:FomalontEtalScoX1}
Fomalont EB, Geldzahler BJ, Bradshaw CF (2001) {Scorpius X-1: The Evolution and
  Nature of the Twin Compact Radio Lobes}. The Astrophysical Journal
  558(1):283--301, \doi{10.1086/322479}

\bibitem[{{Fonseca} et~al.(2021){Fonseca}, {Cromartie}, {Pennucci}, {Ray},
  {Kirichenko}, {Ransom}, {Demorest}, {Stairs}, {Arzoumanian}, {Guillemot},
  {Parthasarathy}, {Kerr}, {Cognard}, {Baker}, {Blumer}, {Brook}, {DeCesar},
  {Dolch}, {Dong}, {Ferrara}, {Fiore}, {Garver-Daniels}, {Good}, {Jennings},
  {Jones}, {Kaspi}, {Lam}, {Lorimer}, {Luo}, {McEwen}, {McKee}, {McLaughlin},
  {McMann}, {Meyers}, {Naidu}, {Ng}, {Nice}, {Pol}, {Radovan},
  {Shapiro-Albert}, {Tan}, {Tendulkar}, {Swiggum}, {Wahl}, and
  {Zhu}}]{bib:FonsecaEtalJ0740}
{Fonseca} E, {Cromartie} HT, {Pennucci} TT, {Ray} PS, {Kirichenko} AY, {Ransom}
  SM, {Demorest} PB, {Stairs} IH, {Arzoumanian} Z, {Guillemot} L,
  {Parthasarathy} A, {Kerr} M, {Cognard} I, {Baker} PT, {Blumer} H, {Brook} PR,
  {DeCesar} M, {Dolch} T, {Dong} FA, {Ferrara} EC, {Fiore} W, {Garver-Daniels}
  N, {Good} DC, {Jennings} R, {Jones} ML, {Kaspi} VM, {Lam} MT, {Lorimer} DR,
  {Luo} J, {McEwen} A, {McKee} JW, {McLaughlin} MA, {McMann} N, {Meyers} BW,
  {Naidu} A, {Ng} C, {Nice} DJ, {Pol} N, {Radovan} HA, {Shapiro-Albert} B,
  {Tan} CM, {Tendulkar} SP, {Swiggum} JK, {Wahl} HM, {Zhu} WW (2021) {Refined
  Mass and Geometric Measurements of the High-mass PSR J0740+6620}. The
  Astrophysical Journal Letters 915(1):L12, \doi{10.3847/2041-8213/ac03b8},
  \eprint{2104.00880}

\bibitem[{Freire(2012)}]{bib:FreireReview}
Freire PCC (2012) {The pulsar population in Globular Clusters and in the
  Galaxy}. Proceedings of the International Astronomical Union
  8(S291):243–250, \doi{10.1017/s1743921312023770}

\bibitem[{Freire et~al.(2017)}]{bib:FreireEtal}
Freire PCC, et~al. (2017) {Long-term observations of the pulsars in 47 Tucanae
  -- II. Proper motions, accelerations and jerks}. Mon Not Roy Astron Soc
  471(1):857--876, \doi{10.1093/mnras/stx1533}, \eprint{1706.04908}

\bibitem[{Freise and Strain(2010)}]{bib:lrre-FreiseStrain}
Freise A, Strain K (2010) {Interferometer Techniques for Gravitational-Wave
  Detection}. Living Rev Rel 13:1, \doi{10.12942/lrr-2010-1},
  \eprint{0909.3661}

\bibitem[{Friedman and Morsink(1998)}]{bib:rmodes3}
Friedman JL, Morsink SM (1998) {Axial instability of rotating relativistic
  stars}. Astrophys J 502:714--720, \doi{10.1086/305920},
  \eprint{gr-qc/9706073}

\bibitem[{{Friedman} and {Schutz}(1978)}]{bib:cfs2}
{Friedman} JL, {Schutz} BF (1978) {Secular instability of rotating Newtonian
  stars.} Astrophys J 222:281--296, \doi{10.1086/156143}

\bibitem[{{Fruchter} et~al.(1988){Fruchter}, {Stinebring}, and
  {Taylor}}]{bib:blackwidowfirst}
{Fruchter} AS, {Stinebring} DR, {Taylor} JH (1988) {A millisecond pulsar in an
  eclipsing binary}. Nature 333(6170):237--239, \doi{10.1038/333237a0}

\bibitem[{Gaensler and Slane(2006)}]{bib:GaenslerSlane}
Gaensler BM, Slane PO (2006) {The evolution and structure of pulsar wind
  nebulae}. Ann Rev Astron Astrophys 44:17--47,
  \doi{10.1146/annurev.astro.44.051905.092528}, \eprint{astro-ph/0601081}

\bibitem[{Galaudage et~al.(2021)Galaudage, Wette, Galloway, and
  Messenger}]{bib:GalaudageEtal}
Galaudage S, Wette K, Galloway DK, Messenger C (2021) {Deep searches for X-ray
  pulsations from Scorpius X-1 and Cygnus X-2 in support of continuous
  gravitational wave searches}. Monthly Notices of the Royal Astronomical
  Society 509(2):1745--1754, \doi{10.1093/mnras/stab3095}

\bibitem[{{Galloway} et~al.(2008){Galloway}, {Muno}, {Hartman}, {Psaltis}, and
  {Chakrabarty}}]{bib:GallowayCygX2Flux}
{Galloway} DK, {Muno} MP, {Hartman} JM, {Psaltis} D, {Chakrabarty} D (2008)
  {Thermonuclear (Type I) X-Ray Bursts Observed by the Rossi X-Ray Timing
  Explorer}. Astrophys J Suppl 179(2):360--422, \doi{10.1086/592044},
  \eprint{astro-ph/0608259}

\bibitem[{Gao et~al.(2020)Gao, Shao, Xu, Sun, Liu, and Xu}]{bib:GaoEtal}
Gao Y, Shao L, Xu R, Sun L, Liu C, Xu RX (2020) {Triaxially deformed freely
  precessing neutron stars: continuous electromagnetic and gravitational
  radiation}. Monthly Notices of the Royal Astronomical Society
  498(2):1826–1838, \doi{10.1093/mnras/staa2476}

\bibitem[{{Geppert} et~al.(1999){Geppert}, {Page}, and
  {Zannias}}]{bib:GeppertEtalAccretion}
{Geppert} U, {Page} D, {Zannias} T (1999) {Submergence and re-diffusion of the
  neutron star magnetic field after the supernova}. Astron Astrophys
  345:847--854

\bibitem[{{Ghosh} and {Lamb}(1979)}]{bib:ghoshlamb}
{Ghosh} P, {Lamb} FK (1979) {Accretion by rotating magnetic neutron stars. III.
  Accretion torques and period changes in pulsating X-ray sources.} Astrophys J
  234:296--316, \doi{10.1086/157498}

\bibitem[{Ghosh et~al.(2019)Ghosh, Berti, Brito, and
  Richartz}]{bib:GhoshEtalSuperradiance}
Ghosh S, Berti E, Brito R, Richartz M (2019) {Follow-up signals from
  superradiant instabilities of black hole merger remnants}. Phys Rev D
  99(10):104030, \doi{10.1103/PhysRevD.99.104030}, \eprint{1812.01620}

\bibitem[{Giliberti and Cambiotti(2022)}]{bib:GilibertiCambiotti}
Giliberti E, Cambiotti G (2022) {Starquakes in millisecond pulsars and
  gravitational waves emission}. Monthly Notices of the Royal Astronomical
  Society 511(3):3365--3376, \doi{10.1093/mnras/stac245}

\bibitem[{Gittins et~al.(2020)Gittins, Andersson, and
  Jones}]{bib:GittinsAnderssonJones}
Gittins F, Andersson N, Jones DI (2020) {Modelling neutron star mountains}. Mon
  Not Roy Astron Soc 500(4):5570--5582, \doi{10.1093/mnras/staa3635},
  \eprint{2009.12794}

\bibitem[{Glampedakis and Gualtieri(2018)}]{bib:GandG}
Glampedakis K, Gualtieri L (2018) {Gravitational waves from single neutron
  stars: an advanced detector era survey}, vol 457, Astrophysics and Space
  Science Library, vol 457. Springer, pp 673--736.
  \doi{10.1007/978-3-319-97616-7\_12}, \eprint{1709.07049}

\bibitem[{Goetz and Riles(2011)}]{bib:twospectmethod}
Goetz E, Riles K (2011) {An all-sky search algorithm for continuous
  gravitational waves from spinning neutron stars in binary systems}. Classical
  and Quantum Gravity 28(21):215006, \doi{10.1088/0264-9381/28/21/215006}

\bibitem[{Goetz and Riles(2016)}]{bib:GoetzRilessftsumming}
Goetz E, Riles K (2016) {Coherently combining data between detectors for
  all-sky semi-coherent continuous gravitational wave searches}. Classical and
  Quantum Gravity 33(8):085007, \doi{10.1088/0264-9381/33/8/085007}

\bibitem[{Goetz et~al.(2021)}]{bib:GoetzEtalO3linelists}
Goetz E, et~al. (2021) {Subtracting Narrow-band Noise from LIGO Strain Data in
  the Third Observing Run}. {LIGO Report T2100200},
  \urlprefix\url{https://dcc.ligo.org/T2100200}

\bibitem[{{Gold}(1968)}]{bib:Gold}
{Gold} T (1968) {Rotating Neutron Stars as the Origin of the Pulsating Radio
  Sources}. Nature 218(5143):731--732, \doi{10.1038/218731a0}

\bibitem[{{Goldreich} and {Julian}(1969)}]{bib:GoldreichJulian}
{Goldreich} P, {Julian} WH (1969) {Pulsar Electrodynamics}. Astrophys J
  157:869, \doi{10.1086/150119}

\bibitem[{Goncharov and Thrane(2018)}]{bib:stochfolding2}
Goncharov B, Thrane E (2018) {All-sky radiometer for narrowband gravitational
  waves using folded data}. Phys Rev D 98:064018,
  \doi{10.1103/PhysRevD.98.064018}

\bibitem[{Gotthelf and Halpern(2009)}]{bib:GotthelfHalpern}
Gotthelf EV, Halpern JP (2009) {Discovery of a Highly Energetic X-ray Pulsar
  Powering HESS J1813-178 in the Young Supernova Remnant G12.82–0.02}. The
  Astrophysical Journal 700(2):L158–L161, \doi{10.1088/0004-637x/700/2/l158}

\bibitem[{{Gottlieb} et~al.(1975){Gottlieb}, {Wright}, and
  {Liller}}]{bib:scox1period}
{Gottlieb} EW, {Wright} EL, {Liller} W (1975) {Optical studies of UHURU
  sources. XI. A probable period for Scorpius X-1 = V818 Scorpii.} Astrophys J
  Lett 195:L33--L35, \doi{10.1086/181703}

\bibitem[{Greco et~al.(2022)}]{bib:GrecoEtalSN1987A}
Greco E, et~al. (2022) {Additional Evidence for a Pulsar Wind Nebula in the
  Heart of SN 1987A from Multiepoch X-Ray Data and MHD Modeling}. Astrophys J
  931(2):132, \doi{10.3847/1538-4357/ac679d}, \eprint{2204.06804}

\bibitem[{Green(2014)}]{bib:GreenSNRCatalog}
Green DA (2014) {A catalogue of 294 Galactic supernova remnants}. Bull Astron
  Soc India 42:47, \eprint{1409.0637}

\bibitem[{Grote and Stadnik(2019)}]{bib:GroteStadnik}
Grote H, Stadnik YV (2019) {Novel signatures of dark matter in
  laser-interferometric gravitational-wave detectors}. Phys Rev Res
  1(3):033187, \doi{10.1103/PhysRevResearch.1.033187}, \eprint{1906.06193}

\bibitem[{Guilet and Müller(2015)}]{bib:MagnetarDynamo1}
Guilet J, Müller E (2015) {Numerical simulations of the magnetorotational
  instability in protoneutron stars – I. Influence of buoyancy}. Monthly
  Notices of the Royal Astronomical Society 450(2):2153--2171,
  \doi{10.1093/mnras/stv727}

\bibitem[{Guo et~al.(2019)Guo, Riles, Yang, and Zhao}]{bib:DPDM_GRYZ}
Guo HK, Riles K, Yang FW, Zhao Y (2019) {Searching for dark photon dark matter
  in LIGO O1 data}. Communications Physics 2(1),
  \doi{10.1038/s42005-019-0255-0}

\bibitem[{Gusakov et~al.(2014)Gusakov, Chugunov, and
  Kantor}]{bib:GusakovChugunovKantor}
Gusakov ME, Chugunov AI, Kantor EM (2014) {Instability windows and evolution of
  rapidly rotating neutron stars}. Phys Rev Lett 112(15):151101,
  \doi{10.1103/PhysRevLett.112.151101}, \eprint{1310.8103}

\bibitem[{Gáspár and Rieke(2020)}]{bib:GasparRiekeFomalhautb}
Gáspár A, Rieke GH (2020) {New HST data and modeling reveal a massive
  planetesimal collision around Fomalhaut}. Proceedings of the National Academy
  of Sciences 117(18):9712–9722, \doi{10.1073/pnas.1912506117}

\bibitem[{Halpern and Gotthelf(2009)}]{bib:HalpernGotthelfCasA}
Halpern JP, Gotthelf EV (2009) {Spin-down Measurement of PSR J1852+0040 in
  Kesteven 79: Central Compact Objects as Anti-magnetars}. The Astrophysical
  Journal 709(1):436–446, \doi{10.1088/0004-637x/709/1/436}

\bibitem[{{Harding} et~al.(1999){Harding}, {Contopoulos}, and
  {Kazanas}}]{bib:HardingContopoulosKazanas}
{Harding} AK, {Contopoulos} I, {Kazanas} D (1999) {Magnetar Spin-Down}.
  Astrophys Jl 525(2):L125--L128, \doi{10.1086/312339},
  \eprint{astro-ph/9908279}

\bibitem[{Haskell and Patruno(2011)}]{bib:HaskellPatruno}
Haskell B, Patruno A (2011) {Spin equilibrium with or without gravitational
  wave emission: the case of XTE J1814-338 and SAX J1808.4-3658}. Astrophys J
  Lett 738:L14, \doi{10.1088/2041-8205/738/1/L14}, \eprint{1106.6264}

\bibitem[{{Haskell} and {Patruno}(2017)}]{bib:HaskellPatrunoJ1023}
{Haskell} B, {Patruno} A (2017) {Are Gravitational Waves Spinning Down PSR
  J1023$+$0038 ?} Phys Rev Lett 119(16):161103,
  \doi{10.1103/PhysRevLett.119.161103}, \eprint{1703.08374}

\bibitem[{Haskell et~al.(2015)Haskell, Priymak, Patruno, Oppenoorth, Melatos,
  and Lasky}]{bib:HaskellEtal}
Haskell B, Priymak M, Patruno A, Oppenoorth M, Melatos A, Lasky PD (2015)
  {Detecting gravitational waves from mountains on neutron stars in the
  Advanced Detector Era}. Mon Not Roy Astron Soc 450(3):2393--2403,
  \doi{10.1093/mnras/stv726}, \eprint{1501.06039}

\bibitem[{Haskell et~al.(2022)Haskell, Antonelli, and
  Pizzochero}]{bib:HaskellEtalPinnedSuperfluid}
Haskell B, Antonelli M, Pizzochero P (2022) {Continuous Gravitational Wave
  Emissions from Neutron Stars with Pinned Superfluids in the Core}. Universe
  8(12):619, \doi{10.3390/universe8120619}, \eprint{2211.15507}

\bibitem[{{Hessels} et~al.(2006){Hessels}, {Ransom}, {Stairs}, {Freire},
  {Kaspi}, and {Camilo}}]{bib:HesselsEtal}
{Hessels} JWT, {Ransom} SM, {Stairs} IH, {Freire} PCC, {Kaspi} VM, {Camilo} F
  (2006) {A Radio Pulsar Spinning at 716 Hz}. Science 311(5769):1901--1904,
  \doi{10.1126/science.1123430}, \eprint{astro-ph/0601337}

\bibitem[{Hewish et~al.(1968)Hewish, Bell, Pilkington, Scott, and
  Collins}]{bib:Hewishetal}
Hewish AR, Bell SJ, Pilkington J, Scott P, Collins R (1968) {Observation of a
  Rapidly Pulsating Radio Source}. Nature 217, \doi{10.1038/217709a0}

\bibitem[{{Heyl}(2002)}]{bib:HeylLMXB}
{Heyl} JS (2002) {Low-Mass X-Ray Binaries May Be Important Laser Interferometer
  Gravitational-Wave Observatory Sources After All}. The Astrophysical Journal
  Letters 574(1):L57--L60, \doi{10.1086/342263}

\bibitem[{Hirakawa et~al.(1978)Hirakawa, Tsubono, and
  Fujimoto}]{bib:EarlyBarLimits}
Hirakawa H, Tsubono K, Fujimoto MK (1978) Search for gravitational radiation
  from the crab pulsar. Phys Rev D 17:1919--1923,
  \doi{10.1103/PhysRevD.17.1919}

\bibitem[{{Ho}(2011)}]{bib:HoAccretion}
{Ho} WCG (2011) {Evolution of a buried magnetic field in the central compact
  object neutron stars}. Mon Not Roy Astron Soc 414(3):2567--2575,
  \doi{10.1111/j.1365-2966.2011.18576.x}, \eprint{1102.4870}

\bibitem[{Ho(2015)}]{bib:emergingBfield}
Ho WCG (2015) {Magnetic field growth in young glitching pulsars with a braking
  index}. Mon Not Roy Astron Soc 452(1):845--851, \doi{10.1093/mnras/stv1339},
  \eprint{1506.03933}

\bibitem[{{Ho}(2016)}]{bib:HoGRB}
{Ho} WCG (2016) {Gravitational waves within the magnetar model of superluminous
  supernovae and gamma-ray bursts}. Mon Not Roy Astron Soc 463(1):489--494,
  \doi{10.1093/mnras/stw2016}, \eprint{1606.00454}

\bibitem[{{Ho} and {Andersson}(2012)}]{bib:HoAnderssonSupercooling}
{Ho} WCG, {Andersson} N (2012) {Rotational evolution of young pulsars due to
  superfluid decoupling}. Nature Physics 8(11):787--789,
  \doi{10.1038/nphys2424}, \eprint{1208.3201}

\bibitem[{{Ho} et~al.(2019){Ho}, {Heinke}, and {Chugunov}}]{bib:HoEtalJ0952}
{Ho} WCG, {Heinke} CO, {Chugunov} AI (2019) {XMM-Newton Detection and Spectrum
  of the Second Fastest Spinning Pulsar PSR J0952-0607}. The Astrophysical
  Journal 882(2):128, \doi{10.3847/1538-4357/ab3578}, \eprint{1905.12001}

\bibitem[{{Ho} et~al.(2020){Ho}, {Espinoza}, {Arzoumanian}, {Enoto}, {Tamba},
  {Antonopoulou}, {Bejger}, {Guillot}, {Haskell}, and {Ray}}]{bib:HoEtalJ0537}
{Ho} WCG, {Espinoza} CM, {Arzoumanian} Z, {Enoto} T, {Tamba} T, {Antonopoulou}
  D, {Bejger} M, {Guillot} S, {Haskell} B, {Ray} PS (2020) {Return of the Big
  Glitcher: NICER timing and glitches of PSR J0537-6910}. Mon Not Roy Astron
  Soc 498(4):4605--4614, \doi{10.1093/mnras/staa2640}, \eprint{2009.00030}

\bibitem[{Ho et~al.(2021)Ho, Zhao, Heinke, Kaplan, Shternin, and
  Wijngaarden}]{bib:HoEtalCasA}
Ho WCG, Zhao Y, Heinke CO, Kaplan DL, Shternin PS, Wijngaarden MJP (2021)
  {X-ray bounds on cooling, composition, and magnetic field of the
  Cassiopeia A neutron star and young central compact objects}. Monthly
  Notices of the Royal Astronomical Society 506(4):5015–5029,
  \doi{10.1093/mnras/stab2081}

\bibitem[{{Hobbs} et~al.(2006){Hobbs}, {Edwards}, and {Manchester}}]{bib:tempo}
{Hobbs} GB, {Edwards} RT, {Manchester} RN (2006) {TEMPO2, a new pulsar-timing
  package - I. An overview}. Monthly Notices of the Royal Astronomical Society
  369(2):655--672, \doi{10.1111/j.1365-2966.2006.10302.x},
  \eprint{astro-ph/0603381}

\bibitem[{Horowitz and Kadau(2009)}]{bib:HorowitzKadau}
Horowitz CJ, Kadau K (2009) {Breaking Strain of Neutron Star Crust and
  Gravitational Waves}. Phys Rev Lett 102:191102,
  \doi{10.1103/PhysRevLett.102.191102}

\bibitem[{Horowitz et~al.(2020)Horowitz, Papa, and
  Reddy}]{bib:HorowitzPapaReddy}
Horowitz CJ, Papa M, Reddy S (2020) {Search for compact dark matter objects in
  the solar system with LIGO data}. Physics Letters B 800:135072,
  \doi{10.1016/j.physletb.2019.135072}

\bibitem[{Hough(1959)}]{bib:houghibm1}
Hough PVC (1959) {Machine analysis of bubble chamber pictures}. In: Conf.
  Proc., vol 590914, pp 554--558

\bibitem[{Hough(1962)}]{bib:houghibm2}
Hough PVC (1962) {Method and Means for Recognizing Complex Patterns}. U.S.
  Patent 3,069,654

\bibitem[{{Hoyle} and {Lyttleton}(1939)}]{bib:bondi1}
{Hoyle} F, {Lyttleton} RA (1939) {The effect of interstellar matter on climatic
  variation}. Proceedings of the Cambridge Philosophical Society 35(3):405,
  \doi{10.1017/S0305004100021150}

\bibitem[{{Hughes}(1980)}]{bib:casabirth}
{Hughes} DW (1980) {Did Flamsteed see the Cassiopeia A supernova?} Nature
  285(5761):132--133, \doi{10.1038/285132a0}

\bibitem[{Hutchins and Jones(2022)}]{bib:HutchinsJones}
Hutchins TJ, Jones DI (2022) {Gravitational radiation from thermal mountains on
  accreting neutron stars: sources of temperature non-axisymmetry}.
  \eprint{2212.07452}

\bibitem[{Huth et~al.(2022)Huth, Pang, Tews, Dietrich, Le~F{\`e}vre, Schwenk,
  Trautmann, Agarwal, Bulla, Coughlin, and Van
  Den~Broeck}]{bib:CombinedEOSAnalysis}
Huth S, Pang PTH, Tews I, Dietrich T, Le~F{\`e}vre A, Schwenk A, Trautmann W,
  Agarwal K, Bulla M, Coughlin MW, Van Den~Broeck C (2022) Constraining
  neutron-star matter with microscopic and macroscopic collisions. Nature
  606(7913):276--280, \doi{10.1038/s41586-022-04750-w}

\bibitem[{Idrisy et~al.(2015)Idrisy, Owen, and Jones}]{bib:IdrisyOwenJones}
Idrisy A, Owen BJ, Jones DI (2015) {R-mode frequencies of slowly rotating
  relativistic neutron stars with realistic equations of state}. Physical
  Review D 91(2), \doi{10.1103/physrevd.91.024001}

\bibitem[{Intini et~al.(2020{\natexlab{a}})Intini, Leaci, Astone, D'Antonio,
  Frasca, La~Rosa, Miller, Palomba, and Piccinni}]{bib:IntiniEtalDopplerVeto}
Intini G, Leaci P, Astone P, D'Antonio S, Frasca S, La~Rosa I, Miller A,
  Palomba C, Piccinni O (2020{\natexlab{a}}) {A Doppler-modulation based veto
  to discard false continuous gravitational-wave candidates}. Class Quant Grav
  37(22):225007, \doi{10.1088/1361-6382/abac43}

\bibitem[{Intini et~al.(2020{\natexlab{b}})Intini, Leaci, Astone, D'Antonio,
  Frasca, La~Rosa, Miller, Palomba, and Piccinni}]{bib:IntiniEtalDoppler}
Intini G, Leaci P, Astone P, D'Antonio S, Frasca S, La~Rosa I, Miller A,
  Palomba C, Piccinni O (2020{\natexlab{b}}) {A Doppler-modulation based veto
  to discard false continuous gravitational-wave candidates}. Class Quant Grav
  37(22):225007, \doi{10.1088/1361-6382/abac43}

\bibitem[{Isi et~al.(2015)Isi, Weinstein, Mead, and Pitkin}]{bib:TGRmethod}
Isi M, Weinstein AJ, Mead C, Pitkin M (2015) {Detecting beyond-Einstein
  polarizations of continuous gravitational waves}. Phys Rev D 91:082002,
  \doi{10.1103/PhysRevD.91.082002}

\bibitem[{Isi et~al.(2019)Isi, Sun, Brito, and Melatos}]{bib:IsiEtalBosons}
Isi M, Sun L, Brito R, Melatos A (2019) {Directed searches for gravitational
  waves from ultralight bosons}. Phys Rev D 99:084042,
  \doi{10.1103/PhysRevD.99.084042}

\bibitem[{Isi et~al.(2020)Isi, Mastrogiovanni, Pitkin, and
  Piccinni}]{bib:IsiEtalSkyShift}
Isi M, Mastrogiovanni S, Pitkin M, Piccinni OJ (2020) {Establishing the
  significance of continuous gravitational-wave detections from known pulsars}.
  Phys Rev D 102(12):123027, \doi{10.1103/PhysRevD.102.123027},
  \eprint{2010.12612}

\bibitem[{{Iyudin} et~al.(1998){Iyudin}, {Sch{\"o}nfelder}, {Bennett},
  {Bloemen}, {Diehl}, {Hermsen}, {Lichti}, {van der Meulen}, {Ryan}, and
  {Winkler}}]{bib:IyudinEtalVelaJr}
{Iyudin} AF, {Sch{\"o}nfelder} V, {Bennett} K, {Bloemen} H, {Diehl} R,
  {Hermsen} W, {Lichti} GG, {van der Meulen} RD, {Ryan} J, {Winkler} C (1998)
  {Emission from $^{44}$Ti associated with a previously unknown Galactic
  supernova}. Nature 396(6707):142--144, \doi{10.1038/24106}

\bibitem[{Jaodand et~al.(2017)Jaodand, Hessels, Archibald, and
  Weltevrede}]{bib:tMSPsystems}
Jaodand A, Hessels JWT, Archibald AM, Weltevrede Pea (2017) {Pulsar
  Astrophysics - The Next 50 Years, IAU Symposium}, vol 337. Cambridge
  University Press

\bibitem[{Jaranowski and Kr\'olak(1999)}]{bib:JK1999}
Jaranowski P, Kr\'olak A (1999) Data analysis of gravitational-wave signals
  from spinning neutron stars. ii. accuracy of estimation of parameters. Phys
  Rev D 59:063003, \doi{10.1103/PhysRevD.59.063003}

\bibitem[{Jaranowski and Kr{\'{o}}lak(2009)}]{bib:JaranowskiKrolaktext}
Jaranowski P, Kr{\'{o}}lak A (2009) {Analysis of Gravitational-Wave Data}.
  Cambridge University Press, Cambridge

\bibitem[{Jaranowski and Kr{\'{o}}lak(2010)}]{bib:gstatisticmethod}
Jaranowski P, Kr{\'{o}}lak A (2010) {Searching for gravitational waves from
  known pulsars using the {$F$} and {$G$} statistics}. Classical and Quantum
  Gravity 27(19):194015, \doi{10.1088/0264-9381/27/19/194015}

\bibitem[{Jaranowski et~al.(1998)Jaranowski, Kr{\'{o}}lak, and
  Schutz}]{bib:JKS}
Jaranowski P, Kr{\'{o}}lak A, Schutz BF (1998) {Data analysis of gravitational
  - wave signals from spinning neutron stars. 1. The Signal and its detection}.
  Phys Rev D 58:063001, \doi{10.1103/PhysRevD.58.063001},
  \eprint{gr-qc/9804014}

\bibitem[{Jasiulek and Chirenti(2017)}]{bib:JasiulekChirenti}
Jasiulek M, Chirenti C (2017) {R-mode frequencies of rapidly and differentially
  rotating relativistic neutron stars}. Phys Rev D 95(6):064060,
  \doi{10.1103/PhysRevD.95.064060}, \eprint{1611.07924}

\bibitem[{Johnson-McDaniel and Owen(2013)}]{bib:JohnsonMcdanielOwen}
Johnson-McDaniel NK, Owen BJ (2013) {Maximum elastic deformations of
  relativistic stars}. Phys Rev D 88:044004, \doi{10.1103/PhysRevD.88.044004},
  \eprint{1208.5227}

\bibitem[{Johnston and Karastergiou(2017)}]{bib:decayinginclination}
Johnston S, Karastergiou A (2017) {Pulsar braking and the P--$\dot{P}$
  diagram}. Mon Not Roy Astron Soc 467(3):3493--3499,
  \doi{10.1093/mnras/stx377}, \eprint{1702.03616}

\bibitem[{Jones and Sun(2021)}]{bib:ViterbiFomalontbO2}
Jones D, Sun L (2021) {Search for continuous gravitational waves from Fomalhaut
  b in the second Advanced LIGO observing run with a hidden Markov model}. Phys
  Rev D 103:023020, \doi{10.1103/PhysRevD.103.023020}

\bibitem[{Jones et~al.(2022)Jones, Sun, Carlin, Dunn, Millhouse, Middleton,
  Meyers, Clearwater, Beniwal, Strang, Vargas, and
  Melatos}]{bib:JonesEtalDoppler}
Jones D, Sun L, Carlin J, Dunn L, Millhouse M, Middleton H, Meyers P,
  Clearwater P, Beniwal D, Strang L, Vargas A, Melatos A (2022) Validating
  continuous gravitational-wave candidates from a semicoherent search using
  doppler modulation and an effective point spread function. Phys Rev D
  106:123011, \doi{10.1103/PhysRevD.106.123011}, \eprint{2203.14468}

\bibitem[{{Jones}(2010)}]{bib:JonesSuperfluid}
{Jones} DI (2010) {Gravitational wave emission from rotating superfluid neutron
  stars}. Mon Not Roy Astron Soc 402(4):2503--2519,
  \doi{10.1111/j.1365-2966.2009.16059.x}, \eprint{0909.4035}

\bibitem[{Jones(2015)}]{bib:JonesParametrization}
Jones DI (2015) {Parameter choices and ranges for continuous gravitational wave
  searches for steadily spinning neutron stars}. Monthly Notices of the Royal
  Astronomical Society 453(1):53--66, \doi{10.1093/mnras/stv1584}

\bibitem[{Jones(2021)}]{bib:JonesFrequency}
Jones DI (2021) {Learning from the Frequency Content of Continuous
  Gravitational Wave Signals}. To appear in "Astrophysics In The XXI Century
  With Compact Stars", World Scientific, eds. Cesar Zen Vasconcellos and
  Fridolin Weber, \eprint{2111.08561}

\bibitem[{Jones and Andersson(2001)}]{bib:JonesAnderssonPrecession1}
Jones DI, Andersson N (2001) {Freely precessing neutron stars: model and
  observations}. Monthly Notices of the Royal Astronomical Society
  324(4):811–824, \doi{10.1046/j.1365-8711.2001.04251.x}

\bibitem[{Jones and Andersson(2002)}]{bib:JonesAnderssonPrecession2}
Jones DI, Andersson N (2002) {Gravitational waves from freely precessing
  neutron stars}. Monthly Notices of the Royal Astronomical Society
  331(1):203–220, \doi{10.1046/j.1365-8711.2002.05180.x}

\bibitem[{Jones et~al.(2005)Jones, Owen, and Whitbeck}]{bib:ptolemymetric}
Jones DI, Owen B, Whitbeck D (2005) {Parameter space metric for combined
  diurnal and orbital motion}. LIGO-T0900500-v1

\bibitem[{Sancho de~la Jordana(2010)}]{bib:SkyHoughHierarchical}
Sancho de~la Jordana L (2010) {Hierarchical Hough all-sky search for periodic
  gravitational waves in LIGO S5 data}. J Phys Conf Ser 228:012004,
  \doi{10.1088/1742-6596/228/1/012004}, \eprint{1001.3754}

\bibitem[{Sancho de~la Jordana and Sintes(2008)}]{bib:SanchodelaJordanaSintes}
Sancho de~la Jordana L, Sintes AM (2008) {A chi**2 veto for continuous wave
  searches}. Class Quant Grav 25:184014, \doi{10.1088/0264-9381/25/18/184014},
  \eprint{0804.1007}

\bibitem[{Jordana-Mitjans et~al.(2022)Jordana-Mitjans, Mundell, Guidorzi,
  Smith, Ramírez-Ruiz, Metzger, Kobayashi, Gomboc, Steele, Shrestha, Marongiu,
  Rossi, and Rothberg}]{bib:Jordana-MitjansEtalGRBRemnant}
Jordana-Mitjans N, Mundell CG, Guidorzi C, Smith RJ, Ramírez-Ruiz E, Metzger
  BD, Kobayashi S, Gomboc A, Steele IA, Shrestha M, Marongiu M, Rossi A,
  Rothberg B (2022) A short gamma-ray burst from a protomagnetar remnant. The
  Astrophysical Journal 939(2):106, \doi{10.3847/1538-4357/ac972b}

\bibitem[{Kalas et~al.(2008)Kalas, Graham, Chiang, Fitzgerald, Clampin, Kite,
  Stapelfeldt, Marois, and Krist}]{bib:KalasEtal}
Kalas P, Graham JR, Chiang E, Fitzgerald MP, Clampin M, Kite ES, Stapelfeldt K,
  Marois C, Krist J (2008) {Optical Images of an Exosolar Planet 25 Light-Years
  from Earth}. Science 322(5906):1345--1348, \doi{10.1126/science.1166609}

\bibitem[{Kantor et~al.(2020)Kantor, Gusakov, and
  Dommes}]{bib:KantorGusakovDommes}
Kantor EM, Gusakov ME, Dommes VA (2020) {Constraining neutron superfluidity
  with $r$-mode physics}. Phys Rev Lett 125(15):151101,
  \doi{10.1103/PhysRevLett.125.151101}, \eprint{2009.12553}

\bibitem[{{Kargaltsev} et~al.(2002){Kargaltsev}, {Pavlov}, {Sanwal}, and
  {Garmire}}]{bib:KargaltsevEtalVelaJr}
{Kargaltsev} O, {Pavlov} GG, {Sanwal} D, {Garmire} GP (2002) {The Compact
  Central Object in the Supernova Remnant G266.2-1.2}. The Astrophysical
  Journal 580(2):1060--1064, \doi{10.1086/343852}, \eprint{astro-ph/0207602}

\bibitem[{{Kashiyama} et~al.(2016){Kashiyama}, {Murase}, {Bartos}, {Kiuchi},
  and {Margutti}}]{bib:KashiyamaEtalGRB}
{Kashiyama} K, {Murase} K, {Bartos} I, {Kiuchi} K, {Margutti} R (2016)
  {Multi-messenger Tests for Fast-spinning Newborn Pulsars Embedded in
  Stripped-envelope Supernovae}. The Astrophysical Journal 818(1):94,
  \doi{10.3847/0004-637X/818/1/94}, \eprint{1508.04393}

\bibitem[{Keitel(2016)}]{bib:lineveto3}
Keitel D (2016) {Robust semicoherent searches for continuous gravitational
  waves with noise and signal models including hours to days long transients}.
  Phys Rev D 93:084024, \doi{10.1103/PhysRevD.93.084024}

\bibitem[{Keitel and Ashton(2018)}]{bib:TransientFstatGPU}
Keitel D, Ashton G (2018) {Faster search for long gravitational-wave
  transients: GPU implementation of the transient $\mathcal F$-statistic}.
  Class Quant Grav 35(20):205003, \doi{10.1088/1361-6382/aade34},
  \eprint{1805.05652}

\bibitem[{Keitel and Prix(2015)}]{bib:lineveto2}
Keitel D, Prix R (2015) {Line-robust statistics for continuous gravitational
  waves: safety in the case of unequal detector sensitivities}. Classical and
  Quantum Gravity 32(3):035004, \doi{10.1088/0264-9381/32/3/035004}

\bibitem[{Keitel et~al.(2014)Keitel, Prix, Papa, Leaci, and
  Siddiqi}]{bib:lineveto1}
Keitel D, Prix R, Papa MA, Leaci P, Siddiqi M (2014) {Search for continuous
  gravitational waves: Improving robustness versus instrumental artifacts}.
  Phys Rev D 89:064023, \doi{10.1103/PhysRevD.89.064023}

\bibitem[{Keitel et~al.(2019)Keitel, Woan, Pitkin, Schumacher, Pearlstone,
  Riles, Lyne, Palfreyman, Stappers, and Weltevrede}]{bib:KeitelEtal}
Keitel D, Woan G, Pitkin M, Schumacher C, Pearlstone B, Riles K, Lyne AG,
  Palfreyman J, Stappers B, Weltevrede P (2019) {First search for long-duration
  transient gravitational waves after glitches in the Vela and Crab pulsars}.
  Phys Rev D 100(6):064058, \doi{10.1103/PhysRevD.100.064058},
  \eprint{1907.04717}

\bibitem[{Keitel et~al.(2021)Keitel, Tenorio, Ashton, and Prix}]{bib:pyFstat}
Keitel D, Tenorio R, Ashton G, Prix R (2021) {PyFstat: a Python package for
  continuous gravitational-wave data analysis}. Journal of Open Source Software
  6(60):3000, \doi{10.21105/joss.03000}

\bibitem[{Kerin and Melatos(2022)}]{bib:KerinMelatosCrustalFailure}
Kerin AD, Melatos A (2022) {Mountain formation by repeated, inhomogeneous
  crustal failure in a neutron star}. Monthly Notices of the Royal Astronomical
  Society 514:1628, \doi{10.1093/mnras/stac1351}

\bibitem[{Knispel and Allen(2008)}]{bib:knispelallen}
Knispel B, Allen B (2008) {Blandford's argument: The strongest continuous
  gravitational wave signal}. Phys Rev D 78:044031,
  \doi{10.1103/PhysRevD.78.044031}

\bibitem[{Kojima(1998)}]{bib:rmodes5}
Kojima Y (1998) {Quasitoroidal oscillations in rotating relativistic stars}.
  Mon Not Roy Astron Soc 293:49--52, \doi{10.1046/j.1365-8711.1998.01119.x},
  \eprint{gr-qc/9709003}

\bibitem[{Konno et~al.(1999)Konno, Obata, and Kojima}]{bib:KonnoObataKojima}
Konno K, Obata T, Kojima Y (1999) {Deformation of relativistic magnetized
  stars}. Astron Astrophys 352:211--216, \eprint{gr-qc/9910038}

\bibitem[{Kramer and Stappers(2015)}]{bib:SKAIncrease}
Kramer M, Stappers B (2015) {Pulsar Science with the SKA}. \eprint{1507.04423}

\bibitem[{Krastev et~al.(2008)Krastev, Li, and Worley}]{bib:KrastevLi}
Krastev PG, Li BA, Worley A (2008) {Nuclear limits on gravitational waves from
  elliptically deformed pulsars}. Phys Lett B 668:1--5,
  \doi{10.1016/j.physletb.2008.07.105}, \eprint{0805.1973}

\bibitem[{Krishnan et~al.(2004)Krishnan, Sintes, Papa, Schutz, Frasca, and
  Palomba}]{bib:houghmethod}
Krishnan B, Sintes AM, Papa MA, Schutz BF, Frasca S, Palomba C (2004) {Hough
  transform search for continuous gravitational waves}. Phys Rev D 70:082001,
  \doi{10.1103/PhysRevD.70.082001}

\bibitem[{Kuwahara and Asada(2022)}]{bib:KuwaharaAsada}
Kuwahara N, Asada H (2022) Earth rotation and time-domain reconstruction of
  polarization states for continuous gravitational waves from known pulsars.
  Phys Rev D 106:024051, \doi{10.1103/PhysRevD.106.024051}, \eprint{2202.00171}

\bibitem[{La~Rosa et~al.(2021)La~Rosa, Astone, D\textquoteright{}Antonio,
  Frasca, Leaci, Miller, Palomba, Piccinni, Pierini, and
  Regimbau}]{bib:LaRosaEtal}
La~Rosa I, Astone P, D\textquoteright{}Antonio S, Frasca S, Leaci P, Miller AL,
  Palomba C, Piccinni OJ, Pierini L, Regimbau T (2021) {Continuous
  Gravitational-Wave Data Analysis with General Purpose Computing on Graphic
  Processing Units}. Universe 7(7):218, \doi{10.3390/universe7070218}

\bibitem[{Lander(2014)}]{bib:Lander2014}
Lander SK (2014) {The contrasting magnetic fields of superconducting pulsars
  and magnetars}. Mon Not Roy Astron Soc 437(1):424--436,
  \doi{10.1093/mnras/stt1894}, \eprint{1307.7020}

\bibitem[{Lander et~al.(2011)Lander, Andersson, and
  Glampedakis}]{bib:LanderEtal2011}
Lander SK, Andersson N, Glampedakis K (2011) {Magnetic neutron star equilibria
  with stratification and type-II superconductivity}. Mon Not Roy Astron Soc
  419:732, \doi{10.1111/j.1365-2966.2011.19720.x}, \eprint{1106.6322}

\bibitem[{{Large} et~al.(1968){Large}, {Vaughan}, and {Mills}}]{bib:largeetal}
{Large} MI, {Vaughan} AE, {Mills} BY (1968) {A Pulsar Supernova Association?}
  Nature 220(5165):340--341, \doi{10.1038/220340a0}

\bibitem[{{Lasky}(2015)}]{bib:Laskyreview}
{Lasky} PD (2015) {Gravitational Waves from Neutron Stars: A Review}. Pubs
  Astron Soc Australia 32:e034, \doi{10.1017/pasa.2015.35}, \eprint{1508.06643}

\bibitem[{Lasky et~al.(2017{\natexlab{a}})Lasky, Leris, Rowlinson, and
  Glampedakis}]{bib:magnetarbrakingindex}
Lasky PD, Leris C, Rowlinson A, Glampedakis K (2017{\natexlab{a}}) {The braking
  index of millisecond magnetars}. Astrophys J Lett 843(1):L1,
  \doi{10.3847/2041-8213/aa79a7}, \eprint{1705.10005}

\bibitem[{Lasky et~al.(2017{\natexlab{b}})Lasky, Sarin, and
  Sammut}]{bib:PostmergerWaveforms}
Lasky PD, Sarin N, Sammut L (2017{\natexlab{b}}) {Long-duration waveform models
  for millisecond magnetars born in binary neutron star mergers}. {LIGO Report
  T1700408}, \urlprefix\url{https://dcc.ligo.org/T1700408}

\bibitem[{{Lattimer} and {Prakash}(2001)}]{bib:LattimerPrakash}
{Lattimer} JM, {Prakash} M (2001) {Neutron Star Structure and the Equation of
  State}. Astrophys J 550(1):426--442, \doi{10.1086/319702},
  \eprint{astro-ph/0002232}

\bibitem[{Lazio and Cordes(1998)}]{bib:LazioCordes}
Lazio TJW, Cordes JM (1998) {Hyperstrong Radio-Wave Scattering in the Galactic
  Center. I. A Survey for Extragalactic Sources Seen Through the Galactic
  Center}. Astrophys J Suppl 118:201, \doi{10.1086/313129},
  \eprint{astro-ph/9804156}

\bibitem[{Leaci(2015)}]{bib:LeaciFilteringDisturbances}
Leaci P (2015) {Methods to filter out spurious disturbances in continuous-wave
  searches from gravitational-wave detectors}. Phys Scripta 90(12):125001,
  \doi{10.1088/0031-8949/90/12/125001}

\bibitem[{Leaci and Prix(2015)}]{bib:StackedFstatScoX1Method}
Leaci P, Prix R (2015) {Directed searches for continuous gravitational waves
  from binary systems: parameter-space metrics and optimal Scorpius X-1
  sensitivity}. Phys Rev D 91(10):102003, \doi{10.1103/PhysRevD.91.102003},
  \eprint{1502.00914}

\bibitem[{Lee(2014)}]{bib:lee2014}
Lee U (2014) {Excitation of a non-radial mode in a millisecond X-ray pulsar XTE
  J1751-305}. Mon Not Roy Astron Soc 442(4):3037--3043,
  \doi{10.1093/mnras/stu1077}, \eprint{1403.3476}

\bibitem[{Levin(1999)}]{bib:rmodeslmxb}
Levin Y (1999) {Runaway heating by R modes of neutron stars in low mass x-ray
  binaries}. Astrophys J 517:328, \doi{10.1086/307196},
  \eprint{astro-ph/9810471}

\bibitem[{Levine and Stebbins(1972)}]{bib:LevineStebbins}
Levine J, Stebbins R (1972) {Upper Limit on the Gravitational Flux Reaching the
  Earth from the Crab Pulsar}. Phys Rev D 6:1465--1468,
  \doi{10.1103/PhysRevD.6.1465}

\bibitem[{{LIGO Scientific Collaboration}(2018)}]{bib:lal}
{LIGO Scientific Collaboration} (2018) {LIGO} {A}lgorithm {L}ibrary -
  {LALS}uite. free software (GPL), \doi{10.7935/GT1W-FZ16}

\bibitem[{Lim and Holt(2019)}]{bib:LimHolt}
Lim Y, Holt JW (2019) {Bayesian modeling of the nuclear equation of state for
  neutron star tidal deformabilities and GW170817}. The European Physical
  Journal A 55(11), \doi{10.1140/epja/i2019-12917-9}

\bibitem[{{Lindblom} and {Detweiler}(1977)}]{bib:cfskiller1}
{Lindblom} L, {Detweiler} SL (1977) {On the secular instabilities of the
  Maclaurin spheroids.} Astrophys J 211:565--567, \doi{10.1086/154964}

\bibitem[{{Lindblom} and {Mendell}(1995)}]{bib:cfskiller2}
{Lindblom} L, {Mendell} G (1995) {Does Gravitational Radiation Limit the
  Angular Velocities of Superfluid Neutron Stars?} Astrophys J 444:804,
  \doi{10.1086/175653}

\bibitem[{Lindblom and Owen(2020)}]{bib:LindblomOwen}
Lindblom L, Owen BJ (2020) {Directed searches for continuous gravitational
  waves from twelve supernova remnants in data from Advanced LIGO's second
  observing run}. Phys Rev D 101(8):083023, \doi{10.1103/PhysRevD.101.083023},
  \eprint{2003.00072}

\bibitem[{Liu and Zou(2022)}]{bib:LiuZhouSNR}
Liu Y, Zou YC (2022) {Directed search for continuous gravitational waves from
  the possible kilonova remnant G4.8+6.2}. Phys Rev D 106(12):123024,
  \doi{10.1103/PhysRevD.106.123024}, \eprint{2211.02855}

\bibitem[{Livas(1989)}]{bib:LivasArticle}
Livas J (1989) {Broadband Search Techniques for Periodic Sources of
  Gravitational Radiation. In: Schutz B.F. (eds) Gravitational Wave Data
  Analysis. NATO ASI Series (Series C: Mathematical and Physical Sciences)},
  vol 253, Springer, Dordrecht, pp 217--238. \doi{10.1007/978-94-009-1185-7}

\bibitem[{Livingstone and Kaspi(2011)}]{bib:Livingstone_Kaspi_2011}
Livingstone MA, Kaspi VM (2011) {Long-Term X-ray Monitoring of the Young Pulsar
  PSR B1509-58}. Astrophys J 742:31, \doi{10.1088/0004-637X/742/1/31},
  \eprint{1110.1312}

\bibitem[{Livingstone et~al.(2007)Livingstone, Kaspi, Gavriil, Manchester,
  Gotthelf, and Kuiper}]{bib:LivingstoneEtal_2007}
Livingstone MA, Kaspi VM, Gavriil FP, Manchester RN, Gotthelf E, Kuiper L
  (2007) {New Phase-coherent Measurements of Pulsar Braking Indices}. Astrophys
  Space Sci 308:317--323, \doi{10.1007/s10509-007-9320-3},
  \eprint{astro-ph/0702196}

\bibitem[{Lorimer(2008)}]{bib:lrre-Lorimer}
Lorimer DR (2008) {Binary and Millisecond Pulsars}. Living Rev Rel 11:8,
  \doi{10.12942/lrr-2008-8}, \eprint{0811.0762}

\bibitem[{Lorimer and Kramer(2005)}]{bib:LorimerKramer}
Lorimer DR, Kramer M (2005) {Handbook of Pulsar Astronomy}. Cambridge
  University Press, Cambridge

\bibitem[{Lower et~al.(2021)Lower, Johnston, Dunn, Shannon, Bailes, Dai, Kerr,
  Manchester, Melatos, Oswald, Parthasarathy, Sobey, and
  Weltevrede}]{bib:LowerEtal}
Lower ME, Johnston S, Dunn L, Shannon RM, Bailes M, Dai S, Kerr M, Manchester
  RN, Melatos A, Oswald LS, Parthasarathy A, Sobey C, Weltevrede P (2021) {The
  impact of glitches on young pulsar rotational evolution}. Monthly Notices of
  the Royal Astronomical Society 508(3):3251--3274,
  \doi{10.1093/mnras/stab2678}

\bibitem[{Lu et~al.(2022)Lu, Wette, Scott, and
  Melatos}]{bib:LuEtalInferringProperties}
Lu N, Wette K, Scott SM, Melatos A (2022) {Inferring neutron star properties
  with continuous gravitational waves}. \eprint{2209.10981}

\bibitem[{Lyne and Graham-Smith(2006)}]{bib:LyneGrahamSmith}
Lyne A, Graham-Smith F (2006) {Pulsar Astronomy}, 3rd edn. Cambridge University
  Press, Cambridge

\bibitem[{Lyne et~al.(2015)Lyne, Jordan, Graham-Smith, Espinoza, Stappers, and
  Weltrvrede}]{bib:LyneEtal_2015}
Lyne A, Jordan C, Graham-Smith F, Espinoza C, Stappers B, Weltrvrede P (2015)
  {45 years of rotation of the Crab pulsar}. Mon Not Roy Astron Soc
  446:857--864, \doi{10.1093/mnras/stu2118}, \eprint{1410.0886}

\bibitem[{Macquart and Kanekar(2015)}]{bib:MacquartEtal}
Macquart JP, Kanekar N (2015) {On Detecting Millisecond Pulsars at the Galactic
  Center}. Astrophys J 805(2):172, \doi{10.1088/0004-637X/805/2/172},
  \eprint{1504.02492}

\bibitem[{Maggiore(2008)}]{bib:Maggioretext1}
Maggiore M (2008) {Gravitational Waves – Volume 1: Theory and Experiments}.
  Oxford University Press, Oxford

\bibitem[{Maggiore(2018)}]{bib:Maggioretext2}
Maggiore M (2018) {Gravitational Waves – Volume 2: Astrophysics and
  Cosmology}. Oxford University Press, Oxford

\bibitem[{Maggiore et~al.(2020)Maggiore, Broeck, Bartolo, Belgacem, Bertacca,
  Bizouard, Branchesi, Clesse, Foffa, García-Bellido, and
  et~al.}]{bib:EinsteinTelescope}
Maggiore M, Broeck CVD, Bartolo N, Belgacem E, Bertacca D, Bizouard MA,
  Branchesi M, Clesse S, Foffa S, García-Bellido J, et~al (2020) {Science case
  for the Einstein telescope}. Journal of Cosmology and Astroparticle Physics
  2020(03):050–050, \doi{10.1088/1475-7516/2020/03/050}

\bibitem[{Manchester and Hobbs(2005)}]{bib:ATNFdb}
Manchester R, Hobbs GB (2005) {The ATNF Pulsar Catalogue}. A. Teoh \&\ M.
  Hobbs, Astronomical Journal, 129, 1993-2006, {URL:
  http://www.atnf.csiro.au/research/pulsar/psrcat/}

\bibitem[{Manchester(2010)}]{bib:ptareview}
Manchester RN (2010) {Detection of Gravitational Waves using Pulsar Timing}.
  \eprint{1004.3602}

\bibitem[{Manchester(2018)}]{bib:velaglitches}
Manchester RN (2018) {Pulsar glitches and their impact on neutron-star
  astrophysics}. \eprint{1801.04332}

\bibitem[{Mauceli et~al.(2000)Mauceli, McHugh, Hamilton, Johnson, and
  Morse}]{bib:cwallegro}
Mauceli E, McHugh M, Hamilton W, Johnson W, Morse A (2000) {Search for periodic
  gravitational radiation with the ALLEGRO gravitational wave detector}.
  \eprint{gr-qc/0007023}

\bibitem[{{McClintock} et~al.(2014){McClintock}, {Narayan}, and
  {Steiner}}]{bib:McClintockEtal}
{McClintock} JE, {Narayan} R, {Steiner} JF (2014) {Black Hole Spin via
  Continuum Fitting and the Role of Spin in Powering Transient Jets}. Space Sci
  Rev 183(1-4):295--322, \doi{10.1007/s11214-013-0003-9}, \eprint{1303.1583}

\bibitem[{McNolty(1973)}]{bib:McNolty}
McNolty F (1973) {Some Probability Density Functions and Their Characteristic
  Functions}. Mathematics of Computation 27(123):495--504,
  \urlprefix\url{http://www.jstor.org/stable/2005656}

\bibitem[{Meadors et~al.(2014)Meadors, Kawabe, and
  Riles}]{bib:MeadorsEtalcleaning}
Meadors GD, Kawabe K, Riles K (2014) {Increasing LIGO sensitivity by
  feedforward subtraction of auxiliary length control noise}. Classical and
  Quantum Gravity 31(10):105014, \doi{10.1088/0264-9381/31/10/105014}

\bibitem[{Meadors et~al.(2016)Meadors, Goetz, and
  Riles}]{bib:twospectdirectedmethod}
Meadors GD, Goetz E, Riles K (2016) {Tuning into Scorpius X-1: adapting a
  continuous gravitational-wave search for a known binary system}. Class Quant
  Grav 33(10):105017, \doi{10.1088/0264-9381/33/10/105017}, \eprint{1512.02105}

\bibitem[{Meadors et~al.(2017)Meadors, Goetz, Riles, Creighton, and
  Robinet}]{bib:twospectS6}
Meadors GD, Goetz E, Riles K, Creighton T, Robinet F (2017) {Searches for
  continuous gravitational waves from Scorpius X-1 and XTE J1751-305 in LIGO's
  sixth science run}. Phys Rev D 95:042005, \doi{10.1103/PhysRevD.95.042005}

\bibitem[{Meadors et~al.(2018)Meadors, Krishnan, Papa, Whelan, and
  Zhang}]{bib:CrossCorrResampling}
Meadors GD, Krishnan B, Papa M, Whelan JT, Zhang Y (2018) {Resampling to
  accelerate cross-correlation searches for continuous gravitational waves from
  binary systems}. Phys Rev D 97(4):044017, \doi{10.1103/PhysRevD.97.044017},
  \eprint{1712.06515}

\bibitem[{Melatos(1997)}]{bib:Melatos}
Melatos A (1997) {Spin-down of an oblique rotator with a current-starved outer
  magnetosphere}. Monthly Notices of the Royal Astronomical Society
  288(4):1049--1059, \doi{10.1093/mnras/288.4.1049}

\bibitem[{Melatos and Payne(2005)}]{bib:MelatosPayne}
Melatos A, Payne DJB (2005) {Gravitational radiation from an accreting
  millisecond pulsar with a magnetically confined mountain}. Astrophys J
  623:1044--1050, \doi{10.1086/428600}, \eprint{astro-ph/0503287}

\bibitem[{{Melatos} et~al.(2015){Melatos}, {Douglass}, and
  {Simula}}]{bib:MelatosEtalSuperfluid}
{Melatos} A, {Douglass} JA, {Simula} TP (2015) {Persistent Gravitational
  Radiation from Glitching Pulsars}. The Astrophysical Journal 807(2):132,
  \doi{10.1088/0004-637X/807/2/132}

\bibitem[{Melatos et~al.(2021)Melatos, Clearwater, Suvorova, Sun, Moran, and
  Evans}]{bib:ViterbiPaperIII}
Melatos A, Clearwater P, Suvorova S, Sun L, Moran W, Evans RJ (2021) {Hidden
  Markov model tracking of continuous gravitational waves from a binary neutron
  star with wandering spin. III. Rotational phase tracking}. Phys Rev D
  104:042003, \doi{10.1103/PhysRevD.104.042003}

\bibitem[{{Melrose} et~al.(2021){Melrose}, {Rafat}, and
  {Mastrano}}]{bib:MelroseEtal}
{Melrose} DB, {Rafat} MZ, {Mastrano} A (2021) {Pulsar radio emission
  mechanisms: a critique}. Monthly Notices of the Royal Astronomical Society
  500(4):4530--4548, \doi{10.1093/mnras/staa3324}, \eprint{2006.15243}

\bibitem[{Mendell and Landry(2005)}]{bib:stackslideimplementation}
Mendell G, Landry M (2005) {StackSlide and Hough Search SNR and Statistics}.
  LIGO Report T050003, \urlprefix\url{https://dcc.ligo.org/T050003}

\bibitem[{Messenger and Woan(2007)}]{bib:sidebandmethod1}
Messenger C, Woan G (2007) {A Fast search strategy for gravitational waves from
  low-mass X-ray binaries}. Class Quant Grav 24:S469--S480,
  \doi{10.1088/0264-9381/24/19/S10}, \eprint{gr-qc/0703155}

\bibitem[{Messenger et~al.(2015)Messenger, Bulten, Crowder, Dergachev,
  Galloway, Goetz, Jonker, Lasky, Meadors, Melatos, Premachandra, Riles,
  Sammut, Thrane, Whelan, and Zhang}]{bib:ScoX1MDC1}
Messenger C, Bulten HJ, Crowder SG, Dergachev V, Galloway DK, Goetz E, Jonker
  RJG, Lasky PD, Meadors GD, Melatos A, Premachandra S, Riles K, Sammut L,
  Thrane EH, Whelan JT, Zhang Y (2015) {Gravitational waves from Scorpius X-1:
  A comparison of search methods and prospects for detection with advanced
  detectors}. Phys Rev D 92:023006, \doi{10.1103/PhysRevD.92.023006}

\bibitem[{{Michel}(1969)}]{bib:Michel}
{Michel} FC (1969) {Relativistic Stellar-Wind Torques}. The Astrophysical
  Journal 158:727, \doi{10.1086/150233}

\bibitem[{{Michel} and {Li}(1999)}]{bib:MichelLi}
{Michel} FC, {Li} H (1999) {Electrodynamics of neutron stars}. Phys Rep
  318(6):227--297, \doi{10.1016/S0370-1573(99)00002-2}

\bibitem[{Michel and Tucker(1969)}]{bib:MichelTucker}
Michel FC, Tucker WH (1969) {Pulsar Emission Mechanism}. Nature
  223(5203):277--279, \doi{10.1038/223277a0}

\bibitem[{{Middleditch} et~al.(2006){Middleditch}, {Marshall}, {Wang},
  {Gotthelf}, and {Zhang}}]{bib:MiddleditchEtalStarquakes}
{Middleditch} J, {Marshall} FE, {Wang} QD, {Gotthelf} EV, {Zhang} W (2006)
  {Predicting the Starquakes in PSR J0537-6910}. The Astrophysical Journal
  652(2):1531--1546, \doi{10.1086/508736}, \eprint{astro-ph/0605007}

\bibitem[{Middleton et~al.(2020)Middleton, Clearwater, Melatos, and
  Dunn}]{bib:ViterbiFiveLMXBsO2}
Middleton H, Clearwater P, Melatos A, Dunn L (2020) {Search for gravitational
  waves from five low mass x-ray binaries in the second Advanced LIGO observing
  run with an improved hidden Markov model}. Physical Review D 102(2),
  \doi{10.1103/physrevd.102.023006}

\bibitem[{Miller et~al.(2018)Miller, Astone, D'Antonio, Frasca, Intini,
  La~Rosa, Leaci, Mastrogiovanni, Muciaccia, Palomba, Piccinni, Singhal, and
  Whiting}]{bib:MillerEtalPostmerger}
Miller A, Astone P, D'Antonio S, Frasca S, Intini G, La~Rosa I, Leaci P,
  Mastrogiovanni S, Muciaccia F, Palomba C, Piccinni OJ, Singhal A, Whiting BF
  (2018) {Method to search for long duration gravitational wave transients from
  isolated neutron stars using the generalized frequency-Hough transform}. Phys
  Rev D 98:102004, \doi{10.1103/PhysRevD.98.102004}

\bibitem[{Miller et~al.(2021{\natexlab{a}})Miller, Clesse, De~Lillo, Bruno,
  Depasse, and Tanasijczuk}]{bib:MillerEtalPBH}
Miller AL, Clesse S, De~Lillo F, Bruno G, Depasse A, Tanasijczuk A
  (2021{\natexlab{a}}) {Probing planetary-mass primordial black holes with
  continuous gravitational waves}. Phys Dark Univ 32:100836,
  \doi{10.1016/j.dark.2021.100836}, \eprint{2012.12983}

\bibitem[{Miller et~al.(2022)Miller, Aggarwal, Clesse, and
  De~Lillo}]{bib:planetaryBBH}
Miller AL, Aggarwal N, Clesse S, De~Lillo F (2022) {Constraints on planetary
  and asteroid-mass primordial black holes from continuous gravitational-wave
  searches}. Phys Rev D 105:062008, \doi{10.1103/PhysRevD.105.062008},
  \eprint{2110.06188}

\bibitem[{Miller et~al.(2019{\natexlab{a}})}]{bib:MillerEtalLongTransient}
Miller AL, et~al. (2019{\natexlab{a}}) {How effective is machine learning to
  detect long transient gravitational waves from neutron stars in a real
  search?} Phys Rev D 100(6):062005, \doi{10.1103/PhysRevD.100.062005},
  \eprint{1909.02262}

\bibitem[{Miller et~al.(2021{\natexlab{b}})}]{bib:MillerEtalDPDM}
Miller AL, et~al. (2021{\natexlab{b}}) {Probing new light gauge bosons with
  gravitational-wave interferometers using an adapted semicoherent method}.
  Phys Rev D 103(10):103002, \doi{10.1103/PhysRevD.103.103002},
  \eprint{2010.01925}

\bibitem[{Miller et~al.(2019{\natexlab{b}})}]{bib:NICERzero}
Miller MC, et~al. (2019{\natexlab{b}}) {{PSR} J0030+0451 Mass and Radius from
  {NICER} Data and Implications for the Properties of Neutron Star Matter}. The
  Astrophysical Journal 887(1):L24, \doi{10.3847/2041-8213/ab50c5}

\bibitem[{{Miller} et~al.(2021)}]{bib:NICERIII}
{Miller} MC, et~al. (2021) {The Radius of PSR J0740+6620 from NICER and
  XMM-Newton Data}. Astrophys J Lett 918(2):L28,
  \doi{10.3847/2041-8213/ac089b}, \eprint{2105.06979}

\bibitem[{Millhouse et~al.(2020)Millhouse, Strang, and
  Melatos}]{bib:ViterbiSNRO2}
Millhouse M, Strang L, Melatos A (2020) {Search for gravitational waves from 12
  young supernova remnants with a hidden Markov model in Advanced LIGO's second
  observing run}. Phys Rev D 102:083025, \doi{10.1103/PhysRevD.102.083025}

\bibitem[{Ming et~al.(2016)Ming, Krishnan, Papa, Aulbert, and
  Fehrmann}]{bib:MingEtalOptimization}
Ming J, Krishnan B, Papa MA, Aulbert C, Fehrmann H (2016) {Optimal directed
  searches for continuous gravitational waves}. Phys Rev D 93(6):064011,
  \doi{10.1103/PhysRevD.93.064011}, \eprint{1510.03417}

\bibitem[{Ming et~al.(2018)Ming, Papa, Krishnan, Prix, Beer, Zhu, Eggenstein,
  Bock, and Machenschalk}]{bib:targetchoice2}
Ming J, Papa MA, Krishnan B, Prix R, Beer C, Zhu SJ, Eggenstein HB, Bock O,
  Machenschalk B (2018) {Optimally setting up directed searches for continuous
  gravitational waves in Advanced LIGO O1 data}. Phys Rev D 97(2):024051,
  \doi{10.1103/PhysRevD.97.024051}, \eprint{1708.02173}

\bibitem[{Ming et~al.(2022)Ming, Papa, Eggenstein, Machenschalk, Steltner,
  Prix, Allen, and Behnke}]{bib:MingEtAlG347}
Ming J, Papa MA, Eggenstein HB, Machenschalk B, Steltner B, Prix R, Allen B,
  Behnke O (2022) {Results From an Einstein@Home Search for Continuous
  Gravitational Waves From G347.3 at Low Frequencies in {LIGO} O2 Data}. The
  Astrophysical Journal 925(1):8, \doi{10.3847/1538-4357/ac35cb}

\bibitem[{Ming et~al.(2019)}]{bib:cwdirectedSNREatHO1}
Ming J, et~al. (2019) {Results from an Einstein@Home search for continuous
  gravitational waves from Cassiopeia A, Vela Jr. and G347.3}. Phys Rev D
  100(2):024063, \doi{10.1103/PhysRevD.100.024063}, \eprint{1903.09119}

\bibitem[{Misner(1972)}]{bib:superradiance2}
Misner CW (1972) {Interpretation of Gravitational-Wave Observations}. Phys Rev
  Lett 28:994--997, \doi{10.1103/PhysRevLett.28.994}

\bibitem[{Misner et~al.(1972)Misner, Thorne, and Wheeler}]{bib:MTWtext}
Misner CW, Thorne K, Wheeler J (1972) {Gravitation}. W.H. Freeman \&\ Company,
  San Francisco

\bibitem[{Modafferi et~al.(2021)Modafferi, Moragues, and
  Keitel}]{bib:ModafferiEtal}
Modafferi LM, Moragues J, Keitel D (2021) {Search for long-duration transient
  gravitational waves from glitching pulsars during LIGO\textemdash{}Virgo
  third observing run}. J Phys Conf Ser 2156(12):012079,
  \doi{10.1088/1742-6596/2156/1/012079}, \eprint{2201.08785}

\bibitem[{Moragues et~al.(2022)Moragues, Modafferi, Tenorio, and
  Keitel}]{bib:MoraguesEtalGlitchDetectionProspects}
Moragues J, Modafferi LM, Tenorio R, Keitel D (2022) {Prospects for detecting
  transient quasi-monochromatic gravitational waves from glitching pulsars with
  current and future detectors}. \doi{10.1093/mnras/stac3665},
  \eprint{2210.09907}

\bibitem[{Morales and Horowitz(2022)}]{bib:MoralesHorowitz}
Morales JA, Horowitz CJ (2022) {Neutron star crust can support a large
  ellipticity}. Mon Not Roy Astron Soc 517(4):5610--5616,
  \doi{10.1093/mnras/stac3058}, \eprint{2209.03222}

\bibitem[{M{\"o}sta et~al.(2015)M{\"o}sta, Ott, Radice, Roberts, Schnetter, and
  Haas}]{bib:MagnetarDynamo2}
M{\"o}sta P, Ott CD, Radice D, Roberts LF, Schnetter E, Haas R (2015) {A
  large-scale dynamo and magnetoturbulence in rapidly rotating core-collapse
  supernovae}. Nature 528(7582):376--379, \doi{10.1038/nature15755}

\bibitem[{Mukherjee et~al.(2018)Mukherjee, Messenger, and
  Riles}]{bib:spinwandering}
Mukherjee A, Messenger C, Riles K (2018) {Accretion-induced spin-wandering
  effects on the neutron star in Scorpius X-1: Implications for continuous
  gravitational wave searches}. Phys Rev D 97(4):043016,
  \doi{10.1103/PhysRevD.97.043016}, \eprint{1710.06185}

\bibitem[{Muno et~al.(2008)Muno, Baganoff, Brandt, Morris, and
  Starck}]{bib:MunoEtal}
Muno MP, Baganoff FK, Brandt WN, Morris MR, Starck JL (2008) {A Catalog of
  Diffuse X-Ray{\textendash}emitting Features within 20 pc of Sagittarius A*:
  Twenty Pulsar Wind Nebulae?} The Astrophysical Journal 673(1):251--263,
  \doi{10.1086/521641}

\bibitem[{Mytidis et~al.(2015)Mytidis, Coughlin, and
  Whiting}]{bib:MytidisEtalRmode}
Mytidis A, Coughlin M, Whiting B (2015) {Constraining the R-mode Saturation
  Amplitude From a Hypothetical Detection of R-mode Gravitational Waves From a
  Newborn Neutron Star: Sensitivity Study}. Astrophys J 810:27,
  \doi{10.1088/0004-637X/810/1/27}, \eprint{1505.03191}

\bibitem[{Mytidis et~al.(2019)Mytidis, Panagopoulos, Panagopoulos, Miller, and
  Whiting}]{bib:MytidisEtalNewborn}
Mytidis A, Panagopoulos AA, Panagopoulos OP, Miller A, Whiting B (2019)
  {Sensitivity study using machine learning algorithms on simulated r-mode
  gravitational wave signals from newborn neutron stars}. Phys Rev D
  99(2):024024, \doi{10.1103/PhysRevD.99.024024}, \eprint{1508.02064}

\bibitem[{{Narayan}(1987)}]{bib:nspopulation}
{Narayan} R (1987) {The Birthrate and Initial Spin Period of Single Radio
  Pulsars}. Astrophys J 319:162, \doi{10.1086/165442}

\bibitem[{Neuhäuser et~al.(2015)Neuhäuser, Hohle, Ginski, Schmidt, Hambaryan,
  and Schmidt}]{bib:NeuhauserEtalFomalhautb}
Neuhäuser R, Hohle MM, Ginski C, Schmidt JG, Hambaryan VV, Schmidt TOB (2015)
  {The companion candidate near Fomalhaut – a background neutron star?}
  Monthly Notices of the Royal Astronomical Society 448(1):376–389,
  \doi{10.1093/mnras/stu2751}

\bibitem[{Neunzert(2019)}]{bib:NeunzertThesis}
Neunzert A (2019) {Searching for Continuous Gravitational Waves from Unknown
  Isolated Neutron Stars in Advanced LIGO Data}. PhD thesis, University of
  Michigan, \urlprefix\url{https://hdl.handle.net/2027.42/151632}

\bibitem[{Ng and Romani(2004)}]{bib:NgRomani}
Ng CY, Romani RW (2004) {Fitting Pulsar Wind Tori}. The Astrophysical Journal
  601(1):479–484, \doi{10.1086/380486}

\bibitem[{Ng et~al.(2021)Ng, Vitale, Hannuksela, and
  Li}]{bib:NgEtalGWTC2BosonConstraints}
Ng KKY, Vitale S, Hannuksela OA, Li TGF (2021) {Constraints on Ultralight
  Scalar Bosons within Black Hole Spin Measurements from the LIGO-Virgo
  GWTC-2}. Phys Rev Lett 126:151102, \doi{10.1103/PhysRevLett.126.151102}

\bibitem[{Nieder et~al.(2019)Nieder, Clark, Bassa, Wu, Singh, Donner, Allen,
  Breton, Dhillon, Eggenstein, Hessels, Kennedy, Kerr, Littlefair, Marsh,
  S{\'{a}}nchez, Papa, Ray, Steltner, and Verbiest}]{bib:nonlvctargetedO2}
Nieder L, Clark CJ, Bassa CG, Wu J, Singh A, Donner JY, Allen B, Breton RP,
  Dhillon VS, Eggenstein HB, Hessels JWT, Kennedy MR, Kerr M, Littlefair S,
  Marsh TR, S{\'{a}}nchez DM, Papa MA, Ray PS, Steltner B, Verbiest JPW (2019)
  {Detection and Timing of Gamma-Ray Pulsations from the 707 Hz Pulsar
  J0952-0607}. The Astrophysical Journal 883(1):42,
  \doi{10.3847/1538-4357/ab357e}

\bibitem[{Nieder et~al.(2020)Nieder, Clark, Kandel, Romani, Bassa, Allen,
  Ashok, Cognard, Fehrmann, Freire, Karuppusamy, Kramer, Li, Machenschalk, Pan,
  Papa, Ransom, Ray, Roy, Wang, Wu, Aulbert, Barr, Beheshtipour, Behnke,
  Bhattacharyya, Breton, Camilo, Choquet, Dhillon, Ferrara, Guillemot, Hessels,
  Kerr, Kwang, Marsh, Mickaliger, Pleunis, Pletsch, Roberts, Sanpa-arsa, and
  Steltner}]{bib:NiederEtalBlackWidow}
Nieder L, Clark CJ, Kandel D, Romani RW, Bassa CG, Allen B, Ashok A, Cognard I,
  Fehrmann H, Freire P, Karuppusamy R, Kramer M, Li D, Machenschalk B, Pan Z,
  Papa MA, Ransom SM, Ray PS, Roy J, Wang P, Wu J, Aulbert C, Barr ED,
  Beheshtipour B, Behnke O, Bhattacharyya B, Breton RP, Camilo F, Choquet C,
  Dhillon VS, Ferrara EC, Guillemot L, Hessels JWT, Kerr M, Kwang SA, Marsh TR,
  Mickaliger MB, Pleunis Z, Pletsch HJ, Roberts MSE, Sanpa-arsa S, Steltner B
  (2020) Discovery of a gamma-ray black widow pulsar by {GPU}-accelerated
  einstein@home. The Astrophysical Journal Letters 902(2):L46,
  \doi{10.3847/2041-8213/abbc02}

\bibitem[{Oliver et~al.(2019)Oliver, Keitel, and
  Sintes}]{bib:OliverKeitelSintes}
Oliver M, Keitel D, Sintes AM (2019) {Adaptive transient Hough method for
  long-duration gravitational wave transients}. Physical Review D 99(10),
  \doi{10.1103/physrevd.99.104067}

\bibitem[{Oppenheimer and Volkoff(1939)}]{bib:OppenheimerVolkoff}
Oppenheimer JR, Volkoff GM (1939) {On Massive Neutron Cores}. Phys Rev
  55:374--381, \doi{10.1103/PhysRev.55.374}

\bibitem[{Osborne and Jones(2020)}]{bib:OsborneJones}
Osborne E, Jones D (2020) {Gravitational waves from magnetically-induced
  thermal neutron star mountains}. Mon Not Roy Astron Soc 494(2):2839--2850,
  \doi{10.1093/mnras/staa858}, \eprint{1910.04453}

\bibitem[{Ostriker et~al.(1970)Ostriker, Rees, and Silk}]{bib:OstrikerReesSilk}
Ostriker JP, Rees MJ, Silk J (1970) {Some Observable Consequences of Accretion
  by Defunct Pulsars}. The Astrophysical Journal Letters 6:179

\bibitem[{Owen(1996)}]{bib:OwenTemplates}
Owen BJ (1996) {Search templates for gravitational waves from inspiraling
  binaries: Choice of template spacing}. Phys Rev D 53:6749--6761,
  \doi{10.1103/PhysRevD.53.6749}

\bibitem[{Owen(2005)}]{bib:OwenElastic}
Owen BJ (2005) {Maximum Elastic Deformations of Compact Stars with Exotic
  Equations of State}. Physical Review Letters 95(21),
  \doi{10.1103/physrevlett.95.211101}

\bibitem[{Owen(2010)}]{bib:Owenalpha}
Owen BJ (2010) {How to adapt broad-band gravitational-wave searches for
  $r$-modes}. Phys Rev D 82:104002, \doi{10.1103/PhysRevD.82.104002}

\bibitem[{Owen et~al.(1998)Owen, Lindblom, Cutler, Schutz, Vecchio, and
  Andersson}]{bib:rmodes4}
Owen BJ, Lindblom L, Cutler C, Schutz BF, Vecchio A, Andersson N (1998)
  {Gravitational waves from hot young rapidly rotating neutron stars}. Phys Rev
  D 58:084020, \doi{10.1103/PhysRevD.58.084020}

\bibitem[{Owen et~al.(2022)Owen, Lindblom, and Pinheiro}]{bib:OwenEtalSn1987A}
Owen BJ, Lindblom L, Pinheiro LS (2022) {First Constraining Upper Limits on
  Gravitational-wave Emission from NS 1987A in SNR 1987A}. Astrophys J Lett
  935(1):L7, \doi{10.3847/2041-8213/ac84dc}, \eprint{2206.01168}

\bibitem[{\"Ozel and Freire(2016)}]{bib:OzelFreire}
\"Ozel F, Freire PCC (2016) {Masses, Radii, and the Equation of State of
  Neutron Stars}. Annual Review of Astronomy and Astrophysics 54(1):401--440,
  \doi{10.1146/annurev-astro-081915-023322}

\bibitem[{{Pacini}(1967)}]{bib:Pacini_1967}
{Pacini} F (1967) {Energy Emission from a Neutron Star}. Nature
  216(5115):567--568, \doi{10.1038/216567a0}

\bibitem[{Pacini(1968)}]{bib:Pacini_1968}
Pacini F (1968) {Rotating Neutron Stars, Pulsars, and Supernova Remnants}.
  Nature 219:145, \doi{10.1038/219145a0}

\bibitem[{Page et~al.(2020)Page, Beznogov, Garibay, Lattimer, Prakash, and
  Janka}]{bib:PageEtalSNR1987A}
Page D, Beznogov MV, Garibay I, Lattimer JM, Prakash M, Janka HT (2020) {NS
  1987A in SN 1987A}. The Astrophysical Journal 898(2):125,
  \doi{10.3847/1538-4357/ab93c2}

\bibitem[{{Palomba}(2000)}]{bib:palomba1}
{Palomba} C (2000) {Pulsars ellipticity revised}. Astron \&\ Astroph
  354:163--168, \eprint{astro-ph/9912356}

\bibitem[{Palomba(2005)}]{bib:palomba2}
Palomba C (2005) {Simulation of a population of isolated neutron stars evolving
  through the emission of gravitational waves}. Mon Not Roy Astron Soc
  359:1150--1164, \doi{10.1111/j.1365-2966.2005.08975.x},
  \eprint{astro-ph/0503046}

\bibitem[{{Palomba}(2012)}]{bib:palombareview}
{Palomba} C (2012) {Searches for continuous gravitational wave signals and
  stochastic backgrounds in LIGO and Virgo data}. arXiv e-prints
  arXiv:1201.3176, \eprint{1201.3176}

\bibitem[{Palomba et~al.(2005)Palomba, Astone, and
  Frasca}]{bib:adaptivefreqhough}
Palomba C, Astone P, Frasca S (2005) {Adaptive Hough transform for the search
  of periodic sources}. Classical and Quantum Gravity 22(18):S1255--S1264,
  \doi{10.1088/0264-9381/22/18/s39}

\bibitem[{Palomba et~al.(2019)}]{bib:PalombaEtalAxion}
Palomba C, et~al. (2019) {Direct constraints on ultra-light boson mass from
  searches for continuous gravitational waves}. Phys Rev Lett 123:171101,
  \doi{10.1103/PhysRevLett.123.171101}, \eprint{1909.08854}

\bibitem[{{Pandharipande} et~al.(1976){Pandharipande}, {Pines}, and
  {Smith}}]{bib:crustdeformation}
{Pandharipande} VR, {Pines} D, {Smith} RA (1976) {Neutron star structure:
  theory, observation, and speculation.} Astrophys J 208:550--566,
  \doi{10.1086/154637}

\bibitem[{Papa et~al.(2020)Papa, Ming, Gotthelf, Allen, Prix, Dergachev,
  Eggenstein, Singh, and Zhu}]{bib:cwdirectedSNREatHO1subthreshold}
Papa MA, Ming J, Gotthelf EV, Allen B, Prix R, Dergachev V, Eggenstein HB,
  Singh A, Zhu SJ (2020) {Search for Continuous Gravitational Waves from the
  Central Compact Objects in Supernova Remnants Cassiopeia A, Vela Jr., and
  G347.3\textendash0.5}. Astrophys J 897(1):22, \doi{10.3847/1538-4357/ab92a6},
  \eprint{2005.06544}

\bibitem[{Papa et~al.(2016)}]{bib:papafollowup}
Papa MA, et~al. (2016) {Hierarchical follow-up of subthreshold candidates of an
  all-sky Einstein@Home search for continuous gravitational waves on LIGO sixth
  science run data}. Phys Rev D 94(12):122006,
  \doi{10.1103/PhysRevD.94.122006}, \eprint{1608.08928}

\bibitem[{{Papaloizou} and {Pringle}(1978)}]{bib:papapringle}
{Papaloizou} J, {Pringle} JE (1978) {Gravitational radiation and the stability
  of rotating stars.} Monthly Notices of the Royal Astronomical Society
  184:501--508, \doi{10.1093/mnras/184.3.501}

\bibitem[{Parthasarathy et~al.(2019)Parthasarathy, Shannon, Johnston, Lentati,
  Bailes, Dai, Kerr, Manchester, Osłowski, Sobey, van Straten, and
  Weltevrede}]{bib:ParthasarathyEtalBrakingIndicesI}
Parthasarathy A, Shannon RM, Johnston S, Lentati L, Bailes M, Dai S, Kerr M,
  Manchester RN, Osłowski S, Sobey C, van Straten W, Weltevrede P (2019)
  {Timing of young radio pulsars – I. Timing noise, periodic modulation, and
  proper motion}. Monthly Notices of the Royal Astronomical Society
  489(3):3810--3826, \doi{10.1093/mnras/stz2383}

\bibitem[{Parthasarathy et~al.(2020)}]{bib:ParthasarathyEtalBrakingIndicesII}
Parthasarathy A, et~al. (2020) {Timing of young radio pulsars II. Braking
  indices and their interpretation}. Mon Not Roy Astron Soc 494(2):2012--2026,
  \doi{10.1093/mnras/staa882}, \eprint{2003.13303}

\bibitem[{Paschalidis and Stergioulas(2017)}]{bib:lrre-PaschalidisStergioulas}
Paschalidis V, Stergioulas N (2017) {Rotating Stars in Relativity}. Living Rev
  Rel 20(1):7, \doi{10.1007/s41114-017-0008-x}, \eprint{1612.03050}

\bibitem[{Patel et~al.(2010)Patel, Siemens, Dupuis, and
  Betzwieser}]{bib:PatelResampling}
Patel P, Siemens X, Dupuis R, Betzwieser J (2010) {Implementation of
  barycentric resampling for continuous wave searches in gravitational wave
  data}. Physical Review D 81(8), \doi{10.1103/physrevd.81.084032}

\bibitem[{Patruno(2017)}]{bib:IGRJ00291Later}
Patruno A (2017) {The Slow Orbital Evolution of the Accreting Millisecond
  Pulsar {IGR} J0029$+$5934}. The Astrophysical Journal 839(1):51,
  \doi{10.3847/1538-4357/aa6986}

\bibitem[{Patruno et~al.(2012)Patruno, Haskell, and
  D'Angelo}]{bib:PatrunoHaskellDangelo}
Patruno A, Haskell B, D'Angelo C (2012) {{Gravitational} {waves} {and} {the}
  {maximum} {spin} {frequency} {of} {neutron} {stars}}. The Astrophysical
  Journal 746(1):9, \doi{10.1088/0004-637x/746/1/9}

\bibitem[{Patruno et~al.(2017)Patruno, Haskell, and
  Andersson}]{bib:PatrunoHaskellAndersson}
Patruno A, Haskell B, Andersson N (2017) {The Spin Distribution of
  Fast-spinning Neutron Stars in Low-mass X-Ray Binaries: Evidence for Two
  Subpopulations}. Astrophys J 850(1):106, \doi{10.3847/1538-4357/aa927a},
  \eprint{1705.07669}

\bibitem[{Pavlov et~al.(2001)Pavlov, Sanwal, Kiziltan, and
  Garmire}]{bib:PavlovEtalVelaJr}
Pavlov GG, Sanwal D, Kiziltan B, Garmire GP (2001) {The compact central source
  in the RX J0852-4622 supernova remnant}. Astrophys J Lett 559:L131,
  \doi{10.1086/323975}, \eprint{astro-ph/0108150}

\bibitem[{{Payne} and {Melatos}(2004)}]{bib:PayneMelatos}
{Payne} DJB, {Melatos} A (2004) {Burial of the polar magnetic field of an
  accreting neutron star - I. Self-consistent analytic and numerical
  equilibria}. Monthly Notices of the Royal Astronomical Society
  351(2):569--584, \doi{10.1111/j.1365-2966.2004.07798.x},
  \eprint{astro-ph/0403173}

\bibitem[{Penrose(1969)}]{bib:penrose1}
Penrose R (1969) {Gravitational collapse: The role of general relativity}. Riv
  Nuovo Cim 1:252--276, \doi{10.1023/A:1016578408204}, [Gen. Rel.
  Grav.34,1141(2002)]

\bibitem[{Piccinni(2022)}]{bib:PiccinniReview}
Piccinni OJ (2022) Status and perspectives of continuous gravitational wave
  searches. Galaxies 10(3), \doi{10.3390/galaxies10030072}

\bibitem[{Piccinni et~al.(2018)Piccinni, Astone, D’Antonio, Frasca, Intini,
  Leaci, Mastrogiovanni, Miller, Palomba, and Singhal}]{bib:BSD}
Piccinni OJ, Astone P, D’Antonio S, Frasca S, Intini G, Leaci P,
  Mastrogiovanni S, Miller A, Palomba C, Singhal A (2018) {A new data analysis
  framework for the search of continuous gravitational wave signals}. Classical
  and Quantum Gravity 36(1):015008, \doi{10.1088/1361-6382/aaefb5}

\bibitem[{Piccinni et~al.(2020)Piccinni, Astone, D'Antonio, Frasca, Intini,
  La~Rosa, Leaci, Mastrogiovanni, Miller, and
  Palomba}]{bib:PiccinniEtalGalacticCenter}
Piccinni OJ, Astone P, D'Antonio S, Frasca S, Intini G, La~Rosa I, Leaci P,
  Mastrogiovanni S, Miller A, Palomba C (2020) {Directed search for continuous
  gravitational-wave signals from the Galactic Center in the Advanced LIGO
  second observing run}. Phys Rev D 101(8):082004,
  \doi{10.1103/PhysRevD.101.082004}, \eprint{1910.05097}

\bibitem[{Pierce et~al.(2018)Pierce, Riles, and Zhao}]{bib:DPDM_PRZ}
Pierce A, Riles K, Zhao Y (2018) {Searching for Dark Photon Dark Matter with
  Gravitational-Wave Detectors}. Physical Review Letters 121(6),
  \doi{10.1103/physrevlett.121.061102}

\bibitem[{Pierini et~al.(2022)Pierini, Astone, Palomba, Nyquist, Dall'Osso,
  D'Antonio, Frasca, La~Rosa, Leaci, Muciaccia, Piccinni, and
  Rei}]{bib:PieriniEtalSignalConfusion}
Pierini L, Astone P, Palomba C, Nyquist A, Dall'Osso S, D'Antonio S, Frasca S,
  La~Rosa I, Leaci P, Muciaccia F, Piccinni OJ, Rei L (2022) Impact of signal
  clusters in wide-band searches for continuous gravitational waves. Phys Rev D
  106:042009, \doi{10.1103/PhysRevD.106.042009}

\bibitem[{Piro et~al.(2017)Piro, Giacomazzo, and Perna}]{bib:PiroEtal}
Piro AL, Giacomazzo B, Perna R (2017) {The Fate of Neutron Star Binary
  Mergers}. The Astrophysical Journal 844(2):L19,
  \doi{10.3847/2041-8213/aa7f2f}

\bibitem[{Pisarski and Jaranowski(2015)}]{bib:PisarskiJaranowski}
Pisarski A, Jaranowski P (2015) {Banks of templates for all-sky narrow-band
  searches of gravitational waves from spinning neutron stars}. Class Quant
  Grav 32(14):145014, \doi{10.1088/0264-9381/32/14/145014}, \eprint{1302.0509}

\bibitem[{Pisarski et~al.(2011)Pisarski, Jaranowski, and
  Pietka}]{bib:PisarskiEtalTemplateBank}
Pisarski A, Jaranowski P, Pietka M (2011) {Banks of templates for directed
  searches of gravitational waves from spinning neutron stars}. Phys Rev D
  83:043001, \doi{10.1103/PhysRevD.83.043001}, \eprint{1010.2879}

\bibitem[{Pitkin et~al.(2011)Pitkin, Reid, Rowan, and
  Hough}]{bib:lrre-Pitkinetal}
Pitkin M, Reid S, Rowan S, Hough J (2011) {Gravitational Wave Detection by
  Interferometry (Ground and Space)}. Living Rev Rel 14:5,
  \doi{10.12942/lrr-2011-5}, \eprint{1102.3355}

\bibitem[{Pitkin et~al.(2015)Pitkin, Gill, Jones, Woan, and
  Davies}]{bib:PitkinEtalTwotones}
Pitkin M, Gill C, Jones DI, Woan G, Davies GS (2015) {First results and future
  prospects for dual-harmonic searches for gravitational waves from spinning
  neutron stars}. Monthly Notices of the Royal Astronomical Society
  453(4):4399--4420, \doi{10.1093/mnras/stv1931}

\bibitem[{Pitkin et~al.(2018)Pitkin, Doolan, McMenamin, and
  Wette}]{bib:PitkinBarycentering}
Pitkin M, Doolan S, McMenamin L, Wette K (2018) {Reduced order modelling in
  searches for continuous gravitational waves – I. Barycentring time delays}.
  Monthly Notices of the Royal Astronomical Society 476(4):4510–4519,
  \doi{10.1093/mnras/sty548}

\bibitem[{Pletsch(2008)}]{bib:Pletsch}
Pletsch HJ (2008) {Parameter-space correlations of the optimal statistic for
  continuous gravitational-wave detection}. Phys Rev D 78:102005,
  \doi{10.1103/PhysRevD.78.102005}

\bibitem[{Pletsch(2010)}]{bib:Pletschmetric}
Pletsch HJ (2010) {Parameter-space metric of semicoherent searches for
  continuous gravitational waves}. Phys Rev D 82:042002,
  \doi{10.1103/PhysRevD.82.042002}, \eprint{1005.0395}

\bibitem[{Pletsch and Allen(2009)}]{bib:PletschAllen}
Pletsch HJ, Allen B (2009) {Exploiting global correlations to detect continuous
  gravitational waves}. Phys Rev Lett 103:181102,
  \doi{10.1103/PhysRevLett.103.181102}, \eprint{0906.0023}

\bibitem[{Pletsch et~al.(2011)Pletsch, Guillemot, Allen, Kramer, Aulbert,
  Fehrmann, Ray, Barr, Belfiore, Camilo, and
  et~al.}]{bib:fermifirsteathdetection}
Pletsch HJ, Guillemot L, Allen B, Kramer M, Aulbert C, Fehrmann H, Ray PS, Barr
  ED, Belfiore A, Camilo F, et~al (2011) {Discovery of Nine Gamma-Ray Pulsars
  in Fermi Large Area Telescope Data using a New Blind Search Method}. The
  Astrophysical Journal 744(2):105, \doi{10.1088/0004-637x/744/2/105}

\bibitem[{Popov et~al.(2015)Popov, Postnov, and
  Shakura}]{bib:PopovISMaccretion}
Popov SB, Postnov KA, Shakura NI (2015) {Settling accretion on to isolated
  neutron stars from interstellar medium}. Monthly Notices of the Royal
  Astronomical Society 447(3):2817--2820, \doi{10.1093/mnras/stu2643}

\bibitem[{Poppenhaeger et~al.(2017)Poppenhaeger, Auchettl, and
  Wolk}]{bib:PoppenhaegerEtal}
Poppenhaeger K, Auchettl K, Wolk SJ (2017) {A test of the neutron star
  hypothesis for Fomalhaut b}. Monthly Notices of the Royal Astronomical
  Society 468(4):4018–4024, \doi{10.1093/mnras/stx565}

\bibitem[{{Potekhin} et~al.(2015){Potekhin}, {Pons}, and
  {Page}}]{bib:NSCooling}
{Potekhin} AY, {Pons} JA, {Page} D (2015) {Neutron Stars{\textemdash}Cooling
  and Transport}. Space Sci Rev 191(1-4):239--291,
  \doi{10.1007/s11214-015-0180-9}, \eprint{1507.06186}

\bibitem[{Premachandra et~al.(2016)Premachandra, Galloway, Casares, Steeghs,
  and Marsh}]{bib:GallowayCygX2Source}
Premachandra SS, Galloway DK, Casares J, Steeghs DT, Marsh TR (2016) {Precision
  Ephemerides For Gravitational Wave Searches: II. Cyg X-2}. Astrophys J
  823(2):106, \doi{10.3847/0004-637X/823/2/106}, \eprint{1604.03233}

\bibitem[{Prix(2007{\natexlab{a}})}]{bib:PrixTemplates1}
Prix R (2007{\natexlab{a}}) {Search for continuous gravitational waves: Metric
  of the multidetector $\mathcal{F}$-statistic}. Phys Rev D 75:023004,
  \doi{10.1103/PhysRevD.75.023004}

\bibitem[{Prix(2007{\natexlab{b}})}]{bib:PrixTemplatesEfficient}
Prix R (2007{\natexlab{b}}) {Template-based searches for gravitational waves:
  efficient lattice covering of flat parameter spaces}. Classical and Quantum
  Gravity 24(19):S481--S490, \doi{10.1088/0264-9381/24/19/s11}

\bibitem[{Prix(2009)}]{bib:Prixreview}
Prix R (2009) {Gravitational Waves from Spinning Neutron Stars}, vol 357,
  Springer International Publishing, pp 651--685.
  \doi{10.1007/978-3-540-76965-1\_24}

\bibitem[{Prix(2018)}]{bib:PrixFstatNote}
Prix R (2018) {The F-statistic and its implementation in
  ComputeFstatistic\_v2}. {LIGO Report T0900149-v6},
  \urlprefix\url{https://dcc.ligo.org/T0900149}

\bibitem[{Prix and Itoh(2005)}]{bib:PrixItoh}
Prix R, Itoh Y (2005) {Global parameter-space correlations of coherent searches
  for continuous gravitational waves}. Classical and Quantum Gravity 22:S1003,
  \doi{10.1088/0264-9381/22/18/S14}

\bibitem[{Prix and Krishnan(2009)}]{bib:PrixKrishnan}
Prix R, Krishnan B (2009) {Targeted search for continuous gravitational waves:
  Bayesian versus maximum-likelihood statistics}. Class Quant Grav 26:204013,
  \doi{10.1088/0264-9381/26/20/204013}, \eprint{0907.2569}

\bibitem[{Prix and Shaltev(2012)}]{bib:PrixShaltev}
Prix R, Shaltev M (2012) {Search for Continuous Gravitational Waves: Optimal
  StackSlide method at fixed computing cost}. Phys Rev D 85:084010,
  \doi{10.1103/PhysRevD.85.084010}, \eprint{1201.4321}

\bibitem[{Prix et~al.(2011)Prix, Giampanis, and
  Messenger}]{bib:PrixGiampanisMessenger}
Prix R, Giampanis S, Messenger C (2011) {Search method for long-duration
  gravitational-wave transients from neutron stars}. Phys Rev D 84:023007,
  \doi{10.1103/PhysRevD.84.023007}

\bibitem[{van~der Putten et~al.(2010)van~der Putten, Bulten, van~den Brand, and
  Holtrop}]{bib:polynomial}
van~der Putten S, Bulten HJ, van~den Brand JFJ, Holtrop M (2010) {Searching for
  gravitational waves from pulsars in binary systems: An all-sky search}.
  Journal of Physics: Conference Series 228:012005,
  \doi{10.1088/1742-6596/228/1/012005}

\bibitem[{Pétri(2019)}]{bib:Petri}
Pétri J (2019) {The illusion of neutron star magnetic field estimates}. Mon
  Not Roy Astron Soc 485(4):4573--4587, \doi{10.1093/mnras/stz711},
  \eprint{1903.01528}

\bibitem[{{Raaijmakers} et~al.(2019){Raaijmakers}, {Riley}, {Watts}, {Greif},
  {Morsink}, {Hebeler}, {Schwenk}, {Hinderer}, {Nissanke}, {Guillot},
  {Arzoumanian}, {Bogdanov}, {Chakrabarty}, {Gendreau}, {Ho}, {Lattimer},
  {Ludlam}, and {Wolff}}]{bib:NICEREOS}
{Raaijmakers} G, {Riley} TE, {Watts} AL, {Greif} SK, {Morsink} SM, {Hebeler} K,
  {Schwenk} A, {Hinderer} T, {Nissanke} S, {Guillot} S, {Arzoumanian} Z,
  {Bogdanov} S, {Chakrabarty} D, {Gendreau} KC, {Ho} WCG, {Lattimer} JM,
  {Ludlam} RM, {Wolff} MT (2019) {A Nicer View of PSR J0030+0451: Implications
  for the Dense Matter Equation of State}. The Astrophysical Journal Letters
  887(1):L22, \doi{10.3847/2041-8213/ab451a}, \eprint{1912.05703}

\bibitem[{{Raaijmakers} et~al.(2021){Raaijmakers}, {Greif}, {Hebeler},
  {Hinderer}, {Nissanke}, {Schwenk}, {Riley}, {Watts}, {Lattimer}, and
  {Ho}}]{bib:NICERV}
{Raaijmakers} G, {Greif} SK, {Hebeler} K, {Hinderer} T, {Nissanke} S, {Schwenk}
  A, {Riley} TE, {Watts} AL, {Lattimer} JM, {Ho} WCG (2021) {Constraints on the
  Dense Matter Equation of State and Neutron Star Properties from NICER's
  Mass-Radius Estimate of PSR J0740+6620 and Multimessenger Observations}.
  Astrophys J Lett 918(2):L29, \doi{10.3847/2041-8213/ac089a},
  \eprint{2105.06981}

\bibitem[{{Radhakrishnan} and
  {Srinivasan}(1982)}]{bib:RadhakrishnanSrinivasanRecycling}
{Radhakrishnan} V, {Srinivasan} G (1982) {On the origin of the recently
  discovered ultra-rapid pulsar}. Current Science 51:1096--1099

\bibitem[{Rajbhandari et~al.(2021)Rajbhandari, Owen, Caride, and
  Inta}]{bib:RajbhandariOwenCarideInta}
Rajbhandari B, Owen BJ, Caride S, Inta R (2021) {First searches for
  gravitational waves from $r$-modes of the Crab pulsar}. Phys Rev D
  104:122008, \doi{10.1103/PhysRevD.104.122008}

\bibitem[{Ravenhall et~al.(1983)Ravenhall, Pethick, and
  Wilson}]{bib:nuclearpasta}
Ravenhall DG, Pethick CJ, Wilson JR (1983) {Structure of Matter below Nuclear
  Saturation Density}. Phys Rev Lett 50:2066--2069,
  \doi{10.1103/PhysRevLett.50.2066}

\bibitem[{Ravi and Lasky(2014)}]{bib:RaviLasky}
Ravi V, Lasky PD (2014) {The birth of black holes: neutron star collapse times,
  gamma-ray bursts and fast radio bursts}. Mon Not Roy Astron Soc
  441(3):2433--2439, \doi{10.1093/mnras/stu720}, \eprint{1403.6327}

\bibitem[{Reardon et~al.(2021)Reardon, Shannon, Cameron, Goncharov, Hobbs,
  Middleton, Shamohammadi, Thyagarajan, Bailes, Bhat, Dai, Kerr, Manchester,
  Russell, Spiewak, Wang, and Zhu}]{bib:ReardonEtalParkes}
Reardon DJ, Shannon RM, Cameron AD, Goncharov B, Hobbs GB, Middleton H,
  Shamohammadi M, Thyagarajan N, Bailes M, Bhat NDR, Dai S, Kerr M, Manchester
  RN, Russell CJ, Spiewak R, Wang JB, Zhu XJ (2021) {The Parkes pulsar timing
  array second data release: timing analysis}. Monthly Notices of the Royal
  Astronomical Society 507(2):2137--2153, \doi{10.1093/mnras/stab1990}

\bibitem[{Reed et~al.(2021)Reed, Deibel, and Horowitz}]{bib:ReedEtal}
Reed BT, Deibel A, Horowitz CJ (2021) {Modeling the Galactic Neutron Star
  Population for Use in Continuous Gravitational-wave Searches}. Astrophys J
  921(1):89, \doi{10.3847/1538-4357/ac1c04}, \eprint{2104.00771}

\bibitem[{{Reed} et~al.(1995){Reed}, {Hester}, {Fabian}, and
  {Winkler}}]{bib:ReedEtalCasA}
{Reed} JE, {Hester} JJ, {Fabian} AC, {Winkler} PF (1995) {The Three-dimensional
  Structure of the Cassiopeia A Supernova Remnant. I. The Spherical Shell}. The
  Astrophysical Journal 440:706, \doi{10.1086/175308}

\bibitem[{Regimbau and de~Freitas~Pacheco(2006)}]{bib:RegimbauDeFreitasPacheco}
Regimbau T, de~Freitas~Pacheco JA (2006) {Gravitational wave background from
  magnetars}. Astron Astrophys 447:1, \doi{10.1051/0004-6361:20053702},
  \eprint{astro-ph/0509880}

\bibitem[{Renaud et~al.(2010)Renaud, Marandon, Gotthelf, Rodriguez, Terrier,
  Mattana, Lebrun, Tomsick, and Manchester}]{bib:RenaudEtal}
Renaud M, Marandon V, Gotthelf EV, Rodriguez J, Terrier R, Mattana F, Lebrun F,
  Tomsick JA, Manchester RN (2010) {{Discovery} {of} a {highly} {energetic}
  {pulsar} {associated} {with} {IGR} J14003{\textendash}6326 {in} {the} {young}
  {uncataloged} {galactic} {supernova} {remnant} G310.6{\textendash}1.6}. The
  Astrophysical Journal 716(1):663--670, \doi{10.1088/0004-637x/716/1/663}

\bibitem[{{Reynolds}(2014)}]{bib:Reynolds}
{Reynolds} CS (2014) {Measuring Black Hole Spin Using X-Ray Reflection
  Spectroscopy}. Space Sci Rev 183(1-4):277--294,
  \doi{10.1007/s11214-013-0006-6}, \eprint{1302.3260}

\bibitem[{Reynolds et~al.(2008)Reynolds, Borkowski, Green, Hwang, Harrus, and
  Petre}]{bib:ReynoldsEtal}
Reynolds SP, Borkowski KJ, Green DA, Hwang U, Harrus I, Petre R (2008) {The
  Youngest Galactic Supernova Remnant: G1.9+0.3}. Astrophys J Lett 680:L41,
  \doi{10.1086/589570}, \eprint{0803.1487}

\bibitem[{Riles(2013)}]{bib:RilesPPNP}
Riles K (2013) {Gravitational Waves: Sources, Detectors and Searches}. Prog
  Part Nucl Phys 68:1--54, \doi{10.1016/j.ppnp.2012.08.001}, \eprint{1209.0667}

\bibitem[{Riles(2017)}]{bib:RilesMPLA}
Riles K (2017) {Recent searches for continuous gravitational waves}. Mod Phys
  Lett A 32(39):1730035, \doi{10.1142/S021773231730035X}, \eprint{1712.05897}

\bibitem[{{Riley} et~al.(2019){Riley}, {Watts}, {Bogdanov}, {Ray}, {Ludlam},
  {Guillot}, {Arzoumanian}, {Baker}, {Bilous}, {Chakrabarty}, {Gendreau},
  {Harding}, {Ho}, {Lattimer}, {Morsink}, and {Strohmayer}}]{bib:NICERanother}
{Riley} TE, {Watts} AL, {Bogdanov} S, {Ray} PS, {Ludlam} RM, {Guillot} S,
  {Arzoumanian} Z, {Baker} CL, {Bilous} AV, {Chakrabarty} D, {Gendreau} KC,
  {Harding} AK, {Ho} WCG, {Lattimer} JM, {Morsink} SM, {Strohmayer} TE (2019)
  {A NICER View of PSR J0030+0451: Millisecond Pulsar Parameter Estimation}.
  The Astrophysical Journal 887(1):L21, \doi{10.3847/2041-8213/ab481c},
  \eprint{1912.05702}

\bibitem[{{Riley} et~al.(2021){Riley}, {Watts}, {Ray}, {Bogdanov}, {Guillot},
  {Morsink}, {Bilous}, {Arzoumanian}, {Choudhury}, {Deneva}, {Gendreau},
  {Harding}, {Ho}, {Lattimer}, {Loewenstein}, {Ludlam}, {Markwardt}, {Okajima},
  {Prescod-Weinstein}, {Remillard}, {Wolff}, {Fonseca}, {Cromartie}, {Kerr},
  {Pennucci}, {Parthasarathy}, {Ransom}, {Stairs}, {Guillemot}, and
  {Cognard}}]{bib:NICERIV}
{Riley} TE, {Watts} AL, {Ray} PS, {Bogdanov} S, {Guillot} S, {Morsink} SM,
  {Bilous} AV, {Arzoumanian} Z, {Choudhury} D, {Deneva} JS, {Gendreau} KC,
  {Harding} AK, {Ho} WCG, {Lattimer} JM, {Loewenstein} M, {Ludlam} RM,
  {Markwardt} CB, {Okajima} T, {Prescod-Weinstein} C, {Remillard} RA, {Wolff}
  MT, {Fonseca} E, {Cromartie} HT, {Kerr} M, {Pennucci} TT, {Parthasarathy} A,
  {Ransom} S, {Stairs} I, {Guillemot} L, {Cognard} I (2021) {A NICER View of
  the Massive Pulsar PSR J0740+6620 Informed by Radio Timing and XMM-Newton
  Spectroscopy}. Astrophys J Lett 918(2):L27, \doi{10.3847/2041-8213/ac0a81},
  \eprint{2105.06980}

\bibitem[{Roberts and van Leeuwen(2013)}]{bib:blackwidowredback}
Roberts MSE, van Leeuwen Je (2013) {Neutron Stars and Pulsars: Challenges and
  Opportunities after 80 years, IAU Symposium}, vol 291. Cambridge University
  Press

\bibitem[{{Romani}(1990)}]{bib:RomaniUnifiedBfieldModel}
{Romani} RW (1990) {A unified model of neutron-star magnetic fields}. Nature
  347(6295):741--743, \doi{10.1038/347741a0}

\bibitem[{Romano and Cornish(2017)}]{bib:lrre-RomanoCornish}
Romano JD, Cornish NJ (2017) {Detection methods for stochastic
  gravitational-wave backgrounds: a unified treatment}. Living Rev Rel 20(1):2,
  \doi{10.1007/s41114-017-0004-1}, \eprint{1608.06889}

\bibitem[{{Roy} et~al.(2012){Roy}, {Gupta}, and
  {Lewandowski}}]{bib:RoyEtal_2012}
{Roy} J, {Gupta} Y, {Lewandowski} W (2012) {Observations of four glitches in
  the young pulsar J1833-1034 and study of its glitch activity}. Monthly
  Notices of the Royal Astronomical Society 424(3):2213--2221,
  \doi{10.1111/j.1365-2966.2012.21380.x}

\bibitem[{Ruderman(1969)}]{bib:Ruderman}
Ruderman M (1969) {Neutron Starquakes and Pulsar Periods}. Nature
  223(5206):597--598, \doi{10.1038/223597b0}

\bibitem[{{Ruderman} and {Sutherland}(1975)}]{bib:RudermanSutherland}
{Ruderman} MA, {Sutherland} PG (1975) {Theory of pulsars: polar gaps, sparks,
  and coherent microwave radiation.} Astrophys J 196:51--72,
  \doi{10.1086/153393}

\bibitem[{Sammut et~al.(2014)Sammut, Messenger, Melatos, and
  Owen}]{bib:sidebandmethod2}
Sammut L, Messenger C, Melatos A, Owen BJ (2014) {Implementation of the
  frequency-modulated sideband search method for gravitational waves from low
  mass X-ray binaries}. Phys Rev D 89(4):043001,
  \doi{10.1103/PhysRevD.89.043001}, \eprint{1311.1379}

\bibitem[{Sanders(2016)}]{bib:SandersThesis}
Sanders J (2016) {Advanced Gravitational Wave Detectors and Detection: Arm
  Length Stabilization and Directed Searches for Isolated Neutron Stars}. PhD
  thesis, University of Michigan,
  \urlprefix\url{https://hdl.handle.net/2027.42/120826}

\bibitem[{Sarin et~al.(2018)Sarin, Lasky, Sammut, and Ashton}]{bib:SarinEtal}
Sarin N, Lasky PD, Sammut L, Ashton G (2018) {X-ray guided gravitational-wave
  search for binary neutron star merger remnants}. Phys Rev D 98(4):043011,
  \doi{10.1103/PhysRevD.98.043011}, \eprint{1805.01481}

\bibitem[{Sarin et~al.(2020)Sarin, Lasky, and Ashton}]{bib:SarinLaskyAshton}
Sarin N, Lasky PD, Ashton G (2020) {Interpreting the X-ray afterglows of
  gamma-ray bursts with radiative losses and millisecond magnetars}. Mon Not
  Roy Astron Soc 499(4):5986--5992, \doi{10.1093/mnras/staa3090},
  \eprint{2008.05745}

\bibitem[{Sathyaprakash and Dhurandhar(1991)}]{bib:SathyaFilterbank}
Sathyaprakash BS, Dhurandhar SV (1991) {Choice of filters for the detection of
  gravitational waves from coalescing binaries}. Phys Rev D 44:3819--3834,
  \doi{10.1103/PhysRevD.44.3819}

\bibitem[{Sathyaprakash and Schutz(2009)}]{bib:lrre-SathyaprakashSchutz}
Sathyaprakash BS, Schutz BF (2009) {Physics, Astrophysics and Cosmology with
  Gravitational Waves}. Living Rev Rel 12:2, \doi{10.12942/lrr-2009-2},
  \eprint{0903.0338}

\bibitem[{Saulson(2017)}]{bib:Saulsontext}
Saulson PR (2017) {Fundamentals of Interferometric Gravitational Wave
  Detectors, 2nd Ed. }. World Scientific, Singapore

\bibitem[{Sauter et~al.(2019)Sauter, Dergachev, and
  Riles}]{bib:SauterBarycentering}
Sauter O, Dergachev V, Riles K (2019) {Efficient estimation of barycentered
  relative time delays for distant gravitational wave sources}. Physical Review
  D 99(4), \doi{10.1103/physrevd.99.044006}

\bibitem[{Schutz(1985)}]{bib:Schutztext}
Schutz BF (1985) {A First Course in General Relativity}. Cambridge University
  Press, Cambridge

\bibitem[{Schutz(1991)}]{bib:SchutzBlairbook}
Schutz BF (1991) {The Detection of Gravitational Waves}, Cambridge University
  Press, Cambridge, p 406

\bibitem[{Serim et~al.(2022)Serim, Serim, and
  Baykal}]{bib:SerimSerimBaykalAccretingPulsarTiming}
Serim D, Serim MM, Baykal A (2022) {Pulse frequency fluctuations of persistent
  accretion powered pulsars}. Mon Not Roy Astron Soc 518(1):1--12,
  \doi{10.1093/mnras/stac3076}, \eprint{2207.00248}

\bibitem[{Seto(2005)}]{bib:SetoWavefrontCurvature}
Seto N (2005) Gravitational wave astrometry for rapidly rotating neutron stars
  and estimation of their distances. Phys Rev D 71:123002,
  \doi{10.1103/PhysRevD.71.123002}

\bibitem[{Shaltev(2016)}]{bib:shaltev}
Shaltev M (2016) {Optimizing the StackSlide setup and data selection for
  continuous-gravitational-wave searches in realistic detector data}. Phys Rev
  D 93(4):044058, \doi{10.1103/PhysRevD.93.044058}, \eprint{1510.06427}

\bibitem[{Shaltev et~al.(2014)Shaltev, Leaci, Papa, and Prix}]{bib:shaltevetal}
Shaltev M, Leaci P, Papa MA, Prix R (2014) {Fully coherent follow-up of
  continuous gravitational-wave candidates: an application to Einstein@Home
  results}. Phys Rev D 89(12):124030, \doi{10.1103/PhysRevD.89.124030},
  \eprint{1405.1922}

\bibitem[{Shapiro and Teukolsky(1983)}]{bib:ShapiroTeukolsky}
Shapiro SL, Teukolsky SA (1983) {Black Holes, White Dwarfs and Neutron Stars :
  The Physics of Compact Objects}. Wiley

\bibitem[{Shklovskii(1970)}]{bib:Shklovskii}
Shklovskii I (1970) {Possible causes of the secular increase in pulsar
  periods}. Sov Astron 13:562

\bibitem[{Siemonsen and East(2020)}]{bib:SiemonsenEast}
Siemonsen N, East WE (2020) {Gravitational wave signatures of ultralight vector
  bosons from black hole superradiance}. Phys Rev D 101(2):024019,
  \doi{10.1103/PhysRevD.101.024019}, \eprint{1910.09476}

\bibitem[{Sieniawska and Bejger(2019)}]{bib:SieniawskaBejgerreview}
Sieniawska M, Bejger M (2019) {Continuous gravitational waves from neutron
  stars: current status and prospects}. Universe 5(11):217,
  \doi{10.3390/universe5110217}, \eprint{1909.12600}

\bibitem[{Sieniawska and Jones(2021)}]{bib:SieniawskaJones}
Sieniawska M, Jones DI (2021) {Gravitational waves from spinning neutron stars
  as not-quite-standard sirens}. Monthly Notices of the Royal Astronomical
  Society 509(4):5179--5187, \doi{10.1093/mnras/stab3315}

\bibitem[{Sieniawska et~al.(2019)Sieniawska, Bejger, and
  Kr{\'{o}}lak}]{bib:SieniawskaBejgerKrolak}
Sieniawska M, Bejger M, Kr{\'{o}}lak A (2019) {Follow-up procedure for
  gravitational wave searches from isolated neutron stars using the time-domain
  \fstatistic\ method}. Classical and Quantum Gravity 36(22):225008,
  \doi{10.1088/1361-6382/ab3ee5}

\bibitem[{Sieniawska et~al.(2022)Sieniawska, Jones, and
  Miller}]{bib:SieniawskaEtalParallax}
Sieniawska M, Jones DI, Miller AL (2022) {Measuring neutron-star distances and
  properties with gravitational-wave parallax}.
  \doi{doi.org/10.1093/mnras/stad624}, \eprint{2212.07506}

\bibitem[{Singh and Papa(2022)}]{bib:SinghPapaLongPeriodBinaries}
Singh A, Papa MA (2022) {Opportunistic search for continuous gravitational
  waves from compact objects in long-period binaries}. \eprint{2208.14117}

\bibitem[{Singh et~al.(2017)Singh, Papa, Eggenstein, and Walsh}]{bib:singh}
Singh A, Papa MA, Eggenstein HB, Walsh S (2017) {Adaptive clustering procedure
  for continuous gravitational wave searches}. Phys Rev D 96(8):082003,
  \doi{10.1103/PhysRevD.96.082003}, \eprint{1707.02676}

\bibitem[{Singh et~al.(2019)Singh, Papa, and Dergachev}]{bib:aeibinarystudy}
Singh A, Papa MA, Dergachev V (2019) {Characterizing the sensitivity of
  isolated continuous gravitational wave searches to binary orbits}. Phys Rev D
  100(2):024058, \doi{10.1103/PhysRevD.100.024058}, \eprint{1904.06325}

\bibitem[{Singh et~al.(2020)Singh, Haskell, Mukherjee, and
  Bulik}]{bib:SinghHaskellMukherjeeBulik}
Singh N, Haskell B, Mukherjee D, Bulik T (2020) {Asymmetric accretion and
  thermal `mountains' in magnetized neutron star crusts}. Mon Not Roy Astron
  Soc 493(3):3866--3878, \doi{10.1093/mnras/staa442}, \eprint{1908.05038}

\bibitem[{Singhal et~al.(2019)}]{bib:SinghalEtalResampling}
Singhal A, et~al. (2019) {A resampling algorithm to detect continuous
  gravitational-wave signals from neutron stars in binary systems}. Class Quant
  Grav 36(20):205015, \doi{10.1088/1361-6382/ab4367}

\bibitem[{Sintes and Krishnan(2006)}]{bib:adaptiveskyhough}
Sintes AM, Krishnan B (2006) {Improved hough search for gravitational wave
  pulsars}. J Phys Conf Ser 32:206--211, \doi{10.1088/1742-6596/32/1/031},
  \eprint{gr-qc/0601081}

\bibitem[{Slane et~al.(1999)Slane, Gaensler, Dame, Hughes, Plucinsky, and
  Green}]{bib:SlaneEtal}
Slane P, Gaensler BM, Dame TM, Hughes JP, Plucinsky PP, Green A (1999)
  {Nonthermal X-Ray Emission from the Shell-Type Supernova Remnant G347.3-0.5}.
  The Astrophysical Journal 525(1):357--367, \doi{10.1086/307893}

\bibitem[{Smith et~al.(2019)Smith, Bruel, Cognard, Cameron, Camilo, Dai,
  Guillemot, Johnson, Johnston, Keith, Kerr, Kramer, Lyne, Manchester, Shannon,
  Sobey, Stappers, and Weltevrede}]{bib:SmithRevisedDeathline}
Smith DA, Bruel P, Cognard I, Cameron AD, Camilo F, Dai S, Guillemot L, Johnson
  TJ, Johnston S, Keith MJ, Kerr M, Kramer M, Lyne AG, Manchester RN, Shannon
  R, Sobey C, Stappers BW, Weltevrede P (2019) {Searching a Thousand Radio
  Pulsars for Gamma-Ray Emission}. The Astrophysical Journal 871(1):78,
  \doi{10.3847/1538-4357/aaf57d}

\bibitem[{Soida et~al.(2003)Soida, Ando, Kanda, Tagoshi, Tatsumi, Tsubono, and
  the TAMA~Collaboration}]{bib:cwtama}
Soida K, Ando M, Kanda N, Tagoshi H, Tatsumi D, Tsubono K, the
  TAMA~Collaboration (2003) {Search for continuous gravitational waves from the
  {SN}1987A remnant using {TAMA}300 data}. Classical and Quantum Gravity
  20(17):S645--S654, \doi{10.1088/0264-9381/20/17/308}

\bibitem[{Spitkovsky(2004)}]{bib:Spitkovsky}
Spitkovsky A (2004) {Electrodynamics of pulsar magnetospheres}. IAU Symp
  218:357, \eprint{astro-ph/0310731}

\bibitem[{Staelin and Reifenstein(1968)}]{bib:staelinreifenstein}
Staelin DH, Reifenstein EC (1968) {Pulsating Radio Sources near the Crab
  Nebula}. Science 162(3861):1481--1483, \doi{10.1126/science.162.3861.1481},
  \eprint{https://science.sciencemag.org/content/162/3861/1481.full.pdf}

\bibitem[{{Starobinski{\v{i}}}(1973)}]{bib:superradiance3}
{Starobinski{\v{i}}} AA (1973) {Amplification of waves during reflection from a
  rotating ``black hole''}. Soviet Journal of Experimental and Theoretical
  Physics 37:28

\bibitem[{Steltner et~al.(2021)Steltner, Papa, Eggenstein, Allen, Dergachev,
  Prix, Machenschalk, Walsh, Zhu, Behnke, and et~al.}]{bib:cwallskyO2EatH}
Steltner B, Papa MA, Eggenstein HB, Allen B, Dergachev V, Prix R, Machenschalk
  B, Walsh S, Zhu SJ, Behnke O, et~al (2021) {Einstein@Home All-sky Search for
  Continuous Gravitational Waves in LIGO O2 Public Data}. The Astrophysical
  Journal 909(1):79, \doi{10.3847/1538-4357/abc7c9}

\bibitem[{Steltner et~al.(2022{\natexlab{a}})Steltner, Menne, Papa, and
  Eggenstein}]{bib:SteltnerEtalClustering}
Steltner B, Menne T, Papa MA, Eggenstein HB (2022{\natexlab{a}})
  Density-clustering of continuous gravitational wave candidates from large
  surveys. Phys Rev D 106:104063, \doi{10.1103/PhysRevD.106.104063}

\bibitem[{Steltner et~al.(2022{\natexlab{b}})Steltner, Papa, and
  Eggenstein}]{bib:SelfGating2}
Steltner B, Papa MA, Eggenstein HB (2022{\natexlab{b}}) {Identification and
  removal of non-Gaussian noise transients for gravitational-wave searches}.
  Phys Rev D 105:022005, \doi{10.1103/PhysRevD.105.022005}

\bibitem[{Stockinger et~al.(2020)}]{bib:StockingerEtal}
Stockinger G, et~al. (2020) {Three-dimensional Models of Core-collapse
  Supernovae From Low-mass Progenitors With Implications for Crab}. Mon Not Roy
  Astron Soc 496(2):2039--2084, \doi{10.1093/mnras/staa1691},
  \eprint{2005.02420}

\bibitem[{Strader et~al.(2019)}]{bib:redbackcollection}
Strader J, et~al. (2019) {Optical spectroscopy and demographics of redback
  millisecond pulsar binaries}. Astrophys J 872(1):42,
  \doi{10.3847/1538-4357/aafbaa}, \eprint{1812.04626}

\bibitem[{Strang et~al.(2021)Strang, Melatos, Sarin, and
  Lasky}]{bib:StrangEtal}
Strang LC, Melatos A, Sarin N, Lasky PD (2021) {Inferring properties of neutron
  stars born in short gamma-ray bursts with a plerion-like X-ray plateau}. Mon
  Not Roy Astron Soc 507(2):2843--2855, \doi{10.1093/mnras/stab2210},
  \eprint{2107.13787}

\bibitem[{Strohmayer and
  Mahmoodifar(2014{\natexlab{a}})}]{bib:strohmayermahmoodifar1}
Strohmayer T, Mahmoodifar S (2014{\natexlab{a}}) {A Non-radial Oscillation Mode
  in an Accreting Millisecond Pulsar?} Astrophys J 784:72,
  \doi{10.1088/0004-637X/784/1/72}, \eprint{1310.5147}

\bibitem[{Strohmayer and
  Mahmoodifar(2014{\natexlab{b}})}]{bib:strohmayermahmoodifar2}
Strohmayer T, Mahmoodifar S (2014{\natexlab{b}}) {Discovery of a Neutron Star
  Oscillation Mode During a Superburst}. Astrophys J Lett 793(2):L38,
  \doi{10.1088/2041-8205/793/2/L38}, \eprint{1409.2847}

\bibitem[{{Sturrock}(1970)}]{bib:Sturrock}
{Sturrock} PA (1970) {Pulsar Radiation Mechanisms}. Nature 227(5257):465--470,
  \doi{10.1038/227465a0}

\bibitem[{Sun(2018)}]{bib:SunThesis}
Sun L (2018) {Hidden Markov model and cross-correlation searches for continuous
  gravitational waves}. PhD thesis, University of Melbourne,
  \urlprefix\url{http://hdl.handle.net/11343/213141}

\bibitem[{Sun and Melatos(2019)}]{bib:SunMelatos}
Sun L, Melatos A (2019) {Application of hidden Markov model tracking to the
  search for long-duration transient gravitational waves from the remnant of
  the binary neutron star merger GW170817}. Phys Rev D 99:123003,
  \doi{10.1103/PhysRevD.99.123003}

\bibitem[{Sun et~al.(2016)Sun, Melatos, Lasky, Chung, and Darman}]{bib:xcorrS5}
Sun L, Melatos A, Lasky PD, Chung CTY, Darman NS (2016) {Cross-correlation
  search for continuous gravitational waves from a compact object in SNR 1987A
  in LIGO Science run 5}. Phys Rev D 94:082004,
  \doi{10.1103/PhysRevD.94.082004}

\bibitem[{Sun et~al.(2018)Sun, Melatos, Suvorova, Moran, and
  Evans}]{bib:ViterbiSNRmethod}
Sun L, Melatos A, Suvorova S, Moran W, Evans R (2018) {Hidden Markov model
  tracking of continuous gravitational waves from young supernova remnants}.
  Physical Review D 97(4), \doi{10.1103/physrevd.97.043013}

\bibitem[{Sun et~al.(2019)Sun, Melatos, and Lasky}]{bib:ViterbiTwoFrequencies}
Sun L, Melatos A, Lasky PD (2019) {Tracking continuous gravitational waves from
  a neutron star at once and twice the spin frequency with a hidden Markov
  model}. Phys Rev D 99:123010, \doi{10.1103/PhysRevD.99.123010}

\bibitem[{Sun et~al.(2020)Sun, Brito, and Isi}]{bib:ViterbiCygX1}
Sun L, Brito R, Isi M (2020) {Search for ultralight bosons in Cygnus X-1 with
  Advanced LIGO}. Physical Review D 101(6), \doi{10.1103/physrevd.101.063020}

\bibitem[{Suvorov and Melatos(2019)}]{bib:SuvorovMelatos}
Suvorov AG, Melatos A (2019) {Relaxation by thermal conduction of a
  magnetically confined mountain on an accreting neutron star}. Mon Not Roy
  Astron Soc 484(1):1079--1099, \doi{10.1093/mnras/sty3518},
  \eprint{1812.10029}

\bibitem[{Suvorova et~al.(2016)Suvorova, Sun, Melatos, Moran, and
  Evans}]{bib:sidebandviterbi}
Suvorova S, Sun L, Melatos A, Moran W, Evans RJ (2016) {Hidden Markov model
  tracking of continuous gravitational waves from a neutron star with wandering
  spin}. Phys Rev D 93(12):123009, \doi{10.1103/PhysRevD.93.123009},
  \eprint{1606.02412}

\bibitem[{Suvorova et~al.(2017)Suvorova, Clearwater, Melatos, Sun, Moran, and
  Evans}]{bib:ViterbiPaperII}
Suvorova S, Clearwater P, Melatos A, Sun L, Moran W, Evans RJ (2017) {Hidden
  Markov model tracking of continuous gravitational waves from a binary neutron
  star with wandering spin. II. Binary orbital phase tracking}. Phys Rev D
  96:102006, \doi{10.1103/PhysRevD.96.102006}

\bibitem[{Suzuki(1995)}]{bib:Suzuki}
Suzuki T (1995) {Search for Continuous Gravitational Wave from Pulsars with
  Resonant Detector}. In: Coccia E (ed) {First Edoardo Amaldi Conference on
  Gravitational Wave Experiments}, World Scientific, p 115

\bibitem[{{Tan} et~al.(2018){Tan}, {Bassa}, {Cooper}, {Dijkema}, {Esposito},
  {Hessels}, {Kondratiev}, {Kramer}, {Michilli}, {Sanidas}, {Shimwell},
  {Stappers}, {van Leeuwen}, {Cognard}, {Grie{\ss}meier}, {Karastergiou},
  {Keane}, {Sobey}, and {Weltevrede}}]{bib:SlowestRadioPulsar}
{Tan} CM, {Bassa} CG, {Cooper} S, {Dijkema} TJ, {Esposito} P, {Hessels} JWT,
  {Kondratiev} VI, {Kramer} M, {Michilli} D, {Sanidas} S, {Shimwell} TW,
  {Stappers} BW, {van Leeuwen} J, {Cognard} I, {Grie{\ss}meier} JM,
  {Karastergiou} A, {Keane} EF, {Sobey} C, {Weltevrede} P (2018) {LOFAR
  Discovery of a 23.5 s Radio Pulsar}. The Astrophysical Journal Letters
  866(1):54, \doi{10.3847/1538-4357/aade88}, \eprint{1809.00965}

\bibitem[{{Tananbaum}(1999)}]{bib:TananbaumCasA}
{Tananbaum} H (1999) {Cassiopeia A}. IAU Circ 7246:1

\bibitem[{Tauris(2012)}]{bib:taurisrldp}
Tauris TM (2012) {Spin-Down of Radio Millisecond Pulsars at Genesis}. Science
  335(6068):561--563, \doi{10.1126/science.1216355}

\bibitem[{{Tauris} and {Konar}(2001)}]{bib:tauriskonar}
{Tauris} TM, {Konar} S (2001) {Torque decay in the pulsar (P,dot \{P\})
  diagram. Effects of crustal ohmic dissipation and alignment}. Astron \&\
  Astroph 376:543--552, \doi{10.1051/0004-6361:20010988},
  \eprint{astro-ph/0101531}

\bibitem[{Taylor(1992)}]{bib:taylorssb}
Taylor JH (1992) {Pulsar Timing and Relativistic Gravity}. Phil Trans A Math
  Phys Eng Sci 341(1660):117--134, \doi{10.1098/rsta.1992.0088}

\bibitem[{Tenorio(2021)}]{bib:TenorioProceedings}
Tenorio R (2021) {An all-sky search in early O3 LIGO data for continuous
  gravitational-wave signals from unknown neutron stars in binary systems}. In:
  {55th Rencontres de Moriond on Gravitation}, \eprint{2105.07455}

\bibitem[{Tenorio et~al.(2021{\natexlab{a}})Tenorio, Keitel, and
  Sintes}]{bib:TenorioEtal}
Tenorio R, Keitel D, Sintes AM (2021{\natexlab{a}}) {Application of a
  hierarchical MCMC follow-up to Advanced LIGO continuous gravitational-wave
  candidates}. Physical Review D 104(8):084012,
  \doi{10.1103/physrevd.104.084012}

\bibitem[{Tenorio et~al.(2021{\natexlab{b}})Tenorio, Keitel, and
  Sintes}]{bib:TenorioKeitelSintesreview}
Tenorio R, Keitel D, Sintes AM (2021{\natexlab{b}}) {Search Methods for
  Continuous Gravitational-Wave Signals from Unknown Sources in the
  Advanced-Detector Era}. Universe 7(12), \doi{10.3390/universe7120474}

\bibitem[{Tenorio et~al.(2021{\natexlab{c}})Tenorio, Keitel, and
  Sintes}]{bib:skyhoughclustering}
Tenorio R, Keitel D, Sintes AM (2021{\natexlab{c}}) {Time-frequency track
  distance for comparing continuous gravitational wave signals}. Phys Rev D
  103(6):064053, \doi{10.1103/PhysRevD.103.064053}, \eprint{2012.05752}

\bibitem[{Tenorio et~al.(2022)Tenorio, Modafferi, Keitel, and
  Sintes}]{bib:TenorioEtalLoudestCandidate}
Tenorio R, Modafferi LM, Keitel D, Sintes AM (2022) {Empirically estimating the
  distribution of the loudest candidate from a gravitational-wave search}. Phys
  Rev D 105(4):044029, \doi{10.1103/PhysRevD.105.044029}, \eprint{2111.12032}

\bibitem[{Thorne(1980)}]{bib:ThorneMultipoles}
Thorne KS (1980) Multipole expansions of gravitational radiation. Rev Mod Phys
  52:299--339, \doi{10.1103/RevModPhys.52.299}

\bibitem[{Thorne(1989)}]{bib:Thorne300}
Thorne KS (1989) {Three Hundred Years of Gravitation}, Cambridge University
  Press, chap~9, p 330

\bibitem[{Thorne and Zytkow(1975)}]{bib:TZO}
Thorne KS, Zytkow AN (1975) {Red giants and supergiants with degenerate neutron
  cores.} The Astrophysical Journal 199:L19--L24, \doi{10.1086/181839}

\bibitem[{{Thorstensen} and {Armstrong}(2005)}]{bib:tMSPfirst2}
{Thorstensen} JR, {Armstrong} E (2005) {Is FIRST J102347.6+003841 Really a
  Cataclysmic Binary?} The Astrophysical Journal 130(2):759--766,
  \doi{10.1086/431326}, \eprint{astro-ph/0504523}

\bibitem[{Thrane et~al.(2011)Thrane, Kandhasamy, Ott, Anderson, Christensen,
  Coughlin, Dorsher, Giampanis, Mandic, Mytidis, and
  et~al.}]{bib:ThraneEtalStamp}
Thrane E, Kandhasamy S, Ott CD, Anderson WG, Christensen NL, Coughlin MW,
  Dorsher S, Giampanis S, Mandic V, Mytidis A, et~al (2011) {Long
  gravitational-wave transients and associated detection strategies for a
  network of terrestrial interferometers}. Physical Review D 83(8),
  \doi{10.1103/physrevd.83.083004}

\bibitem[{Thrane et~al.(2015)Thrane, Mitra, Christensen, Mandic, and
  Ain}]{bib:stochfolding}
Thrane E, Mitra S, Christensen N, Mandic V, Ain A (2015) {All-sky, narrowband,
  gravitational-wave radiometry with folded data}. Phys Rev D 91:124012,
  \doi{10.1103/PhysRevD.91.124012}

\bibitem[{Tiwari et~al.(2015)Tiwari, Drago, Frolov, Klimenko, Mitselmakher,
  Necula, Prodi, Re, Salemi, Vedovato, and Yakushin}]{bib:TiwariEtalRegression}
Tiwari V, Drago M, Frolov V, Klimenko S, Mitselmakher G, Necula V, Prodi G, Re
  V, Salemi F, Vedovato G, Yakushin I (2015) {Regression of environmental noise
  in {LIGO} data}. Classical and Quantum Gravity 32(16):165014,
  \doi{10.1088/0264-9381/32/16/165014}

\bibitem[{Torres et~al.(2008)Torres, Jonker, Steeghs, Roelofs, Bloom, Casares,
  Falco, Garcia, Marsh, Mendez, Miller, Nelemans, and
  Rodriguez-Gil}]{bib:IGRJ00291Outburst}
Torres MAP, Jonker PG, Steeghs D, Roelofs GHA, Bloom JS, Casares J, Falco EE,
  Garcia MR, Marsh TR, Mendez M, Miller JM, Nelemans G, Rodriguez-Gil P (2008)
  {Observations of the 599 Hz Accreting X-Ray Pulsar {IGR} J00291$+$5934 during
  the 2004 Outburst and in Quiescence}. The Astrophysical Journal
  672(2):1079--1090, \doi{10.1086/523831}

\bibitem[{Treves et~al.(2000)Treves, Turolla, Zane, and Colpi}]{bib:TrevesEtal}
Treves A, Turolla R, Zane S, Colpi M (2000) {Isolated neutron stars: accretors
  and coolers}. Publications of the Astronomical Society of the Pacific
  112(769):297

\bibitem[{Tse et~al.(2019)}]{bib:squeezingO3}
Tse M, et~al. (2019) {Quantum-Enhanced Advanced LIGO Detectors in the Era of
  Gravitational-Wave Astronomy}. Phys Rev Lett 123:231107,
  \doi{10.1103/PhysRevLett.123.231107}

\bibitem[{Tsukada et~al.(2019)Tsukada, Callister, Matas, and
  Meyers}]{bib:TsukadaEtal2019}
Tsukada L, Callister T, Matas A, Meyers P (2019) {First search for a stochastic
  gravitational-wave background from ultralight bosons}. Phys Rev D 99:103015,
  \doi{10.1103/PhysRevD.99.103015}

\bibitem[{Tsukada et~al.(2021)Tsukada, Brito, East, and
  Siemonsen}]{bib:TsukadaEtal2021}
Tsukada L, Brito R, East WE, Siemonsen N (2021) {Modeling and searching for a
  stochastic gravitational-wave background from ultralight vector bosons}. Phys
  Rev D 103:083005, \doi{10.1103/PhysRevD.103.083005}

\bibitem[{Ushomirsky et~al.(2000)Ushomirsky, Cutler, and
  Bildsten}]{bib:UshomirskyEtal}
Ushomirsky G, Cutler C, Bildsten L (2000) {Deformations of accreting neutron
  star crusts and gravitational wave emission}. Mon Not Roy Astron Soc 319:902,
  \doi{10.1046/j.1365-8711.2000.03938.x}, \eprint{astro-ph/0001136}

\bibitem[{Vajente et~al.(2020)Vajente, Huang, Isi, Driggers, Kissel,
  Szczepańczyk, and Vitale}]{bib:VajenteEtalcleaning}
Vajente G, Huang Y, Isi M, Driggers J, Kissel J, Szczepańczyk M, Vitale S
  (2020) {Machine-learning nonstationary noise out of gravitational-wave
  detectors}. Physical Review D 101(4), \doi{10.1103/physrevd.101.042003}

\bibitem[{Valluri et~al.(2021)Valluri, Dergachev, Zhang, and
  Chishtie}]{bib:CWFourierTransform}
Valluri SR, Dergachev V, Zhang X, Chishtie FA (2021) {Fourier transform of the
  continuous gravitational wave signal}. Phys Rev D 104:024065,
  \doi{10.1103/PhysRevD.104.024065}

\bibitem[{Van Den~Broeck(2005)}]{bib:VanDenBroeck}
Van Den~Broeck C (2005) {The gravitational wave spectrum of non-axisymmetric,
  freely precessing neutron stars}. Classical and Quantum Gravity
  22(9):1825--1839, \doi{10.1088/0264-9381/22/9/022}

\bibitem[{Vermeulen et~al.(2021)Vermeulen, Relton, Grote, Raymond, Affeldt,
  Bergamin, Bisht, Brinkmann, Danzmann, Doravari, Kringel, Lough, L{\"u}ck,
  Mehmet, Mukund, Nadji, Schreiber, Sorazu, Strain, Vahlbruch, Weinert, Willke,
  and Wittel}]{bib:GEOscalar}
Vermeulen SM, Relton P, Grote H, Raymond V, Affeldt C, Bergamin F, Bisht A,
  Brinkmann M, Danzmann K, Doravari S, Kringel V, Lough J, L{\"u}ck H, Mehmet
  M, Mukund N, Nadji S, Schreiber E, Sorazu B, Strain KA, Vahlbruch H, Weinert
  M, Willke B, Wittel H (2021) {Direct limits for scalar field dark matter from
  a gravitational-wave detector}. Nature 600(7889):424--428,
  \doi{10.1038/s41586-021-04031-y}

\bibitem[{Vicer{\'{e}} and Yvert(2016)}]{bib:vicereautocorr}
Vicer{\'{e}} A, Yvert M (2016) {An autocorrelation method to detect periodic
  gravitational waves from neutron stars in binary systems}. Classical and
  Quantum Gravity 33(16):165006, \doi{10.1088/0264-9381/33/16/165006}

\bibitem[{Viets and Wade(2021)}]{bib:VietsWadeNoiseSubtraction}
Viets A, Wade M (2021) {Subtracting Narrow-band Noise from LIGO Strain Data in
  the Third Observing Run}. {LIGO Report T2100058},
  \urlprefix\url{https://dcc.ligo.org/T2100058}

\bibitem[{Vigelius and Melatos(2010)}]{bib:vigeliusmelatos}
Vigelius M, Melatos A (2010) {Gravitational-wave spin-down and stalling lower
  limits on the electrical resistivity of the accreted mountain in a
  millisecond pulsar}. Astrophys J 717:404--410,
  \doi{10.1088/0004-637X/717/1/404}, \eprint{1005.2257}

\bibitem[{{Viterbi}(1967)}]{bib:ViterbiOriginal}
{Viterbi} A (1967) {Error bounds for convolutional codes and an asymptotically
  optimum decoding algorithm}. IEEE Transactions on Information Theory
  13(2):260--269

\bibitem[{Wade et~al.(2012)Wade, Siemens, Kaplan, Knispel, and
  Allen}]{bib:Wadeetal}
Wade L, Siemens X, Kaplan DL, Knispel B, Allen B (2012) {Continuous
  gravitational waves from isolated Galactic neutron stars in the advanced
  detector era}. Phys Rev D 86:124011, \doi{10.1103/PhysRevD.86.124011}

\bibitem[{{Wagoner}(1984)}]{bib:wagoner}
{Wagoner} RV (1984) {Gravitational radiation from accreting neutron stars}.
  Astrophys J 278:345--348, \doi{10.1086/161798}

\bibitem[{Walsh et~al.(2019)Walsh, Wette, Papa, and
  Prix}]{bib:WalshEtaltemplates}
Walsh S, Wette K, Papa MA, Prix R (2019) {Optimizing the choice of analysis
  method for all-sky searches for continuous gravitational waves with
  Einstein@Home}. Physical Review D 99(8), \doi{10.1103/physrevd.99.082004}

\bibitem[{Walsh et~al.(2016)}]{bib:allskymdc}
Walsh S, et~al. (2016) {Comparison of methods for the detection of
  gravitational waves from unknown neutron stars}. Phys Rev D 94(12):124010,
  \doi{10.1103/PhysRevD.94.124010}, \eprint{1606.00660}

\bibitem[{Wang et~al.(2018)Wang, Steeghs, Galloway, Marsh, and
  Casares}]{bib:WangEtalScoX1}
Wang L, Steeghs D, Galloway DK, Marsh T, Casares J (2018) {Precision
  Ephemerides for Gravitational-wave Searches – III. Revised system
  parameters of Sco X-1}. Monthly Notices of the Royal Astronomical Society
  478(4):5174--5183, \doi{10.1093/mnras/sty1441}

\bibitem[{Watts et~al.(2008)Watts, Krishnan, Bildsten, and
  Schutz}]{bib:WattsEtal}
Watts A, Krishnan B, Bildsten L, Schutz BF (2008) {Detecting gravitational wave
  emission from the known accreting neutron stars}. Mon Not Roy Astron Soc
  389:839--868, \doi{10.1111/j.1365-2966.2008.13594.x}, \eprint{0803.4097}

\bibitem[{{Weltevrede} et~al.(2011){Weltevrede}, {Johnston}, and
  {Espinoza}}]{bib:WeltevredeEtal_2011}
{Weltevrede} P, {Johnston} S, {Espinoza} CM (2011) {The glitch-induced identity
  changes of PSR J1119-6127}. Monthly Notices of the Royal Astronomical Society
  411(3):1917--1934

\bibitem[{Wette(2012)}]{bib:WetteEstimation}
Wette K (2012) {Estimating the sensitivity of wide-parameter-space searches for
  gravitational-wave pulsars}. Phys Rev D 85:042003,
  \doi{10.1103/PhysRevD.85.042003}, \eprint{1111.5650}

\bibitem[{Wette(2014)}]{bib:WetteTemplates1}
Wette K (2014) {Lattice template placement for coherent all-sky searches for
  gravitational-wave pulsars}. Phys Rev D 90:122010,
  \doi{10.1103/PhysRevD.90.122010}

\bibitem[{Wette(2015)}]{bib:WetteTemplates2}
Wette K (2015) {Parameter-space metric for all-sky semicoherent searches for
  gravitational-wave pulsars}. Phys Rev D 92:082003,
  \doi{10.1103/PhysRevD.92.082003}

\bibitem[{Wette(2016)}]{bib:WetteTemplates3}
Wette K (2016) {Empirically extending the range of validity of parameter-space
  metrics for all-sky searches for gravitational-wave pulsars}. Phys Rev D
  94(12):122002, \doi{10.1103/PhysRevD.94.122002}, \eprint{1607.00241}

\bibitem[{Wette(2021)}]{bib:WetteGeometry}
Wette K (2021) {Geometric Approach to Analytic Marginalisation of the
  Likelihood Ratio for Continuous Gravitational Wave Searches}. Universe
  7(6):174, \doi{10.3390/universe7060174}, \eprint{2104.14829}

\bibitem[{Wette and Prix(2013)}]{bib:WettePrix}
Wette K, Prix R (2013) {Flat parameter-space metric for all-sky searches for
  gravitational-wave pulsars}. Phys Rev D 88:123005,
  \doi{10.1103/PhysRevD.88.123005}

\bibitem[{Wette et~al.(2008)Wette, Owen, Allen, Ashley, Betzwieser,
  Christensen, Creighton, Dergachev, Gholami, Goetz, Gustafson, Hammer, Jones,
  Krishnan, Landry, Machenschalk, McClelland, Mendell, Messenger, Papa, Patel,
  Pitkin, Pletsch, Prix, Riles, de~la Jordana, Scott, Sintes, Trias, Whelan,
  and Woan}]{bib:cwcasamethod}
Wette K, Owen BJ, Allen B, Ashley M, Betzwieser J, Christensen N, Creighton TD,
  Dergachev V, Gholami I, Goetz E, Gustafson R, Hammer D, Jones DI, Krishnan B,
  Landry M, Machenschalk B, McClelland DE, Mendell G, Messenger CJ, Papa MA,
  Patel P, Pitkin M, Pletsch HJ, Prix R, Riles K, de~la Jordana LS, Scott SM,
  Sintes AM, Trias M, Whelan JT, Woan G (2008) {Searching for gravitational
  waves from Cassiopeia A with {LIGO}}. Classical and Quantum Gravity
  25(23):235011, \doi{10.1088/0264-9381/25/23/235011}

\bibitem[{Wette et~al.(2018)Wette, Walsh, Prix, and Papa}]{bib:WetteTemplates4}
Wette K, Walsh S, Prix R, Papa M (2018) {Implementing a semicoherent search for
  continuous gravitational waves using optimally constructed template banks}.
  Physical Review D 97(12), \doi{10.1103/physrevd.97.123016}

\bibitem[{Wette et~al.(2021)Wette, Dunn, Clearwater, and
  Melatos}]{bib:WetteEtalDeep}
Wette K, Dunn L, Clearwater P, Melatos A (2021) {Deep exploration for
  continuous gravitational waves at 171–172 Hz in LIGO second observing run
  data}. Physical Review D 103(8), \doi{10.1103/physrevd.103.083020}

\bibitem[{Whelan et~al.(2014)Whelan, Prix, Cutler, and
  Willis}]{bib:WhelanEtalCoordinates}
Whelan JT, Prix R, Cutler CJ, Willis JL (2014) {New Coordinates for the
  Amplitude Parameter Space of Continuous Gravitational Waves}. Class Quant
  Grav 31:065002, \doi{10.1088/0264-9381/31/6/065002}, \eprint{1311.0065}

\bibitem[{Whelan et~al.(2015)Whelan, Sundaresan, Zhang, and
  Peiris}]{bib:xcorrmethod3}
Whelan JT, Sundaresan S, Zhang Y, Peiris P (2015) {Model-Based
  Cross-Correlation Search for Gravitational Waves from Scorpius X-1}. Phys Rev
  D 91:102005, \doi{10.1103/PhysRevD.91.102005}, \eprint{1504.05890}

\bibitem[{Whitbeck(2006)}]{bib:Whitbeck}
Whitbeck DM (2006) {Observational Consequences of Gravitational Wave Emission
  From Spinning Compact Sources}. PhD thesis, Penn State U.

\bibitem[{Williams and Schutz(2000)}]{bib:WilliamsSchutzDemod}
Williams PR, Schutz BF (2000) {An Efficient matched filtering algorithm for the
  detection of continuous gravitational wave signals}. AIP Conf Proc
  523(1):473--476, \doi{10.1063/1.1291918}, \eprint{gr-qc/9912029}

\bibitem[{Woan et~al.(2018)Woan, Pitkin, Haskell, Jones, and
  Lasky}]{bib:WoanEtalMSP}
Woan G, Pitkin MD, Haskell B, Jones DI, Lasky PD (2018) {Evidence for a Minimum
  Ellipticity in Millisecond Pulsars}. The Astrophysical Journal 863(2):L40,
  \doi{10.3847/2041-8213/aad86a}

\bibitem[{Worley et~al.(2008)Worley, Krastev, and Li}]{bib:UncertainIzz}
Worley A, Krastev PG, Li BA (2008) {Nuclear Constraints on the Moments of
  Inertia of Neutron Stars}. The Astrophysical Journal 685(1):390--399,
  \doi{10.1086/589823}

\bibitem[{Yamamoto and Tanaka(2021)}]{bib:YamamotoTanaka}
Yamamoto TS, Tanaka T (2021) {Use of an excess power method and a convolutional
  neural network in an all-sky search for continuous gravitational waves}. Phys
  Rev D 103(8):084049, \doi{10.1103/PhysRevD.103.084049}, \eprint{2011.12522}

\bibitem[{Yao et~al.(2017)Yao, Manchester, and Wang}]{bib:YMWModel}
Yao JM, Manchester RN, Wang N (2017) {A {New} {Electron}-{Density} {Model}
  {for} {Estimation} {of} {Pulsar} {and} {FRB} {Distances}}. The Astrophysical
  Journal 835(1):29, \doi{10.3847/1538-4357/835/1/29}

\bibitem[{Yim and Jones(2020)}]{bib:YimJones}
Yim G, Jones DI (2020) {Transient gravitational waves from pulsar post-glitch
  recoveries}. Mon Not Roy Astron Soc 498(3):3138--3152,
  \doi{10.1093/mnras/staa2534}, \eprint{2007.05893}

\bibitem[{Yoshida et~al.(2005)Yoshida, Yoshida, and
  Eriguchi}]{bib:YoshidaYoshidaEriguchi}
Yoshida S, Yoshida S, Eriguchi Y (2005) {R-mode oscillations of rapidly
  rotating barotropic stars in general relativity: analysis by the relativistic
  Cowling approximation}. Monthly Notices of the Royal Astronomical Society
  356(1):217–224, \doi{10.1111/j.1365-2966.2004.08436.x}

\bibitem[{Yoshino and Kodama(2014)}]{bib:YoshinoKodama2014}
Yoshino H, Kodama H (2014) {Gravitational radiation from an axion cloud around
  a black hole: Superradiant phase}. PTEP 2014:043E02,
  \doi{10.1093/ptep/ptu029}, \eprint{1312.2326}

\bibitem[{Yoshino and Kodama(2015)}]{bib:YoshinoKodama2015}
Yoshino H, Kodama H (2015) {The bosenova and axiverse}. Classical and Quantum
  Gravity 32(21):214001, \doi{10.1088/0264-9381/32/21/214001}

\bibitem[{Yunes et~al.(2022)Yunes, Miller, and Yagi}]{bib:YunesMiller}
Yunes N, Miller MC, Yagi K (2022) Gravitational-wave and x-ray probes of the
  neutron star equation of state. Nature Reviews Physics 4(4):237--246,
  \doi{10.1038/s42254-022-00420-y}

\bibitem[{{Zel'dovich}(1971)}]{bib:superradiance1}
{Zel'dovich} YB (1971) {Generation of Waves by a Rotating Body}. Soviet Journal
  of Experimental and Theoretical Physics Letters 14:180

\bibitem[{Zhang et~al.(2000)Zhang, Harding, and
  Muslimov}]{bib:ZhangHardingMuslimov}
Zhang B, Harding AK, Muslimov AG (2000) {Radio pulsar death line revisited: Is
  PSR J2144-3933 anomalous?} Astrophys J Lett 531:L135--L138,
  \doi{10.1086/312542}, \eprint{astro-ph/0001341}

\bibitem[{Zhang et~al.(2021)Zhang, Papa, Krishnan, and
  Watts}]{bib:O2CrossCorrAEI}
Zhang Y, Papa MA, Krishnan B, Watts AL (2021) {Search for Continuous
  Gravitational Waves from Scorpius X-1 in LIGO O2 Data}. The Astrophysical
  Journal Letters 906(2):L14, \doi{10.3847/2041-8213/abd256}

\bibitem[{Zhu et~al.(2016)Zhu, Papa, Eggenstein, Prix, Wette, Allen, Bock,
  Keitel, Krishnan, Machenschalk, Shaltev, and Siemens}]{bib:directedEatHS6}
Zhu SJ, Papa MA, Eggenstein HB, Prix R, Wette K, Allen B, Bock O, Keitel D,
  Krishnan B, Machenschalk B, Shaltev M, Siemens X (2016) {Einstein@Home search
  for continuous gravitational waves from Cassiopeia A}. Phys Rev D 94:082008,
  \doi{10.1103/PhysRevD.94.082008}

\bibitem[{Zhu et~al.(2017)Zhu, Papa, and Walsh}]{bib:ZhuEtalDoppler}
Zhu SJ, Papa MA, Walsh S (2017) {New veto for continuous gravitational wave
  searches}. Phys Rev D 96(12):124007, \doi{10.1103/PhysRevD.96.124007},
  \eprint{1707.05268}

\bibitem[{Zhu et~al.(2020)Zhu, Baryakhtar, Papa, Tsuna, Kawanaka, and
  Eggenstein}]{bib:ZhuEtalBosonCWSignal}
Zhu SJ, Baryakhtar M, Papa MA, Tsuna D, Kawanaka N, Eggenstein HB (2020)
  {Characterizing the continuous gravitational-wave signal from boson clouds
  around Galactic isolated black holes}. Physical Review D 102(6),
  \doi{10.1103/physrevd.102.063020}

\bibitem[{{Zimmermann}(1978)}]{bib:ZimmermannBuriedBfield}
{Zimmermann} M (1978) {Revised estimate of gravitational radiation from Crab
  and Vela pulsars}. Nature 271(5645):524--525, \doi{10.1038/271524a0}

\bibitem[{Zimmermann and Szedenits(1979)}]{bib:zimmermannszedenits}
Zimmermann M, Szedenits E (1979) {Gravitational waves from rotating and
  precessing rigid bodies: Simple models and applications to pulsars}. Phys Rev
  D 20:351--355, \doi{10.1103/PhysRevD.20.351}

\bibitem[{Zweizig and Riles(2021)}]{bib:SelfGating1}
Zweizig J, Riles K (2021) {Information on self-gating of $h(t)$ used in O3a
  continuous-wave searches}. LIGO Report T2000384,
  \urlprefix\url{https://dcc.ligo.org/T2000384}

\end{thebibliography}

\end{document}